\documentclass[12pt,titlepage]{article}

\usepackage[default]{jasa_harvard}    
\usepackage{JASA_manu}

\newcommand {\ctn}{\citeasnoun} 
\usepackage{graphicx,subfigure,amsmath,latexsym,amssymb}
\usepackage{float,epsfig,multirow,rotating,times}
\usepackage{upgreek,wrapfig}
\usepackage{comment}
\usepackage{pgfplots}
\usetikzlibrary{shapes, calc, shapes, arrows}

\newcommand{\ba}{\boldsymbol{a}}

\newcommand{\bTheta}{\boldsymbol{\Theta}}

\newcommand{\bSigma}{\boldsymbol{\Sigma}}

\newcommand{\bpi}{\boldsymbol{\pi}}

\newcommand{\bmu}{\boldsymbol{\mu}}
\newcommand{\bnu}{\boldsymbol{\nu}}
\newcommand{\btau}{\boldsymbol{\tau}}

\newcommand{\bOmega}{\boldsymbol{\Omega}}
\newcommand{\bomega}{\boldsymbol{\omega}}

\newcommand{\bI}{\boldsymbol{I}}

\newcommand{\bL}{\boldsymbol{L}}

\newcommand{\bS}{\boldsymbol{S}}

\newcommand{\bx}{\bm{x}}

\newcommand{\by}{\boldsymbol{y}}

\newcommand{\bz}{\boldsymbol{z}}

\newcommand{\bzero}{\boldsymbol{0}}

\newcommand{\bj}{{\boldsymbol{j}}}
\newcommand{\bone}{\boldsymbol{1}}

\newtheorem{theorem}{Theorem}

\newtheorem{definition}[theorem]{Definition}

\newcommand{\statesp}{\ensuremath{\mathcal X}}
\newcommand{\Y}{\ensuremath{\mathcal Y}}
\newcommand{\D}{\ensuremath{\mathcal D}}

\newcommand{\e}{\ensuremath{\epsilon}}

\newcommand{\supr}[2]{{#1}^{(#2)}}

\newcommand{\be}{\boldsymbol\e}
\newcommand{\bm}{\mathbf}
\renewcommand{\t}{\ensuremath{\theta}}

\newcommand{\topline}{\hrule height 1pt width \textwidth \vspace*{2pt}}
\newcommand{\botline}{\vspace*{2pt}\hrule height 1pt width \textwidth \vspace*{4pt}}
\newtheorem{algo}{Algorithm} 

\numberwithin{equation}{section}
\numberwithin{algo}{section}
\numberwithin{table}{section}
\numberwithin{figure}{section}

\begin{document}

\normalsize

\title{\vspace{-0.8in}
Transdimensional Transformation based Markov Chain Monte Carlo}
\author{Moumita Das and Sourabh Bhattacharya\thanks{
Moumita Das is a PhD student and Sourabh Bhattacharya 
is an Associate Professor in Interdisciplinary Statistical Research Unit, Indian Statistical
Institute, 203, B. T. Road, Kolkata 700108.
Corresponding e-mail: sourabh@isical.ac.in.}}
\date{\vspace{-0.5in}}
\maketitle%

\begin{abstract}

Variable dimensional problems, where not only the parameters, but also the number of parameters
are random variables, pose serious challenge to Bayesians. Although
in principle the Reversible Jump Markov Chain Monte Carlo (RJMCMC) methodology
is a response to such challenges,
the dimension-hopping strategies need not be always convenient for practical implementation, particularly because
efficient ``move-types" having reasonable acceptance rates are often difficult to devise.

In this article, we propose and develop a novel and general dimension-hopping MCMC methodology that can update
all the parameters as well as the number of parameters simultaneously using simple deterministic
transformations of some low-dimensional (often one-dimensional) random variable. This methodology,
which has been inspired by Transformation based MCMC (TMCMC) of \ctn{Dutta14}, 
facilitates great speed in terms of computation time and provides reasonable acceptance rates and mixing properties. 
Quite importantly, our approach provides a natural way to automate the move-types in variable
dimensional problems. We refer to this methodology as Transdimensional Transformation based Markov Chain Monte Carlo (TTMCMC). 
Comparisons with RJMCMC in gamma and normal mixture examples demonstrate
far superior performance of TTMCMC in terms of mixing, acceptance rate, computational speed and automation.
Furthermore, we demonstrate good performance of TTMCMC in multivariate normal mixtures, even for dimension
as large as $20$. To our knowledge, there exists no application of RJMCMC for such high-dimensional mixtures.

As by-products of our effort on the development of TTMCMC, we propose a novel methodology to summarize
the posterior distributions of the mixture densities, providing a way to obtain the mode
of the posterior distribution of the densities and the associated highest posterior density
credible regions. Based on our method we also propose a criterion to assess convergence
of variable-dimensional algorithms.   
These methods of summarization and convergence assessment are applicable to general problems, not just to
mixtures.
\\[2mm]

\end{abstract}

\tableofcontents

\section{{\bf Introduction}}
\label{sec:intro}

Markov chain Monte Carlo (MCMC) is known to have revolutionized Bayesian computation. 
In modern times, it is often required to analyze high-dimensional, complex data, and the Bayesian paradigm, with the
MCMC machinery, provides an ideal package to the statistical scientist for the purpose. As is to be
anticipated, to simulate from complex Bayesian posteriors, development of quite sophisticated MCMC methods were necessary,
and various approaches based on component-wise and joint updating of the parameters, such as
the adaptive direction sampling (\ctn{Gilks94}), 
the multiple-try Metropolis method (\ctn{Liu00}), 
the auxiliary variable approach (\ctn{Storvik11}),
parallel MCMC methods (\ctn{Martino16}), 
have emerged in response to the needs of the modern Bayesian.

However, the above methods are appropriate when the number of parameters is known in advance. 
When one of the unknown parameters is the number of parameters itself, then none of the traditional MCMC methods are  
applicable, irrespective of how sophisticated they are. 
Indeed, simultaneous inference on both model and parameter space
is an issue that is fundamental to modern statistical practice (\ctn{Sisson05}). Examples
of such problems arise in mixture analysis where the parameters associated with the mixture components
as well as the number of mixture components are unknown (see, for example, \ctn{Richardson97});
in change point analysis where the locations and the number of change points are unknown (see, for example, \ctn{Green95});
in variable selection problems where the number of covariates and the associated coefficients are unknown
(\ctn{Dellaportas02}, \ctn{Dellaportas99});
in spline smoothing where the location and the number of knots are unknown (see \ctn{Denison98} for instance); 
in continuous wavelet representation of unknown functions with a finite, but unknown number of wavelet basis functions
and the corresponding parameters (\ctn{Chu09}); 
in autoregressive time series models where the order of the 
autoregression and the associated
parameters are unknown (\ctn{Vermaak04}); in factor analysis where the dimension of the 
latent factor loading matrix and the associated
parameters are unknown (\ctn{Lopes04}); 
in spatial point processes where the locations and the number of points
are random (see \ctn{Moller04}); 
to name only a few. 

A general MCMC strategy which can explore variable dimensional
spaces by jumping between different dimensions has been proposed by \ctn{Green95}, and is 
well-known as Reversible Jump MCMC (RJMCMC). The versatility of the methodology
is well-reflected in the large varieties of variable-dimensional problems to which it has been applied; indeed,
all the aforementioned examples make use of RJMCMC.
However, one difficulty is frequently encountered when designing reversible jump algorithms is the construction 
of efficient proposals. 
Typically, dimension jumping moves in reversible jump samplers exhibit much lower acceptance rate than in 
fixed-dimensional moves.
\ctn{Awadhi04} observed that models with multimodal distributions yield particularly low acceptance rates. 
There have been many attempts of creating automatic RJMCMC samplers which also maintain high acceptance
rates; see, for example, \ctn{Brooks03}, \ctn{Robert03}, \ctn{Green03}, \ctn{Godsill03}, \ctn{Robert04}, 
\ctn{Sisson05}, \ctn{Fan11} and the references therein. However, in spite of the commendable attempts, 
these ideas are perhaps relevant in quite specific
models with several restrictive assumptions; see \ctn{Robert04}, \ctn{Sisson05}, \ctn{Fan11}. 

The issues discussed above point towards the need to develop general and natural move types that can 
change dimensions as well as update the other (within model) parameters simultaneously, 
while maintaining reasonable acceptance rates and mixing properties. 
In this regard, the transformation based MCMC (TMCMC) approach of 
\ctn{Dutta14} in the fixed dimensional set-up provides the necessary motivation. The key concept
of TMCMC is to propose a move-type from a set of available move-types, 
simulate a single, one-dimensional random variable from some arbitrary distribution
and propose simple deterministic transformations to all the parameters using the one-dimensional random
variable, within the proposed move-type. In this article we show that the same concept of deterministic 
transformations of a single 
random variable can be exploited to construct, for any general variable dimensional problem, 
a generic and effective dimension-hopping sampler which can change dimensions and update all the
parameters of the proposed model in a single block while maintaining reasonable acceptance rates and
mixing properties.
We refer to this general variable dimensional MCMC sampler as Transdimensional Transformation based
Markov Chain Monte Carlo (TTMCMC).  

\subsection{{\bf Overview of contributions and organisation of this paper}}
\label{subsec:contributions}

Before a formal introduction of TTMCMC, it is necessary to provide a brief overview of the basic concept
of TMCMC. We do this in Section \ref{sec:tmcmc}. 

We introduce TTMCMC in Section \ref{sec:ttmcmc}, and in Section \ref{sec:jump_more} we 
extend our proposed methodology to more general situations where one wishes to jump more than one dimension
at a time. 
That TTMCMC thus developed closely qualifies as an automatic variable dimensional sampler,
is argued in Section \ref{sec:automation}. 

Although our proposed sampler is quite general and readily applicable to all transdimensional sampling
frameworks, for the purpose of illustration and comparison with RJMCMC we restrict ourselves to gamma and 
normal mixture problems with unknown number of components. 
In this regard, in Section \ref{sec:gamma_mixtures} we first conduct four simulation experiments with
gamma mixtures with true number of components being 1, 2, 3 and 4, respectively. 
In Section \ref{sec:normal_mixtures} 
we provide details regarding applications of our methods to analyse 
three well-studied real data sets, namely, the enzyme, acidity and the galaxy data (see \ctn{Richardson97}, for instance). 
In Section \ref{sec:mvn_mixtures} we demonstrate the application of TTMCMC in mixtures of multivariate normal densities.
In particular, we consider three simulation studies for dimensions $3$, $10$ and $20$. 

We show that the simplest possible TTMCMC algorithm, 
which is based on additive transformations, puts up excellent performance in all the examples, even in all the
multivariate scenarios, providing ample support to our claim of automation. Also interestingly, the TTMCMC applications are able
to capture very precise information regarding the number of mixture components, for both simulated and real data sets. 
None of the previous methods
(see \ctn{Richardson97} and the references therein)
were able to capture so precise information as TTMCMC. Moreover, there possibly does not exist any RJMCMC algorithm
that works for multivariate mixtures with dimension as high as $20$. Hence, from the high-dimensional perspective, 
TTMCMC is clearly far ahead of RJMCMC.

For the gamma mixtures and the normal mixtures associated with the real data applications we compare 
additive TTMCMC with the closest RJMCMC analogue of additive TTMCMC, 
based on random walk proposals.
This RJMCMC algorithm seems to be the more natural, intuitive and computationally far simpler alternative to
the random walk-motivated ``automatic generic transdimensional RJMCMC sampler" proposed in \ctn{Green03}. 
Indeed, the approach of \ctn{Green03} is appropriate only when a small set of models is considered in the 
variable-dimensional problem, and as such not a viable option for our normal mixtures with maximum of $30$
components; see Section \ref{subsec:random_walk_rjmcmc} for details.

Unfortunately, the random walk RJMCMC algorithm analogue of additive TTMCMC fails to 
produce satisfactory results in a way that even convergence is not assured in any of the examples.
In particular, with the same scales of additive TTMCMC, random walk RJMCMC yields extremely poor acceptance rate
in general.
Moreover, the RJMCMC-based posterior of the number of components tends to assign 
higher posterior probabilities to implausibly large values, 
clearly indicating lack of convergence. We argue that the same issue persists
with general RJMCMC algorithms. 
This suggests that complex and difficult-to-implement algorithms with extremely large convergence time 
are required for RJMCMC to yield sensible results, and that there is no default choice of such algorithms. 
On the other hand, the potentiality of additive TTMCMC 
in conjunction with the results of our experiments demonstrate that additive TTMCMC is close to 
qualifying as the default variable-dimensional algorithm, even for large dimensions.


We summarize our work and make concluding remarks in Section \ref{sec:conclusion}.
Additional details are provided in the supplement \ctn{Das14b}, whose sections 
have the prefix ``S-" when referred to in this paper.  

\section{{\bf A brief overview of the key idea of TMCMC}}
\label{sec:tmcmc}

In order to obtain a valid algorithm based on transformations, \ctn{Dutta14} design appropriate move
types so that detailed balance and irreducibility hold. We first illustrate the basic idea of transformation
based moves with a simple example. Given that we are in the current state $x$, we may
propose the ``forward move" $x'=x+\e$, where
$\e>0$ is a simulation from some arbitrary density $\varrho(\cdot)$ which is supported on the positive part
of the real line. To move back to $x$ from $x'$, we need to apply the ``backward transformation" $x'-\e$.
In general, given $\e$
and the current state $x$, we shall denote the forward transformation by $T(x,\e)$, and
the backward transformation by $T^b(x,\e)$.
For fixed $\e$ the forward and backward transformations must be one-to-one and onto,
and must satisfy $T^b(T(x,\e),\e)=x=T(T^b(x,\e),\e)$; see \ctn{Dutta14} for a detailed discussion
regarding these. 

The simple idea discussed above has been generalized to the multi-dimensional situation by
\ctn{Dutta14}. Remarkably, for any dimension, the moves can be constructed by simple deterministic
transformations of the one-dimensional random variable $\e$, which is simulated from any arbitrary
distribution on some relevant support. We provide some examples of such moves in the next section
after introducing some necessary notation borrowed from \ctn{Dutta14}.

\subsection{{\bf Notation}}
\label{subsec:notation}
Suppose that $\statesp$ is a $k$-dimensional space of the form $\statesp = \prod_{i=1}^k \statesp_i$ 
so that $T = (T_1,\ldots,T_k)$ where each $T_i : \statesp_i \times \mathcal D \to \statesp_i$, for some 
set $\mathcal D$, are the component-wise transformations. 
Let $\bz=(z_1,\ldots,z_k)$ be a vector of indicator variables, where, 
for $i=1,\ldots,k$, 
$z_i=1$ and $z_i=-1$ indicate, respectively, application of forward transformation and  
backward transformation to $x_i$, and let $z_i=0$ denote no change to $x_i$. This ``no change" step
is sufficient to ensure irreducibility of TMCMC in non-additive transformations; see \ctn{Dutta14}.
Given any such indicator vector $\bz$, let us define
$ T_{\bz} = (g_{1,z_1},g_{2,z_2},\ldots,g_{k,z_k})$
where 
\[ g_{i,z_i} = \left\{ \begin{array}{ccc}
                  T_{i}^b & \textrm{ if } & z_i=-1 \\ 
		  x_i   & \textrm{ if } & z_i=0 \\
		  T_{i} & \textrm{ if } & z_i=1. 
                 \end{array}
\right.\]
Corresponding to any given $\bz$, we also define the following `conjugate' vector
$\bz^c=(z^c_1,z^c_2,\ldots,z^c_k)$, where $$z^c_i=-z_i.$$ 
With this definition of $\bz^c$, $T_{\bz^c}$ can be interpreted as the conjugate of $T_{\bz}$.

Since $3^k$ values of $\bz$ are possible, it is clear that $T$, via $\bz$,
induces $3^k$ many types of `moves' of the forms $\{T_{\bz_i};i=1,\ldots,3^k\}$ on the state-space. 
Suppose now that there is a subset $\Y$ of $\D$ such that the sets $T_{\bz_i}(\bm x,\Y)$ and 
$T_{\bz_j}(\bm x,\Y)$ are disjoint for every $\bz_i \ne \bz_j$. In fact, $\Y$ denotes the support
of the distribution $\varrho(\cdot)$ from which $\e$ is simulated. 
This mutual exclusiveness is required to satisfy the detailed balance property; see \ctn{Dutta14} for the details.
Thus, although $\D$ denotes the actual range of values that $\e$ can assume in principle, for implementation
of TMCMC we must restrict the support of $\e$ to $\Y$.

\subsection{{\bf Examples of transformations on two-dimensional state-space using single $\e$}}
Although for the sake of illustration we provide below examples pertaining to two-dimensional cases
it is important to remark at the outset that these examples can be easily generalized to any dimension; 
see \ctn{Dutta14}.

\begin{enumerate}


\item {\it Additive transformation:} Suppose $\statesp = \D=\mathbb R^2$. With two
positive scale parameters $a_1$ and $a_2$, we can then consider the following
additive transformation:
$T_{(1,1)}(\bm x,\e) = (x_1 + a_1\e,x_2+a_2\e)$, 
$T_{(-1,1)}(\bm x,\e) = (x_1 - a_1\e,x_2 + a_2\e)$, 
$T_{(1,-1)}(\bm x,\e) = (x_1 + a_1\e,x_2 - a_2\e)$ and 
$T_{(-1,-1)}(\bm x,\e)= (x_1 - a_1\e,x_2 - a_2\e)$. 
We set $\Y = (0,\infty)$. 

\item {\it Multiplicative transformation:} Suppose $\statesp = \D = \mathbb R^2$.
Then we may consider the following multiplicative transformation:
$T_{(1,1)}(\bm x,\e) = (x_1\e , x_2\e)$,
$T_{(-1,1)}(\bm x,\e) = (x_1/\e,x_2\e)$, 
$T_{(1,-1)}(\bm x,\e) = (x_1\e,x_2/\e)$,
$T_{(-1,-1)}(\bm x,\e) = (x_1/\e,x_2/\e)$,
$T_{(1,0)}(\bm x,\e) = (x_1\e,x_2)$,
$T_{(1,0)}(\bm x,\e) = (x_1\e,x_2)$,
$T_{(-1,0)}(\bm x,\e) = (x_1/\e,x_2)$,
$T_{(0,1)}(\bm x,\e) = (x_1,x_2\e)$,
$T_{(0,-1)}(\bm x,\e) = (x_1,x_2/\e)$,
$T_{(0,0)}(\bm x,\e) = (x_1,x_2)$.
We choose $\Y = \left\{(-1,1)-\{0\}\right\}$.

\item {\it Additive-multiplicative transformation:} 
It is possible to combine additive and multiplicative transformations, but here we need
at least two $\e$'s, one for the additive, and another for the multiplicative transformation.
For instance, if $\statesp = \D = \mathbb R^2$, then we may consider the following moves:
$T_{(1,1)}(\bm x, \e_1,\e_2) = (x_1 + \e_1, x_2\e_2)$,
$T_{(-1,1)}(\bm x, \e_1,\e_2) = (x_1 - \e_1,x_2\e_2)$, 
$T_{(1,-1)}(\bm x, \e_1,\e_2) = (x_1 + \e_1,x_2/\e_2)$, 
$T_{(-1,-1)}(\bm x,\e_1,\e_2) = (x_1 - \e_1,x_2/\e_2)$,
$T_{(1,0)}(\bm x,\e_1,\e_2) = (x_1 + \e_1,x_2)$,
$T_{(-1,0)}(\bm x,\e_1,\e_2) = (x_1 - \e_1,x_2)$,
$T_{(0,1)}(\bm x,\e_1,\e_2) = (x_1,x_2\e_2)$,
$T_{(0,-1)}(\bm x,\e_1,\e_2) = (x_1,x_2/\e_2)$,
$T_{(0,0)}(\bm x,\e_1,\e_2) = (x_1,x_2)$.
We let $\Y = (0,\infty) \times \left\{(-1,1)-\{0\}\right\}$. 
Although this example uses two $\e$'s for two dimensions, it is important to note that for any dimension higher than two,
at most two $\e$'s will be required for validity of additive-multiplicative TMCMC, one for the additive part
and another for the multiplicative part, irrespective of the dimensionality. Thus, the minimum effective dimensionality
of additive TMCMC and multiplicative TMCMC is $1$, while that of additive-multiplicative TMCMC in this setting is $2$, for 
any dimensionality greater than one.


\end{enumerate}

The key observation underlying the above examples is that it is always possible to construct 
valid transformations in high-dimensional spaces
using combinations of appropriate transformations on one-dimensional spaces.
These transformations and the underlying principle remain valid even in TTMCMC. 

\subsection{{\bf The general form of the TMCMC algorithm}}
\label{subsec:general_form_tmcmc}

For a $k~(\geq 1)$-dimensional target distribution, with current state $\bx=(x_1,\ldots,x_k)$,
\ctn{Dutta14} apply forward and backward transformations to $x_i$ with probabilities $p_i$ and $q_i$, respectively
and keep $x_i$ unchanged with probability $1-p_i-q_i$, for $i=1,\ldots,k$. Thus, $\bz$ can now be interpreted as a random vector
such that for $i=1,\ldots,k$, $z_i\in\{-1,0,1\}$ with probabilities $q_i,1-p_i-q_i,p_i$, respectively.
Thus, we simulate $z_i \sim Multinomial(1;p_i,q_i,1-p_i-q_i)$ independently for $i=1,\ldots,k$, 
draw $\e \sim \varrho(\cdot)$, and form the proposed move $\bx\mapsto \bx'=T_{\bz}(\bx,\e)$,
which is accepted with probability
\begin{equation}
\alpha(\bm x, \e) = \min\left(1, \dfrac{P(\bz^c)}{P(\bz)} ~\dfrac{\pi(\bm x')}{\pi(\bm x)} 
 ~\left|\frac{\partial (T_{\bz}(\bm x, \e),\e)}{\partial(\bm x, \e)}\right| \right),
\label{eq:acc_tmcmc}
\end{equation}
 where 
 \[\dfrac{P(\bz^c)}{P(\bz)}=\underset{\{i_1:z_{i_1}=-1\}}\prod \frac{p_{i_1}}{q_{i_1}}
 \underset{\{i_2:z_{i_2}=1\}}\prod \frac{q_{i_2}}{p_{i_2}}.\]
Note that the acceptance ratio is {\it always} independent of the proposal density $\varrho$. 

The redundant move-type $\bx\mapsto\bx$ has positive probability of occurrence, and hence \ctn{Dutta14}
suggest rejection of this move whenever it appears. That is, sampling of $\bz$ is to be continued until
at least one $z_i\neq 0$. This rejection sampling of $\bz$ is very efficient since the rejection region
is a singleton and has very small probability, particularly in high dimensions. The normalizing constant that arises
because of this truncation cancels in the acceptance ratio of TMCMC, as shown in \ctn{Dutta14}.

\section{{\bf TTMCMC for updating the dimension and the parameters in a single block using deterministic
transformations of a single random variable}}
\label{sec:ttmcmc}

First we illustrate the main idea of TTMCMC informally using the additive transformation. 

\subsection{{\bf Illustration of the key idea of TTMCMC with a simple example}}
\label{subsec:ttmcmc_simple_example}

Assume that the current state is $\bm x=(x_1,x_2)\in\mathbb R^2$. We first randomly select
$u=(u_1,u_2,u_3)\sim Multinomial (1;w_b,w_d,w_{nc})$, where $w_b,w_d,w_{nc}~(>0)$ such that $w_b+w_d+w_{nc}=1$ are
the probabilities of birth, death, and no-change moves, respectively. That is, if $u_1=1$, then we increase
the dimensionality from 2 to 3; if $u_2=1$, then we decrease the dimensionality from 2 to 1, and if $u_3=1$, then
we keep the dimensionality unchanged. 
In the latter case, when the dimensionality is unchanged, the acceptance probability remains the same as in
TMCMC, given by (\ref{eq:acc_tmcmc}). 

If $u_1=1$, we can increase the dimensionality by first selecting one of $x_1$ and $x_2$
with probability $1/2$; for the sake of clarity, we assume that $x_1$ has been selected, 
Here, as in TMCMC, we draw $\e\sim \varrho(\cdot)$, where $\varrho(\cdot)$ is supported on the positive part
of the real line, and draw $z_2$ where 
$z_2=1$ with probability $p_2$ 
and $z_2=-1$ with probability
$1-p_2$. Also, as before, $\bz^c=(z^c_1,z^c_2)$ is the conjugate of $\bz$, where 
$z^c_i=-z_i$.
We then construct the move-type $T_{b,\bz}(\bm x,\e)=(x_1+a_1\e,x_1-a_1\e,x_2+z_2a_2\e)$
$=(g_{1,z_1=1}(x_1,\e),~g_{1,z^c_1=-1}(x_1,\e),~g_{2,z_2}(x_2,\e))$, say. 
We re-label $\bm x'=T_{b,\bz}(\bm x,\e)=(x_1+a_1\e,x_1-a_1\e,x_2+z_2a_2\e)$ as $(x'_1,x'_2,x'_3)$.
Thus, $T_{b,\bz}(\bm x,\e)$ increases the dimension from 2 to 3. 

We accept this birth move with probability
\begin{align}
a_b(\bm x,\e)
&=\min\left\{1, \frac{1}{3}\times\frac{w_d}{w_b}\times\frac{p^{I_{\{1\}}(z^c_2)}_2q^{I_{\{-1\}}(z^c_2)}_2}
{p^{I_{\{1\}}(z_2)}_2q^{I_{\{-1\}}(z_2)}_2}\right.\notag\\
&\left.\quad\quad\quad\quad\times\frac{\pi(x_1+a_1\e,~x_1-a_1\e,~x_2+z_2a_2\e)}{\pi(x_1,x_2)}\times
\left|\frac{\partial (T_{b,\bz}(\bm x, \e))}{\partial(\bm x, \e)}\right|
\right\}.
\label{eq:acc_birth}
\end{align}
In (\ref{eq:acc_birth}),
\begin{align}
& \left|\frac{\partial (T_{b,\bz}(\bm x, \e))}{\partial(\bm x, \e)}\right|
 =\left|\frac{\partial (x_1+a_1\e,x_1-a_1\e,x_2+z_2a_2\e)}{\partial(x_1,x_2, \e)}\right|
 =\left|\left(\begin{array}{ccc}
1 & 1 & 0 \\
0 & 0 & 1 \\
a_1 & -a_1 & z_2a_2 \\
 \end{array}
 \right)\right|=2a_1.
 \label{eq:Jacobian_birth}
\end{align}

Now let us illustrate the problem of returning to $\bm =(x_1,x_2)~(\in\mathbb R^2)$ from 
$T_{b,\bz}(\bm x,\e)=(x_1+a_1\e,x_1-a_1\e,x_2+z_2a_2\e)~(\in\mathbb R^3)$.
For our purpose, we can select $x_1+a_1\e$ with probability $1/3$; then select $x_1-a_1\e$ from
the remaining two elements with probability $1/2$, and form the average $x^*_1=((x_1+a_1\e)+(x_1-a_1\e))/2=x_1$.
For non-additive transformations we can consider the averages of the backward moves of 
each of the selected elements. Even in this additive transformation example, after simulating $\e$ as before 
we can consider 
the respective backward moves of $x_1+a_1\e$ and $x_1-a_1\e$, both yielding $x_1$, and then take
the average denoted by $x^*_1$. For the remaining element $x_2+z_2a_2\e$, we need to simulate
$z^c_2$ and then consider the move $(x_2+z_2a_2\e)+z^c_2a_2\e=x_2$. Thus, we can return to $(x_1,x_2)$ using
this strategy.

Letting $\bm x'=(x'_1,x'_2,x'_3)$, 
and denoting the average
involving the first two elements by $x^*_1$, 
the death move is then given by $\bm x''=T_{d,\bz}(\bm x',\e)=(x^*_1,x'_3+z^c_2a_2\e)$
$=(\frac{x'_1+x'_2}{2},x'_3+z^c_2a_2\e)$.
Now observe that for returning to $(x'_1,x'_2)$ from $x^*_1$,
we must have ${x_{1}}^*+a_1\e^*=x'_1$ and ${x_{1}}^*-a_1\e^*=x'_2$, which yield $\e^*=(x'_1-x'_2)/2a_1$.
Hence, the Jacobian associated with the death move in this case is given by
\begin{align}
 \left|\frac{\partial \left(T_{d,\bz}(\bm x', \e),\e^*,\e\right)}{\partial(\bm x', \e)}\right|
 = \left|\frac{\partial \left(\frac{x'_1+x'_2}{2},x'_3+z^c_2a_2\e,\frac{x'_1-x'_2}{2a_1},\e\right)}
 {\partial(x'_1,x'_2,x'_3, \e)}\right|
 &=\left|\left(\begin{array}{cccc}
\frac{1}{2} & 0  & \frac{1}{2a_1} & 0\\
\frac{1}{2} & 0 & -\frac{1}{2a_1} & 0 \\
0 & 1 & 0 & 0 \\
0 & z^c_2a_2 & 0 & 1\\
 \end{array}
 \right)\right|=\frac{1}{2a_1}.
 \label{eq:Jacobian_death}
\end{align}
We accept this death move with probability
 \begin{align}
 a_d(\bm x'', \e, \e^*) &= 
 \min\left\{1,3\times\frac{w_b}{w_d}\times\dfrac{P(\bz^c)}{P(\bz)} ~\dfrac{\pi(\bm x'')}{\pi(\bm x')} 
 ~\left|\frac{\partial (T_{d,\bz}(\bm x', \e),\e^*,\e)}{\partial(\bm x',\e)}\right| \right\}\notag\\
 &= \min\left\{1,3\times\frac{w_b}{w_d}\times\dfrac{p^{I_{\{1\}}(z^c_2)}_2q^{I_{\{-1\}}(z^c_2)}_2}
 {p^{I_{\{1\}}(z_2)}_2q^{I_{\{-1\}}(z_2)}_2} \times\dfrac{\pi(\bm x'')}{\pi(\bm x')} 
 \times\frac{1}{2a_1}\right\}.
 \label{eq:acc_death}
 \end{align}

In the general situation, we shall make the birth, death and no-change probabilities $w_b$, $w_d$, $w_{nc}$
depend upon the current dimension $k$, and denote them by $w_{b,k}$, $w_{d,k}$ and $w_{nc,k}$, respectively,
satisfying $w_{b,k}+w_{d,k}+w_{nc,k}=1$ for every $k\geq 1$. Note that when the current dimension $k=1$,
then $w_{d,k}=0$, as $k\geq 1$. 
Similarly, if in some cases there is reason to assume that the number of parameters can not exceed some
finite quantity denoted by $k_{max}$, then $w_{b,k_{max}}=0$.

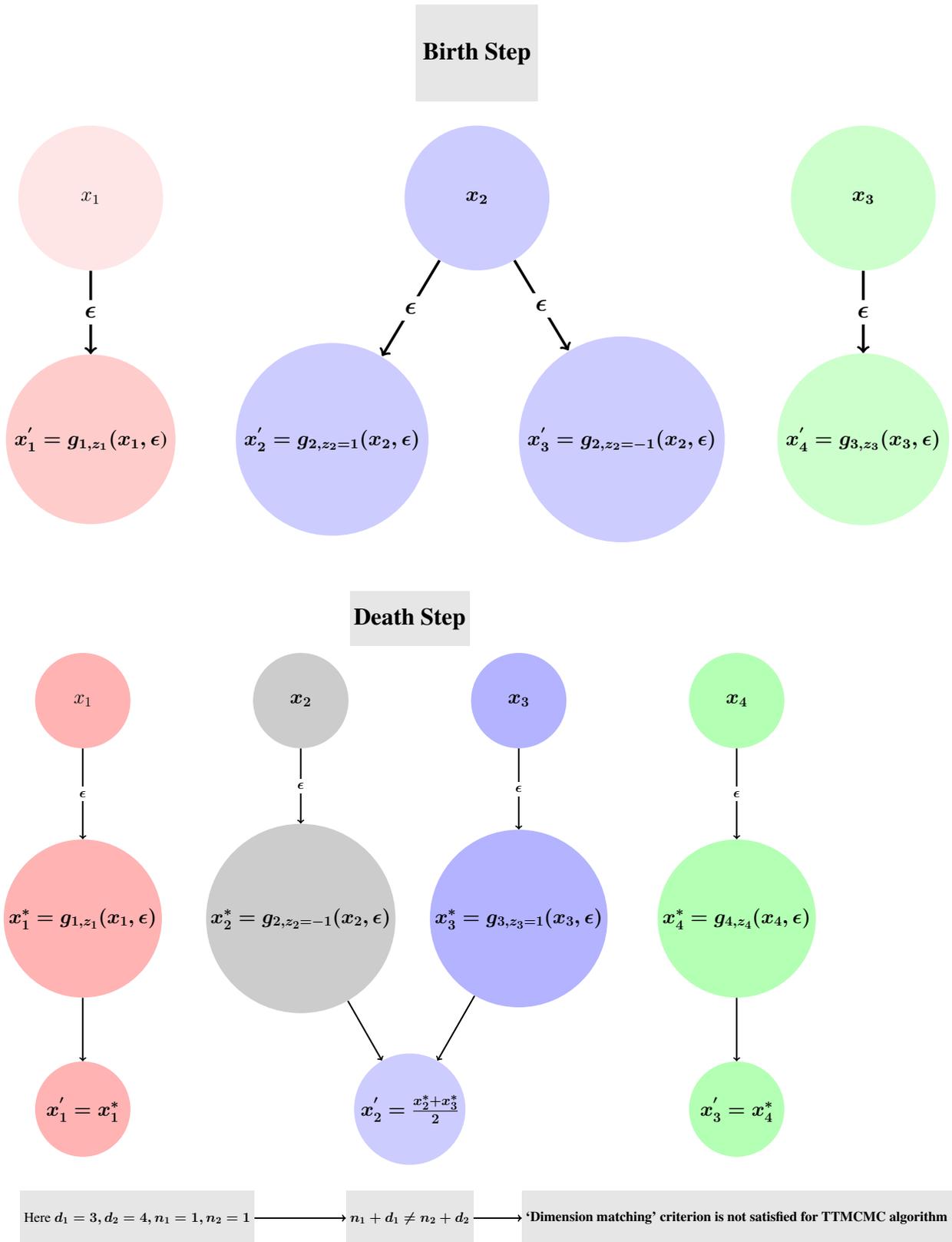
\begin{figure}[H]
\subfigure
\centering
    \resizebox{\textwidth}{!}{
        \begin{tikzpicture}[font=\boldmath]
        \node[rectangle,fill=gray!20,minimum width=2cm,minimum height=2cm] (12) at (0,75) {\large \bf Birth Step};
        \node[circle,fill=red!10,minimum size=3cm,font=\bf] (1) at (-8,72) {$x_1$};
        \node[circle,fill=blue!20,minimum size=3cm] (2) at (0,72) {$x_2$};
        \node[circle,fill=green!20,minimum size=3cm] (3) at (8,72) {$x_3$};
        \node[circle,fill=red!20,minimum size=3cm] (4) at (-8,67) {$x^{'}_1=g_{1,z_1}(x_1,\epsilon$)};
        \node[circle,fill=blue!20,minimum size=3cm] (5) at (-3,67) {$x^{'}_2=g_{2,z_2=1}(x_2,\epsilon)$};
        \node[circle,fill=blue!20,minimum size=3cm] (6) at (3,67) {$x^{'}_3=g_{2,z_2=-1}(x_2,\epsilon)$};
        \node[circle,fill=green!20,minimum size=3cm] (7) at (8,67) {$x^{'}_4=g_{3,z_3}(x_3,\epsilon)$};
        
        
        \foreach \from/\to in {1/4,2/5,2/6,3/7}
        {\draw[->,ultra thick, fill=blue] (\from) to node[fill=white] {\large{$\epsilon$}} (\to);
        }
        
        \end{tikzpicture}

    }
    \hspace{2mm}

   \subfigure
   \centering
    \resizebox{\textwidth}{!}{
        \begin{tikzpicture}[font=\boldmath]
       
        \node[rectangle,fill=gray!20,minimum width=2cm,minimum height=2cm,font=\bf] (12) at (28,65) {\Huge \bf Death Step};
        \node[circle,fill=red!30,minimum size=3.5cm,font=\bfseries] (1) at (16,62) {\LARGE$x_1$};
        \node[circle,fill=gray!40,minimum size=3.5cm] (2) at (24,62) {\LARGE$x_2$};
        \node[circle,fill=blue!30,minimum size=3.5cm] (3) at (32,62) {\LARGE$x_3$};
        \node[circle,fill=green!30,minimum size=3.5cm] (4) at (40,62) {\LARGE$x_4$};
        \node[circle,fill=red!30,minimum size=3.5cm] (5) at (16,54) {\LARGE$x^{*}_1=g_{1,z_1}(x_1,\epsilon)$};
        \node[circle,fill=gray!40,minimum size=3.5cm] (6) at (24,54) {\LARGE$x^{*}_2=g_{2,z_2=-1}(x_2,\epsilon)$};
        \node[circle,fill=blue!30,minimum size=3.5cm] (7) at (32,54) {\LARGE$x^{*}_3=g_{3,z_3=1}(x_3,\epsilon)$};
        \node[circle,fill=green!30,minimum size=3.5cm](8) at (40,54) {\LARGE$x^{*}_4=g_{4,z_4}(x_4,\epsilon)$};
        \node[circle,fill=red!30,minimum size=3.5cm] (9) at (16,47) {\LARGE$x^{'}_1=x^{*}_1$};
        \node[circle,fill=blue!20,minimum size=3.5cm] (10) at (28,47) {\LARGE$x^{'}_2=\frac{x^*_{2}+x^*_{3}}{2}$};
        \node[circle,fill=green!30,minimum size=3.5cm] (11) at (40,47) {\LARGE$x^{'}_3=x^{*}_4$};
        \node[rectangle,fill=gray!20,minimum width=3cm,minimum height=2cm] (12) at (18,43) {\large Here $d_1=3, d_2=4,n_1=1,n_2=1$};
        \node[rectangle,fill=gray!20,minimum width=3cm,minimum height=2cm] (13) at (28,43) { \large $n_1+d_1\neq n_2+d_2$};
        \node[rectangle,fill=gray!20,minimum width=4cm,minimum height=2cm] (14) at (40,43) {\large \bf `Dimension matching' criterion is not satisfied for TTMCMC algorithm};
       
       \foreach \from/\to in {1/5,2/6,3/7,4/8}
        {\draw[->,ultra thick, fill=blue] (\from) to node[fill=white] {\large{$\epsilon$}} (\to);
        }
        {\draw[->,ultra thick, fill=blue] (5) -- (9);
          \draw[->,ultra thick, fill=blue] (6) -- (10);
          \draw[->,ultra thick, fill=blue] (7) -- (10);
           \draw[->,ultra thick, fill=blue] (8) -- (11);
           \draw[->,ultra thick, fill=blue] (12) -- (13);
           \draw[->,ultra thick, fill=blue] (13) -- (14);

        }
        
        \end{tikzpicture}
    }
     \caption{\bf Illustration of TTMCMC algorithm for jumping between dimension $3$ and $4$.}

\label{fig:diagram_TTMCMC}
\end{figure}

Figure \ref{fig:diagram_TTMCMC} illustrates the idea of TTMCMC schematically, and compares it with the RJMCMC principle, shown diagrammatically in Figure \ref{fig:diagram_RJMCMC}. 
As illustrated, for RJMCMC, the necessary `` dimension matching'' criterion is satisfied, but 
the criterion is not satisfied, indeed, not necessary, for TTMCMC.

\hspace{2mm}

\begin{figure}[H]
    \centering
    \resizebox{\textwidth}{!}{
        \begin{tikzpicture}[font=\boldmath]
         \node[rectangle,fill=gray!20,minimum size=2cm,font=\bf] (12) at (48,55){\bf Birth Step};
        \node[circle,fill=red!30,minimum size=3cm] (1) at (40,50) {$x_1$};
        \node[circle,fill=blue!30,minimum size=3cm] (2) at (48,50) {$x_2$};
        \node[circle,fill=green!30,minimum size=3cm] (3) at (56,50) {$x_3$};
        \node[circle,fill=red!30,minimum size=3cm] (4) at (40,46) {$x^{'}_1=x_1\pm\epsilon_1$};
        \node[circle,fill=blue!30,minimum size=3cm] (5) at (45,46) {$x^{'}_2=x_2+\epsilon_2$};
        \node[circle,fill=blue!30,minimum size=3cm] (6) at (51,46) {$x^{'}_3=x_2-\epsilon_2$};
        \node[circle,fill=green!30,minimum size=3cm] (7) at (56,46) {$x^{'}_4=x_3\pm\epsilon_3$};

        \foreach \from/\to in {1/4,2/5,2/6,3/7}
        {\draw[->,ultra thick, fill=blue] (\from) to node[fill=white] {\large{$\epsilon_{\from}$}} (\to);
        }
        \end{tikzpicture}
    }
    
   \hspace{2mm}
    

    \centering
    \resizebox{\textwidth}{!}{
        \begin{tikzpicture}[font=\boldmath]
         \node[rectangle,fill=gray!20,minimum size=6cm,font=\bf] (12) at (80,24) {\Huge \bf Death Step};
        \node[circle,fill=red!30,minimum size=7.5cm] (1) at (64,15) {\LARGE $x_1$};
        \node[circle,fill=gray!40,minimum size=7.5cm] (2) at (74,15) {\LARGE $x_2$};
        \node[circle,fill=blue!30,minimum size=7.5cm] (3) at (84,15) {\LARGE $x_3$};
        \node[circle,fill=green!30,minimum size=7.5cm] (4) at (94,15) {\LARGE $x_4$};
        \node[circle,fill=red!40,minimum size=7.5cm] (5) at (64,2) {\LARGE$x^{'}_1=x_1\pm\epsilon_1$};
        \node[circle,fill=blue!20,minimum size=7.5cm] (6) at (79,2) {\LARGE $x^{'}_2=\frac{x_3+x_4}{2}$};
        \node[circle,fill=green!20,minimum size=7.5cm] (7) at (94,2) {\LARGE $x^{'}_3=x_4\pm\epsilon_4$};
        \node[rectangle,fill=gray!20,minimum size=5.5cm] (8) at (64,-6) {\LARGE Here $d_1=3, d_2=4,n_1=3,n_2=2$};
        \node[rectangle,fill=gray!20,minimum size=5.5cm] (9) at (78,-6) {\LARGE $n_1+d_1= n_2+d_2$};
        \node[rectangle,fill=gray!20,minimum size=5.5cm] (10) at (94,-6) {\LARGE \bf `Dimension matching' criterion is satisfied for RJMCMC algorithm};

        {\draw[->,ultra thick, fill=blue] (2) -- (6);
         \draw[->,ultra thick] (3) -- (6);
          \draw[->,ultra thick, fill=blue] (8) -- (9);
          \draw[->,ultra thick, fill=blue] (9) -- (10);
        }
         \foreach \from/\to in {1/5,4/7}
        {\draw[->,ultra thick, fill=blue] (\from) to node[fill=white] {\Huge $\epsilon_{\from}$} (\to);
        }

        \end{tikzpicture}
    }
    \caption{\bf Illustration of RJMCMC algorithm for jumping between dimension $3$ and $4$.}
\label{fig:diagram_RJMCMC}
\end{figure}

\subsection{{\bf General TTMCMC algorithm for jumping one dimension at a time}}
\label{subsec:ttmcmc_general}

We now provide the TTMCMC algorithm in the general case, as follows.

\begin{algo}\label{algo:ttmcmc} \topline General TTMCMC algorithm based on a single $\e$.
\botline \normalfont \ttfamily
\begin{itemize}
 \item Let the initial value be $\supr{\bm x}{0}\in\mathbb R^k$. 
 \item For $t=0,1,2,\ldots$
\begin{enumerate}
 \item Generate $u=(u_1,u_2,u_3)\sim Multinomial (1;w_{b,k},w_{d,k},w_{nc,k})$.
 \item If $u_1=1$ (increase dimension), then
 \begin{enumerate}
 \item Randomly select a co-ordinate from $\bm x^{(t)}=(x^{(t)}_1,\ldots,x^{(t)}_k)$ assuming uniform 
 probability $1/k$ for each co-ordinate.
 Let $j$ denote the chosen co-ordinate.
 \item Generate $\e \sim \varrho(\cdot)$ and for $i=1,\ldots,k;~i\neq j$ simulate  
 $$z_i \sim Multinomial(1;p_i,q_i,1-p_i-q_i)$$ independently. 
 \item Propose the following birth move: 
 \begin{align} 
 \bm x' &= T_{b,\bz}(\supr{\bm x}{t}, \e)=(g_{1,z_1}(x^{(t)}_1,\e),\ldots,g_{j-1,z_{j-1}}(x^{(t)}_{j-1},\e),\notag\\
& g_{j,{z_j=1}}(x^{(t)}_j,\e),g_{j,{z^c_j=-1}}(x^{(t)}_j,\e),g_{j+1,z_{j+1}}(x^{(t)}_{j+1},\e),\ldots,
g_{k,z_k}(x^{(t)}_k,\e)).\notag
\end{align}
Re-label the elements of $\bm x'$ as $(x'_1,x'_2,\ldots,x'_{k+1})$.
\item Calculate the acceptance probability of the birth move $\bm x'$:
 \begin{align}
 a_b(\supr{\bm x}{t}, \e) &= 
 \min\left\{1, \frac{1}{k+1}\times\frac{w_{d,k+1}}{w_{b,k}}\times\dfrac{P_{(j)}(\bz^c)}{P_{(j)}(\bz)} 
 ~\dfrac{\pi(\bm x')}{\pi(\supr{\bm x}{t})} 
 ~\left|\frac{\partial (T_{b,\bz}(\supr{\bm x}{t}, \e))}{\partial(\supr{\bm x}{t}, \e)}\right| \right\},\notag
 \end{align}
 where
 \[ 
P_{(j)}(\bz)=\prod_{i\neq j=1}^kp^{I_{\{1\}}(z_i)}_iq^{I_{\{-1\}}(z_i)}_i,
 \]
 and
 \[ 
P_{(j)}(\bz^c)=\prod_{i\neq j=1}^kp^{I_{\{1\}(z^c_i)}}_iq^{I_{\{-1\}}(z^c_i)}_i.
 \]

\item Set \[ \supr{\bm x}{t+1}= \left\{\begin{array}{ccc}
 \bm x' & \mbox{ with probability } & a_b(\supr{\bm x}{t},\e) \\
 \supr{\bm x}{t}& \mbox{ with probability } & 1 - a_b(\supr{\bm x}{t},\e).
\end{array}\right.\]
\end{enumerate}
\item If $u_2=1$ (decrease dimension), then
 \begin{enumerate}
 \item Generate $\e \sim \varrho(\cdot)$. 
 \item Randomly select co-ordinate $j$ with probability $1/k$, and randomly select co-ordinate $j'$ from the remaining
 co-ordinates with probability $1/(k-1)$. Let $x^*_j=\left(g_{j,z^c_j=-1}(x_j,\e)+g_{j',z_{j'}=1}(x_{j'},\e)\right)/2$; 
 replace the co-ordinate $x_j$
 drawn first by the average $x^*_j$, and delete $x_{j'}$.
 \item Simulate $\bz$ by generating independently, for $i=1,\ldots,k$, but $i\neq j,j'$, 
 $z_i \sim Multinomial(1;p_i,q_i,1-p_i-q_i)$.
 For $i\neq j,j'$, apply the transformation $x'_i=g_{i,z_i}(x^{(t)}_i,\e)$.  
 \item Propose the following death move:
 \begin{align} 
 \bm x' &= T_{d,\bz}(\supr{\bm x}{t}, \e)=(g_{1,z_1}(x^{(t)}_1,\e),\ldots,g_{j-1,z_{j-1}}(x^{(t)}_{j-1},\e),x^*_j,
g_{j+1,z_{j+1}}(x^{(t)}_{j+1},\e),\notag\\
&\ldots,g_{j'-1,z_{j'-1}}(x^{(t)}_{j'-1},\e),
g_{j'+1,z_{j'+1}}(x^{(t)}_{j'+1},\e),
\ldots,g_{k,z_k}(x^{(t)}_k,\e)).\notag
\end{align}
Re-label the elements of $\bm x'$ as $(x'_1,x'_2,\ldots,x'_{k-1})$.
 \item Solve for $\e^*$ from the equations $g_{j,z_j=1}(x^*_j,\e^*)=x_j$ and $g_{j,z^c_{j}=-1}(x^*_j,\e^*)=x_{j'}$
and express $\e^*$ in terms of $x_j$ and $x_{j'}$.
\item Calculate the acceptance probability of the death move:
 \begin{align}
 a_d(\supr{\bm x}{t}, \e, \e^*) &= 
 \min\left\{1,k\times\frac{w_{b,k-1}}{w_{d,k}}\times\dfrac{P_{(j,j')}(\bz^c)}{P_{(j,j')}(\bz)} 
 ~\dfrac{\pi(\bm x')}{\pi(\supr{\bm x}{t})} 
 ~\left|\frac{\partial (T_{d,\bz}(\supr{\bm x}{t}, \e),\e^*,\e)}{\partial(\supr{\bm x}{t},\e)}\right| \right\},\notag
 \end{align}
 where
 \[ 
P_{(j,j')}(\bz)=\prod_{i\neq j,j'=1}^kp^{I_{\{1\}}(z_i)}_iq^{I_{\{-1\}}(z_i)}_i,
 \]
 and
 \[ 
P_{(j,j)}(\bz^c)=\prod_{i\neq j,j'=1}^kp^{I_{\{1\}(z^c_i)}}_iq^{I_{\{-1\}}(z^c_i)}_i.
 \]
\item Set \[ \supr{\bm x}{t+1}= \left\{\begin{array}{ccc}
 \bm x' & \mbox{ with probability } & a_d(\supr{\bm x}{t},\e,\e^*) \\
 \supr{\bm x}{t}& \mbox{ with probability } & 1 - a_d(\supr{\bm x}{t},\e,\e^*).
\end{array}\right.\]
 \end{enumerate}
\item If $u_3=1$ (dimension remains unchanged), then implement steps (1), (2), (3) of Algorithm 3.1 
of \ctn{Dutta14}. 
\end{enumerate}
\item End for
\end{itemize}
\botline \rmfamily
\end{algo}
In Sections S-1 and S-2 of the supplement we provide the proofs of detailed balance and ergodicity (irreducibility
and aperiodicity) of the above TTMCMC method.

\subsubsection{{\bf Observations regarding Algorithm 3.1}}
\label{subsubsec:obs_ttmcmc}
\begin{itemize}
\item Note that the acceptance
probabilities are independent of the proposal density $\varrho(\cdot)$ irrespective of its form, just as in TMCMC. 
The reason is that in TTMCMC we simulate $\epsilon\sim \varrho$, for some
appropriate density $\varrho$, for increasing, as well as for
decreasing dimension (see the proof of detailed balance
in Section S-1 for the precise details). In other words, the ``dimension-matching" criterion of RJMCMC 
is not required for TTMCMC. Indeed, recall that, to accomplish the birth step in RJMCMC one needs 
to simulate an $\epsilon$, but in the death step two randomly chosen components are averaged to reduce
the dimension, and no simulation of $\epsilon$ is done. 
As such, in RJMCMC the dimension-matching criterion is responsible for the 
presence of the proposal density in the acceptance ratio. 

\item Consequently, it is not possible to interpret TTMCMC as a special case of RJMCMC. 
Also, neither is RJMCMC a special case of TTMCMC, even though in fixed-dimensional problems,
TMCMC with additive transformations contains the random walk Metropolis algorithm as a special case when
as many $\e$'s as the number of variables to be updated are used for TMCMC. 

\item Independence of the acceptance ratio of the proposal density $\varrho$ has pleasing consequences
for TTMCMC in the sense that for any finite TTMCMC sample (which is always the case in practice), 
the possible bias in the acceptance 
probabilities of birth and death moves due to 
involvement of $\varrho$ is absent. Since for RJMCMC this is not the case, the performance may be seriously affected.
For instance, if $\varrho$ is strictly bounded above by $1$, then the birth move will have significantly greater acceptance
probability than the death move. The advantage of TTMCMC and disadvantage of RJMCMC in this regard are clearly 
reflected in all our experiments that we report in this article. 
\item In the acceptance
probabilities, $\dfrac{P_{(j)}(\bz^c)}{P_{(j)}(\bz)}=1$ and $\dfrac{P_{(j,j')}(\bz^c)}{P_{(j,j')}(\bz)}=1$ if 
$p_i=q_i$ for each $i$. This results in simplification of the acceptance ratio computation.
The birth, death and the no-change probabilities given by $w_{b,k}$, $w_{d,k}$ and $w_{nc,k}$
can also be chosen to be equal for every $k>1$, which will result in further simplification
of the computation of the acceptance ratio.

\item
In our algorithm, the new variables created from one variable are never ``necessarily adjacent".
Even in the case of adjacency, our method does absolutely fine; indeed, for the death step,
we only need to have appropriate positive probability of selecting 
the two variables for combining them into one (or deleting one) such that the detailed balance holds. Specifically, 
suppose that we create adjacent variables in the birth move. Then, in the corresponding death move we will 
choose adjacent pairs with appropriate probability and combine them into one. 
Alternatively, one may select two variables, but should reject the entire death move 
if the selected variables are not adjacent. In fact, the issue of adjacency is nothing specific to TTMCMC, 
and can be handled by RJMCMC as well as by TTMCMC. 

\end{itemize}

\subsection{{\bf Structured dependence within the moves}}
\label{subsec:dependent_z}

In Algorithm \ref{algo:ttmcmc} we have assumed that for $i\in\{1,\ldots,k\}\backslash\{j\}$ and 
for $i\in\{1,\ldots,k\}\backslash\{j,j'\}$ (accordingly as the move-type is birth move or death move), 
$z_i$ are independently simulated in every iteration. 
Although the co-ordinate-wise moves are dependent since the same $\e$ is used for updating them, 
more flexible and structured dependence can be induced within the moves in the TTMCMC context. 
Such structured dependence allows
for selecting the co-ordinate-wise forward or backward transformations in ways that take account
of the posterior correlation between the parameters, thus facilitating more efficient moves.

Briefly, at each iteration, for $i=1,\ldots,k$, we can reparameterize $p_i$ and $q_i$ as
\begin{align}	
p_i &= \frac{\exp(\psi_{1i})}{\sum_{j=1}^3\exp(\psi_{ji})};
\quad\quad q_i = \frac{\exp(\psi_{2i})}{\sum_{j=1}^3\exp(\psi_{ji})};
\quad\quad 1-p_i-q_i = \frac{\exp(\psi_{3i})}{\sum_{j=1}^3\exp(\psi_{ji})},
\label{eq:repara}
\end{align}
where, for $j=1,2,3$,
\begin{equation}
(\psi_{j1},\psi_{j2},\ldots,\psi_{jk})\sim N_k\left(\bmu_j,\bSigma_j\right)
\label{eq:psi}
\end{equation}
independently, where $\left(\bmu_j,\bSigma_j\right);~j=1,2,3$ may be estimated from a pilot run of TMCMC
with the dimensionality fixed at $k=k_{max}$.
Specifically, from a pilot run of TMCMC with $p_i=q_i$, for each variable $x_i$, $i=1,\ldots,k_{max}$, we may   
consider the three empirical means of $x_i$ associated with $z_i=1$, $-1$ and $0$, 
as good candidates for the $i$-th components of $\bmu_1$, $\bmu_2$ and $\bmu_3$, respectively. 
For the covariance matrices $\bSigma_j$, the empirical estimates of the covariances between $x_i$
and $x_j$ associated with $(z_i=1,z_j=1)$, $(z_i=-1,z_j=-1)$, and $(z_i=0,z_j=0)$
may be considered as the $(i,j)$-th elements of $\bSigma_1$, $\bSigma_2$ and $\bSigma_3$, respectively.
The above strategy yields three $k_{max}$-dimensional vectors $\tilde\bmu_j$; $j=1,2,3$, and
three $k_{max}\times k_{max}$-dimensional covariance matrices $\tilde\bSigma_j$; $j=1,2,3$.
The required $k$-dimensional $\bmu_j$ and $k\times k$-dimensional $\bSigma_j$ are then simply 
relevant sub-vectors and sub-matrices of $\tilde\bmu_j$ and $\tilde\bSigma_j$ respectively.

At each iteration of TTMCMC we then first simulate $(\psi_{j1},\psi_{j2},\ldots,\psi_{jk});~j=1,2,3$
using (\ref{eq:psi}), obtain $\{p_i,q_i,1-p_i-q_i;~i=1,\ldots,k\}$ using (\ref{eq:repara}); then
given $\{p_i,q_i,1-p_i-q_i;~i=1,\ldots,k\}$ we simulate $z_i\sim Multinomial(1;p_i,q_i,1-p_i-q_i)$ 
independently as before, where
$i\in\{1,\ldots,k\}\backslash\{j\}$ or $i\in\{1,\ldots,k\}\backslash\{j,j'\}$.

As in the case of TMCMC, it can be easily verified that our modified TTMCMC algorithm with this
hierarchical dependence structure for the distribution of $\bz$ satisfies detailed balance.

\section{{\bf Jumping more than one dimensions at a time}}
\label{sec:jump_more}

We now consider the situations where instead of jumping one dimension, one wishes to jump several dimensions
at a time.
That is, we now consider the more general framework where $\bx=(x_1,\ldots,x_k)\in\mathbb R^k$ and that we wish to 
increase the dimension to $k+m$, 
or to decrease the dimension from $k+m$ to $k$, where 
$1\leq m\leq k$. It follows that TTMCMC can jump from $k$ to $2k$ dimensions and from $2k$ to $k$ dimensions at the maximum.
RJMCMC does not have such restriction, but jumping many dimensions at a time will only add to the general inefficiency of RJMCMC.

For an illustrative TTMCMC example where jumping more than one dimension is desired, assume that $k=3$ and $m=2$, so that 
it is required to jump from $\mathbb R^3$ to
$\mathbb R^5$. For simplicity, we illustrate with the additive transformation. 
One may anticipate that this can be accomplished by simulating a single positive $\e\sim \varrho(\cdot)$, 
selecting, say, $x_1$ and $x_2$ at random without replacement from $\bx=(x_1,x_2,x_3)$, simulating $z_3$, 
and then constructing the birth move $\bx'=T_{b,z_2}(\bx,\e)=(x_1+a_1\e,x_1-a_1\e,x_2+a_2\e,x_2-a_2\e,x_3+z_3a_3\e)
=(x'_1,x'_2,x'_3,x'_4,x'_5)$.
However, for this move, the dimension of $(\bx,\e)=(x_1,x_2,x_3,\e)$ is 4,
while that of $\bx'=(x'_1,x'_2,x'_3,x'_4,x'_5)$ is 5. In other words, 
the Jacobian 
$\left|\frac{\partial (T_{b,\bz}(\bm x, \e))}{\partial(\bm x, \e)}\right|$
is not well-defined.

To get past the above difficulty with dimensions, we need to simulate two $\e$'s from $\varrho(\cdot)$: $\e_1$ for
splitting $x_1$ into $x_1+a_1\e_1$ and $x_1-a_1\e_1$, and $\e_2$ for splitting $x_2$ into $x_2+a_2\e_2$
and $x_2-a_2\e_2$, and also to update $x_3$ to $x_3+z_3a_3\e_2$ ($\e_1$ can also be used to update $x_3$). Hence the birth move takes the form
$\bx'=T_{b,z_3}(\bx,\e_1,\e_2)=(x_1+a_1\e_1,x_1-a_1\e_1,x_2+a_2\e_2,x_2-a_2\e_2,x_3+z_3a_3\e_2)=(x'_1,x'_2,x'_3,x'_4,x'_5)$.
Now the dimensions of both $\bx'=(x'_1,x'_2,x'_3,x'_4,x'_5)$ and $(\bx,\e_1,\e_2)=(x_1,x_2,x_3,\e_1,\e_2)$
are the same and equals 5; hence the Jacobian 
$$\left|\frac{\partial (T_{b,z_3}(\bx, \e_1,\e_2))}{\partial(\bx, \e_1,\e_2)}\right|
=\left|\frac{\partial (x_1+a_1\e_1,x_1-a_1\e_1,x_2+a_2\e_2,x_2-a_2\e_2,x_3+z_3a_3\e_2)}
{\partial(x_1,x_2,x_3, \e_1,\e_2)}\right|=4a_1a_2,$$
is well-defined.
The acceptance probability of the birth move in this example is given by
\begin{align}
a_b(\bm x,\e_1,\e_2)
&=\min\left\{1, \frac{1}{(3+2)(3+1)}\times\frac{w_{d,5}}{w_{b,3}}\times\frac{p^{I_{\{1\}}(z^c_3)}_3q^{I_{\{-1\}}(z^c_3)}_3}
{p^{I_{\{1\}}(z_3)}_3q^{I_{\{-1\}}(z_3)}_3}
\times\frac{\pi(\bx')}
{\pi(\bx)}\times
\left|\frac{\partial (T_{b,z_3}(\bm x, \e_1,\e_2))}{\partial(\bm x, \e_1,\e_2)}\right|
\right\}\notag\\
&=\min\left\{1, \frac{1}{20}\times\frac{w_{d,5}}{w_{b,3}}\times\frac{p^{I_{\{1\}}(z^c_3)}_3q^{I_{\{-1\}}(z^c_3)}_3}
{p^{I_{\{1\}}(z_3)}_3q^{I_{\{-1\}}(z_3)}_3}
\frac{\pi(\bx')}{\pi(\bx)}\times
4a_1a_2
\right\}.\notag\\
\label{eq:acc_birth2}
\end{align}

For the corresponding death move, that is, for moving from $\bx'=(x'_1,x'_2,x'_3,x'_4,x'_5)$
to $\bx''=T_{d,\bz}(\bx',\e_1)=(\frac{x'_1+x'_2}{2},\frac{x'_3+x'_4}{2},x'_5+z^c_3a_3\e_1)=(x''_1,x''_2,x''_3)$, 
we must have, for the reverse of this death move,
$x''_1+a_1\e^*_1=x'_1$, $x''_1-a_1\e^*_1=x'_2$, $x''_2+a_2\e^*_2=x'_3$, 
$x''_2-a_2\e^*_2=x'_4$.
The first two equations yield $\e^*_1=\frac{x'_1-x'_2}{2a_1}$ and the last two equations yield 
$\e^*_2=\frac{x'_3-x'_4}{2a_2}$.
The Jacobian is given by
\begin{align}
&\left|\frac{\partial (T_{d,z_3}(\bx', \e_1);\e^*_1,\e^*_2,\e_1)}{\partial(\bx', \e_1)}\right|
=\left|\frac{\partial \left(\frac{x'_1+x'_2}{2},\frac{x'_3+x'_4}{2},x'_5+z^c_3a_3\e_1,\frac{x'_1-x'_2}{2a_1},
\frac{x'_3-x'_4}{2a_2},\e_1\right)}{\partial\left(x'_1,x'_2,x'_3,x'_4,x'_5, \e_1\right)}\right|=\frac{1}{4a_1a_2}.
\end{align}
We accept this death move with probability
 \begin{align}
 a_d(\bm x'', \e_1,\e^*_1,\e^*_2) &= 
 \min\left\{1,5\times 4\times\frac{w_{b,3}}{w_{d,5}}\times\dfrac{P(\bz^c)}{P(\bz)} ~\dfrac{\pi(\bm x'')}{\pi(\bm x')} 
 ~\left|\frac{\partial (T_{d,z_3}(\bx', \e_1);\e^*_1,\e^*_2,\e_1)}{\partial(\bx', \e_1)}\right| \right\}\notag\\
 &= \min\left\{1,20\times\frac{w_{b,3}}{w_{d,5}}\times\dfrac{p^{I_{\{1\}}(z_3)}_3q^{I_{\{-1\}}(z_3)}_3}
 {p^{I_{\{1\}}(z^c_3)}_3q^{I_{\{-1\}}(z^c_3)}_3} \times\dfrac{\pi(\bm x'')}{\pi(\bm x')} 
 \times\frac{1}{4a_1a_2}\right\}.
 \label{eq:acc_death2}
 \end{align}
We illustrate the idea of this algorithm in Figure \ref{fig:diagram_morethanone} 
diagrammatically for the ease of understanding. 
 \begin{figure}[]
 
    \centering
    \resizebox{\textwidth}{!}{
        \begin{tikzpicture}[font=\boldmath]
        \node[rectangle,fill=gray!20,minimum size=3cm] (12) at (-4,55) {\large \bf Birth Step};
        \node[circle,fill=red!20,minimum size=4cm,font=\bf] (1) at (-12,50) {$x_1$};
        \node[circle,fill=blue!20,minimum size=4cm] (2) at (0,50) {$x_2$};
        \node[circle,fill=green!20,minimum size=4cm] (3) at (8,50) {$x_3$};
        \node[circle,fill=red!20,minimum size=4cm] (4) at (-16,43) {$x^{'}_1=g_{1,z_1=1}(x_1,\epsilon_1$)};
        \node[circle,fill=red!20,minimum size=4cm] (5) at (-8,43) {$x^{'}_2=g_{1,z_1=-1}(x_1,\epsilon_1$)};
        \node[circle,fill=blue!20,minimum size=3.5cm] (6) at (-3,43) {$x^{'}_3=g_{2,z_2=1}(x_2,\epsilon_2)$};
        \node[circle,fill=blue!20,minimum size=3.5cm] (7) at (3,43) {$x^{'}_4=g_{2,z_2=-1}(x_2,\epsilon_2)$};
        \node[circle,fill=green!20,minimum size=4cm] (8) at (8,43) {$x^{'}_5=g_{3,z_3}(x_3,\epsilon_2)$};

        
        \foreach \from/\to in {1/4,1/5,2/6,2/7}
        {\draw[->, ultra thick, fill=blue] (\from) to node[fill=white] {\large{$\epsilon_\from$}} (\to);
        }
        \foreach \from/\to in {3/8}
        {\draw[->,ultra thick, fill=blue] (\from) to node[fill=white] {\large{$\epsilon_2$}} (\to);
 
        }

        \end{tikzpicture}

    }


    \centering
    \resizebox{\textwidth}{!}{
        \begin{tikzpicture}[font=\boldmath]
       
        \node[rectangle,fill=gray!20,minimum size=4cm,font=\bf] (12) at (32,60) {\Huge \bf Death Step};
        \node[circle,fill=red!30,minimum size=6.5cm,font=\bfseries] (1) at (16,52) {\LARGE$x_1$};
        \node[circle,fill=yellow!50,minimum size=6.5cm] (2) at (24,52) {\LARGE$x_2$};
        \node[circle,fill=blue!20,minimum size=6.5cm] (3) at (32,52) {\LARGE$x_3$};
        \node[circle,fill=gray!30,minimum size=6.5cm] (4) at (40,52) {\LARGE$x_4$};
        \node[circle,fill=green!30,minimum size=6.5cm] (5) at (48,52) {\LARGE$x_5$};

        \node[circle,fill=red!30,minimum size=6.5cm] (6) at (16,43) {\LARGE$x^{*}_1=g_{1,z_1=-1}(x_1,\epsilon_1)$};
        \node[circle,fill=yellow!50,minimum size=6.5cm] (7) at (24,43) {\LARGE$x^{*}_2=g_{2,z_2=1}(x_2,\epsilon_1)$};
        \node[circle,fill=blue!20,minimum size=6.5cm] (8) at (32,43) {\LARGE$x^{*}_3=g_{3,z_3=-1}(x_3,\epsilon_2)$};
        \node[circle,fill=gray!30,minimum size=6.5cm](9) at (40,43) {\LARGE$x^{*}_4=g_{4,z_4=1}(x_4,\epsilon_2)$};
        \node[circle,fill=green!30,minimum size=6.5cm] (10) at (48,43) {\LARGE$x^{*}_5=g_{5,z_5}(x_5,\epsilon_2)$};
        \node[circle,fill=orange!40,minimum size=6.5cm] (11) at (20,35) {\LARGE $x^{'}_1=\frac{(x^{*}_1+x^{*}_2)}{2}$};
         \node[circle,fill=blue!15,minimum size=6.5cm] (12) at (36,35) {\LARGE $x^{'}_2=\frac{(x^{*}_3+x^{*}_4)}{2}$};
        \node[circle,fill=green!30,minimum size=6.5cm] (13) at (48,35) {\LARGE$x^{'}_3=x^{*}_5$};
        \node[rectangle,fill=gray!20,minimum width =5cm,minimum height=2cm] (14) at (18,30) {\large Here $d_1=3, d_2=5,n_1=2,n_2=2$};
        \node[rectangle,fill=gray!20,minimum width =4cm,minimum height=2cm] (15) at (28,30) { \large $n_1+d_1\neq n_2+d_2$};
        \node[rectangle,fill=gray!20,minimum width =7cm,minimum height=2cm] (16) at (40,30) {\large \bf `Dimension matching' criterion is not satisfied for TTMCMC algorithm};
       
       \foreach \from/\to in {1/6,2/7}
        {\draw[->,ultra thick, fill=blue] (\from) to node[fill=white] {\Large{$\epsilon_1$}} (\to);
        }
         \foreach \from/\to in {3/8,4/9,5/10}
        {\draw[->,ultra thick, fill=blue] (\from) to node[fill=white] {\Large{$\epsilon_2$}} (\to);
        }
      
        {\draw[->,ultra thick, fill=blue] (6) -- (11);
          \draw[->,ultra thick, fill=blue] (7) -- (11);
          \draw[->,ultra thick, fill=blue] (8) -- (12);
           \draw[->,ultra thick, fill=blue] (9) -- (12);
           \draw[->,ultra thick, fill=blue] (10) -- (13);
           \draw[->,ultra thick, fill=blue] (14) -- (15);
           \draw[->,ultra thick,scale=0.5, fill=blue] (15) -- (16);

        }
        
        \end{tikzpicture}
    }
     \caption{ Illustration of TTMCMC algorithm for jumping more than one dimension.}
\label{fig:diagram_morethanone}
\end{figure}

Thus, in general, for moving from dimension $k$ to dimension $k+m$, we need to simulate $\e_1,\ldots\e_m$ for
updating $\bx=(x_1,\ldots,x_k)$ to $\bx'=(x'_1,x'_2,\ldots,x'_k,x'_{k+1},\ldots,x'_{k+m})$.
The associated general TTMCMC algorithm for jumping $m$ dimensions  is provided as Algorithm S-3.1 of Section S-3,
and the proof of its detailed balance is provided in Section S-4. 

In variable dimensional problems such as mixtures, changing the dimension of one set of parameters necessitates
changing the dimensions of the other sets of parameters. Thus, more than one dimension must be changed at a time,
while the parameters are inter-related. We provide the details and the relevant algorithm (Algorithm S-5.1) 
in Section S-5 of the supplement. Indeed, for our mixture applications of TTMCMC, we implement Algorithm S-5.1,
choosing the additive transformation.

Note that exactly as discussed in Section \ref{subsec:dependent_z} we can incorporate a hierarchical
dependence structure on the distribution of $\bz$ in Algorithms S-3.1 and S-5.1, 
which does not hamper the detailed balance condition.

\section{{\bf TTMCMC: towards automation}}
\label{sec:automation}
Algorithms \ref{algo:ttmcmc}, S-3.1 and S-5.1 provide concrete ways to
implement our TTMCMC procedure, in general variable dimensional problems. 
Below we detail the manyfold advantages of TTMCMC, which point towards the fact that TTMCMC 
is close to qualifying as an automatic sampler in variable dimensional problems.


\subsection{{\bf Reasonably high acceptance rate}}
\label{subsec:acceptance_rate}

The additive and the
multiplicative transformations, and combinations of them can be effectively utilized, in conjunction
with just a few, fixed number of $\e$'s, to 
accomplish transdimensional movement. The methodology reduces the variable dimensional problem
to effectively fixed dimensional, indexed by a fixed and small number of $\e$'s. The fixed
and low-dimensional nature of $\e$ (or the set $\{\e_1,\ldots,\e_m\}$) ensures reasonably high acceptance rate. 
Indeed, for high-dimensional proposals, 
with high probability at least one component would be ill-proposed,
which would render the acceptance probability extremely small, even in fixed-dimensional cases. In the context
of TMCMC, theoretical and empirical results are provided in \ctn{Dutta14}, \ctn{Dey14a}, \ctn{Dey14b}.
Our experiments in this paper 
provide ample support to our claim of adequate acceptance rate of TTMCMC.

\subsection{{\bf Good mixing properties in high-dimensional and multimodal cases}}
\label{subsec:mixing}

\ctn{Dutta14} discussed that in one-dimensional situations, TMCMC reduces to a Metropolis-Hastings algorithm with 
a specialized mixture proposal density, and hence, is expected to explore multimodal target densities quite
efficiently (see \ctn{Guan07}, for example). In higher dimensions, due to singularity, 
the proposal does not admit a Lebesgue-measure-dominated mixture density form directly, 
but since the method employs similar principles,
good convergence properties of TMCMC are to be expected for high-dimensional multimodal targets as well. 
Since TTMCMC samplers are also based on the same principles of deterministic transformations and construction of move types 
within each of the birth, death and no-change move types, good convergence properties are expected 
when the target density is multimodal for each dimension. 
In the context of TMCMC, \ctn{Dey14}, \ctn{Dey14a} and \ctn{Dey14b} demonstrate far superior mixing of TMCMC
compared to random walk Metropolis-Hastings. The results of our TTMCMC applications 
reported in this paper provide ample support to this discussion.

\subsection{{\bf Applicability to all variable dimensional problems}}
\label{subsec:applicability}

The construction of TTMCMC sampler does not require any assumptions regarding the model, such as
existence of moments or unimodality. Note that in the attempts made so far for constructing
generic RJMCMC samplers, these assumptions are quite crucial; see \ctn{Sisson05}, \ctn{Fan11} for comprehensive
discussions regarding these assumptions. 
So, for the construction of TTMCMC sampler for switching between two models, namely, from $\mathcal{M}_{k}$ 
to $\mathcal{M}_{k'}$, we only need to determine if some sets of parameters are related and 
decide on the number of parameters to be added or deleted, 
in a single step. Accordingly we will choose one of the above mentioned algorithms and update all the 
parameters in a single block. 
Hence, our proposed sampler is very much applicable to any variable dimensional problem.


\subsection{{\bf Default TTMCMC algorithm and its tuning}}
\label{subsec:optimal_scaling}
In order to design efficient MCMC algorithms it has become standard practice to
tune the proposals. For the default, random walk proposals, this is synonymous with choosing
the scales optimally. \ctn{Dutta14} recommended additive TMCMC as the default TMCMC proposal
since this transformation requires much smaller number of move-types and the corresponding acceptance
probability has a simple form in that it is free of the Jacobian of transformations. 
Already \ctn{Dey14a}, \ctn{Dey14b} have developed some theory 
on optimal scaling in the context of additive TMCMC. 
In keeping with \ctn{Dutta14} we advocate additive TTMCMC as the default TTMCMC sampler, which
again requires specification of the scaling constants.
In this regard, in Section S-8.2 of the supplement 
we propose a convergence diagnostic that 
is generally applicable.
Guided by our proposed convergence diagnostic it is possible to find the 
appropriate value of scaling constants. Instances of the idea are illustrated
in Sections \ref{sec:gamma_mixtures} and \ref{sec:normal_mixtures}. The results
of our experiments demonstrate great ease of implementation and excellent performance
of the default additive TTMCMC sampler in all the examples. 
Further experiments with additive TTMCMC, conducted by these authors and their colleagues 
in challenging, high-dimensional spatio-temporal problems (see, for example, \ctn{Das14}), 
variable-selection problems, (high-dimensional) curve-fitting problems also yielded excellent results.
Thus, it seems that additive TTMCMC is close towards the kind of automation that we desire.
%


\section{{\bf Simulation studies with mixtures of gamma distributions with unknown number of components}}
\label{sec:gamma_mixtures}

\ctn{Wiper01} implement RJMCMC in mixtures of gamma distributions of the form $\mathcal G\left(\nu,\frac{\nu}{\mu}\right)$,
where by $\mathcal G\left(a,b\right)$ we mean a gamma distribution with mean $a/b$ and variance $a/b^2$.
In other words, \ctn{Wiper01} consider the following mixture density for $y>0$:
\begin{equation}
f(y\vert\bnu_k,\bmu_k,\bpi_k,k)=\sum_{j=1}^k\pi_j\frac{\left(\nu_j/\mu_j\right)^{\nu_j}}{\Gamma(\nu_j)} 
y^{\nu_j-1}\exp\left(-\frac{\nu_j}{\mu_j}y\right),
\label{eq:gammamix}
\end{equation}
where $\bnu_k=(\nu_1,\ldots,\nu_k)$, $\bmu_k=(\mu_1,\ldots,\mu_k)$, and $\bpi_k=(\pi_1,\ldots,\pi_k)$.
Given $k>0$, for each $j$, $\nu_j>0$, $\mu_j>0$, $0<\pi_j<1$ such that
$\sum_{j=1}^k\pi_j=1$. We assume $k$ to be unknown, so that the dimension of the model (that is,
the number of the component parameters) is unknown and considered random.

\subsection{{\bf Prior structure}}
\label{subsec:gammamix_prior}

\ctn{Wiper01} assumed the following prior structure given $k$:
\begin{align}
\bpi_k&\sim\mathcal D(1,\ldots,1);\label{eq:prior_pi_gammamix}\\
\nu_j &\stackrel{iid}{\sim}\mathcal E(100);~j=1,\ldots,k;\label{eq:prior_nu_gammamix}\\ 
\mu^{-1}_j &\stackrel{iid}{\sim}\mathcal G\left(1,1\right);~j=1,\ldots,k,\label{eq:prior_mu_gammamix} 
\end{align}
such that $\mu_1<\cdots<\mu_k$.
In (\ref{eq:prior_pi_gammamix}), $\mathcal D(1,\ldots,1)$ denotes the Dirichlet distribution
with all the parameters equal to $1$, and in (\ref{eq:prior_nu_gammamix}), $\mathcal E(100)$ stands for the 
exponential distribution with mean $100$.
As regards $k$, \ctn{Wiper01} consider the discrete uniform distribution on $\{1,\ldots,10\}$.

For the implementation purpose, we reparameterize $\nu_j$ and $\mu_j$
as $\exp(\nu^*_j)$ and $\exp(\mu^*_j)$, where 
$\nu^*_j\sim\mathcal \log\left(\mbox{Exponential}(100)\right)$ and 
$\left(\mu^*_j\right)^{-1}\sim\mathcal \log\left(\mathcal G\left(1,1\right)\right)$. Since $-\infty<\nu^*_j<\infty$
and $-\infty<\mu^*_j<\infty$, this reparameterization frees the parameter space from any restrictions, allowing TTMCMC 
to move freely, while keeping the original prior distributions intact. 
We denote $(\nu^*_1,\ldots,\nu^*_k)$ by $\bnu^*_k$ and $(\mu^*_1,\ldots,\mu^*_k)$ by $\bmu^*_k$.

For $\bpi$ we propose the following prior based on reparameterization:
for $j=1,\ldots,k$,
\begin{align}
\pi_j=\frac{\exp\left(\omega_j\right)}{\sum_{\ell=1}^k\exp\left(\omega_j\right)};\quad
\omega_1,\ldots,\omega_k\stackrel{iid}{\sim}N\left(\mu_{\omega},\sigma^2_{\omega}\right),
\label{eq:prior_pi}
\end{align}
where $\omega_j\stackrel{iid}\sim\log\left(\mathcal G(1,1)\right)$, so that the prior (\ref{eq:prior_pi_gammamix})
remains intact. Thus, we need to update $\bomega_k=(\omega_1,\ldots,\omega_k)$, instead of $\bpi$, using TTMCMC. 

\subsection{{\bf Label switching}}
\label{subsec:label_switching}
A brief account of the so-called ``label-switching problem" associated with identifiability
of mixtures is provided in Section S-6 of the supplement.
In this article our goal is to demonstrate TTMCMC with inference regarding posterior distributions of
densities. Since inference on densities is not affected by label switching, the problem of label switching is not
of much importance in our context. Moreover, we argue in Section S-6 that 
identifiability in the mixture context is not generally desirable.
However, since \ctn{Wiper01} enforced the restriction $\mu_1<\cdots<\mu_k$ in an attempt to mitigate
identifiability problems, for fair comparison we also impose the same restriction.

\subsection{{\bf Posterior summary}}
\label{subsec:posterior_summary}
An important aspect to any Bayesian analysis is summarization of the posterior in the sense of obtaining
a measure of central tendency and appropriate credible regions. Here we are interested in the posterior
distribution of the entire mixture density, induced by the posterior of the unknown number of parameters.
Thus, we need a measure of central tendency for the set of mixture densities supported by the posterior,
and appropriately constructed credible regions. Indeed, in Section S-7 of the supplement, we develop
a methodology for obtaining the modal mixture density associated with the posterior, along with the desired
credible regions and highest posterior density (HPD) credible regions. In the context of our experiments we shall display
the modal mixture densities and several other mixture densities falling within the 95\% HPD regions.

\subsection{{\bf Convergence diagnostics}}
\label{subsec:convergence_diagnostics}
Convergence assessment even in fixed-dimensional set-ups is a difficult proposition; in variable-dimensional problems,
the challenges increase manyfold. We provide a briefing on these in Section S-8.1 of the supplement.
As an attempt to make some progress on convergence assessment in variable-dimensional problems we propose a
convergence diagnostic in Section S-8.2 of the supplement, which is based on the methodology
for summarizing the posterior. In a nutshell, we obtain 95\% (or any other desired) credible regions
from the first and second halves of a complete run of TTMCMC, and then obtain the minimum increments of the radii
required for the credible regions to contain one another; small values of the increments indicate convergence
of TTMCMC. Not only do we assess convergence of TTMCMC with this method, we exploit this idea to select the scales
of the additive transformation that we employ for the illustrations. 

\subsection{{\bf General TTMCMC strategy for our experiments}}
\label{subsec:general_ttmcmc}
We conduct four simulation studies, with data generated from the same 1-component, 2-component, 3-component and 4-component
gamma mixtures as considered by \ctn{Wiper01} and apply TTMCMC and compare our results with those obtained by the
RJMCMC algorithm of \ctn{Wiper01}. In particular,
we apply Algorithm S-3.1,
updating $(k,\bnu^*,\bmu^*,\bomega)$ simultaneously in a single block 
using the additive transformation; we choose the proposal density to be
$\varrho(\epsilon)\equiv N(\epsilon:0,1)\mathbb I_{(0,\infty)}(\epsilon)$, where
$N(\epsilon:0,1)$ denotes the normal density with mean $0$, variance $1$, and evaluated at $\epsilon$;
$\mathbb I_{(0,\infty)}(\cdot)$ denotes the indicator function for the set $(0,\infty)$. 
For every iteration of TTMCMC we choose equal move-type probabilities 
of birth, death and no-change strategies. Also, for the underlying additive transformation, we choose
equal probabilities of forward and backward transformations.
The forms of the Jacobian for the birth and the death moves are given by 
$8a_{\nu^*_j}a_{\mu^*_j}a_{\omega_j}$ and $(8a_{\nu^*_j}a_{\mu^*_j}a_{\omega_j})^{-1}$ respectively, where
$a_{\nu^*_j}$, $a_{\mu^*_j}$ and $a_{\omega_j}$ are the scales for additive TTMCMC updating of
$\nu^*_j$, $\mu^*_j$ and $\omega_j$ respectively. 
We base the choices of these scales on the convergence diagnostic proposed in Section S-8.2 of the supplement.
The experimental details are provided in the context-specific applications.
All our codes are written in C and implemented on a 32 bit, dual core ($2.53$ GHz $\times 2$) laptop 
with 2.8 GiB memory. However, for high-dimensional multivariate experiments we implemented our C codes
on a VMWare.

\subsection{{\bf An RJMCMC algorithm based on random walk proposals}}
\label{subsec:random_walk_rjmcmc}
Since, in this paper, we apply additive TTMCMC to our examples, it makes sense to compare our TTMCMC results 
with those obtained by the RJMCMC algorithm based on random walk, which is the closest to additive TTMCMC among all RJMCMC
algorithms. 
Recall that random walk involves additive transformations of the same form as additive TTMCMC, but with independent
jump sizes for every variable, unlike TTMCMC. Also, unlike TTMCMC, the acceptance ratios for the birth 
and death moves involves products of the densities $\varrho(u_i)\equiv N(u_i:0,1)\mathbb I_{(0,\infty)}(u_i);~i=1,2,3$, 
corresponding 
to the birth proposals for $(\nu^*_1,\tau^*_1,\omega_1)$. 
Since the proposals of additive TTMCMC and random walk have the same additive form, the variabilities of the
jump sizes of the competing proposals are not expected to be different. This is confirmed by the optimal
scaling theory of TMCMC developed by \ctn{Dey14a}, where it is shown that the optimal scales of additive
TMCMC and random walk are the same. Hence, in this work, we choose the same scales of random walk  RJMCMC as additive TTMCMC.

The main difference between our random walk RJMCMC and 
the proposal of \ctn{Green03} is that the latter is deterministic
unless movement to a higher dimension is attempted; the moves also involve dimension-specific mean vectors and
covariance matrices, which are to be estimated from the dimension-specific posteriors. Even for moderate
number of models this is a difficult and computationally burdensome proposition; see \ctn{Fan11} for example.
Indeed, as stressed in \ctn{Green03}, the approach is unlikely to be useful for more than a small set of models.

However, for all our examples related to the gamma mixture, our random walk RJMCMC had very small overall acceptance rate,
and completely failed to change the dimension in any such example.
Hence, we do not provide further details regarding the performance of the random walk RJMCMC in gamma mixtures. 
In the normal mixture context, random walk RJMCMC performed somewhat better, although still not at all
satisfactorily. Since this algorithm fails even in univariate contexts, we do not pursue this for the multivariate 
situations.

\subsection{{\bf First simulation study with data generated from a one-component gamma mixture}}
\label{subsec:1comp}

Following \ctn{Wiper01} we generate $400$ realizations 
from $\mathcal Gamma(3,3)$, and model the realized data with the gamma mixture of the form (\ref{eq:gammamix}). Assuming 
the same prior structure described in Section \ref{subsec:gammamix_prior}, we then simulate from the resulting
variable-dimensional posterior using TTMCMC. 

For implementing TTMCMC it is necessary to select the scales  $a_{\nu^*_j},a_{\tau^*_j},a_{\omega_j}$
appropriately for each $j=1,\ldots,k$. 
Rather than selecting the scales in order to optimize the acceptance rate (see \ctn{Dey14a}
for optimal scaling theory in the context of additive TMCMC), here we
choose the scales by directly quantifying convergence of the TTMCMC chain using
the convergence diagnostic procedure proposed in Section S-8.2 of the supplement. 
We experimented by setting, for every $j=1,\ldots,k$, the scale values
$a_{\nu^*_j}=a_{\nu^*}$; $a_{\mu^*_j}=a_{\mu^*}$, and $a_{\omega_j}=a_{\omega}$, with 
$a_{\nu^*},a_{\tau^*},a_{\omega}$ being one of the trial values $0.05,0.1,0.12,0.15,0.20,0.25,0.50$.
With every trial value, we ran our TTMCMC algorithm for
a burn-in of $750,000$ iterations, and
a further $1,500,000$ iterations, storing one in $150$ iterations, thus obtaining
a total of $10,000$ realizations from the posterior distribution.
For each trial run
we assessed convergence of our TTMCMC chain using the method proposed in Section S-8.2. 
We divided our TTMCMC samples into two parts, one part consisting of the first $5,000$ realizations
and the other part containing the next $5,000$ realizations. Constructing the approximate
95\% credible regions as prescribed, we then obtained the minimum increment, $\eta_1$, of the radius 
of the first credible region such that the increased first credible region wholly contains the second credible region.
Similarly, we obtained $\eta_2$, the radius increment associated with the second credible region.
Small values of $\eta_1$ and $\eta_2$ indicate convergence of the algorithm.
We selected that set of trial values of the scales
which yielded the smallest $\eta_1$ and $\eta_2$ among the trial runs. Indeed, the smallest
$\eta_1$ and $\eta_2$ turned out to be 
$\eta_1=0.041460$ and $\eta_2=0.027130$, which corresponded to
$a_{\nu}=a_{\mu}=0.5$ and $a_{\omega}=1.5$. Hence, we report our results with respect to these trial values. 
Moreover, since
both these quantities are small, we conclude that convergence has taken place appropriately.
We remark here that the rather long burn-in that we had considered was unnecessary, as further experiments 
showed that the chain converged in far less number
of iterations. But we feel it is a good practice to allow large enough burn-in when it is feasible computationally.
The overall acceptance rate, evaluated empirically, turned out to be $0.036596$. 
The birth, death, and no-change rates are $0.004206$, $0.053131$ and $0.067679$, respectively.
Our TTMCMC implementation with the scales selected as above took 10 minutes and 57 seconds. 

\begin{figure}
\centering
\subfigure[Trace plot of $k$.]{ \label{fig:gamma1comp_k}
\includegraphics[width=7cm,height=6cm]{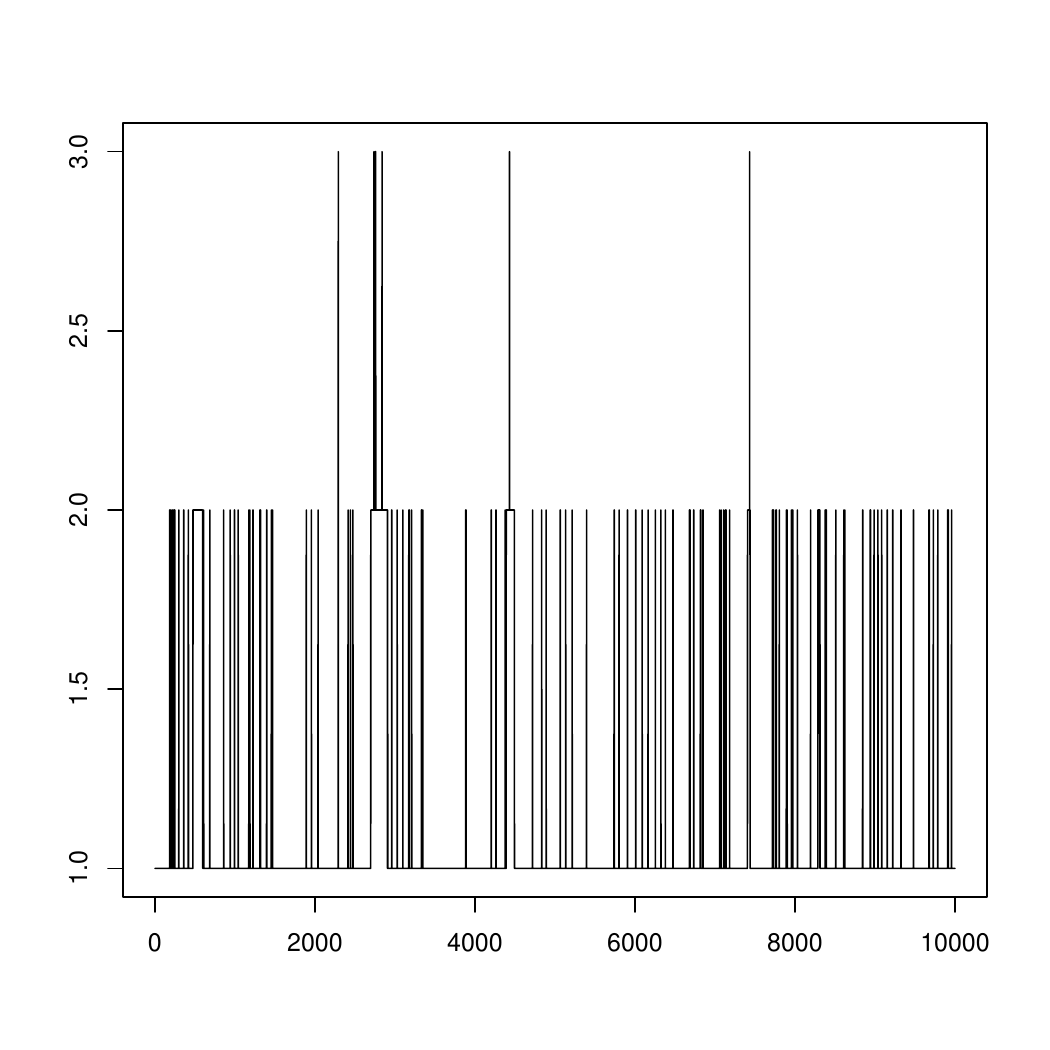}}
\hspace{2mm}
\subfigure[Trace plot of $\nu^*_1$.]{ \label{fig:gamma1comp_nu}
\includegraphics[width=7cm,height=6cm]{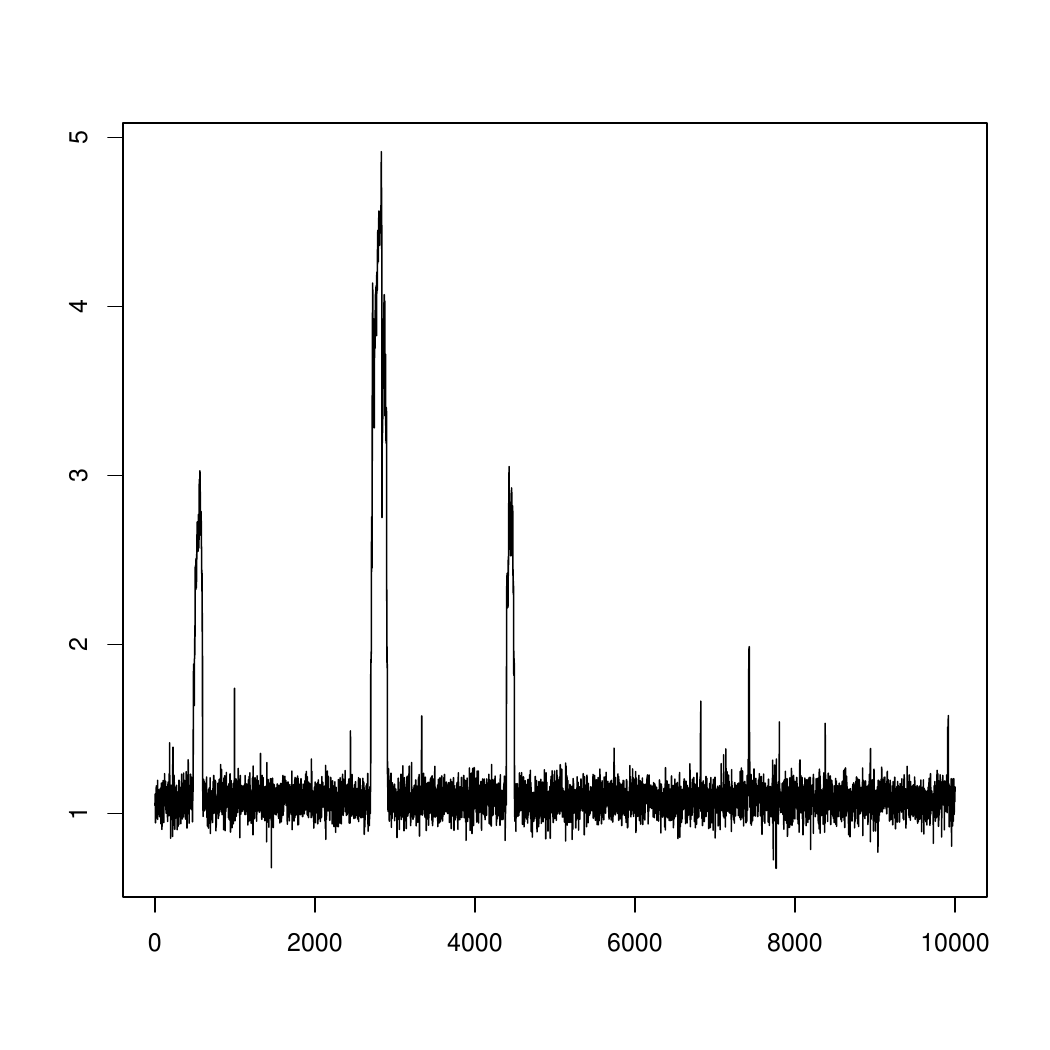}}\\
\vspace{2mm}
\subfigure[Trace plot of $\mu^*_1$.]{ \label{fig:gamma1comp_mu}
\includegraphics[width=7cm,height=6cm]{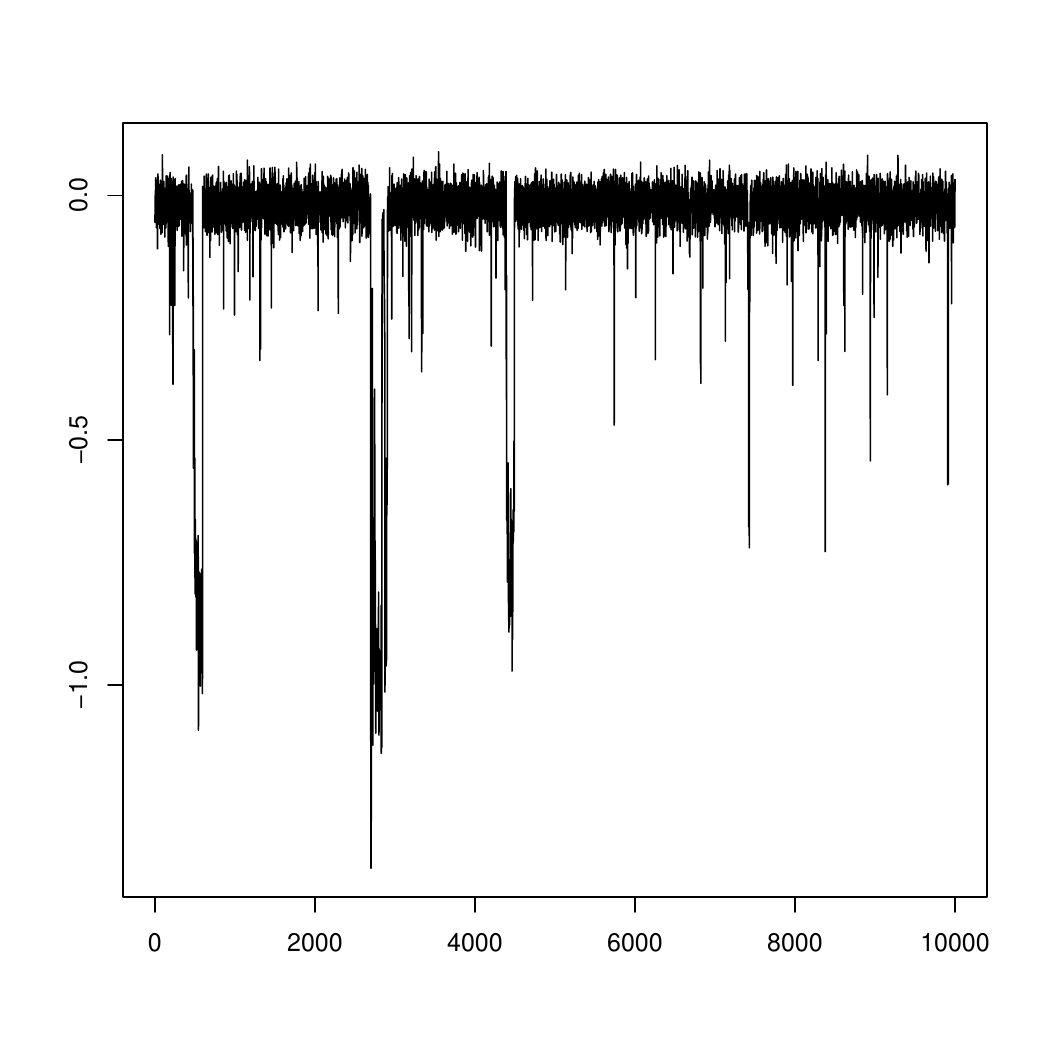}}
\hspace{2mm}
\subfigure[Trace plot of $\omega_1$.]{ \label{fig:gamma1comp_w}
\includegraphics[width=7cm,height=6cm]{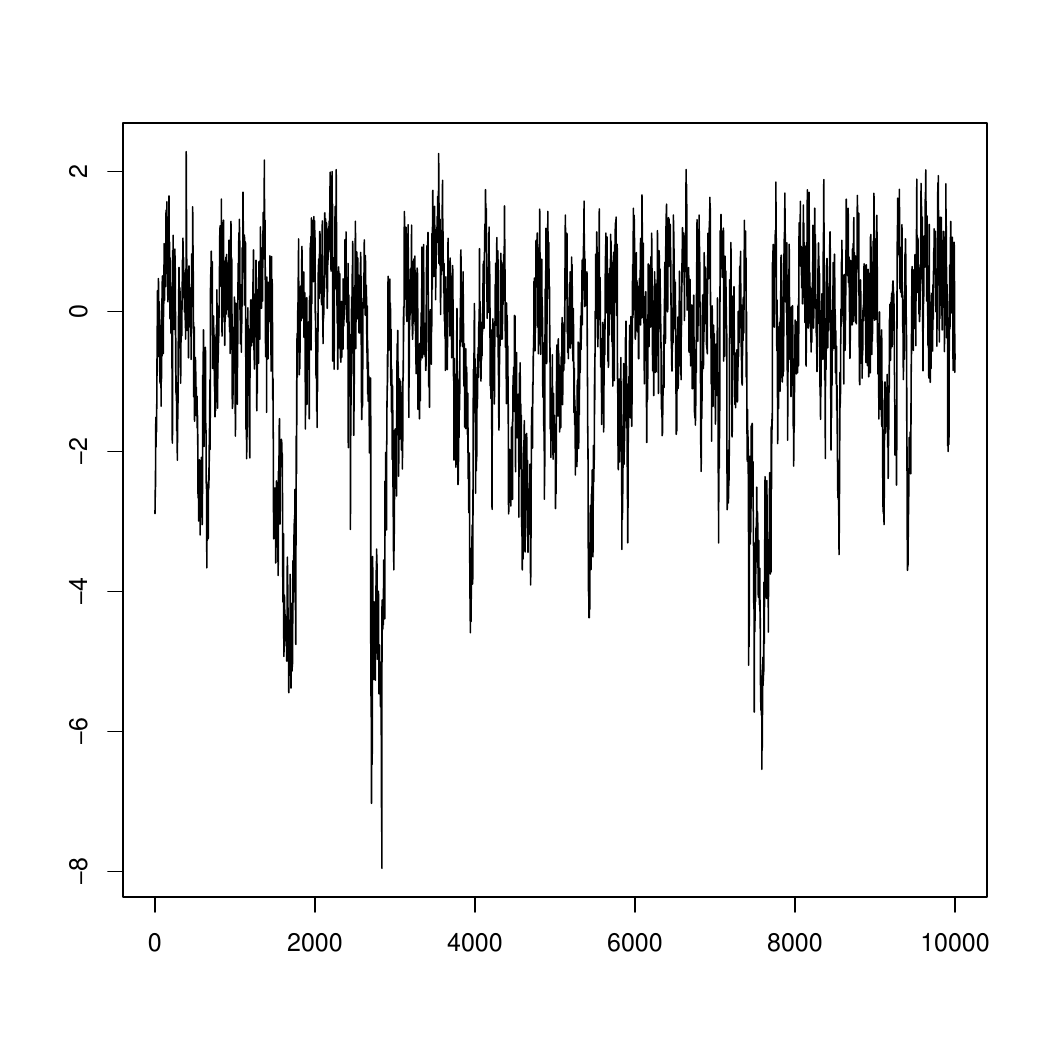}}
\caption{{\bf TTMCMC for 1-component gamma mixture:} Trace plots of $k$, $\nu^*_1$, $\mu^*_1$ and $\omega_1$.} 
\label{fig:gamma1comp_trace_plots}
\end{figure}

The trace plots of $k$, $\nu^*_1$, $\mu^*_1$ and $\omega_1$, provided in Figure \ref{fig:gamma1comp_trace_plots},
exhibits quite adequate mixing properties consistent with our more formal test of convergence.
Also very encouragingly, the posterior distribution of $k$ gives probabilities 
$0.9344$, $0.0649$ and $0.0007$ to $k=1,2,3$ respectively, heavily supporting the true, single-component
gamma mixture. Since the data size is rather large, such high support to the truth is expected.
Indeed, with further simulation studies we demonstrate in Section S-9.1 of the supplement, that as the data size
increases, the posterior distribution of $k$ concentrates around the truth, namely, $k=1$.

\begin{figure}
\includegraphics[width=7in,height=6.5in]{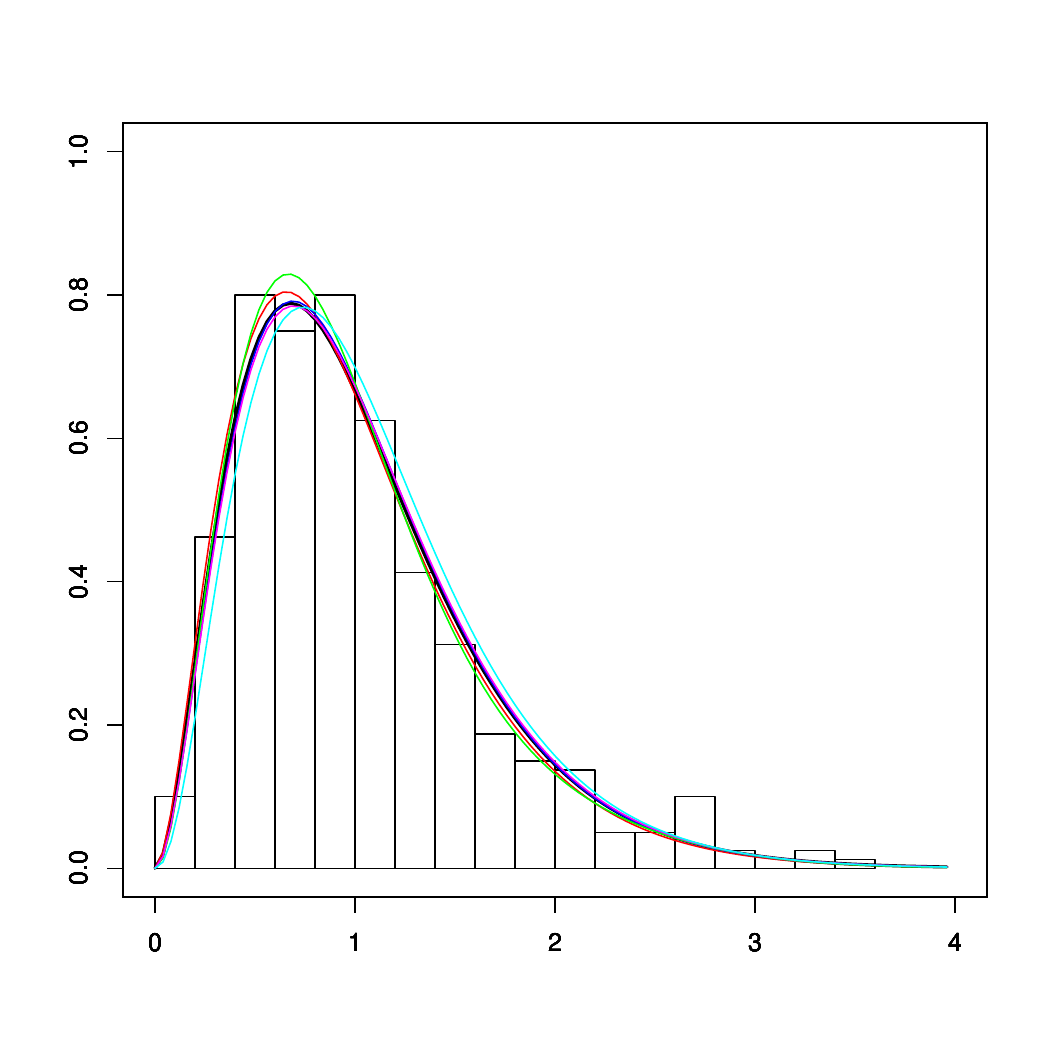}
\caption{{\bf TTMCMC for 1-component gamma mixture:} Goodness of fit of the posterior distribution of 
densities (coloured curves) to the simulated data (histogram). 
The thick black curve is the modal density and the other coloured curves are some densities contained in the 95\% HPD.}
\label{fig:gamma1comp_hpd}
\end{figure}

Figure \ref{fig:gamma1comp_hpd} shows the modal density (thick, black curve), along with some 
other densities within the 95\% HPD region overlapped on the histogram of the simulated data. Excellent
fit of the posterior distribution of the densities to the data is indicated by the diagram.

\subsubsection{{\bf Comparison with the results obtained by Wiper et al. (2001)}} 
\label{subsubsec:1comp_wiper}
In sharp contrast with our TTMCMC results, \ctn{Wiper01}, using an RJMCMC algorithm that is very similar to that proposed
by \ctn{Richardson97} for normal mixtures, obtained a posterior distribution that supports all possible
values of $k\in\{1,\ldots,10\}$. In particular, their posterior probabilities of $k=1,2,3,4,5$ turned out to be 
$0.41$, $0.24$, $0.12$, $0.08$ and $0.05$, respectively, with other values of $k$ having posterior 
probabilities less than $0.03$. In other words, driven by RJMCMC, the true value $k=1$ received lower posterior support, 
in comparison with our TTMCMC based posterior. This performance can possibly be attributed to the 
$\mathcal G(5,5)$ proposal density they used for their dimension-changing
move. Since this density is uniformly less than one and features in the acceptance ratio, heavy bias towards large
values of $k$ is to be expected as per our discussion in the third point following Algorithm \ref{algo:ttmcmc}.
Thus, there seems to be good reasons to suspect the convergence of the RJMCMC algorithm in this case. In fact,
as we shall show, the same issue hinders convergence of the RJMCMC algorithms for the remaining experiments as well. 

It is important to remark in this context that the actual mixture density can be approximated well in spite of poor
mixing, provided that $k$ takes on large values with significant posterior probabilities. Therefore fitting the actual
density alone can be very misleading as a criterion of assessment of variable-dimensional algorithms, particularly
for RJMCMC algorithms, because of their inherent bias towards large values of $k$ in any practical implementation.
In all the four simulation examples considered by \ctn{Wiper01}, the actual densities are well-approximated
by RJMCMC, but in all the cases, large values of $k$ seemed to play vital important roles in this regard. 
Such an issue is clearly of more concern in real data cases where the truth is unknown. 
As we demonstrate with TTMCMC in the supplement with the real galaxy data example of \ctn{Richardson97}, their prior structure
perhaps actually supports unimodal density, while the histogram is highly multimodal. However, because of large values
of $k$ supported by RJMCMC, the approximated density seems to appear as a good fit.

\subsection{{\bf Second simulation study with data generated from a two-component gamma mixture}}
\label{subsec:2comp}

Following \ctn{Wiper01} we now generate $400$ realizations 
from the two-component mixture $0.1\times\mathcal G(9,27)+0.9\times\mathcal G(90,270)$. 

\begin{figure}
\centering
\subfigure[Trace plot of $k$.]{ \label{fig:gamma2comp_k}
\includegraphics[width=7cm,height=6cm]{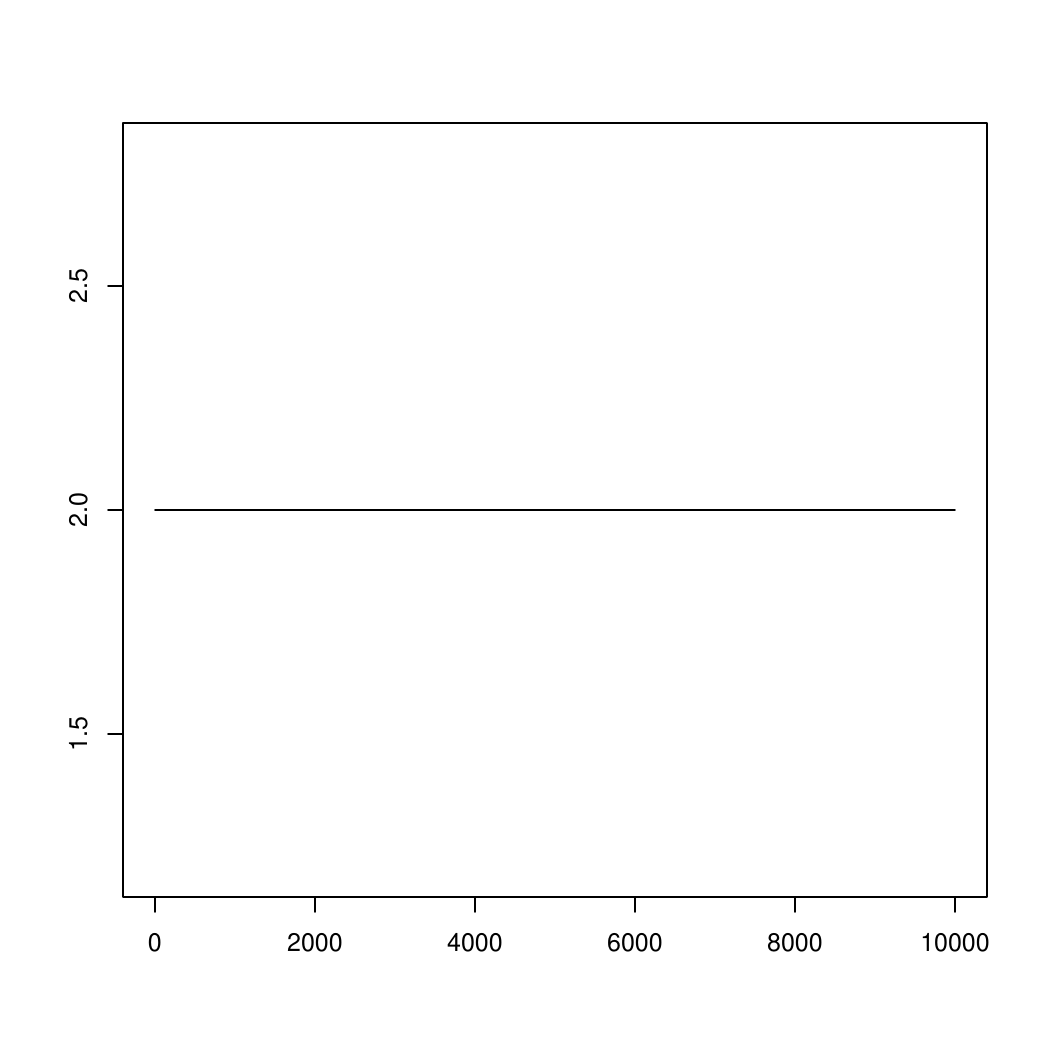}}
\hspace{2mm}
\subfigure[Trace plot of $\nu^*_1$.]{ \label{fig:gamma2comp_nu}
\includegraphics[width=7cm,height=6cm]{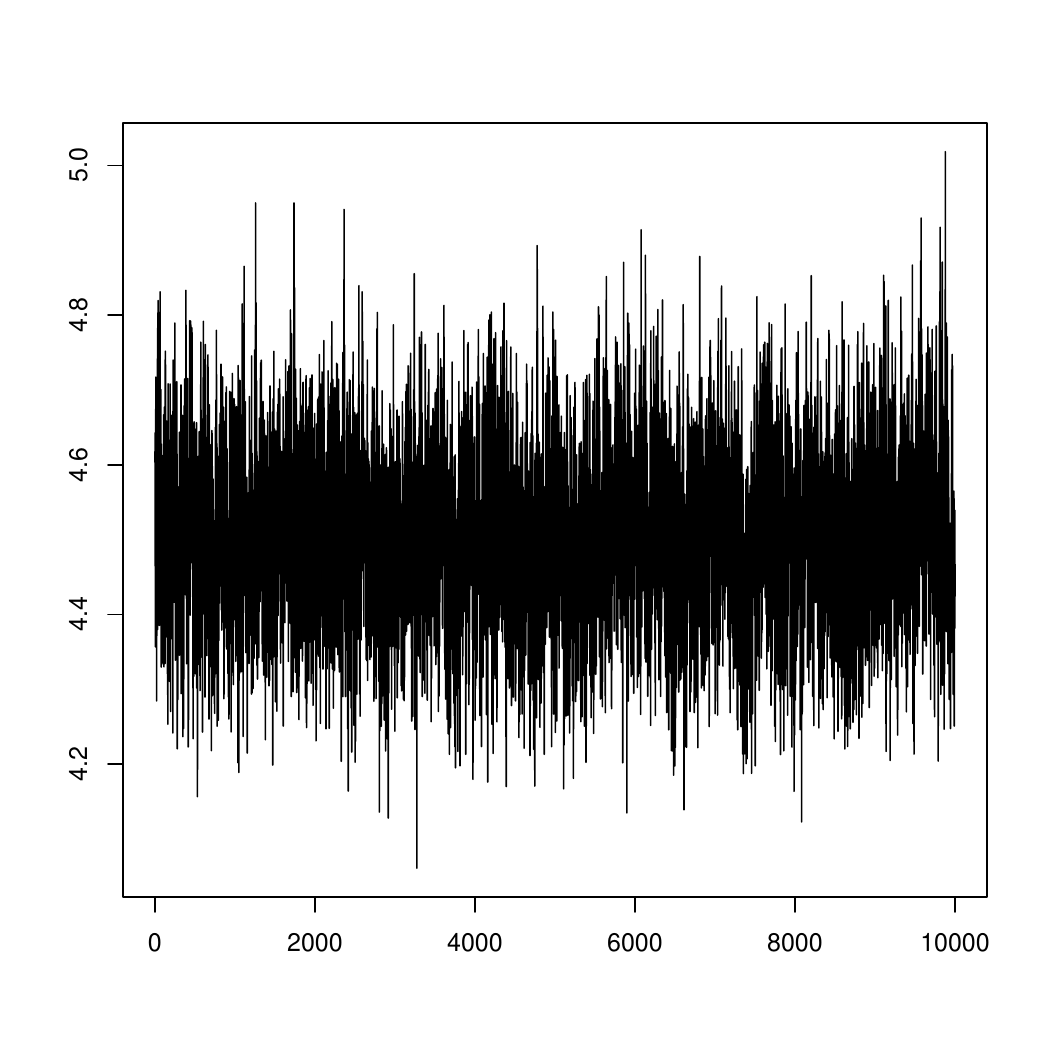}}\\
\vspace{2mm}
\subfigure[Trace plot of $\mu^*_1$.]{ \label{fig:gamma2comp_mu}
\includegraphics[width=7cm,height=6cm]{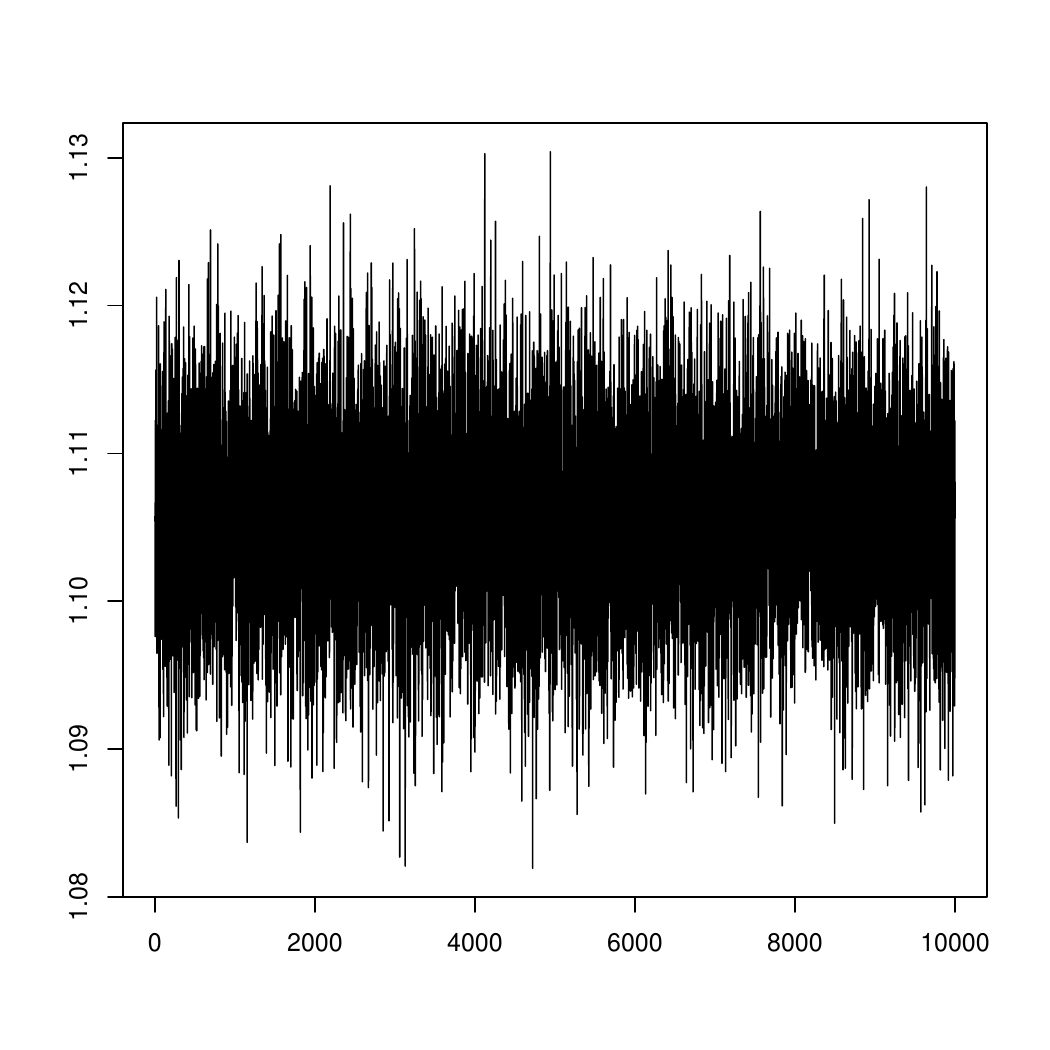}}
\hspace{2mm}
\subfigure[Trace plot of $\omega_1$.]{ \label{fig:gamma2comp_w}
\includegraphics[width=7cm,height=6cm]{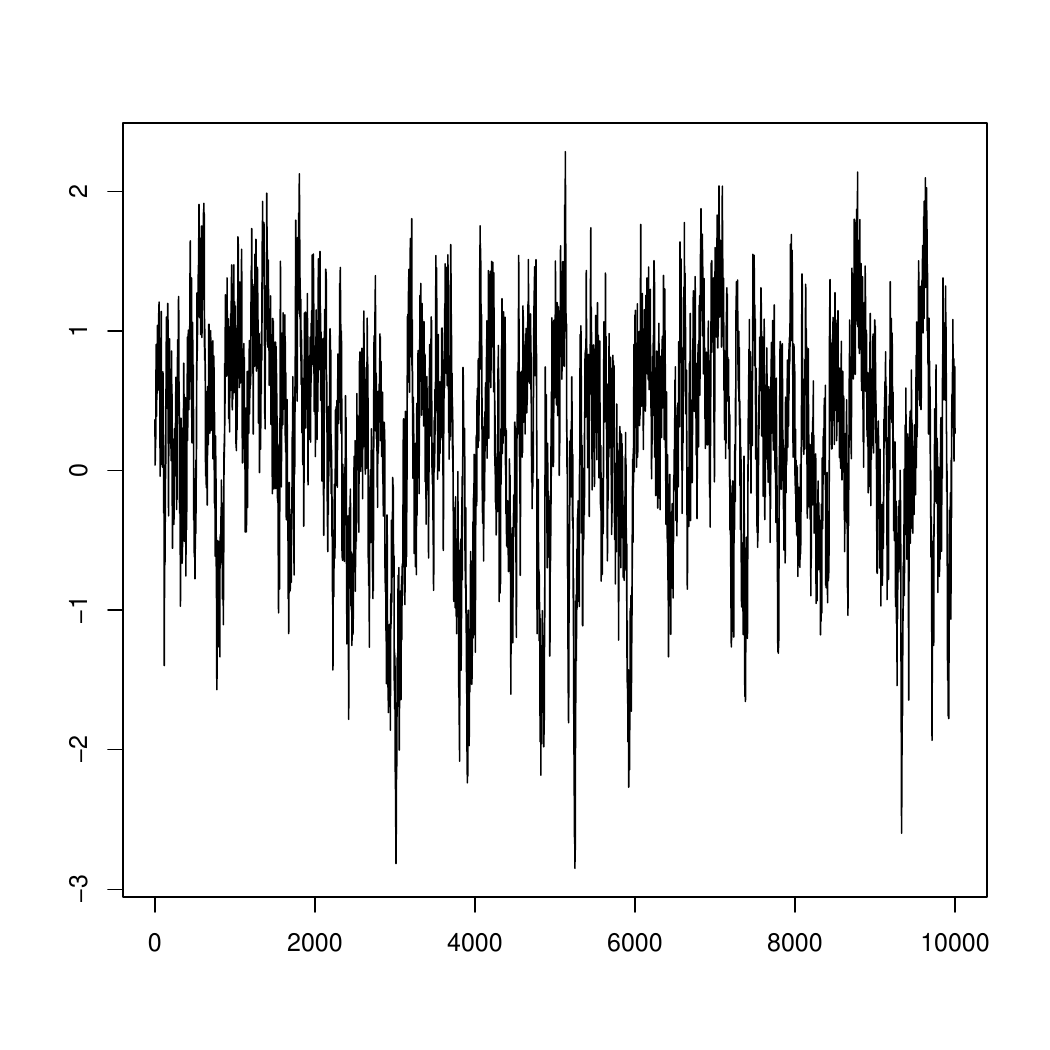}}
\caption{{\bf TTMCMC for 2-component gamma mixture:} Trace plots of $k$, $\nu^*_1$, $\mu^*_1$ and $\omega_1$.} 
\label{fig:gamma2comp_trace_plots}
\end{figure}

In this case, for TTMCMC implementation we obtained $a_{\nu^*}=0.05$; $a_{\mu^*}=0.005$, 
and $a_{\omega}=0.05$ using our convergence
diagnostic procedure. We set a considerably large burn-in time of 30,00,000 iterations as convergence seemed
to be somewhat slow compared to the one-component example. We stored
one in $150$ iterations of a further run of $1,500,000$ iterations, so that, as before we stored 
a total of $10,000$ realizations from the posterior distribution.
This took 31 minutes 6 seconds and yielded an overall acceptance rate $0.229365$. 
Also, the birth, death and no-change rates are $0.000017$, $0.000022$ and $0.688025$, respectively.
In this case, we obtained $\eta_1=0.28333$ and $\eta_2=0.30828$, which are reasonably small, providing
reasonably strong evidence in support of convergence of our TTMCMC chain. This is further supported strongly by
the visual information carried by the trace plots of $k$, $\nu^*_1$, $\mu^*_1$ and $\omega_1$, 
shown in Figure \ref{fig:gamma2comp_trace_plots}.

Interestingly, after burn-in, TTMCMC gives full mass to $2$ components, thus completely
supporting the truth. However, as demonstrated in Section S-9 of the supplement with simulation studies
for different data sizes (see Section S-9.2 for simulations with this 2-component mixture), 
it is possible that the actual posterior distribution of 
$k$ gives ``almost" point mass
to $k=2$, such that with probability close to zero some other components may also occur, but might have been missed
by us in this case due to the finite run length of our algorithm.

\begin{figure}
\includegraphics[width=7in,height=6.5in]{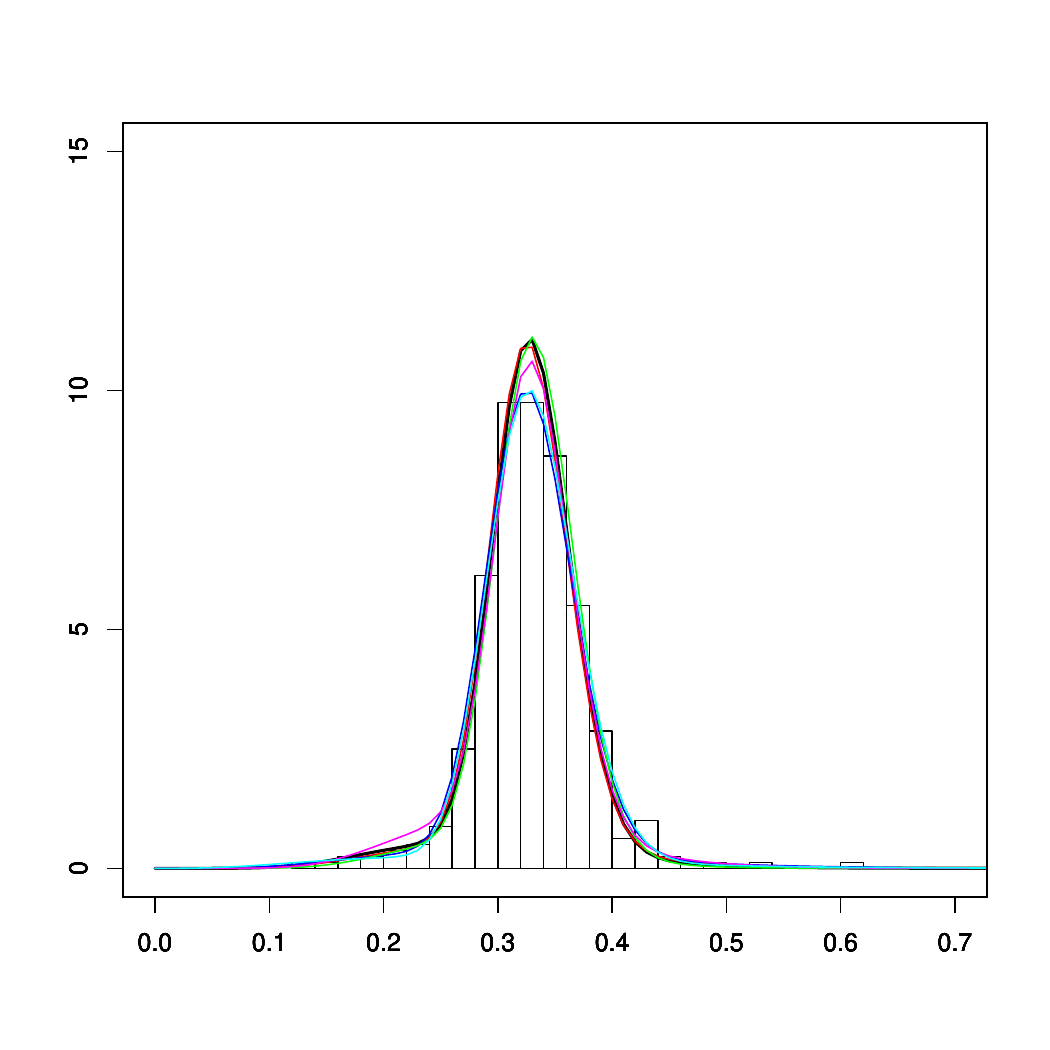}
\caption{{\bf TTMCMC for 2-component gamma mixture:} Goodness of fit of the posterior distribution of 
densities (coloured curves) to the simulated data (histogram). 
The thick black curve is the modal density and the other coloured curves are some densities contained in the 95\% HPD.}
\label{fig:gamma2comp_hpd}
\end{figure}

As before, Figure \ref{fig:gamma2comp_hpd} shows excellent
fit of the posterior distribution of the densities to the simulated data.

\subsubsection{{\bf Comparison with the results obtained by Wiper at al. (2001)}} 
\label{subsubsec:2comp_wiper}
As to be anticipated, bias towards
large values of $k$ continued in this example. Indeed, although \ctn{Wiper01} 
obtained $k=2$ as the mode of their RJMCMC based posterior of $k$, they also found that their RJMCMC algorithm
yielded the posterior probability about $0.01$ for $k=1$, and supported other larger values of $k$.
Thus, compared to TTMCMC, which identifies the truth very precisely, RJMCMC manages to facilitate 
only weak inference because of its lack of convergence.

\subsection{{\bf Third simulation study with data generated from a three-component gamma mixture}}
\label{subsec:3comp}

Here we generate $400$ realizations 
from the three-component mixture $0.2\times\mathcal G(40,20)+0.6\times\mathcal G(6,1)+0.2\times \mathcal G(200,20)$,
following \ctn{Wiper01}. 

\begin{figure}
\centering
\subfigure[Trace plot of $k$.]{ \label{fig:gamma3comp_k}
\includegraphics[width=7cm,height=6cm]{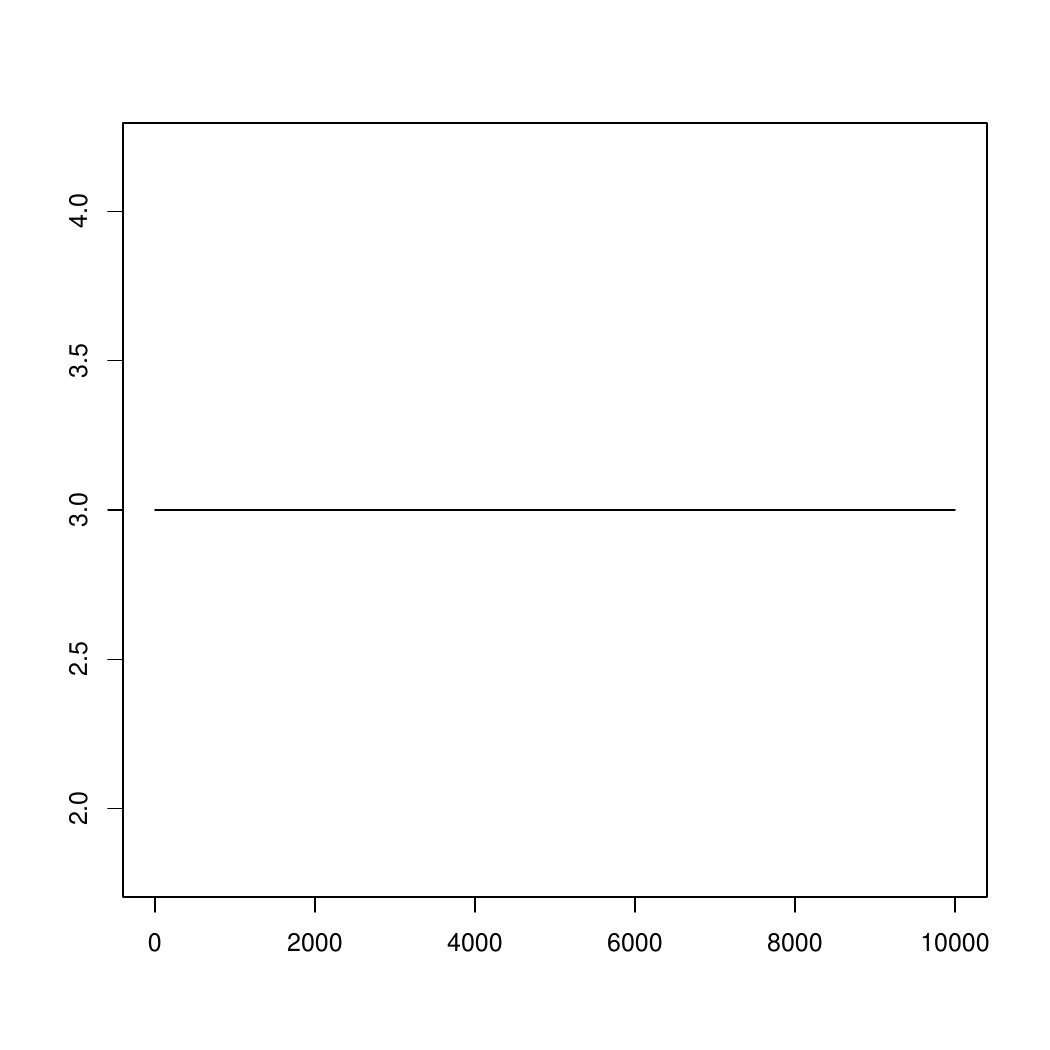}}
\hspace{2mm}
\subfigure[Trace plot of $\nu^*_1$.]{ \label{fig:gamma3comp_nu}
\includegraphics[width=7cm,height=6cm]{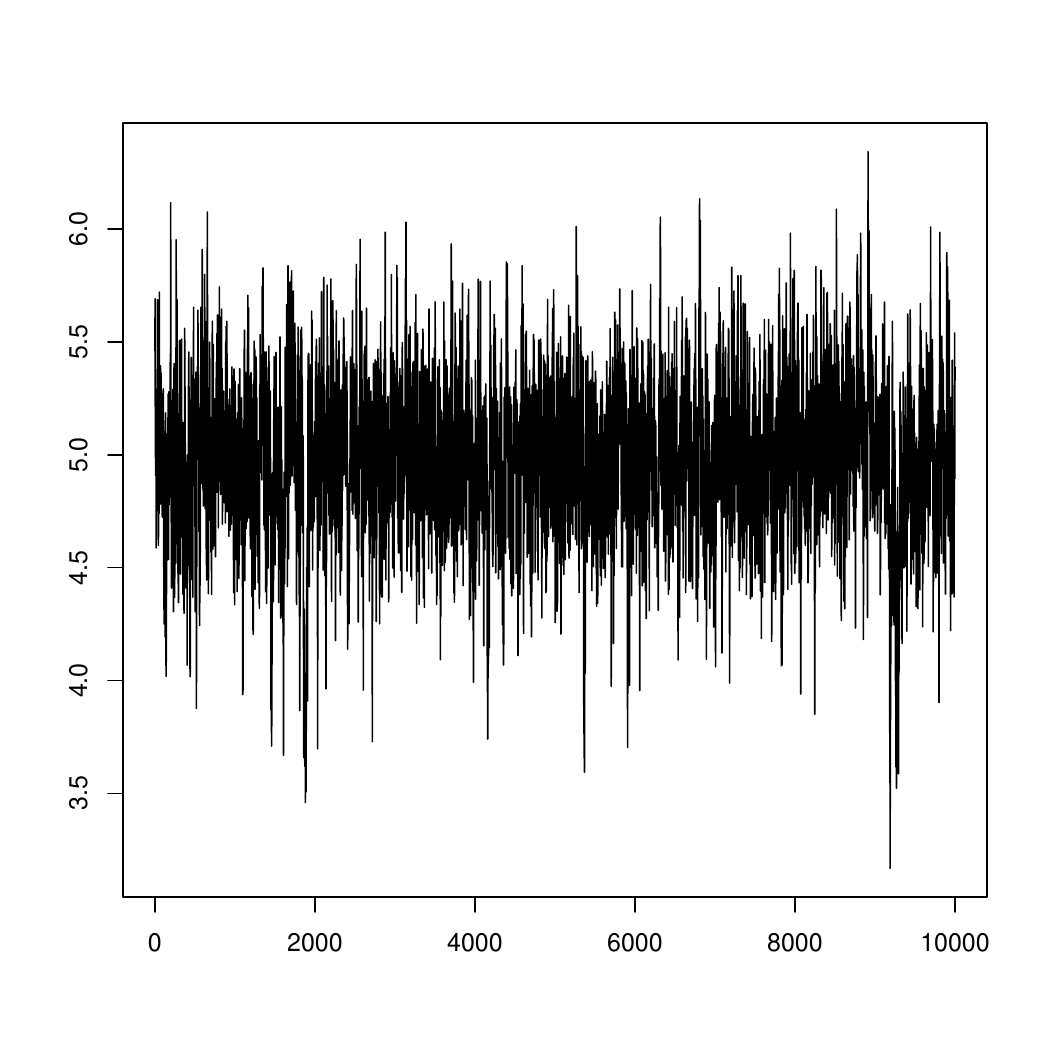}}\\
\vspace{2mm}
\subfigure[Trace plot of $\mu^*_1$.]{ \label{fig:gamma3comp_mu}
\includegraphics[width=7cm,height=6cm]{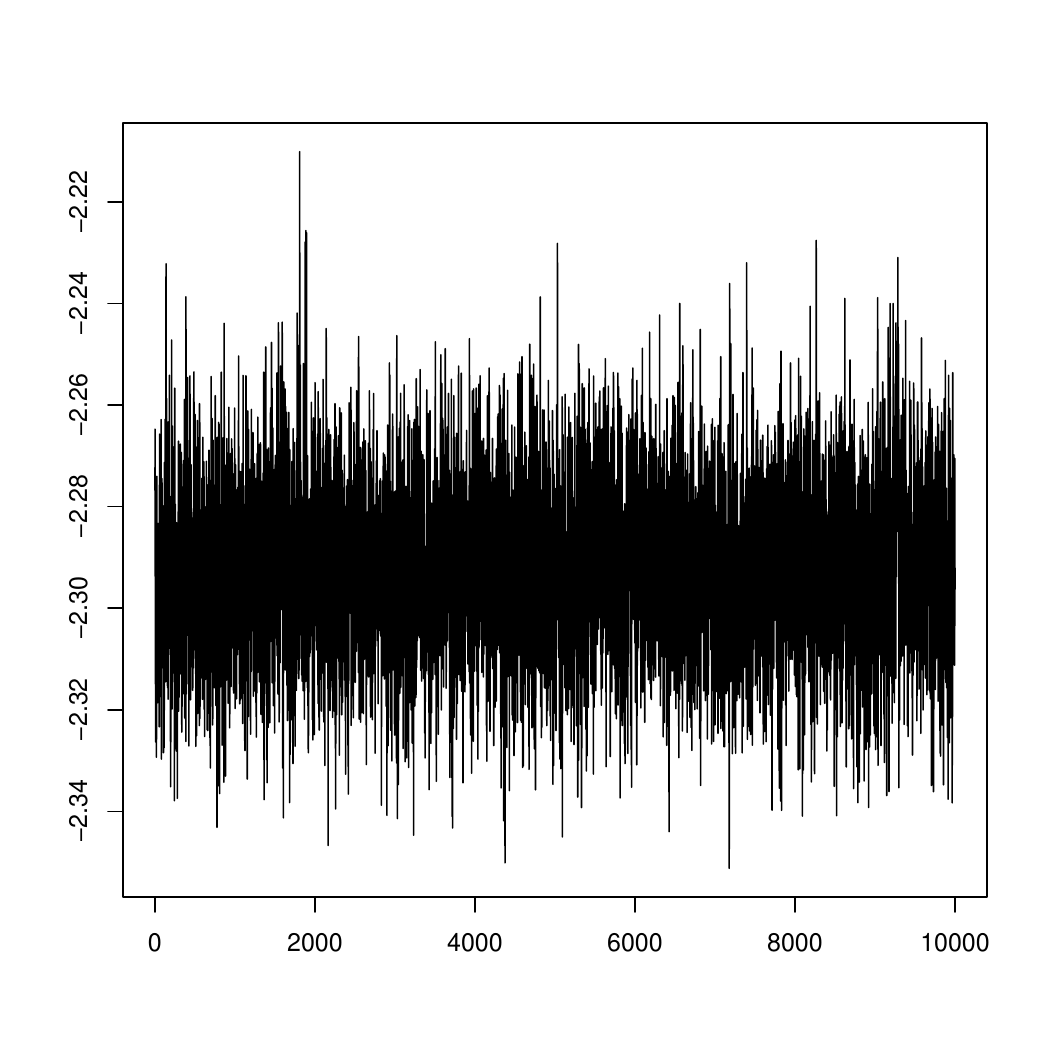}}
\hspace{2mm}
\subfigure[Trace plot of $\omega_1$.]{ \label{fig:gamma3comp_w}
\includegraphics[width=7cm,height=6cm]{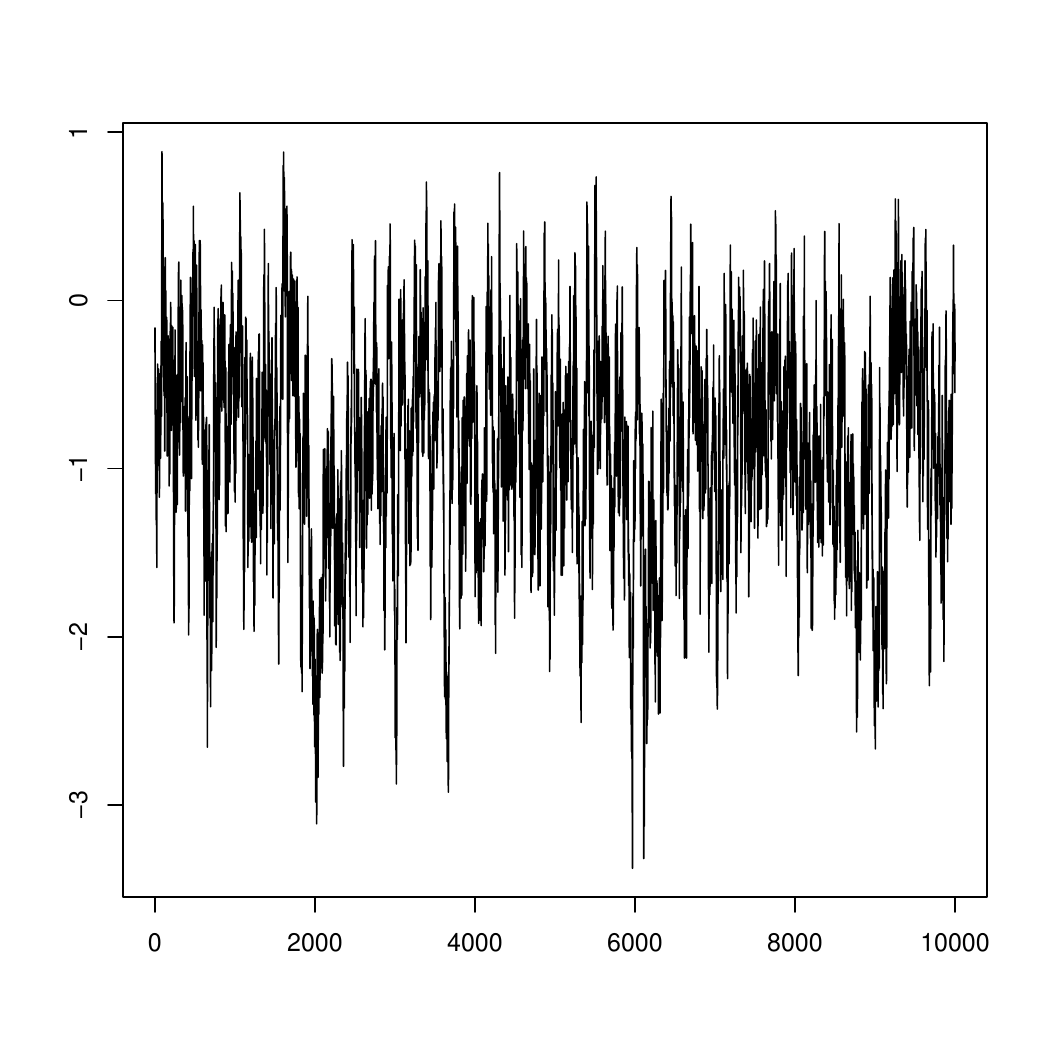}}
\caption{{\bf TTMCMC for 3-component gamma mixture:} Trace plots of $k$, $\nu^*_1$, $\mu^*_1$ and $\omega_1$.} 
\label{fig:gamma3comp_trace_plots}
\end{figure}

Again we obtained $a_{\nu^*}=0.05$; $a_{\mu^*}=0.005$, and $a_{\omega}=0.05$ using our convergence
diagnostic procedure. Here a burn-in of 15,00,000 iterations turned out to be more than sufficient. As before we stored
$10,000$ realizations from the posterior distribution out of a further $1,500,000$ iterations after the burn-in
with a thinning of size $150$. The overall acceptance rate was $0.240443$ and the time taken was 36 minutes and 5 seconds.
The birth, death and no-change rates are $0.00001$, $0.000017$ and $0.720547$, respectively.
As regards the convergence diagnostic, $\eta_1=0.01602$ and $\eta_2= 0.01757$, which are both small enough to let us
conclude that the TTMCMC chain has converged very well. The trace plots displayed in Figure \ref{fig:gamma3comp_trace_plots}
completely support our conclusion regarding convergence.

Again, the posterior distribution of $k$ completely supports the truth, giving full mass to $3$, which, in this example, is 
the correct number of components. The simulation study in Section S-9.3 of the supplement demonstrates that 
it is possible that here TTMCMC has missed $k=4$, which might have occurred with extremely small probability.

\begin{figure}
\includegraphics[width=7in,height=6.5in]{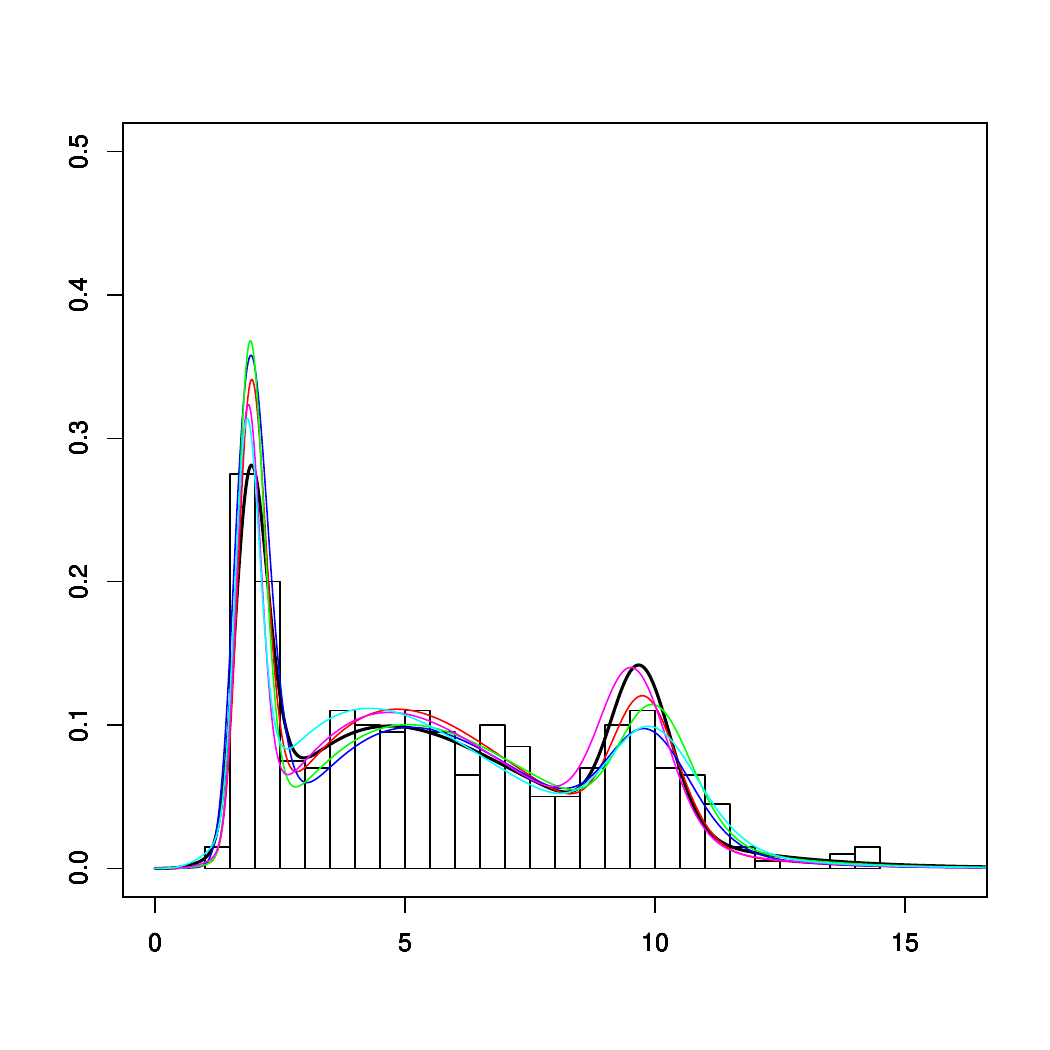}
\caption{{\bf TTMCMC for 3-component gamma mixture:} Goodness of fit of the posterior distribution of 
densities (coloured curves) to the simulated data (histogram). 
The thick black curve is the modal density and the other coloured curves are some densities contained in the 95\% HPD.}
\label{fig:gamma3comp_hpd}
\end{figure}

As to be expected, Figure \ref{fig:gamma3comp_hpd} confirms excellent
fit of the posterior distribution of the densities to the simulated data.

\subsubsection{{\bf Comparison with the results obtained by Wiper at al. (2001)}} 
\label{subsubsec:3comp_wiper}
Specific RJMCMC based results pertaining to the three component mixture are not provided in \ctn{Wiper01},
but larger values of $k$ compared to the truth, are certain to occur with significant probabilities.

\subsection{{\bf Fourth simulation study with data generated from a four-component gamma mixture}}
\label{subsec:4comp}

For the final simulation study with gamma mixtures, following \ctn{Wiper01} we generate $400$ realizations 
from the four-component mixture 
$0.25\times\mathcal G(200,100)+0.25\times\mathcal G(400,100)+0.25\times \mathcal G(600,100)
+0.25\times \mathcal G(800,100)$.

\begin{figure}
\centering
\subfigure[Trace plot of $k$.]{ \label{fig:gamma4comp_k}
\includegraphics[width=7cm,height=6cm]{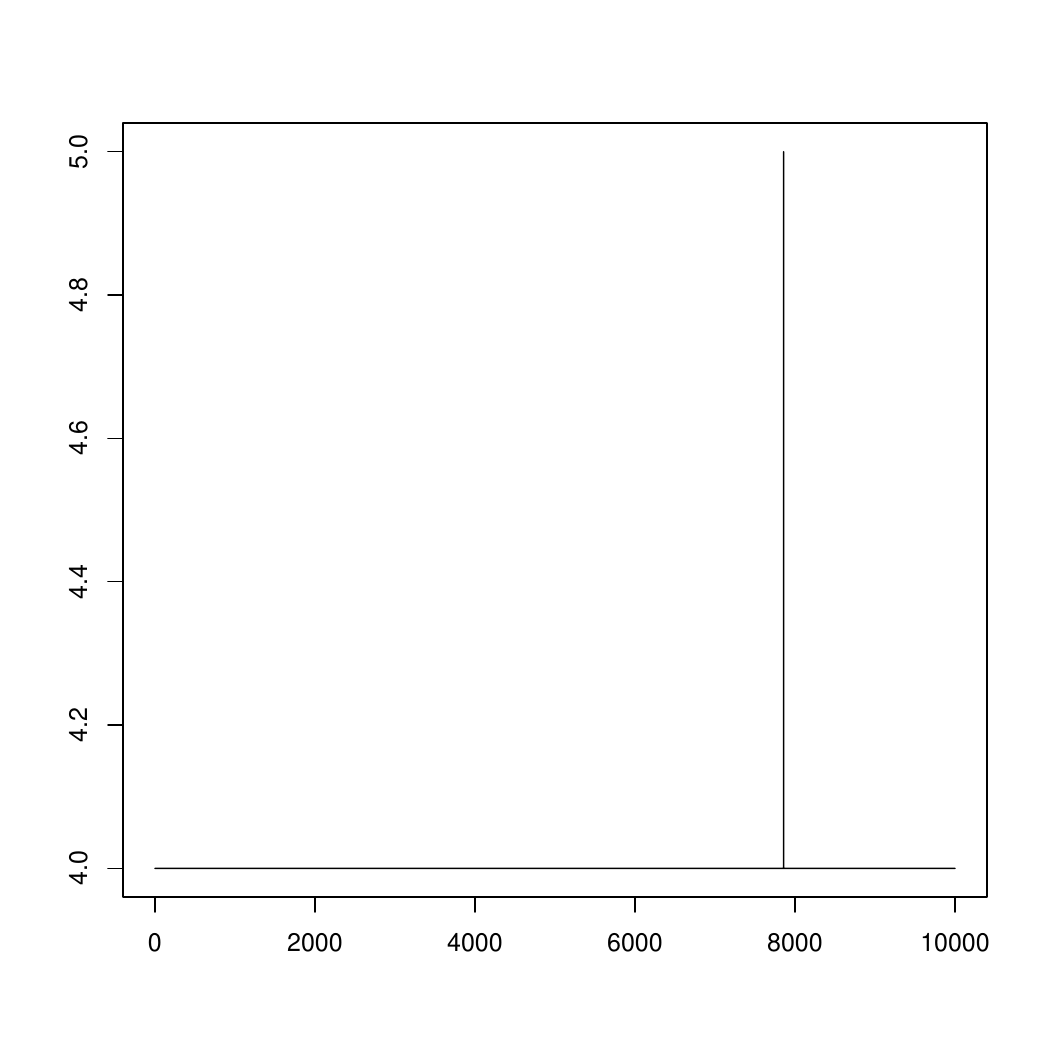}}
\hspace{2mm}
\subfigure[Trace plot of $\nu^*_1$.]{ \label{fig:gamma4comp_nu}
\includegraphics[width=7cm,height=6cm]{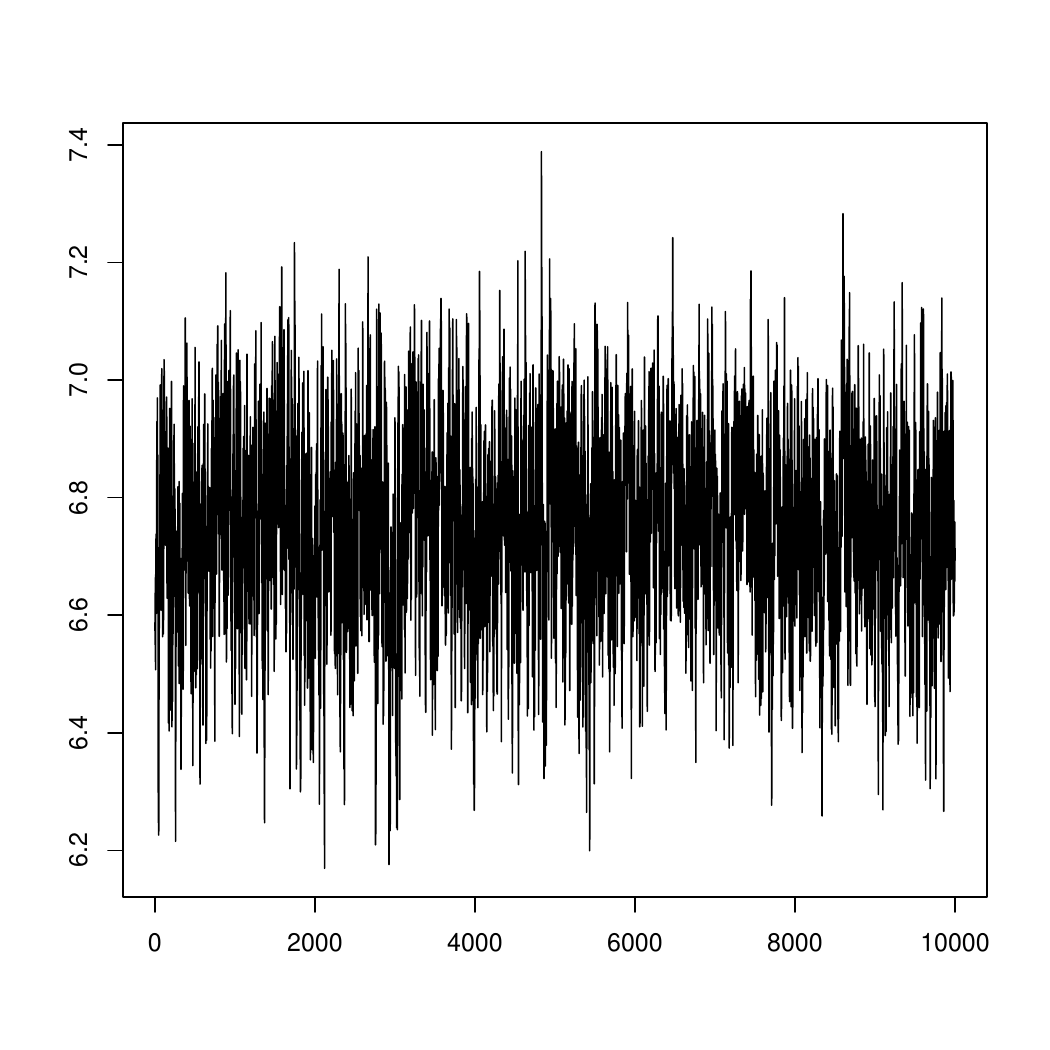}}\\
\vspace{2mm}
\subfigure[Trace plot of $\mu^*_1$.]{ \label{fig:gamma4comp_mu}
\includegraphics[width=7cm,height=6cm]{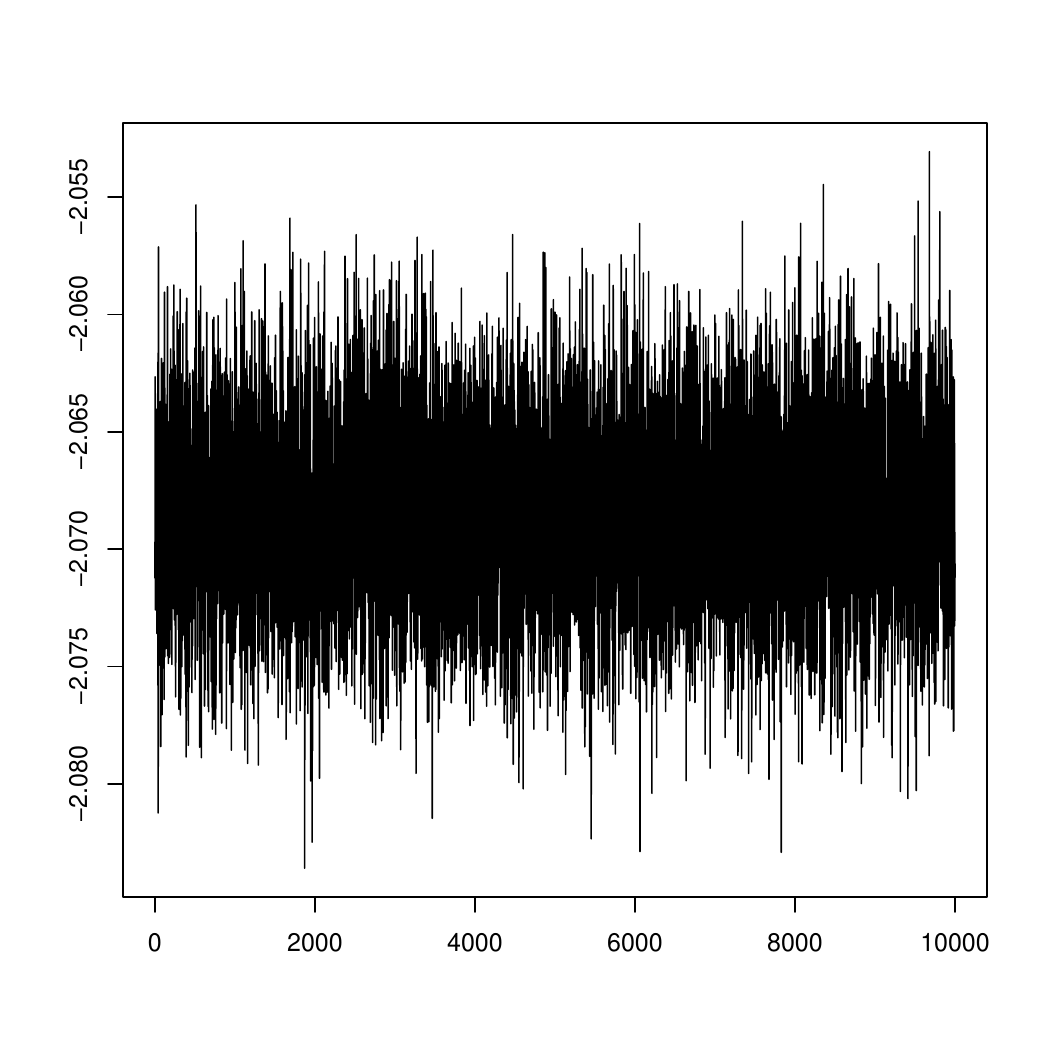}}
\hspace{2mm}
\subfigure[Trace plot of $\omega_1$.]{ \label{fig:gamma4comp_w}
\includegraphics[width=7cm,height=6cm]{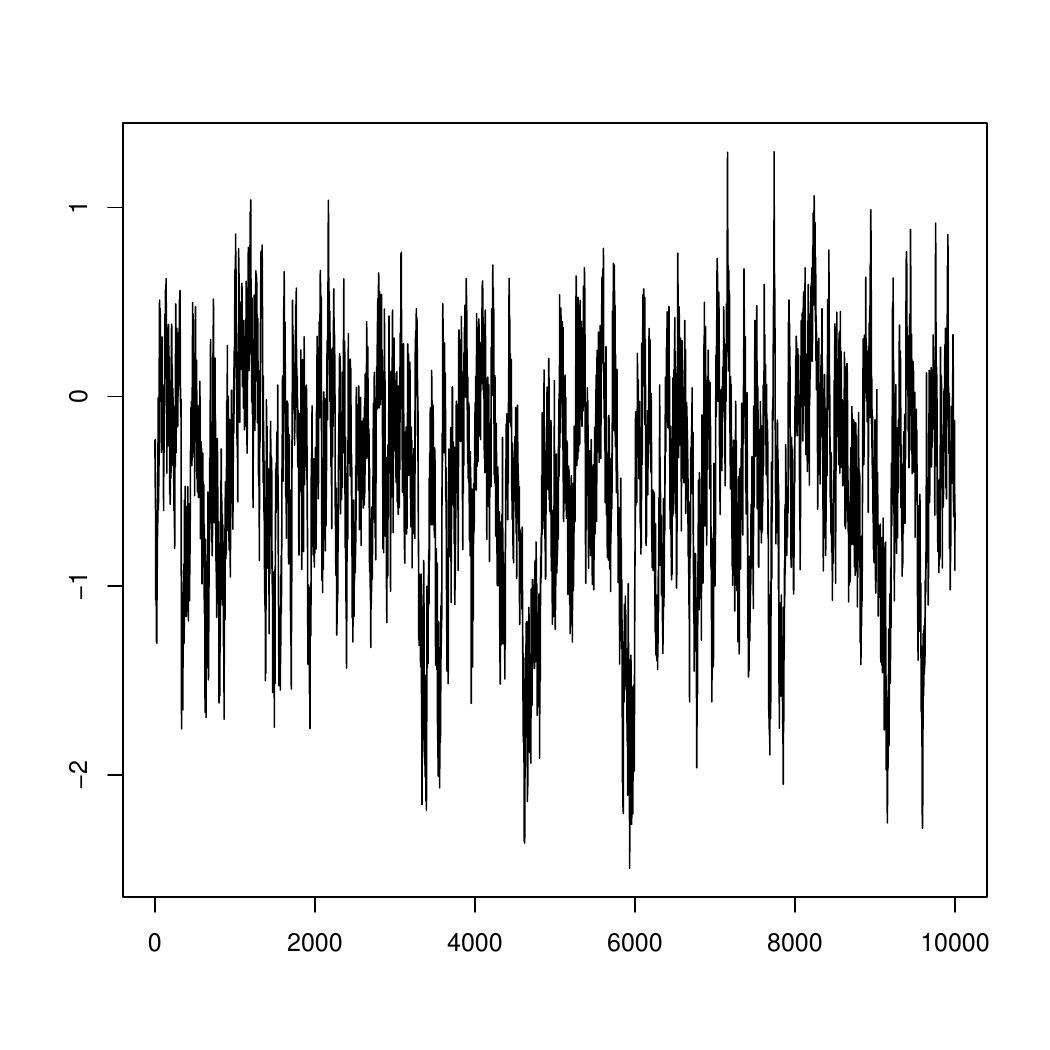}}
\caption{{\bf TTMCMC for 4-component gamma mixture:} Trace plots of $k$, $\nu^*_1$, $\mu^*_1$ and $\omega_1$.} 
\label{fig:gamma4comp_trace_plots}
\end{figure}

Here we obtained $a_{\nu^*}=0.05$; $a_{\mu^*}=0.005$, and $a_{\omega}=0.12$, with a burn-in of 15,00,000 iterations 
and with respect to $10,000$ realizations from the posterior distribution stored as before after burn-in 
with a thinning of size $150$. The time to implement TTMCMC was 40 minutes and 35 seconds and we obtained an 
overall acceptance rate $0.117432$. The birth, death and no-change rates are $0.000417$, $0.000401$ and
$0.345363$, respectively.
That the chain converged reasonably well can be inferred since
$\eta_1=0.08154$ and $\eta_2=0.11839$ are both reasonably small.
As before, the trace plots displayed in Figure \ref{fig:gamma4comp_trace_plots}
confirm our conclusion regarding convergence.

Here the posterior distribution of $k$ gives almost full mass to the truth $k=4$, and seems to be consistent
with the further simulation study conducted in Section S-9.4 of the supplement, considering a data of size $1000$.

\begin{figure}
\includegraphics[width=7in,height=6.5in]{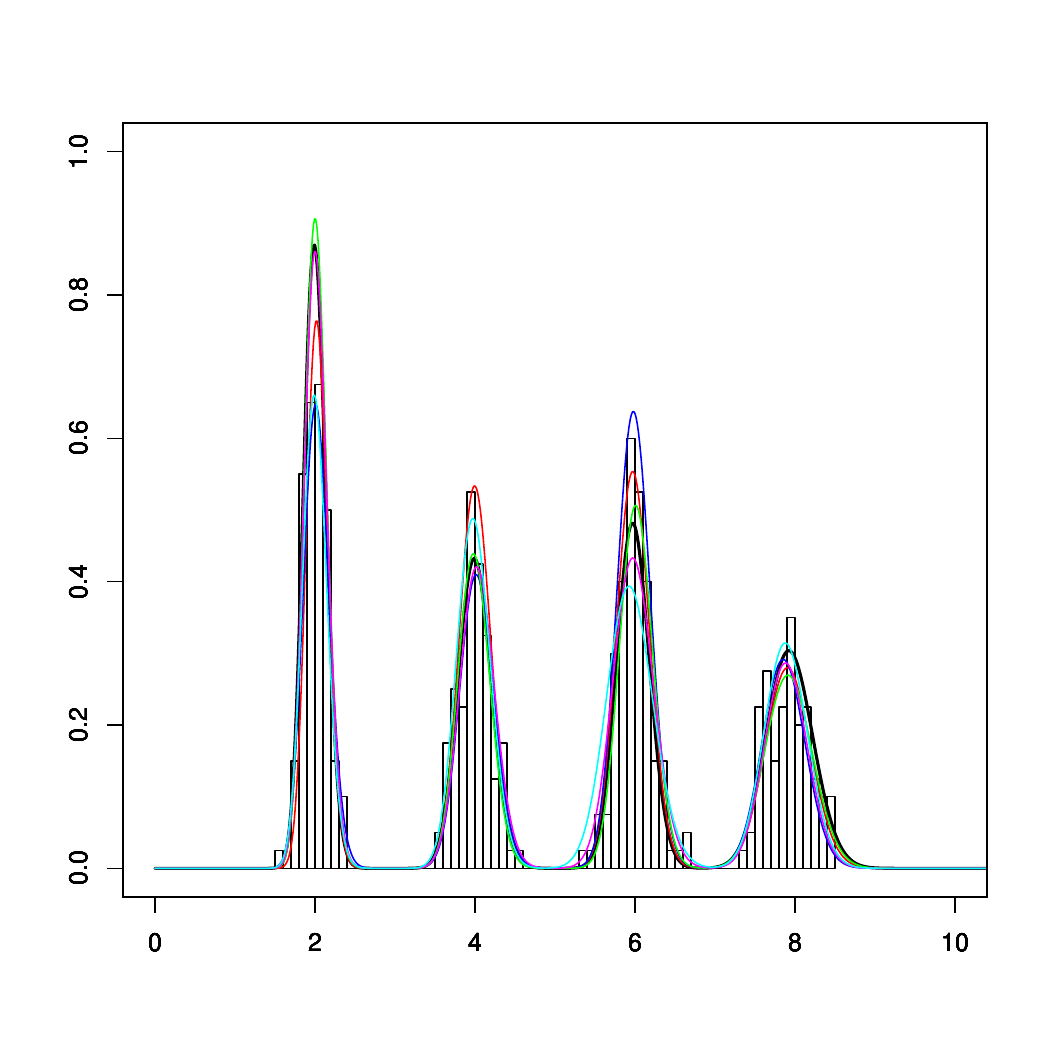}
\caption{{\bf TTMCMC for 4-component gamma mixture:} Goodness of fit of the posterior distribution of 
densities (coloured curves) to the simulated data (histogram). 
The thick black curve is the modal density and the other coloured curves are some densities contained in the 95\% HPD.}
\label{fig:gamma4comp_hpd}
\end{figure}

As before, Figure \ref{fig:gamma4comp_hpd} shows that excellent
fit of the posterior distribution of the densities to the simulated data has been achieved.

\subsubsection{{\bf Comparison with the results obtained by Wiper at al. (2001)}} 
\label{subsubsec:4comp_wiper}
Even for this 4-component example specific RJMCMC based results are not provided in \ctn{Wiper01},
but as in the other RJMCMC based examples, larger values of $k$ compared to the truth, 
are certain to occur with significant probabilities.

\section{{\bf Comparison of TTMCMC and RJMCMC in the normal mixture set up with unknown number of components}}
\label{sec:normal_mixtures}

We now illustrate TTMCMC on normal mixture models with unknown number of components
with application to the well-studied enzyme, acidity and the galaxy data sets. 
\ctn{Richardson97} modeled these data sets using parametric normal mixtures and 
applied RJMCMC for Bayesian inference. On the other hand, \ctn{Bhattacharya08} (see also \ctn{Escobar95}) 
proposed a semi parametric normal mixture model based on Dirichlet process
and used Gibbs sampler for Bayesian inference. 

\subsection{{\bf Normal mixture}}
\label{subsec:normix}

Let the data points $y_1,\ldots,y_n$ be independently and identically distributed ($iid$) 
as the normal mixture of the following form: for $i=1,\ldots,n$
\begin{equation}
f(y_i\vert\bnu_k,\btau_k,\bpi_k,k)=\sum_{j=1}^k\pi_j\sqrt{\frac{\tau_j}{2\pi}}\exp\left\{-\frac{\tau_j}{2}(y_i-\nu_j)^2\right\},
\label{eq:normix}
\end{equation}
where $\bnu_k=(\nu_1,\ldots,\nu_k)$, $\btau_k=(\tau_1,\ldots,\tau_k)$, and $\bpi_k=(\pi_1,\ldots,\pi_k)$.
Given $k>0$, for each $j$, $-\infty<\nu_j<\infty$, $\tau_j>0$, $0<\pi_j<1$ such that
$\sum_{j=1}^k\pi_j=1$. As before, we assume that $k$ is unknown.

\subsection{{\bf Prior structure}}
\label{subsec:normix_prior}

Note that the semi parametric mixture model of \ctn{Bhattacharya08} can be viewed as a parametric
model when the scale parameter associated with the base distribution of the Dirichlet process
prior tends to infinity. Hence, from that perspective, the base distributions of $\nu_j$ and $\tau_j$
may be regarded as the respective priors for our current parametric mixture context. Thus, 
motivated by \ctn{Bhattacharya08}, we consider
the following prior for $\bnu$ and $\btau$:
\begin{align}
[\tau_j] &\sim\mathcal G\left(\frac{s}{2},\frac{S}{2}\right);\label{eq:prior_tau}\\
[\nu_j\vert\tau_j] &\sim N\left(\nu_0,\frac{\psi}{\tau_j}\right).\label{eq:prior_nu}
\end{align}
In the above, 
$N\left(\mu,\sigma^2\right)$ denotes the normal distribution with mean $\mu$ and variance $\sigma^2$.
Specifications of the values of the hyperparameters $s, S, \nu_0, \psi$ are discussed in the context of the applications. 

Analogous to the gamma mixture context here we reparameterize $\tau_j$
as $\exp(\tau^*_j)$, where $\tau^*_j\sim\mathcal \log\left(\mathcal G(s/2,S/2)\right)$. 
We denote $(\tau^*_1,\ldots,\tau^*_k)$ by $\btau^*_k$.

For $\bpi$ we propose the same reparameterization (\ref{eq:prior_pi}).
In this case, we consider two kinds of priors on $\bomega$. One is 
$\omega_j\sim N\left(\mu_{\omega},\sigma^2_{\omega}\right)$, and the other is
$\omega_j\sim\log\left(\mathcal G(\alpha_j,1)\right)$ independently, for $j=1,\ldots,k$, where
$\alpha_j>0;~j=1,\ldots,k$.
Note that, for the normal prior on $\omega_j$, the induced prior on $\bpi$ is not the traditional
Dirichlet distribution, while the second prior implies that $\bpi\sim\mathcal D(\alpha_1,\ldots,\alpha_k)$. 

As regards the prior on $k$, we consider the uniform distribution on $\{1,2,\ldots,30\}$, the 
truncated Poisson distribution on $\{1,2,\ldots,30\}$ and the discretized normal
with mean $\mu_k$ and variance $\sigma^2_k$ on $\{1,2,\ldots,30\}$ (that is, the normal
density with mean $\mu_k$ and variance $\sigma^2_k$ evaluated and re-normalized on $\{1,2,\ldots,30\}$
to render it a discrete probability mass function).


We fit normal mixture models to each of the three data sets -- enzyme, acidity, and galaxy, 
using the general TTMCMC strategy provided in Section \ref{subsec:general_ttmcmc}.
The details are provided in the context-specific applications.

We compare the performance of additive TTMCMC with random walk RJMCMC, which is analogous to additive TTMCMC
but with independent jump-sizes for every co-ordinate and with the proposal density associated with the birth move
incorporated within the acceptance ratio, unlike TTMCMC; see Section \ref{subsec:random_walk_rjmcmc}.

Our main aim is to demonstrate that the 
simplest version of TTMCMC, namely, TTMCMC with
the additive transformation, is efficient enough for adequately exploring the complicated mixture-based posteriors
in all the three applications, while the corresponding RJMCMC version, composed of random walk based moves, 
fails miserably.

Specific details of inference and implementation 
of our methodologies follow. 

\subsection{{\bf Enzyme data}}
\label{subsec:enzyme}
Following \ctn{Bhattacharya08} we set $s = 4.0$; $S = 2 \times (0.2/1.22) = 0.3278689$; 
$\nu_0 = 1.45$; $\psi= 33.3$. Rather than assuming $\omega_j\sim\log\left(\mathcal G(\alpha_j,1)\right)$
which induce the traditional Dirichlet distribution for $\bpi$, here
we assume that $\omega_j\sim N\left(\mu_{\omega_j},\sigma^2_{\omega_j}\right)$, with
$\mu_{\omega}=0$ and $\sigma^2_{\omega}=0.25$. We chose somewhat small variance
to reflect our belief that $\omega_j$'s are relatively close to constant, so that {\it a priori} the
mixing probabilities $\bpi$ are approximately the same. We specify the uniform distribution
on $\{1,\ldots,30\}$ as the prior on $k$.


As in the gamma mixture set-up we experimented by setting, for every $j=1,\ldots,k$, the scale values
$a_{\nu^*_j}=a_{\nu^*}$; $a_{\tau^*_j}=a_{\tau^*}$, and $a_{\omega_j}=a_{\omega}$, with 
$a_{\nu^*},a_{\tau^*},a_{\omega}$ being one of the trial values $0.05$, $0.1$, $0.12$, $0.15$, $0.20$, $0.25$, $0.50$.
We considered a burn-in of $375,000$ iterations and a further $15,00,000$ iterations, storing as before one in $150$ iterations
to obtain $10,000$ realizations from the posterior.
Here $\eta_1$ and $\eta_2$ turned out to be 
$\eta_1=0.07291$ and $\eta_2=0.039230$, which corresponded to
$a_{\nu}=a_{\tau}=a_{\omega}=0.05$. The results we report are with respect to these trial values. 
Since both $\eta_1$ and $\eta_2$ are small, we conclude that convergence has taken place appropriately.
The overall acceptance rate, evaluated empirically, turned out to be $0.05284032$, and the birth, death, no-change
rates are $0.000306$, $0.000304$ and $0.157810$, respectively.
Our TTMCMC implementation with the scales selected as above took 2 minutes and 56 seconds. 

\begin{figure}
\centering
\subfigure[Trace plot of $k$.]{ \label{fig:enzyme_k}
\includegraphics[width=7cm,height=6cm]{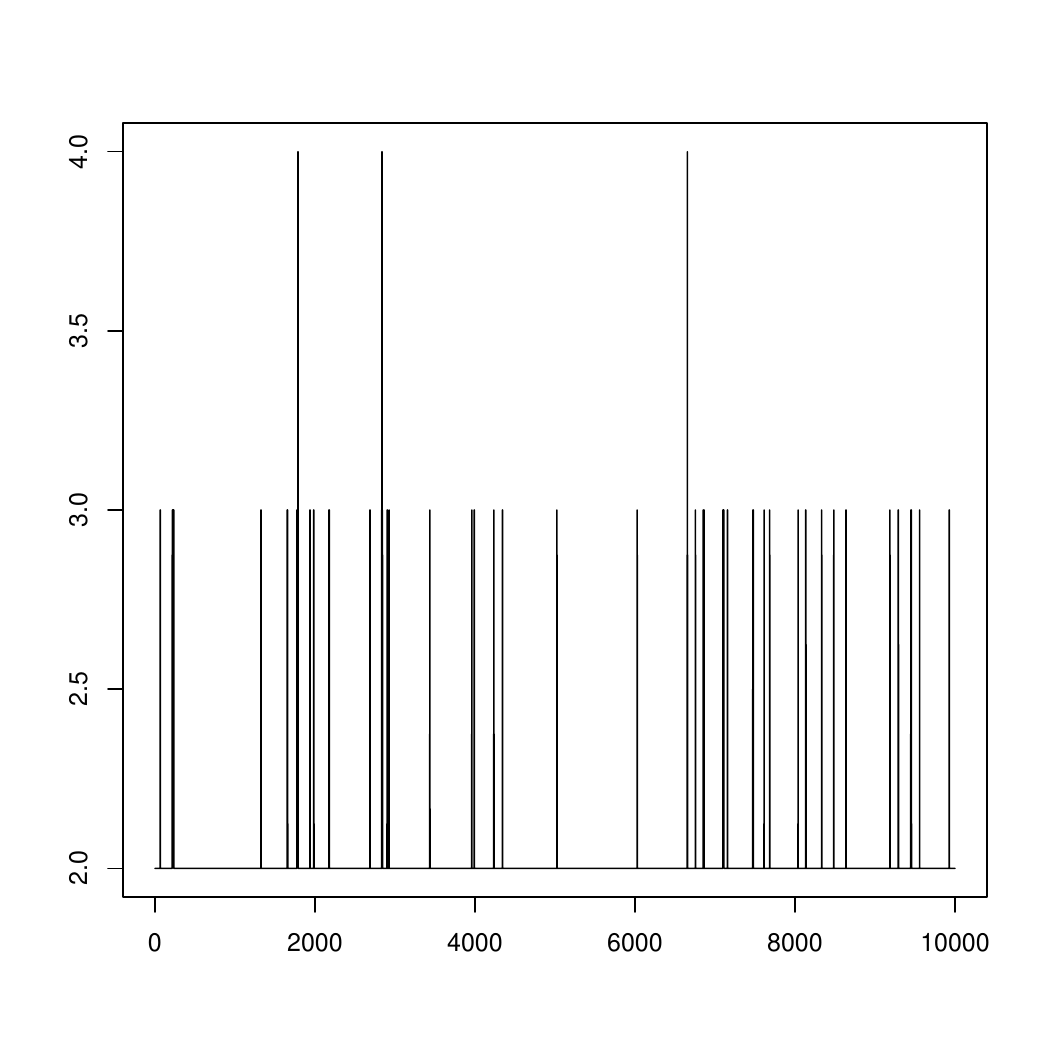}}
\hspace{2mm}
\subfigure[Trace plot of $\nu^*_1$.]{ \label{fig:enzyme_nu}
\includegraphics[width=7cm,height=6cm]{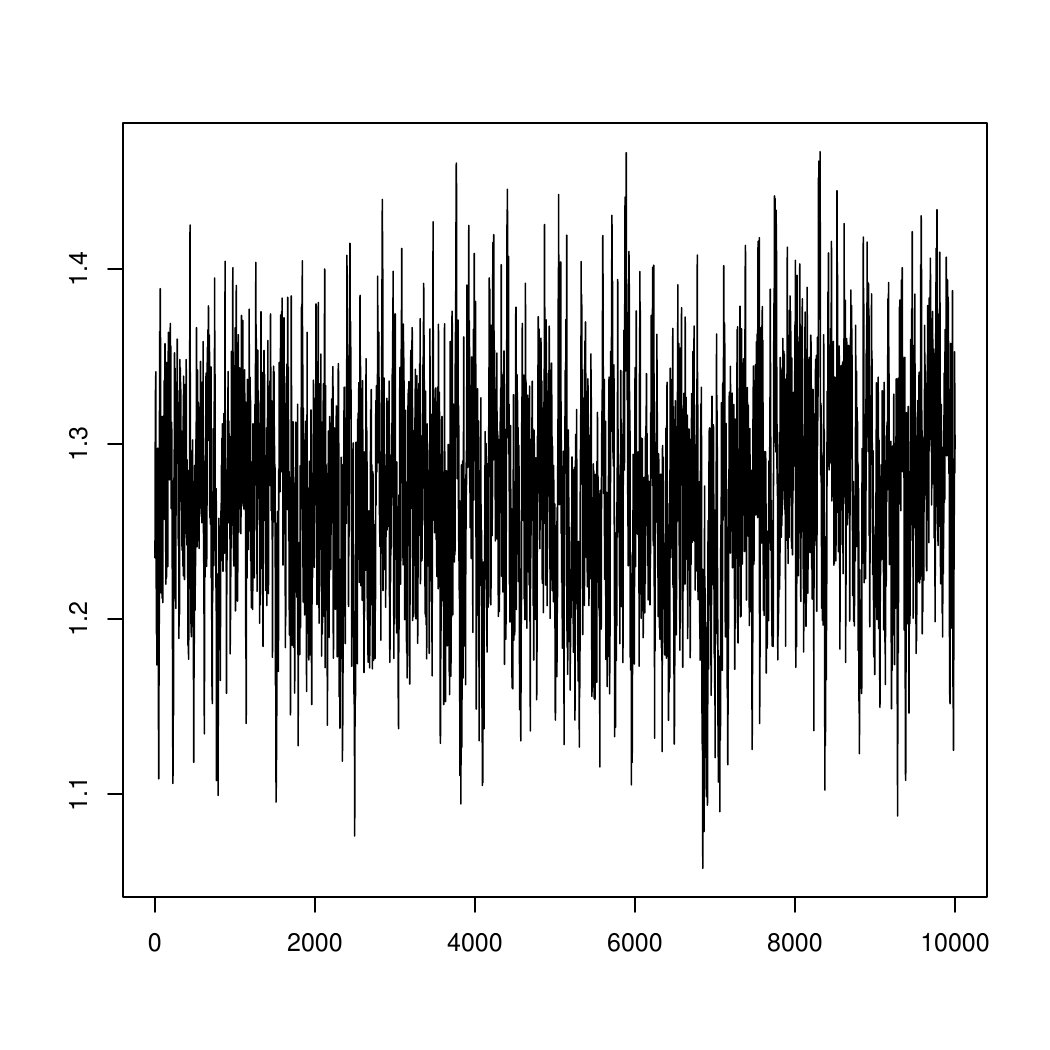}}\\
\vspace{2mm}
\subfigure[Trace plot of $\tau^*_1$.]{ \label{fig:enzyme_tau}
\includegraphics[width=7cm,height=6cm]{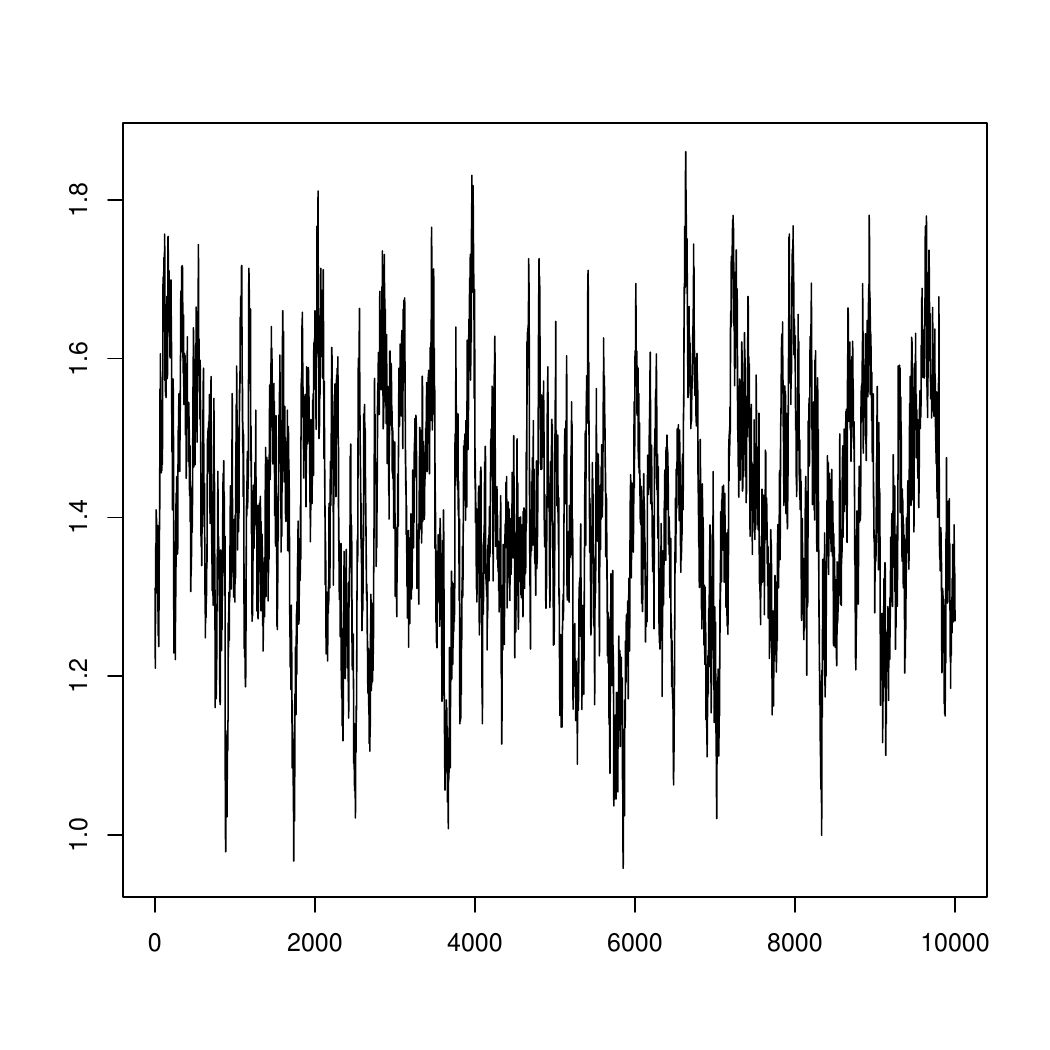}}
\hspace{2mm}
\subfigure[Trace plot of $\omega_1$.]{ \label{fig:enzyme_w}
\includegraphics[width=7cm,height=6cm]{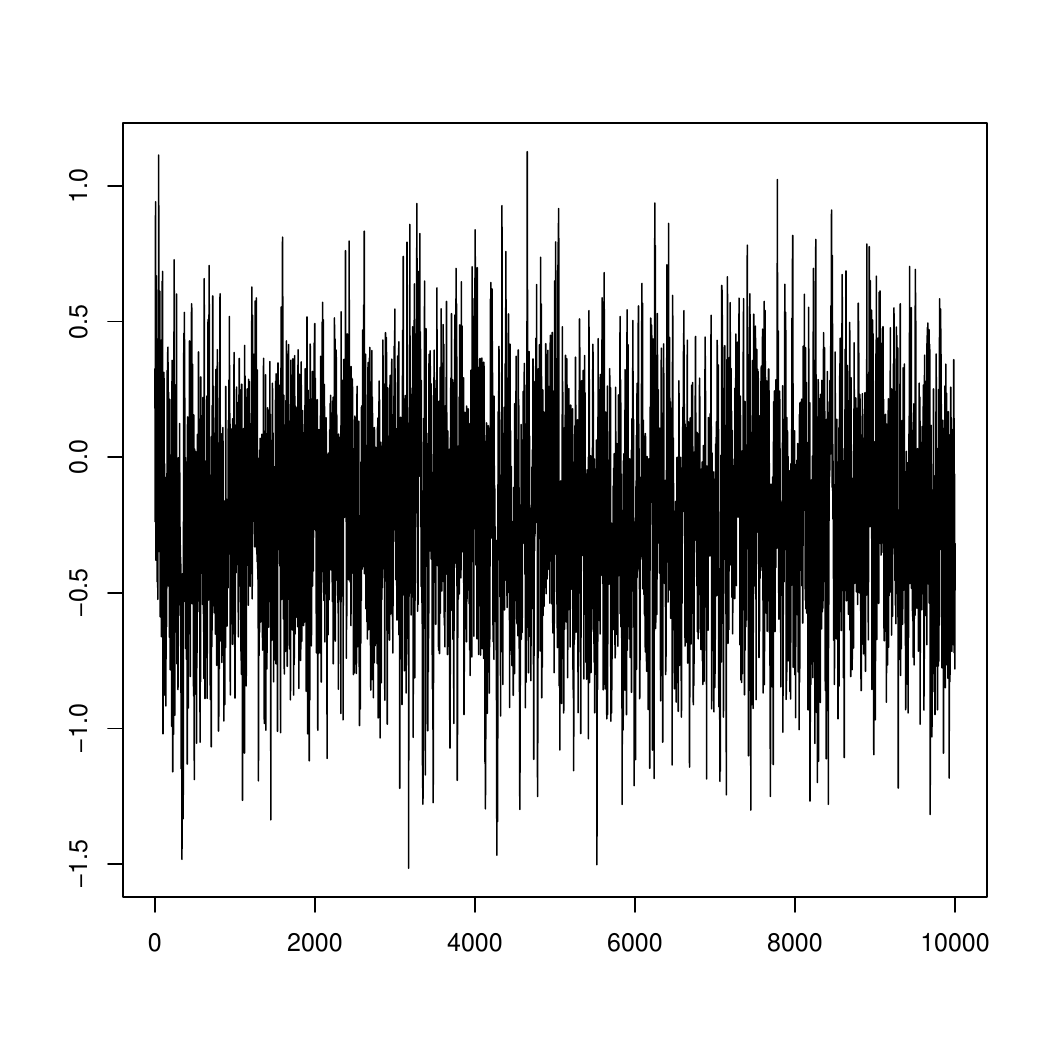}}
\caption{{\bf TTMCMC for the enzyme data:} Trace plots of $k$, $\nu^*_1$, $\tau^*_1$ and $\omega_1$.} 
\label{fig:enzyme_trace_plots}
\end{figure}

We also verified convergence of our TTMCMC chain with informal trace plots.
Figure \ref{fig:enzyme_trace_plots} displays the trace plots of $k$, $\nu^*_1$, $\tau^*_1$ and $\omega_1$.
As seen in panel (a) of Figure \ref{fig:enzyme_trace_plots} the posterior distribution of $k$ 
placed highest mass on 2 components (posterior probability 0.986), followed by 3 components
(posterior probability 0.0137), and then by 4 components (probability 0.0003).
In other words, our Bayesian analysis strongly supports bimodality.
Indeed, the information regarding bimodality is particularly
strong thanks to the small range on which the data are supported and the large size of the data
(the data set contains $245$ observations on an effective support $(0,3)$). 
Panels (b), (c) and (d) of Figure \ref{fig:enzyme_trace_plots} show adequate mixing properties
of the chain. Thus, the mixing information provided by these trace plots supports the conclusion
obtained by our proposed credible region based convergence assessment method.

\begin{figure}
\includegraphics[width=7in,height=6.5in]{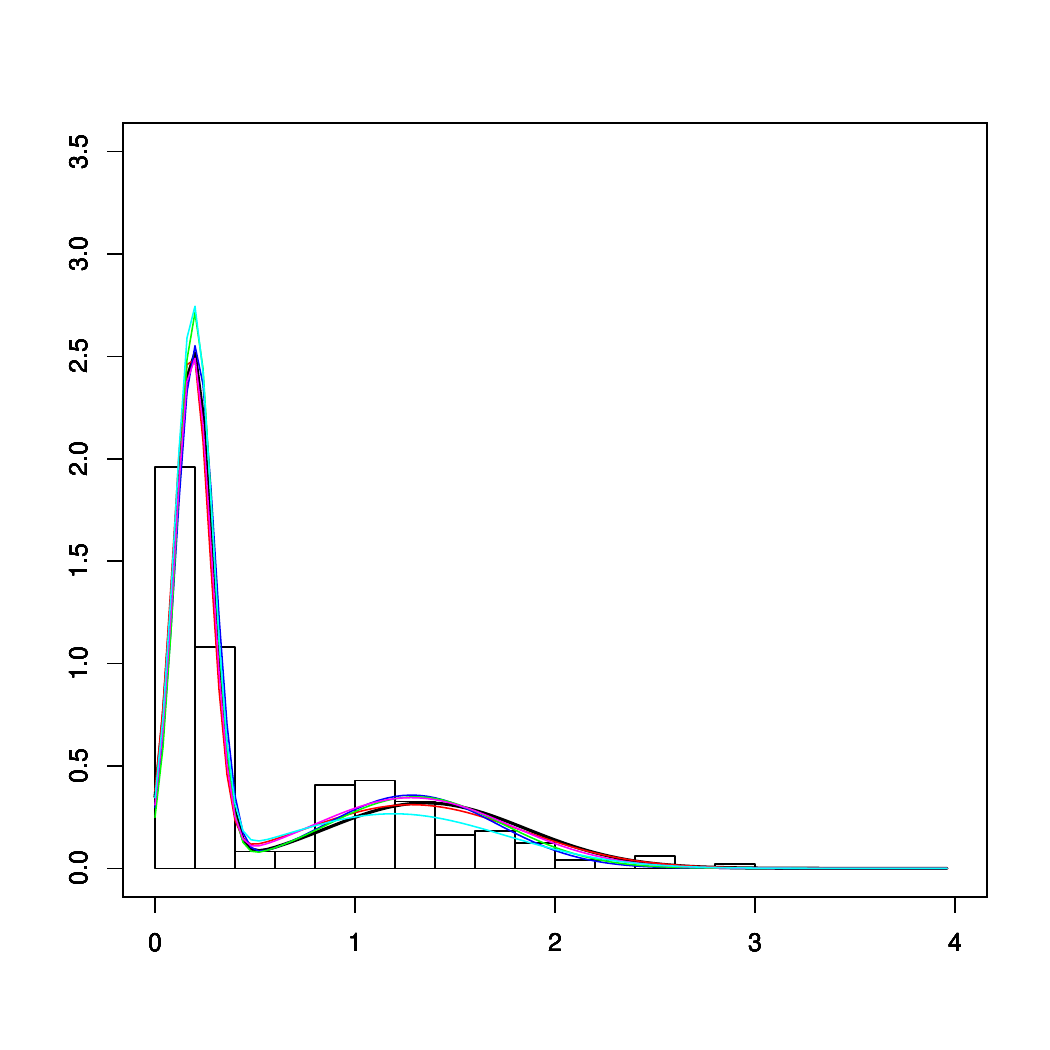}
\caption{{\bf TTMCMC for the enzyme data:} Goodness of fit of the posterior distribution of 
densities (coloured curves) to the observed 
data (histogram). 
The thick black
curve is the modal density and the other coloured curves are some densities contained in the 95\% HPD.}
\label{fig:enzyme_hpd}
\end{figure}

Figure \ref{fig:enzyme_hpd} shows excellent fit of the posterior distribution of the densities 
to the data.

\subsection{{\bf Acidity data}}
\label{subsec:acidity}

Again following \ctn{Bhattacharya08} we set $s = 4.0$; $S = 2 \times (0.2/0.573) = 0.6980803$; 
$\nu_0 = 5.02$; $\psi= 33.3$. 
Here also we assume that $\omega_j\sim N\left(\mu_{\omega_j},\sigma^2_{\omega_j}\right)$, with
$\mu_{\omega}=0$ and $\sigma^2_{\omega}=0.25$. 
As before we put the uniform prior distribution
on $\{1,\ldots,30\}$ on $k$.

Following the convergence diagnostic method detailed above for choosing appropriate scales
here we obtain $a_{\nu^*_j}=a_{\tau^*_j}=a_{\omega^*_j}=0.05$ for $j=1,\ldots,k$.
For these scales we obtained $\eta_1=0.0049$ and $\eta_2=0.0080$, which are very
small, indicating very good convergence.

With the chosen scales our implementation took $1$ minute and $43$ seconds to yield
$10,000$ realizations following a burn-in of $300,000$ iterations, after storing
one in 150 iterations out of further $15,00,000$ iterations after the burn-in period.
The overall acceptance rate turned out to be $0.198572$, and the birth, death, no-change rates turned out
to be $0.000795$, $0.000842$ and $0.593601$, respectively.

\begin{figure}
\centering
\subfigure[Trace plot of $k$.]{ \label{fig:acidity_k}
\includegraphics[width=7cm,height=6cm]{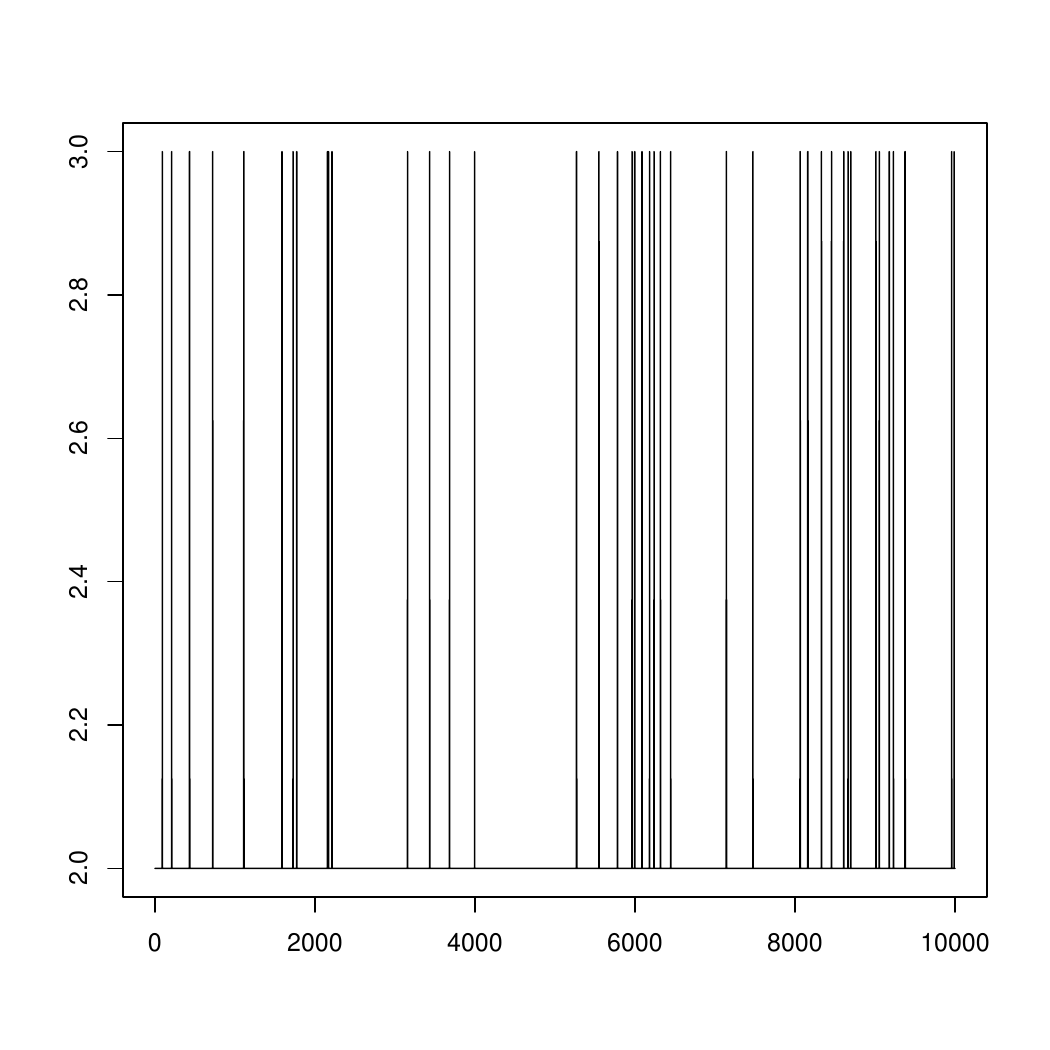}}
\hspace{2mm}
\subfigure[Trace plot of $\nu^*_1$.]{ \label{fig:acidity_nu}
\includegraphics[width=7cm,height=6cm]{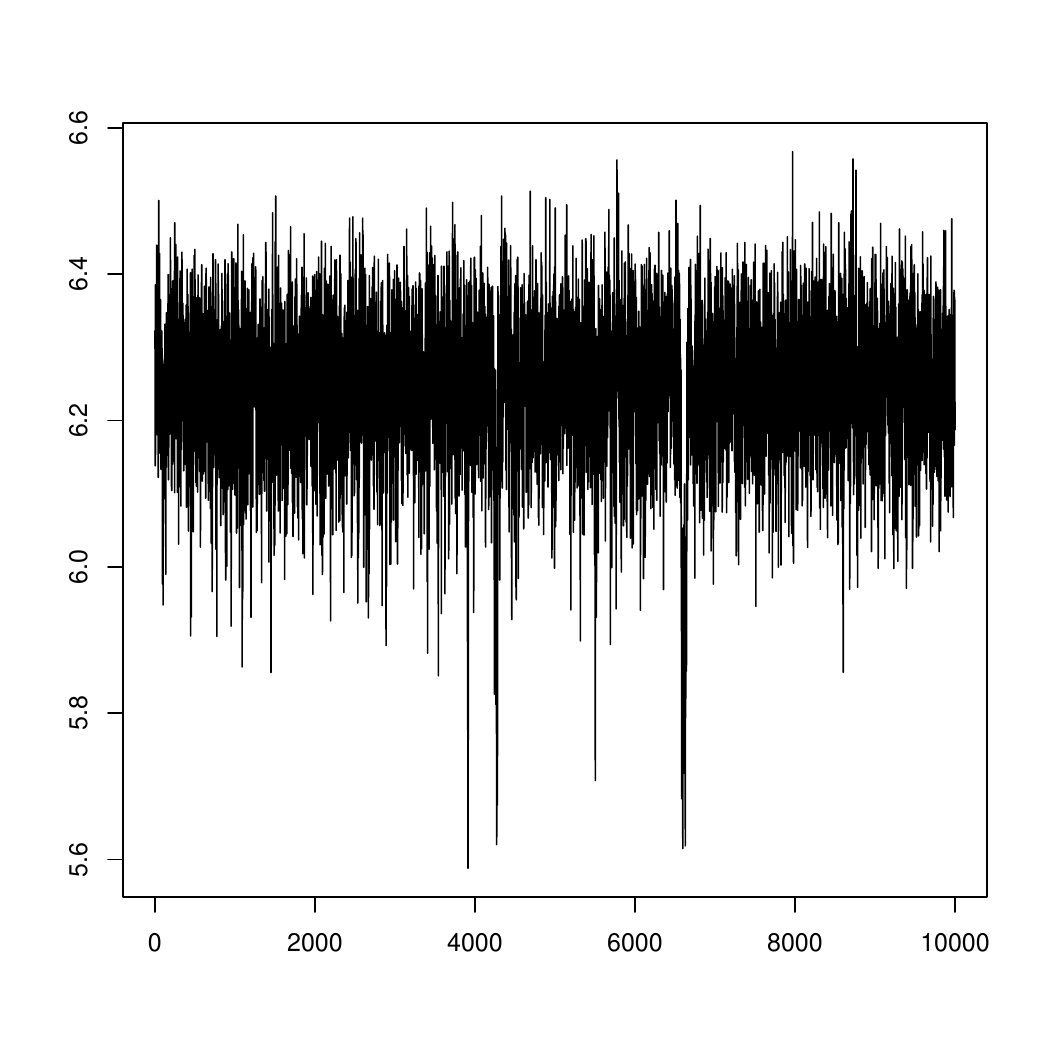}}\\
\vspace{2mm}
\subfigure[Trace plot of $\tau^*_1$.]{ \label{fig:acidity_tau}
\includegraphics[width=7cm,height=6cm]{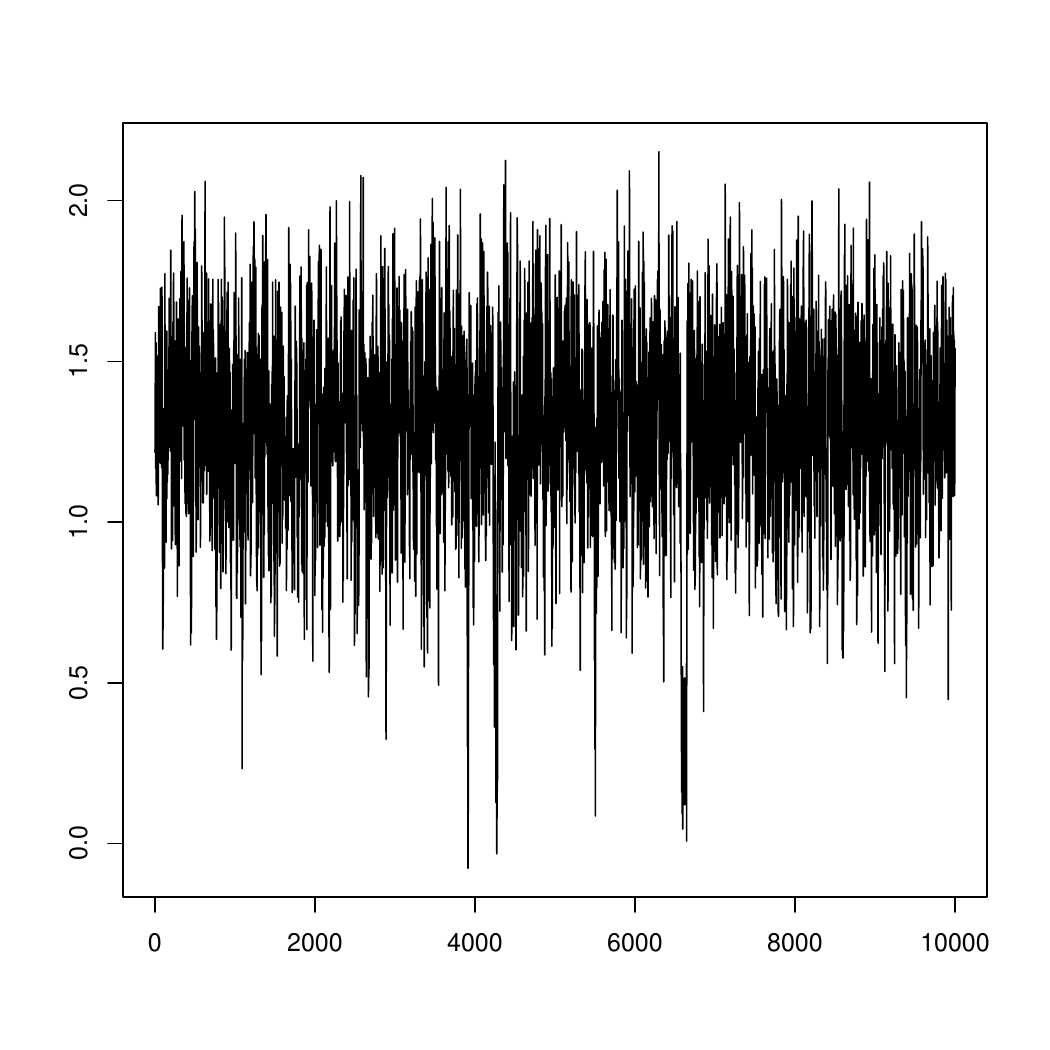}}
\hspace{2mm}
\subfigure[Trace plot of $\omega_1$.]{ \label{fig:acidity_w}
\includegraphics[width=7cm,height=6cm]{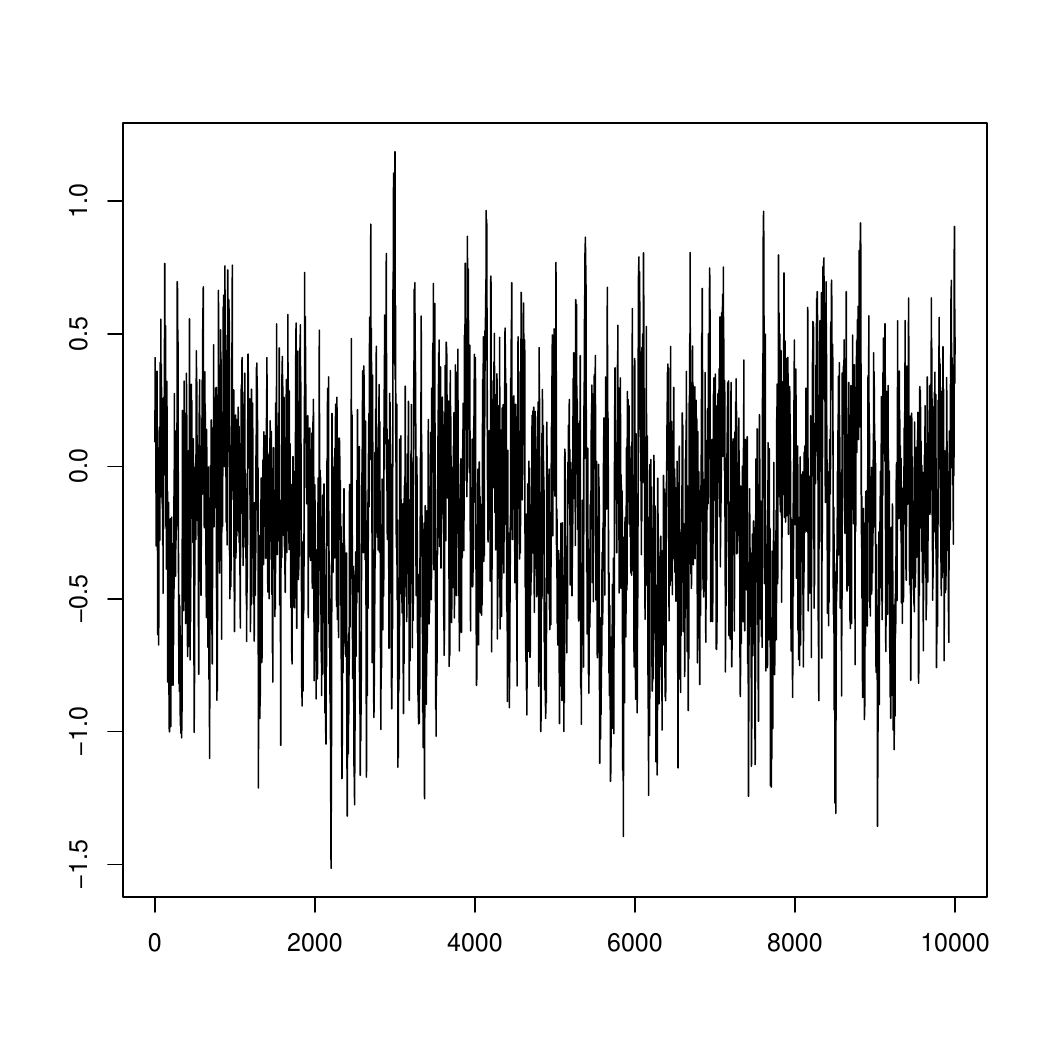}}
\caption{{\bf TTMCMC for the acidity data:} Trace plots of $k$, $\nu^*_1$, $\tau^*_1$ and $\omega_1$.} 
\label{fig:acidity_trace_plots}
\end{figure}

The trace plots of $k,\nu^*_1,\tau^*_1$ and $\omega_1$, shown in Figure \ref{fig:acidity_trace_plots},
again indicate quite good mixing properties and are consistent with the conclusions of our proposed 
credible region based convergence
assessment criterion.

With our prior structure here the posterior distribution of $k$ 
again strongly favoured 2 and 3 components, with $k=2$ receiving significantly larger posterior mass
0.9941 compared to the posterior probability of $k=3$.
The reason for the strong support for bimodality can be attributed to the large size of the data
contained in the relatively small interval $(2,8)$.

\begin{figure}
\includegraphics[width=7in,height=6.5in]{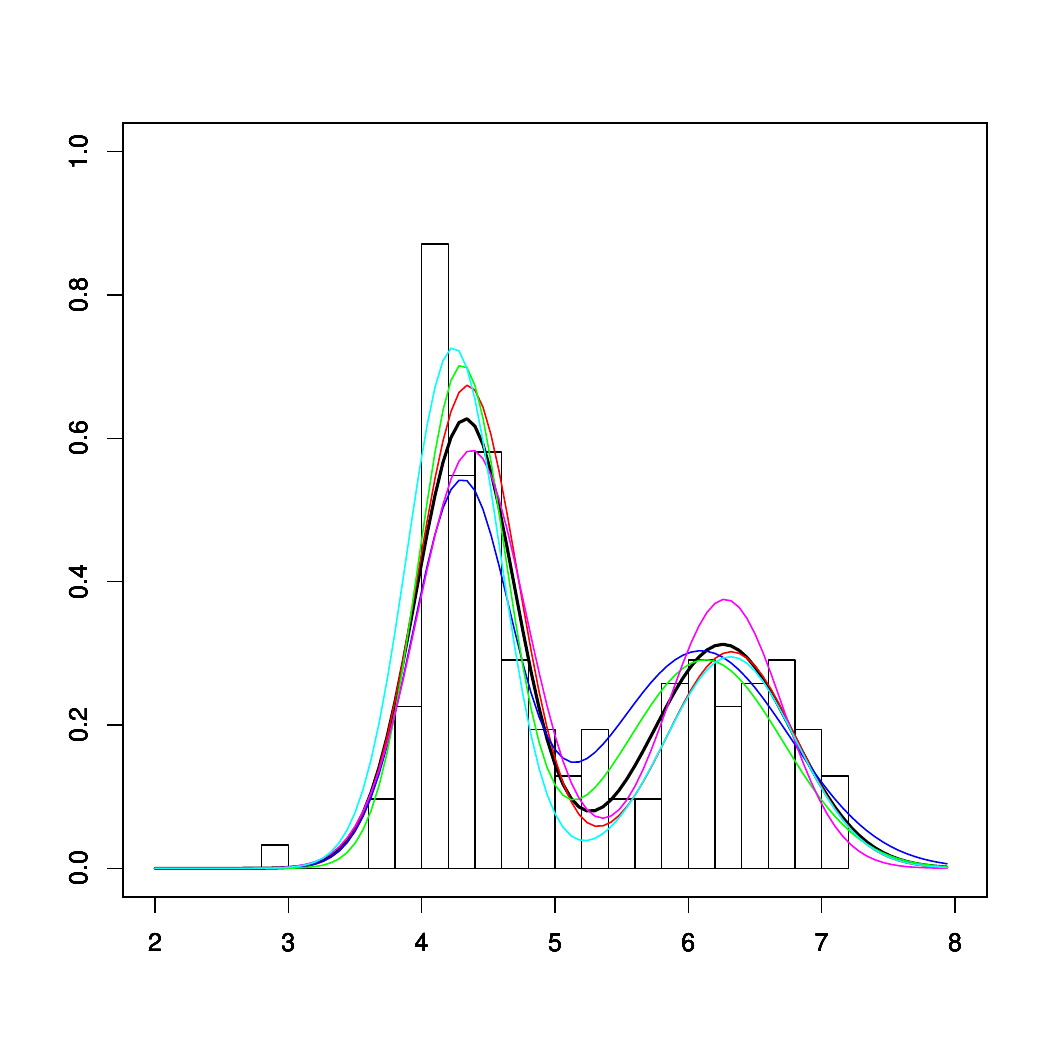}
\caption{{\bf TTMCMC for the acidity data:} Goodness of fit of the posterior distribution of densities (coloured curves) 
to the observed data (histogram). The thick black curve is the modal density and the other coloured curves are 
some densities contained in the 95\% HPD.}
\label{fig:acidity_hpd}
\end{figure}

The modal density and sample densities falling in the 95\% HPD region, overlapped on
the histogram of the observed data are shown in Figure \ref{fig:acidity_hpd}. 
Once again, good fit to the data is indicated.

\subsection{{\bf Galaxy data}}
\label{subsec:galaxy}

In contrast with the previous two cases of the enzyme and the acidity data, 
the galaxy data, which is much more sparse 
and seems to exhibit far greater number of modes, seems to be much more challenging to analyze.
Thus, we consider a somewhat different prior structure to reflect our beliefs regarding the Bayesian mixture analysis.

Here, following \ctn{Bhattacharya08} we set $s = 4.0$; $S = 2$; 
$\nu_0 = 20$; $\psi=33.3$.  However, unlike the previous
two cases here we assume that $\omega_j\sim \log\left(\mathcal G(5,1)\right)$, 
so that $\bpi$ follows the Dirichlet distribution with all the parameters equal to $5$.
The prior mean and mode of $\pi_j$ associated with this Dirichlet distribution 
are $1/k$ and the variance is $(k-1)/6k^2$. Note that the mean and the variance
of the uniform Dirichlet distribution, which corresponds to taking all the parameters equal to $1$,
are $1/k$ and $(k-1)/\{k(k+1)\}$, respectively. Hence, for large $k$, the variance of our
prior distribution is about $1/6$ times that of the uniform Dirichlet. This lesser variability
ensures that the minor local modes receive non-negligible prior weights, and hence makes sense
in this galaxy data scenario.
As regards the prior on $k$, here we choose a discretized normal distribution
on $\{1,\ldots,30\}$ with mean $15$ and variance $50$.
This reflects our belief that although all the values in $\{1,\ldots,30\}$ receive significant
prior masses, relatively large number of components is preferable in this application where many
local modes are exhibited by the data.

In this application, following the previous convergence diagnostic method, we found the 
appropriate scales to be
$a_{\nu^*_j}=a_{\tau^*_j}=a_{\omega_j}=1$ for $j=1,\ldots,k$.
These scales correspond to $\eta_1=0.01657$ and $\eta_2=0.01039$, which 
indicate good convergence.
Here the overall acceptance rate, computed over $18,00,000$ iterations, turned out to be $0.036388$,
while the birth, death and no-change rates are $0.007517$, $0.007559$ and $0.094195$, respectively.

The implementation of TTMCMC in this application took $6$ minutes and $33$ seconds to yield
$10,000$ realizations after discarding a burn-in of $300,000$ iterations, and then storing
one iteration in every 150 iterations out of further $15,00,000$ iterations following the burn-in period.

\begin{figure}
\centering
\subfigure[Trace plot of $k$.]{ \label{fig:galaxy_k}
\includegraphics[width=7cm,height=6cm]{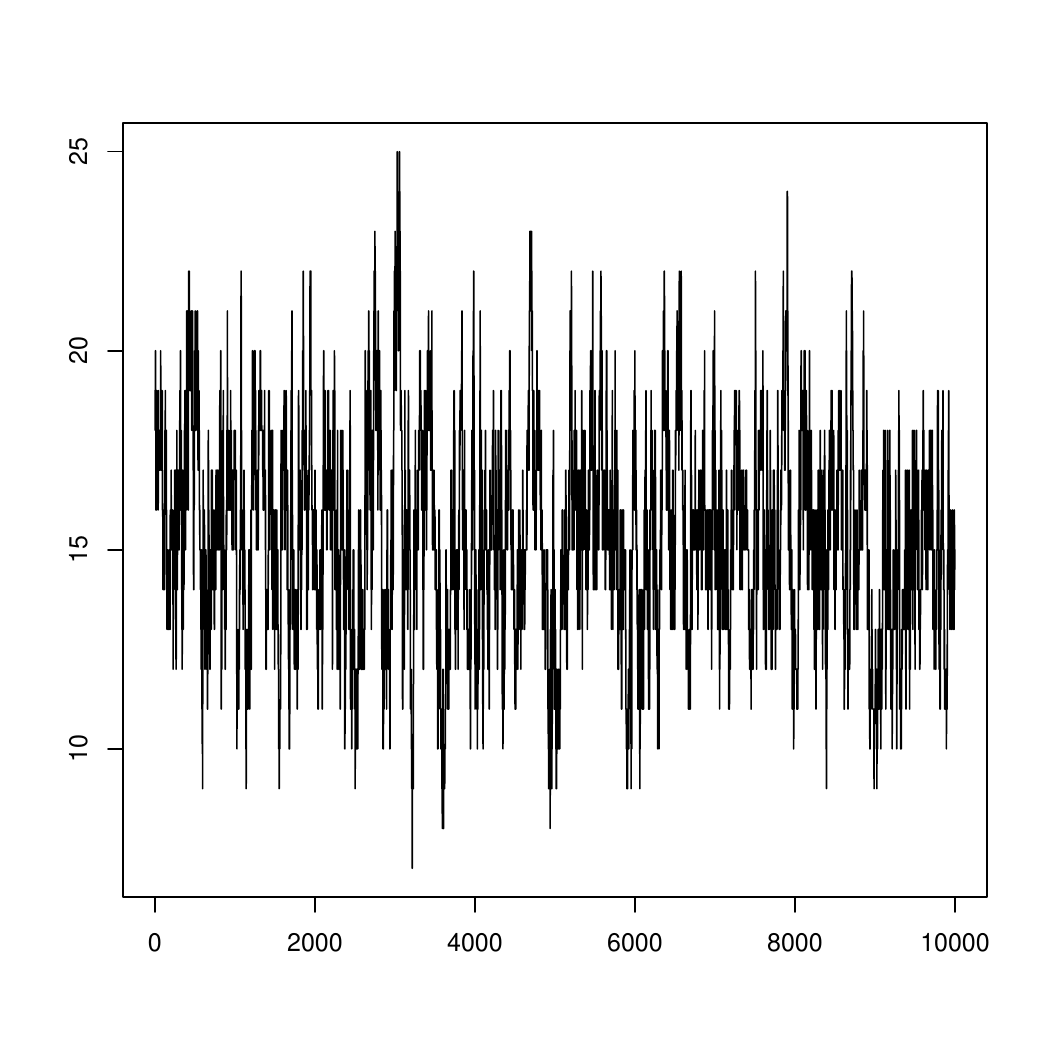}}
\hspace{2mm}
\subfigure[Trace plot of $\nu^*_1$.]{ \label{fig:galaxy_nu}
\includegraphics[width=7cm,height=6cm]{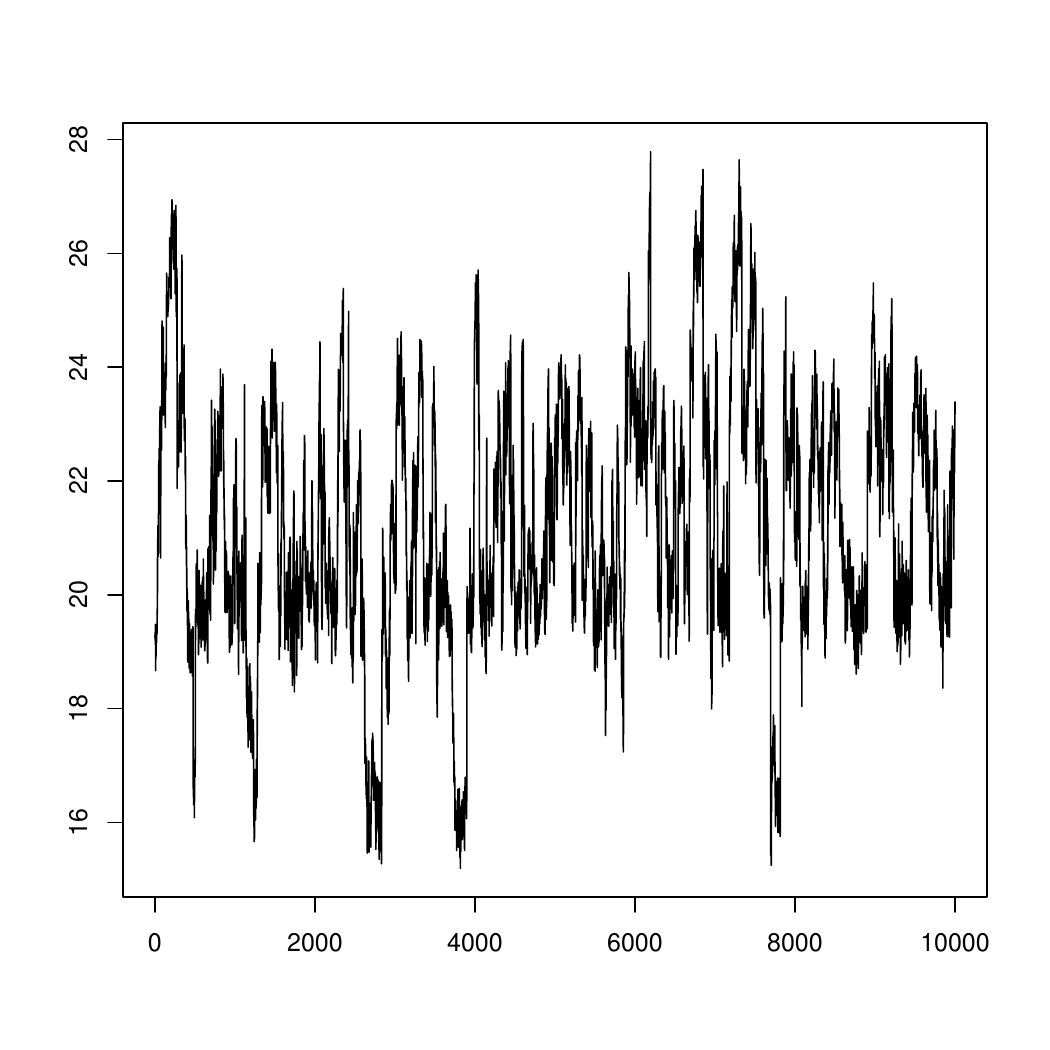}}\\
\vspace{2mm}
\subfigure[Trace plot of $\tau^*_1$.]{ \label{fig:galaxy_tau}
\includegraphics[width=7cm,height=6cm]{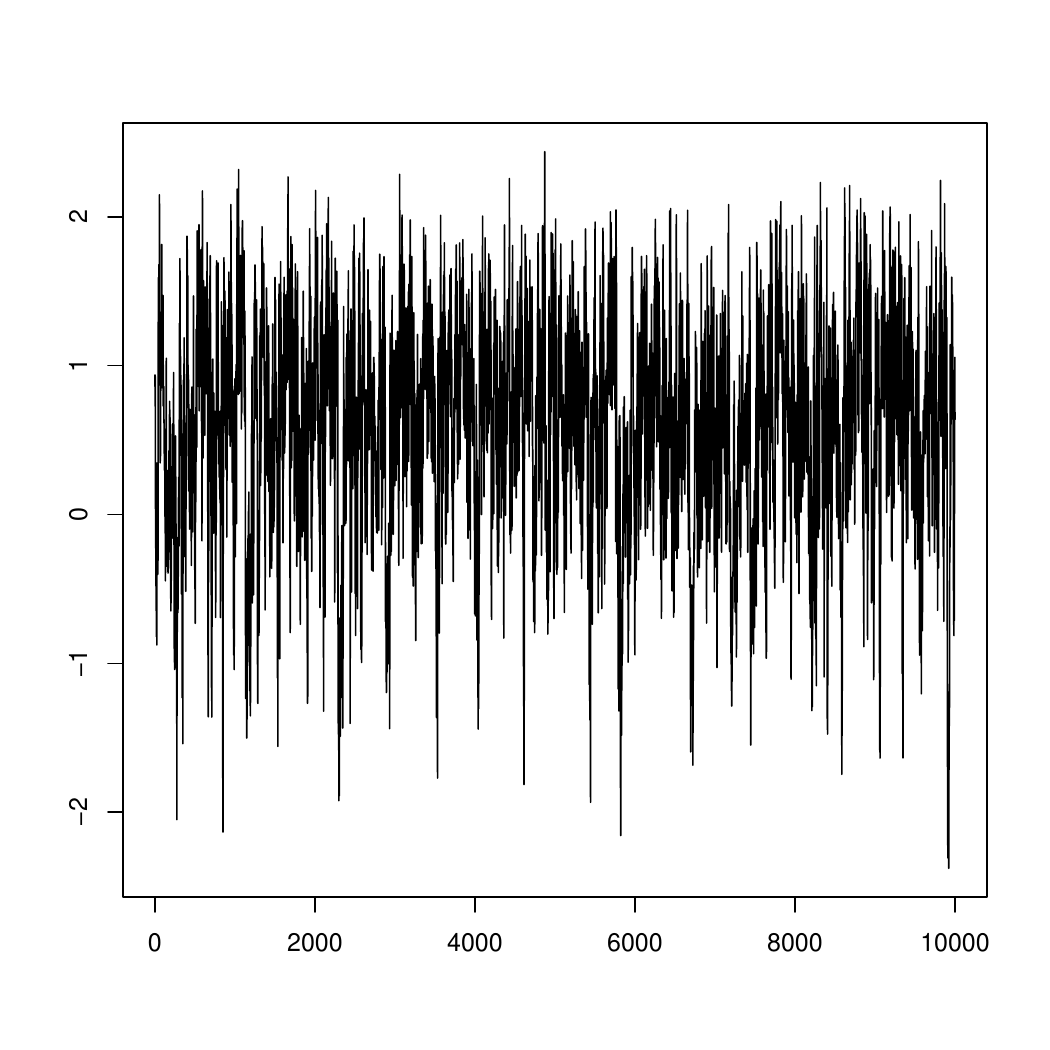}}
\hspace{2mm}
\subfigure[Trace plot of $\omega_1$.]{ \label{fig:galaxy_w}
\includegraphics[width=7cm,height=6cm]{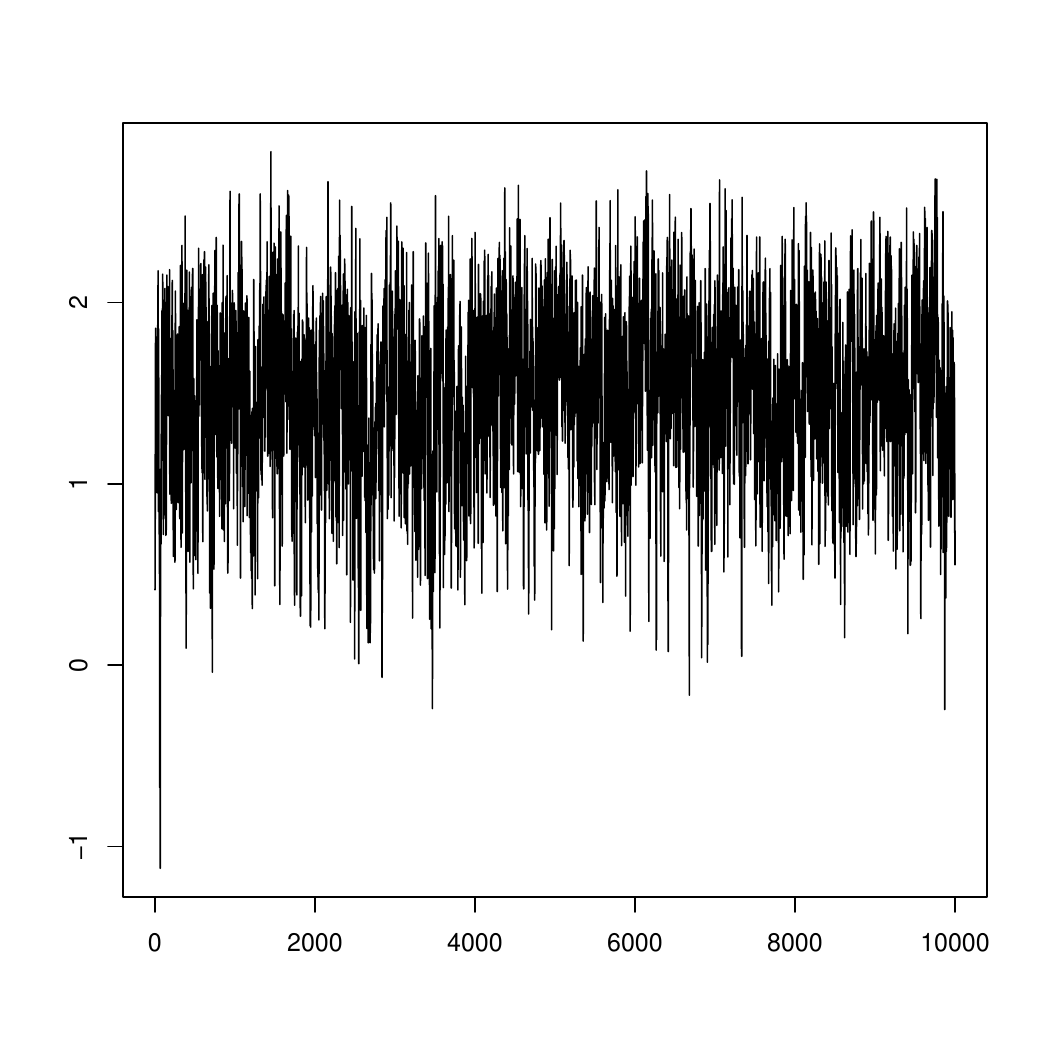}}
\caption{{\bf TTMCMC for the galaxy data:} Trace plots of $k$, $\nu^*_1$, $\tau^*_1$ and $\omega_1$. Good mixing
behavior of the TTMCMC chain is exhibited by the above panels.} 
\label{fig:galaxy_trace_plots}
\end{figure}

Note that, even in this challenging galaxy data application, the trace plots turned out to be quite reasonable, as shown in
Figure \ref{fig:galaxy_trace_plots}. Thus, reasonable overall mixing behavior of the TTMCMC chain is indicated
by the trace plots, consistent with the results of our credible region based convergence assessment criterion.

In this problem the posterior distribution of $k$ turned out to be much more variable
than in the previous two cases. Here $k\in\{7,8,9,10,11,12,13,14,15,16,17,18,19,20,21,22,23,24,25\}$
with respective probabilities $\{0.0002, 0.0005,
0.0059, 0.0191, 0.0455, 0.0784, 0.1044, 0.1371, 0.1596,\\ 0.1457, 0.1115, 0.0869,
0.0513, 0.0277, 0.0128, 0.0097, 0.0018, 0.0016, 0.0003\}$. 
Thus most of the possible values of $k$ received positive posterior masses. It is also difficult
to single out any particular value of $k$ that is very strongly favoured by the posterior, unlike
the previous two applications.

\begin{figure}
\includegraphics[width=7in,height=6.5in]{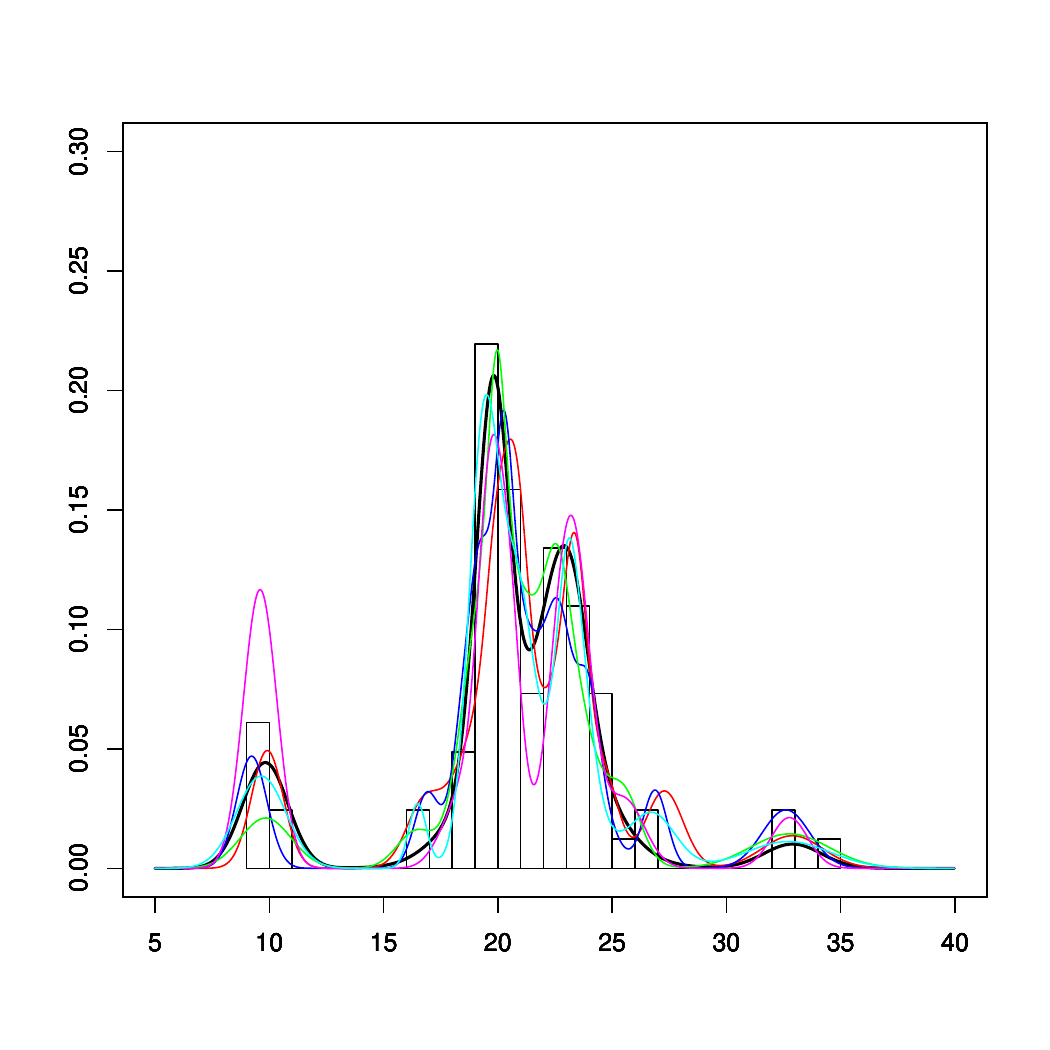}
\caption{{\bf TTMCMC for the galaxy data:} Goodness of fit of the posterior distribution of densities 
(coloured curves) to the observed data (histogram). The thick black
curve is the modal density and the other coloured curves are some densities contained in the 95\% HPD.}
\label{fig:galaxy_hpd}
\end{figure}

Figure \ref{fig:galaxy_hpd} depicts the modal density and sample densities falling in the 95\% HPD region, 
overlapped on
the histogram of the observed data. 
The fit to the data seems to be quite encouraging with the sample densities capturing even the minor
modes located at the extreme ends of the support of the data.

\subsection{{\bf Comparison of TTMCMC with random walk RJMCMC with respect to the three real data sets}}
\label{subsec:compare_ttmcmc_rjmcmc}
To save space, we have provided the details of the comparisons in Section S-10 of the supplement.
Briefly, in all the three examples, random walk RJMCMC places much higher posterior mass to large number
of components that are very implausible. The reason for this can be attributed to the product
of the left truncated standard normal densities that features in the denominator of the acceptance ratio
of the birth move of RJMCMC; since the aforementioned densities are bounded above by $1$, this makes
the acceptance rate for the birth move exceeding large, which, in effect, seriously slows down convergence.
In addition, for the somewhat challenging galaxy data set, the random walk RJMCMC chain has extremely poor acceptance
rate, and the chain hardly moved. Recall that this was the case for all the four gamma mixture examples as well. 
Thus, random walk RJMCMC completely fails to act as the default RJMCMC algorithm.  

\subsection{{\bf Relevance of autocorrelation plots for convergence diagnosis in variable dimensions}}
\label{subsec:autocorrelations}
Convergence assessment with the help of autocorrelations is not always appropriate in variable dimensional MCMC algorithms. 
Since there is no fixed Euclidean structure, parameters may not retain the same meaning throughout the iterations. 
To proceed with autocorrelation plots, it is necessary to focus attention on those parameters 
which retain constant interpretation across all models. In the mixture case the number of components may be considered. 
In this regard, the autocorrelation plots presented in Figure S-4 of the supplement 
reveal far superior mixing of the $k$-chain obtained by our TTMCMC sampler compared to 
random walk RJMCMC for all the three real data sets. 
In particular, for the galaxy data set, the RJMCMC based autocorrelations are simply hopeless!

\subsection{{\bf Comparison between TTMCMC and RJMCMC when the prior of Richardson and Green (1997) is considered}}
\label{subsec:further_comparisons}

Further comparisons between TTMCMC and RJMCMC with respect to the prior structure and the algorithm of \ctn{Richardson97}, 
are provided in Section S-10 of the supplement, in the context of the challenging galaxy data. 
We argue that actually their prior structure, where
$\btau$ are made dependent in a way that they are approximately of the same size,
is not expected to provide good fit to the observed histogram, but the large number of components supported by
their algorithm as a result of its inherent bias as discussed, create the appearance of good fit. 
We further argue that the prior structure of \ctn{Cappe03}, which is essentially the prior of \ctn{Richardson97}
but $\btau$ are independent {\it a priori}, is a more appropriate prior for capturing the varieties of modes
in the galaxy data.

\section{{\bf TTMCMC for multivariate normal mixtures}}
\label{sec:mvn_mixtures}
We now consider $iid$ $p$-variate data 
$\left\{\by_i=\left(y_{i1},\ldots,y_{ip}\right)^T;i=1,\ldots,n\right\}$ arising from the $p$-variate normal mixture 
having the following density when the number of components is $k$: for $i=1,\ldots,n$,
\begin{equation}
f(\by_i|\bTheta_k)
=\sum_{j=1}^k\pi_j\frac{1}{\left(2\pi\right)^{p/2}\left|\bSigma_j\right|^{\frac{1}{2}}}
\exp\left\{-\frac{1}{2}\left(\by_i-\bmu_j\right)^T\bSigma^{-1}\left(\by_i-\bmu_j\right)\right\},
\label{eq:mvn_mixture}
\end{equation}
where $\bTheta_k=\left\{\bmu_1,\ldots,\bmu_k,\bSigma_1,\ldots,\bSigma_k,\pi_1,\ldots,\pi_k\right\}$.

Letting $\bar\by$ denote the $p$-dimensional sample mean vector and $\bS=\mbox{diag}\left\{s^2_1,\ldots,s^2_p\right\}$
denote the diagonal matrix with the sample variances in the diagonal, we transform the data $y_i$, following \ctn{Dell06},
to $\bS^{-1/2}\left(\by_i-\bar\by\right)$, once the data are generated. 

\subsection{{\bf Prior structure}}
\label{subsec:mvn_prior}
Following \ctn{Dell06}, we assume that {\it a priori}
\begin{equation}
[\bmu_j|\bSigma_j]\sim N_p\left(\bzero,\bSigma_j\right),
\label{eq:mvn_mu_prior}
\end{equation}
a $p$-variate normal with mean $\bzero$ and covariance matrix $\bSigma_j$.
We also assume following \ctn{Dell06} that
\begin{equation}
[\bSigma_j]\sim W^{-1}\left(p+1,\bOmega\right),
\label{eq:mvn_sigma_prior}
\end{equation}
an inverse-Wishart distributon with $(p+1)$ degrees of freedom and diagonal matrix $\bOmega$.
However, instead of considering the gamma prior on the diagonal elements of $\bOmega$ as in \ctn{Dell06},
we set all the diagonal elements equal to $1$. This we do to avoid oversmoothness induced by the dependence structure between
the $\bSigma_j; j=1,\ldots,k$, and to facilitate adaptive learning from the data. Recall that
(see Section \ref{subsec:further_comparisons}) a similar issue 
of oversmoothness seems to render the prior of \ctn{Richardson97} less appropriate for capturing the varieties of modes
as compared to the prior of \ctn{Cappe03}, in the univariate normal mixture case.

As before, we consider a discrete uniform prior for $k$ on $\{1,2,\ldots,30\}$. Here we remark that although
\ctn{Dell06} also report a discrete uniform prior on $k$, they did not specify the range.

\subsection{{\bf TTMCMC strategy for multivariate situations}}
\label{subsec:mvn_ttmcmc_algo}

As before we reparameterize $\pi_j$ as $\exp(w_j)/\sum_{i=1}^k\exp(w_i)$. As for $\bSigma_j$, we consider
the Cholesky decomposition $\bSigma_j=\bL\bL^T$, where $\bL=(L_{rs})_{r,s=1,\ldots,p}$ is the appropriate 
lower triangular matrix.
Thus, there are $1+p+p(p+1)/2$ number of parameters to be split in any given birth move
given that the $j$-th mixture component is chosen; $w_j$, the $p$ components
of $\bmu_j=\left(\mu_{j1},\ldots,\mu_{jp}\right)^T$ 
and $p(p+1)/2$ non-zero elements of $\bL_j$. Thus, we need $1+p+p(p+1)/2$ $\epsilon$'s
to define our additive TTMCMC move types. The Jacobian of the birth move is given 
$2^{1+p+p(p+1)/2}\times a_{w_j}\times\prod_{r=1}^pa_{\mu_{jr}}\prod_{r\geq s=1}^pa_{L_{jrs}}$,
where $a_{\mu_{jr}}$ is the scale for the additive transformation of the $r$-th component of $\bmu_j$
and $a_{L_{jrs}}$ is the same for the $(r,s)$-th element of $\bL_{jrs}$, where $r\geq s$. The Jacobian
for the death move is the inverse of that of the birth move with the relevant scale values. 
We reject the entire move if any of the diagonal elements of $\bL$ becomes negative.

\subsection{{\bf Simulation experiment with $p=3$}}
\label{subsec:simstudy_3d}
Following \ctn{Dell06} we set generate $80$, $100$ and $100$ data points from 3-variate normal distributions
with means $\bmu_1=(6,4,2)^T$, $\bmu_2=(-11,-4,-1)^T$, $\bmu_3=(-7,-11,-5)^T$ and covariance matrices
$\bSigma_1=\left(\begin{array}{ccc} 3 & 2 & 1\\ 2 & 5 & 0\\ 1 & 0 & 4\end{array}\right)$,
$\bSigma_2=\left(\begin{array}{ccc} 2 & -1.5 & 1\\ -1.5 & 5 & 2\\ 1 & 2 & 3\end{array}\right)$,
$\bSigma_3=\left(\begin{array}{ccc} 5 & -1 & 1\\ -1 & 4 & -2\\ 1 & -2 & 3\end{array}\right)$,
respectively, and fit our 3-variate mixture model to the data assuming unknown number of components.

\begin{figure}
\centering
\subfigure[Trace plot of $k$.]{ \label{fig:3d_k}
\includegraphics[width=7cm,height=6cm]{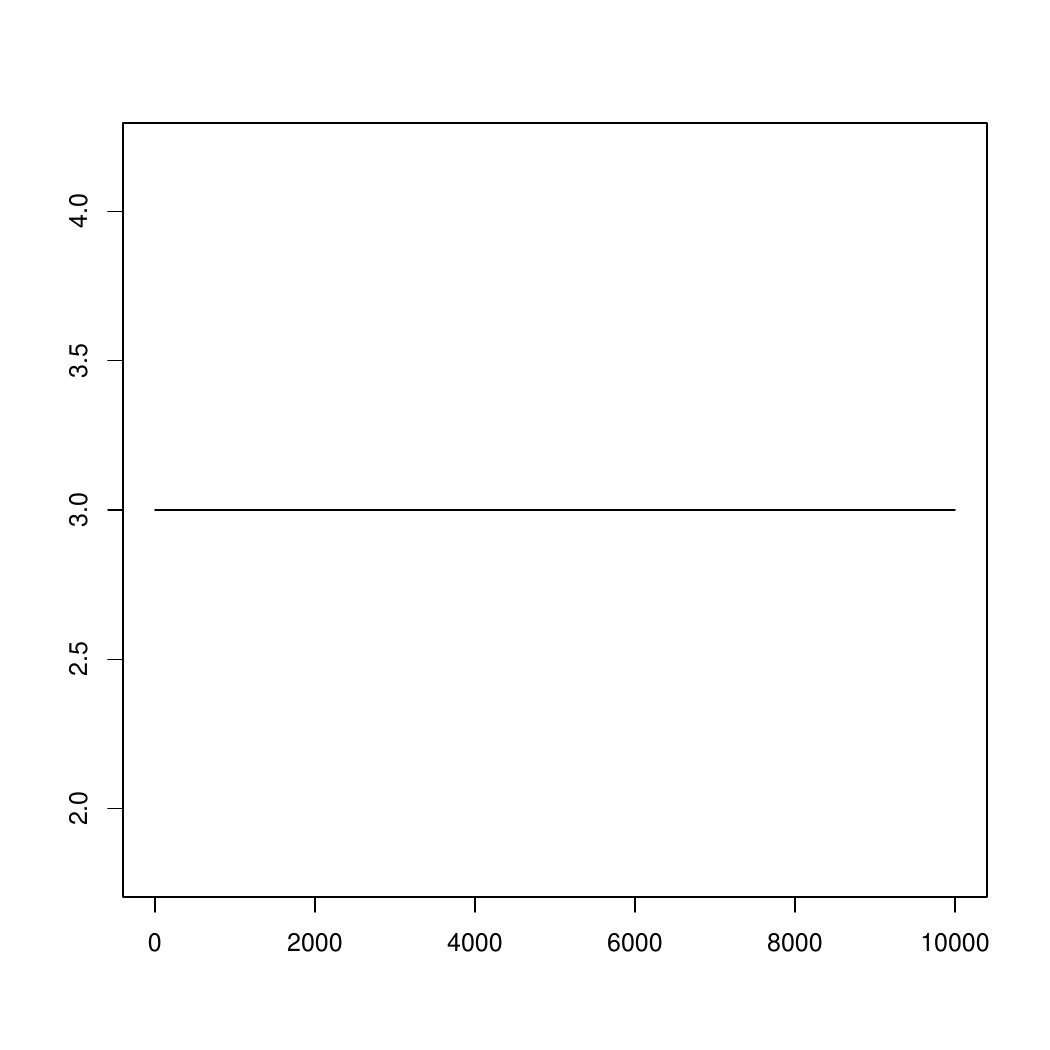}}
\hspace{2mm}
\subfigure[Trace plot of $\mu_{11}$.]{ \label{fig:3d_nu}
\includegraphics[width=7cm,height=6cm]{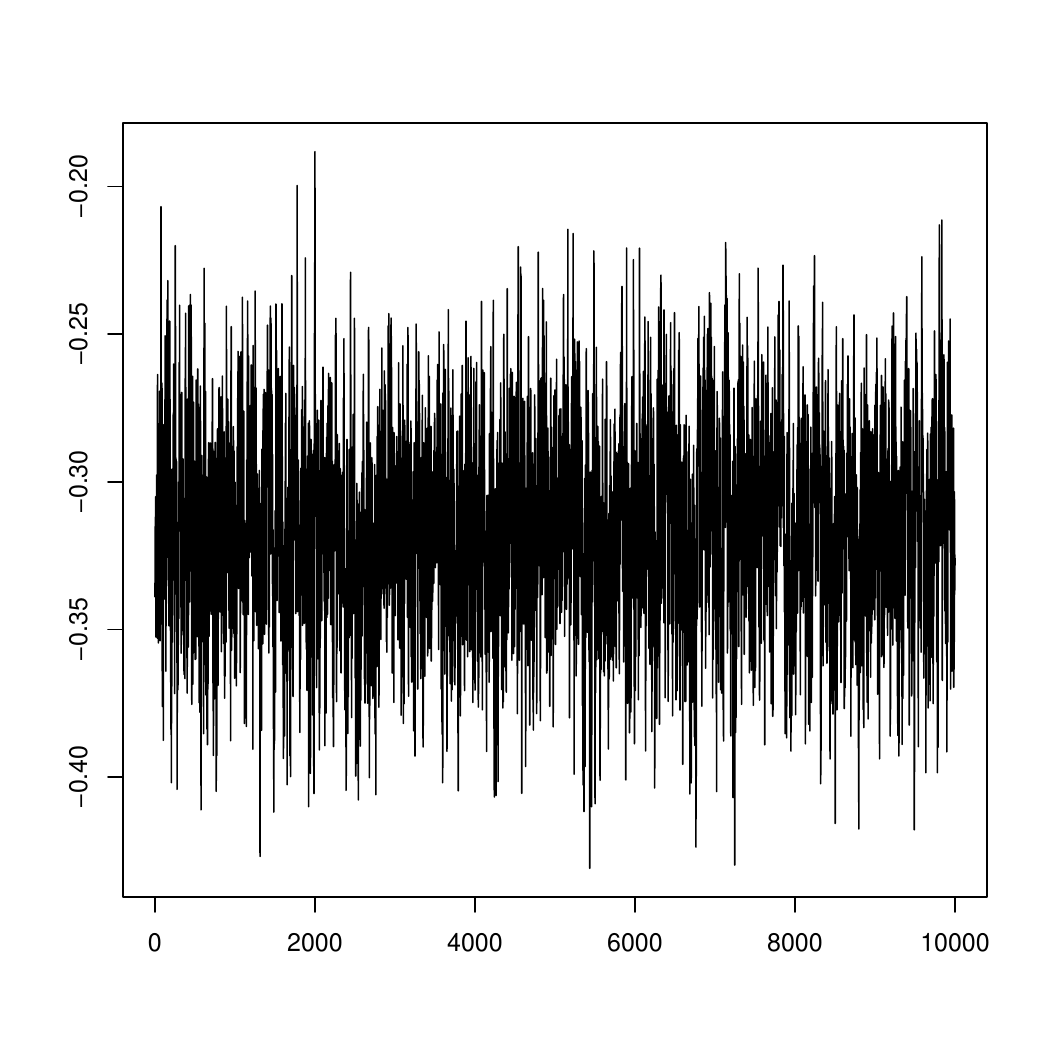}}\\
\vspace{2mm}
\subfigure[Trace plot of $L_{11}$.]{ \label{fig:3d_tau}
\includegraphics[width=7cm,height=6cm]{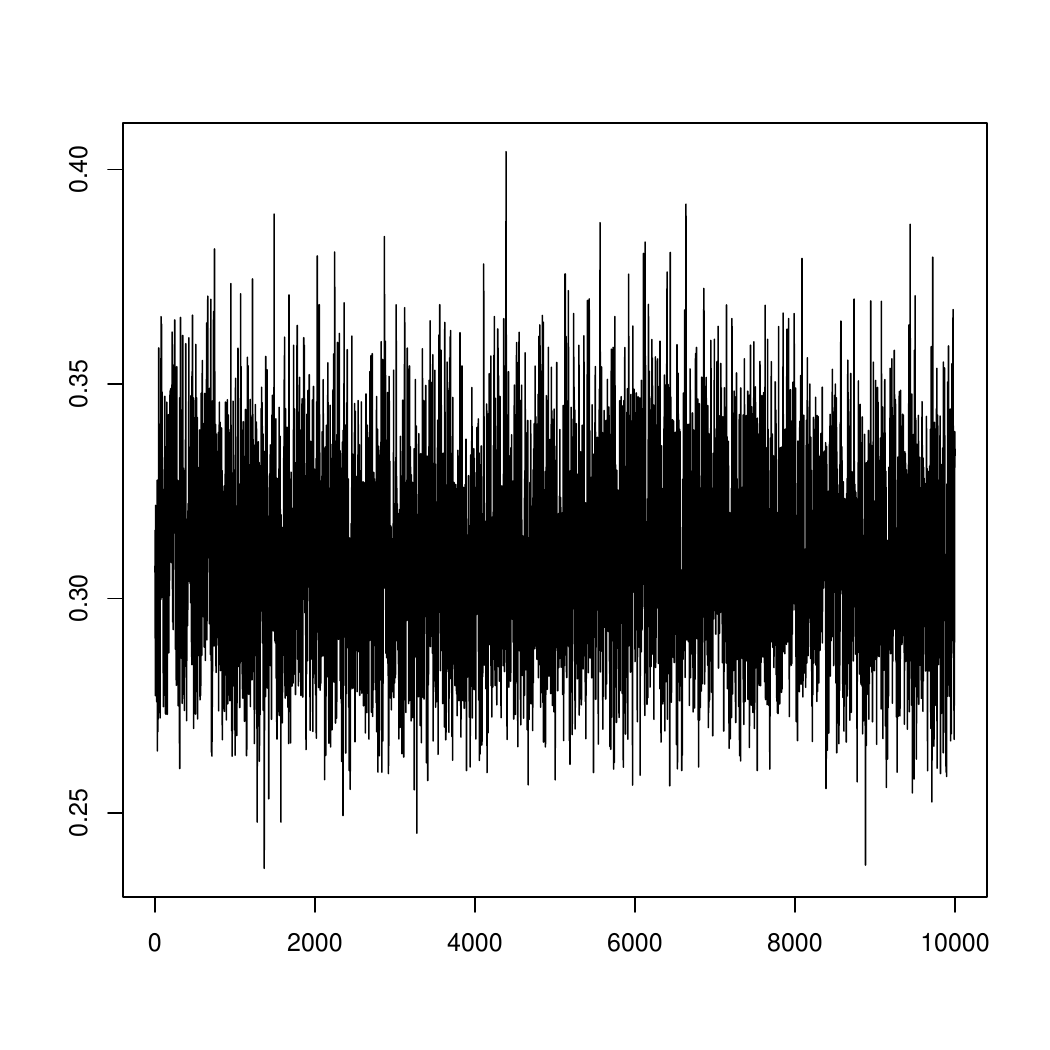}}
\hspace{2mm}
\subfigure[Trace plot of $\omega_1$.]{ \label{fig:3d_w}
\includegraphics[width=7cm,height=6cm]{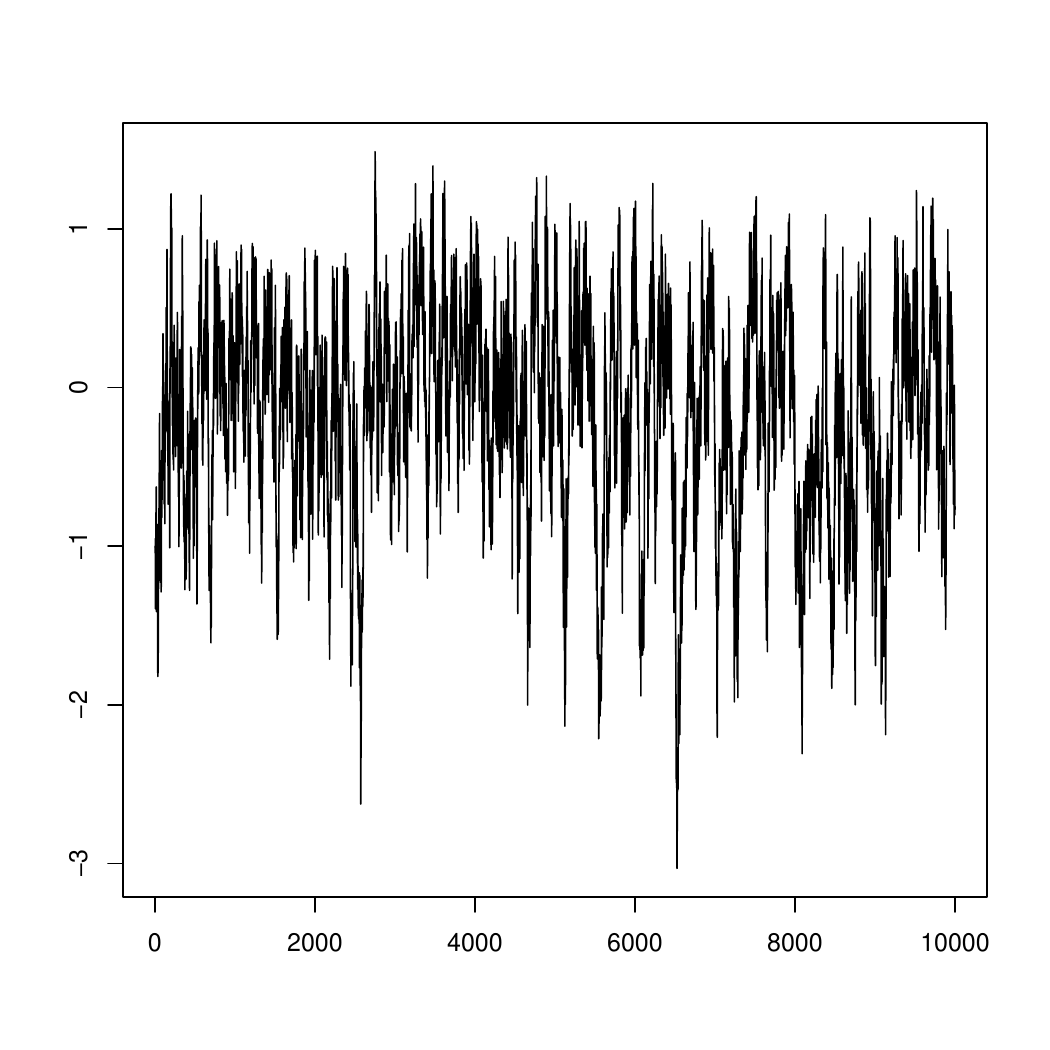}}
\caption{{\bf TTMCMC for 3-dimensional case:} Trace plots of $k$, $\mu_{11}$, $L_{11}$ and $\omega_1$. Good mixing
behavior of the TTMCMC chain is exhibited by the above panels.} 
\label{fig:3d_trace_plots}
\end{figure}

Considering a burn-in of 3,000,000 iterations, we ran the TTMCMC algorithm for a further 3,000,000 iterations,
storing one in 300 iterations, to obtain 10,000 realizations from the posterior. The implementation
took 49 minutes and 14 seconds on our laptop.
The overall acceptance rate turned out to be $0.038231$ when the scales of the additive transformations are set to be
$0.05$ and $0.05$ for the means and the elements of the Cholesky factors, and $0.5$ for the weights.  
The birth, death and the no-change rates are $0.000002$, $0.000015$ and $0.114539$, respectively.
The trace plots shown in Figure \ref{fig:3d_trace_plots} confirm excellent convergence properties
of our algorithm, even in the multivariate case. Importantly, we obtained point mass at the true number
of mixture components (as before, we do not rule out the possibility of missing some component other than $3$
in our finite TTMCMC run). In contrast, \ctn{Dell06} report 6 models associated with $k=1,\ldots,6$, with 3 components
receiving $0.9493$ posterior probability.

\begin{figure}
\includegraphics[width=7in,height=6.5in]{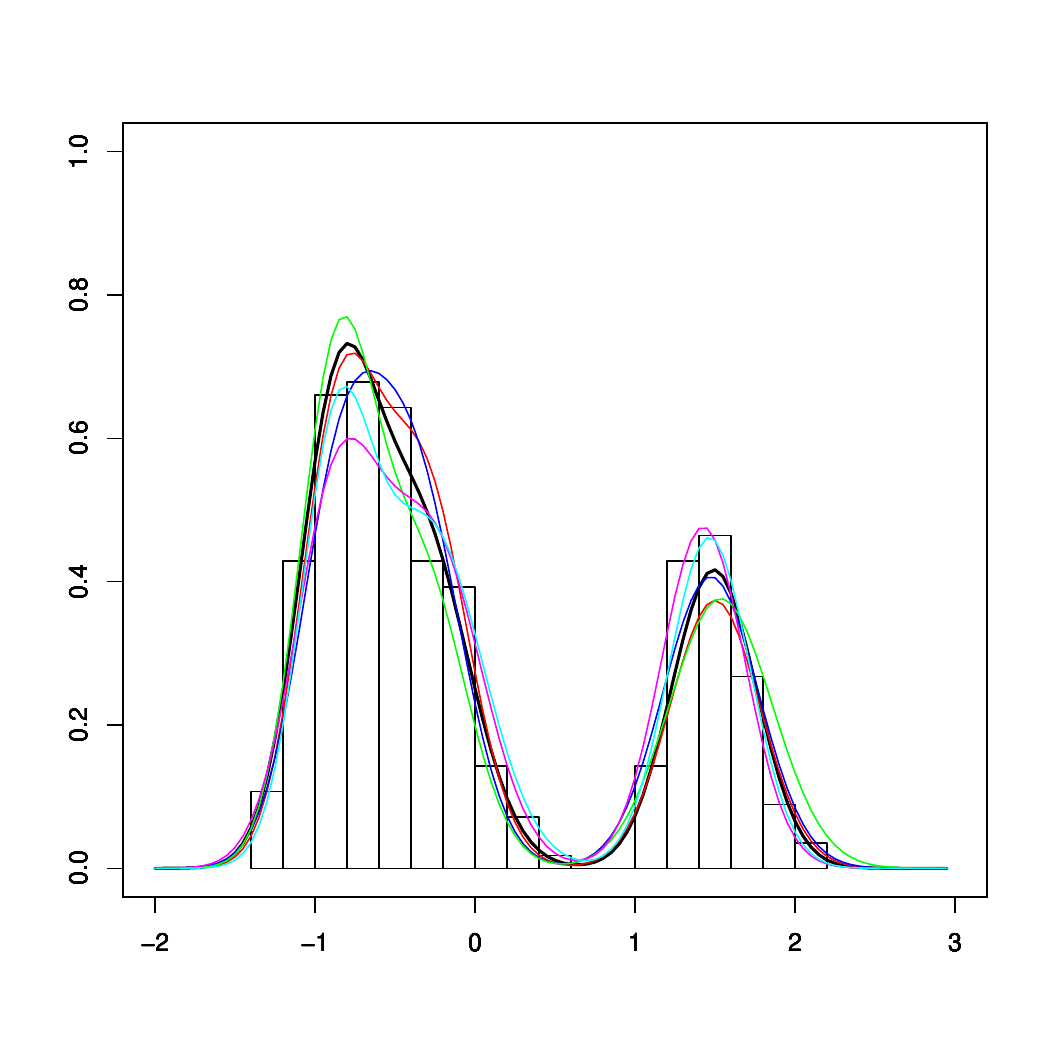}
\caption{{\bf TTMCMC for 3-dimensional case:} Goodness of fit of the posterior distribution of densities 
(coloured curves) to the histogram of the first component of the observed data $\left\{y_{i1};i=1,\ldots,285\right\}$. 
The thick black curve is the modal density and the other coloured curves are some densities contained in the 95\% HPD.}
\label{fig:3d_hpd}
\end{figure}

Figure \ref{fig:3d_hpd} depicts the modal density and sample densities falling in the 95\% HPD region, 
overlapped on the histogram of the first component $\left\{y_{i1};i=1,\ldots,n\right\}$ (here $n=280$) of the observed data. 
Excellent fit to the data is clearly indicated.

\subsection{{\bf Simulation experiment with $p=10$}}
\label{subsubsec:simstudy_10d}
We now consider application of TTMCMC to mixtures of $p=10$ dimensional multivariate normals. Specifically,
we first generate two mean vectors $\bmu_1$ and $\bmu_2$ from two $10$-dimensional, normal distributions 
$N_{10}\left(4\bone_{10},\bI_{10}\right)$ and $N_{10}\left(-5\bone_{10},\bI_{10}\right)$,
where, for any integer $p\geq 1$, $\bone_{p}$ is a $d$-component vector with each component $1$, and $\bI_{p}$ 
is the identity matrix of order $p$. Corresponding to the mean vectors $\bmu_1$ and $\bmu_2$, we specify
covariance matrices $\bSigma_1$ and $\bSigma_2$ of the following form: the off-diagonal elements are
given by $\sigma^2_j\rho$ and the diagonal elements are all equal to $\sigma^2_j$, for $j=1,2$.
For our illustration we consider $\sigma^2_1=4$, $\sigma^2_2=3$ and $\rho=0.5$.

We then generate $300$ realizations from $N_{10}\left(\bmu_1,\bSigma_1\right)$ and
$300$ realizations from $N_{10}\left(\bmu_2,\bSigma_2\right)$, which constitute our data set
$\{\by_1,\ldots,\by_{600}\}$ of size $600$.

We use the same TTMCMC algorithm as in the 3-dimensional experiment, 
but as to be anticipated for
higher dimensions, the convergence was slower compared to the 3-dimensional example. 
To improve mixing, we employed the following strategy. At the end of each iteration $t\geq 1$, we simulated 
$r^{(t)}\sim N(0,1)$ and proposed the further additive transformation $\bTheta^{(t)}\mapsto\bTheta^{(t)}+\ba r^{(t)}$, where 
$\bTheta^{(t)}$ denotes the stage of the parameters at iteration $t$, and $\ba$ denotes the vector of scaling
constants for the additive transformation. We then calculated the acceptance probability of this proposal in the usual
TMCMC set-up to either accept the new proposal $\bTheta^{(t)}+\ba r^{(t)}$ or to remain at $\bTheta^{(t)}$.
Such a strategy has also been employed by \ctn{Sabya13} to improve mixing in the context of palaeoclimate modeling. 
The strategy is akin to the so-called generalized Gibbs/MH methods in fixed-dimensional set-ups have the potential
of improving mixing (see, for example, \ctn{Liu00}, \ctn{Liu01}; see also \ctn{Liu99}). Further details can be found
in the supplement of \ctn{Dutta14}.

For our purpose, we chose the scales of the additive transformation associated with the original TTMCMC to be relatively large;
$0.5$ for the means, $0.05$ for the Cholesky components and $1.5$ for the weights, while for the mixing improvement
step we chose the scales to be $1/10$-th of the above scales. This ensures relatively small acceptance rate but large moves
for the original TTMCMC steps but much higher acceptance rate at the mixing improvement step.

However, in spite of the above strategy, the mixing improvement was not dramatic in our case, 
and still a considerably long run was necessary. As such, we discarded
the first $3\times 10^7$ iterations, and stored one in $300$ iterations out of the next $12\times 10^7$ iterations
to store $4\times 10^5$ iterations. We applied further thinning of size 40 to the stored samples, finally storing
$10,000$ iterations. The entire procedure took about 68 hours on our VMWare.
The overall acceptance rate, birth rate, death rate and the no-change rates in this implementation
are $0.008173$, $0.00014$, $0.00037$ and $0.023949$, respectively.

\begin{figure}
\centering
\subfigure[Trace plot of $k$.]{ \label{fig:10d_k}
\includegraphics[width=7cm,height=6cm]{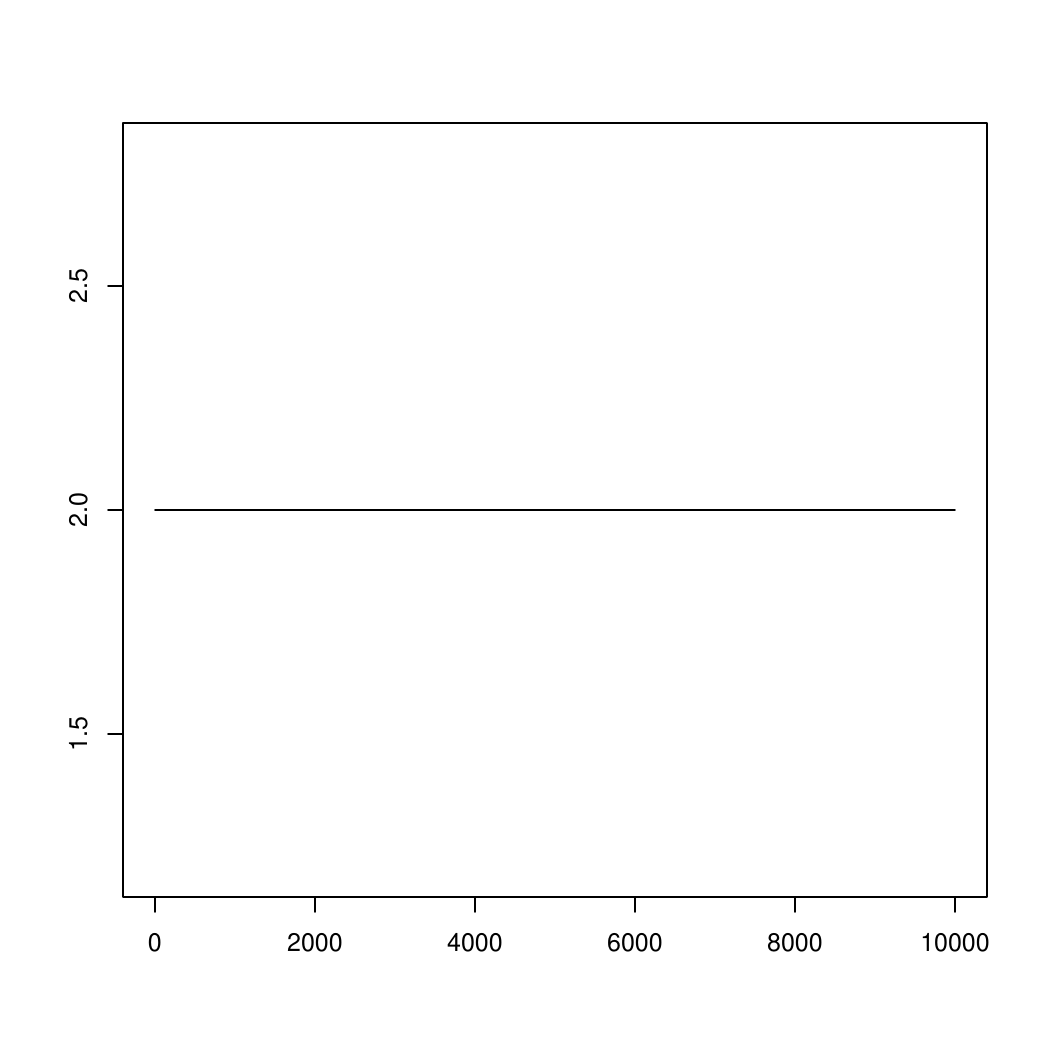}}
\hspace{2mm}
\subfigure[Trace plot of $\mu_{11}$.]{ \label{fig:10d_nu}
\includegraphics[width=7cm,height=6cm]{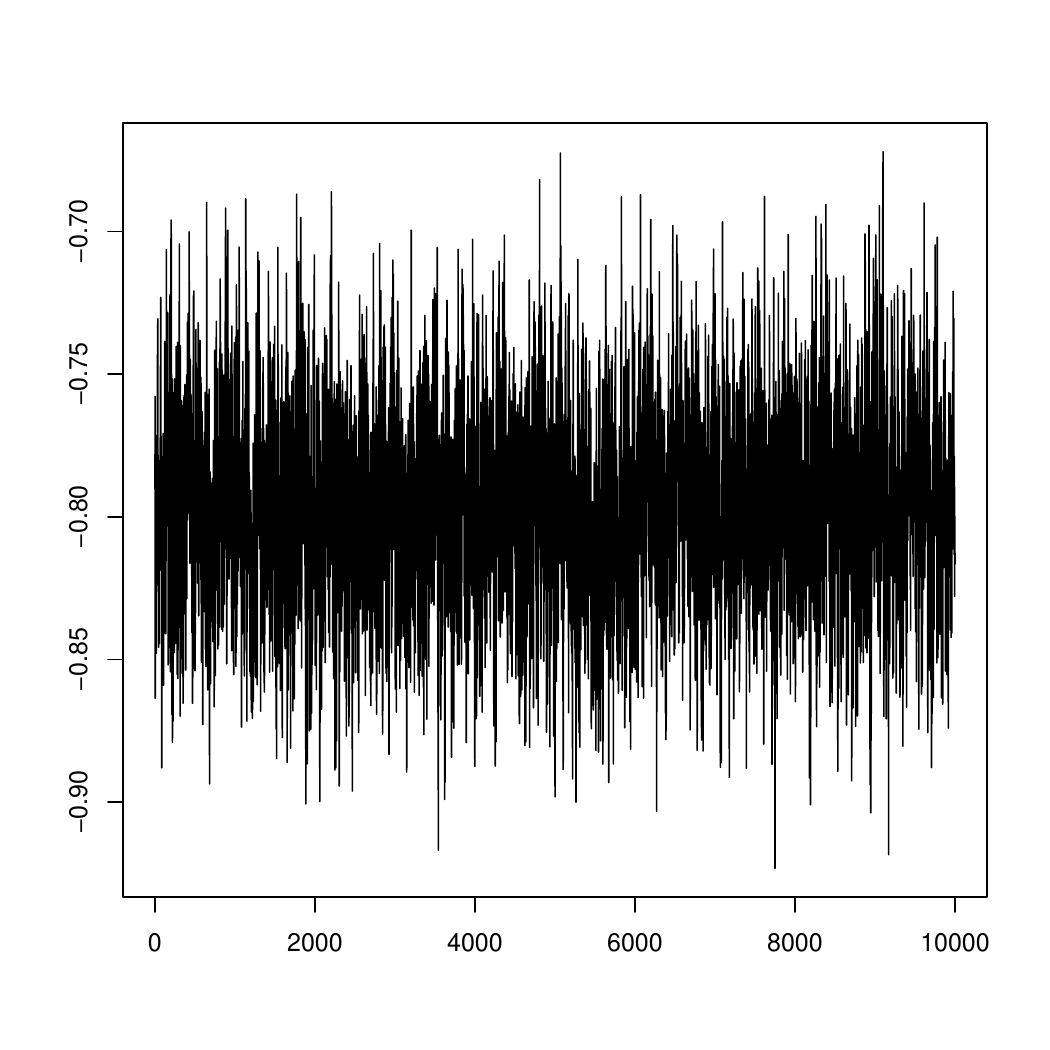}}\\
\vspace{2mm}
\subfigure[Trace plot of $L_{11}$.]{ \label{fig:10d_tau}
\includegraphics[width=7cm,height=6cm]{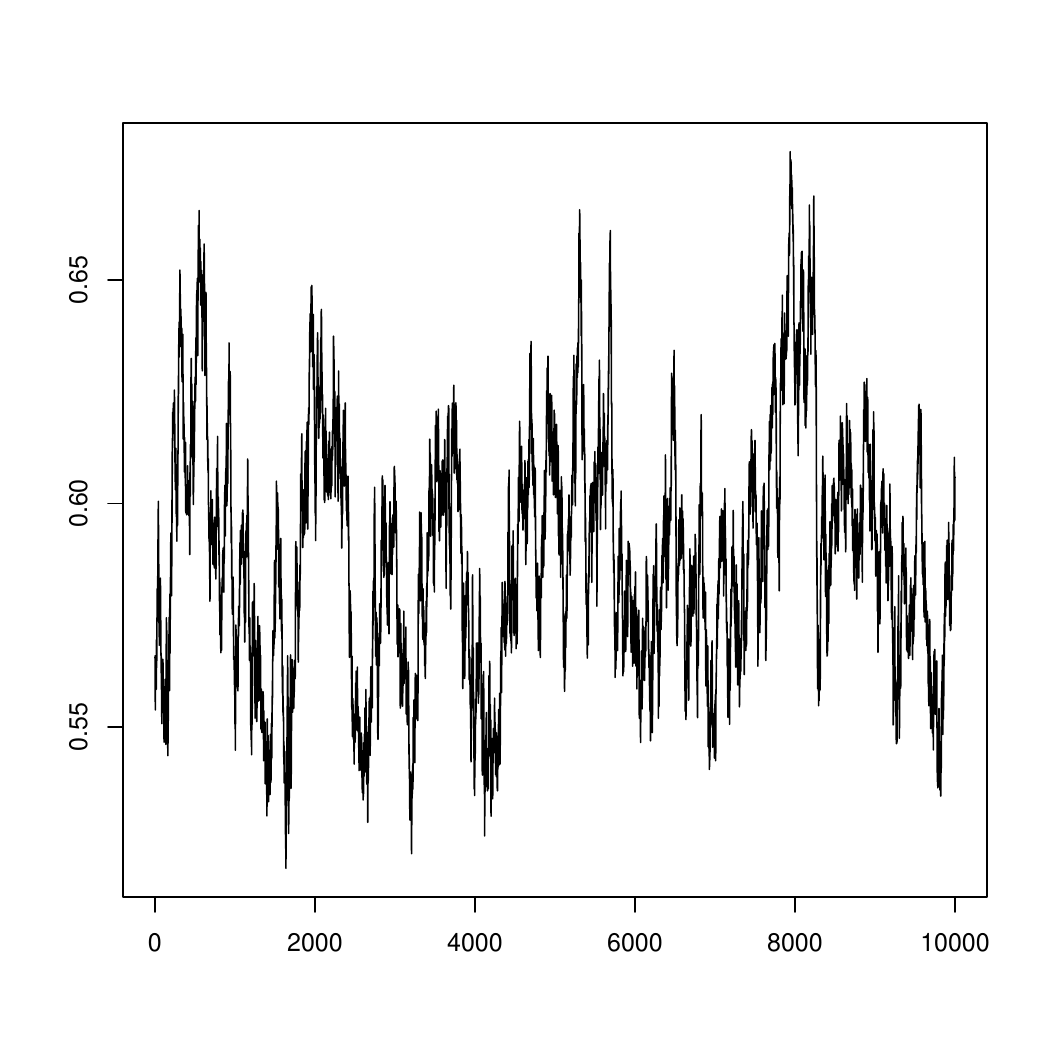}}
\hspace{2mm}
\subfigure[Trace plot of $\omega_1$.]{ \label{fig:10d_w}
\includegraphics[width=7cm,height=6cm]{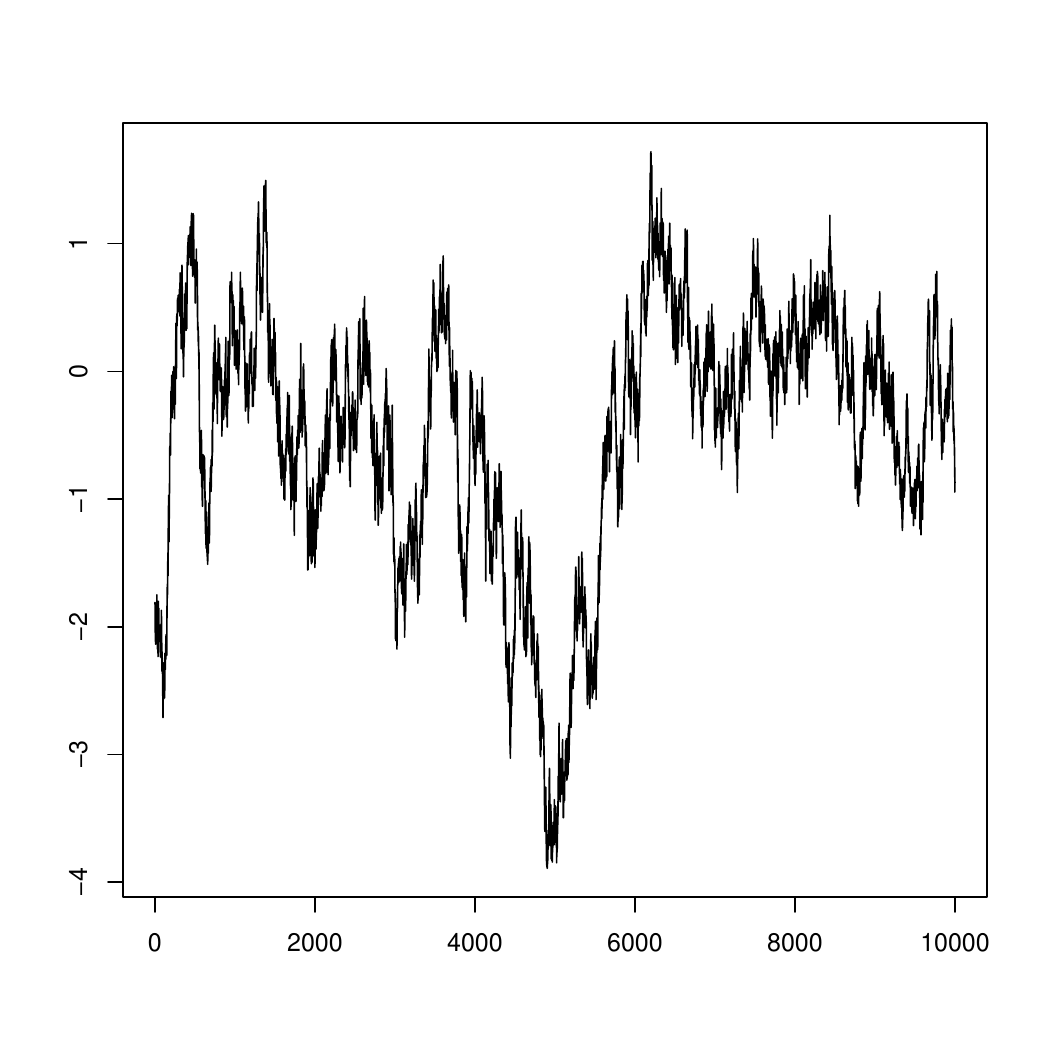}}
\caption{{\bf TTMCMC for 10-dimensional case:} Trace plots of $k$, $\mu_{11}$, $L_{11}$ and $\omega_1$. Adequate mixing
behavior of the TTMCMC chain is exhibited by the above panels.} 
\label{fig:10d_trace_plots}
\end{figure}

\begin{figure}
\includegraphics[width=7in,height=6.5in]{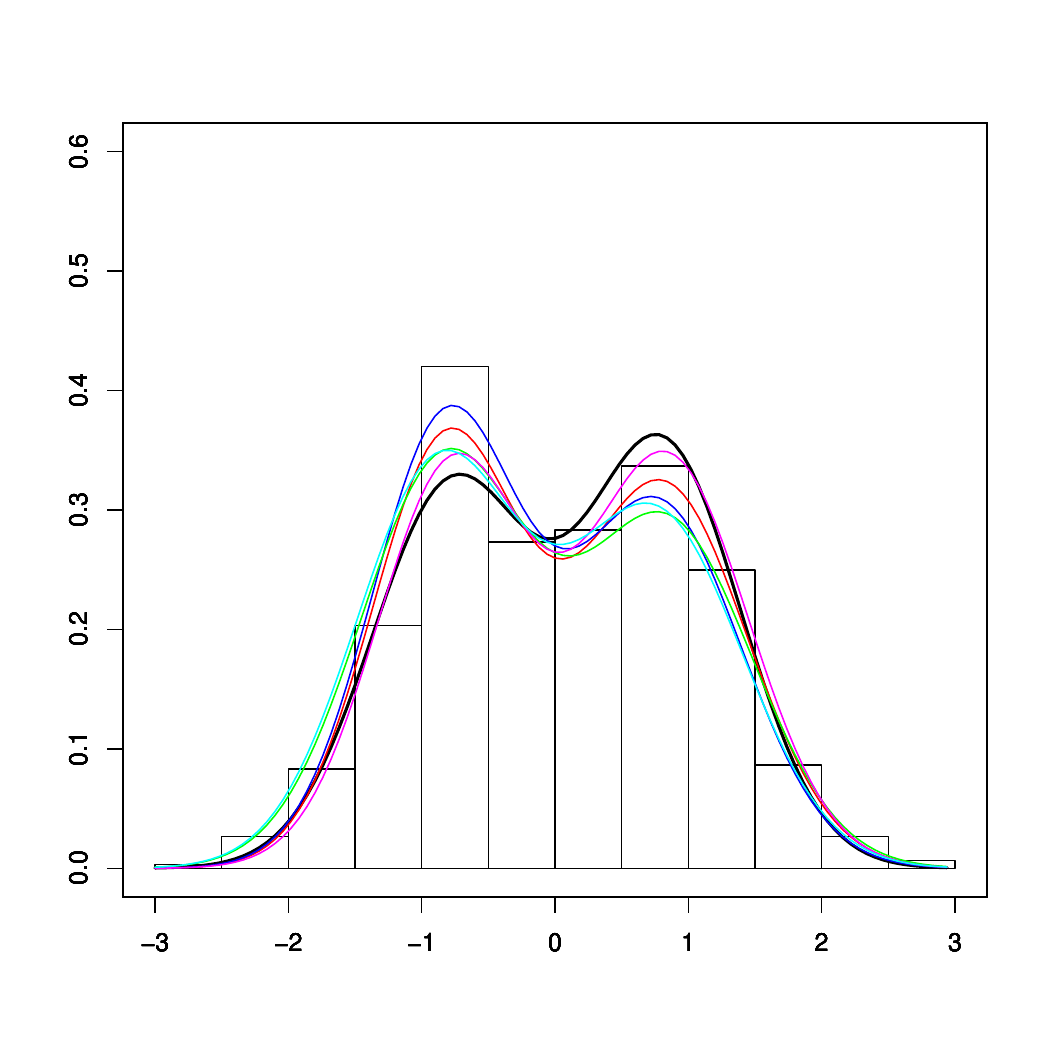}
\caption{{\bf TTMCMC for 10-dimensional case:} Goodness of fit of the posterior distribution of densities 
(coloured curves) to the histogram of the first component of the observed data $\left\{y_{i1};i=1,\ldots,600\right\}$. 
The thick black curve is the modal density and the other coloured curves are some densities contained in the 95\% HPD.}
\label{fig:10d_hpd}
\end{figure}

The trace plots and the goodness of fit (for the first co-ordinate of the $10$-dimensional data) diagram
shown in Figures \ref{fig:10d_trace_plots} and \ref{fig:10d_hpd} vindicate satisfactory performance of our
method, in spite of high dimensionality. Importantly, the correct number of components, namely, $k=2$ has been
identified correctly. 

\subsection{{\bf Simulation experiment with $p=20$}}
\label{subsec:simstudy_20d}
\begin{figure}
\centering
\subfigure[Trace plot of $k$.]{ \label{fig:20d_k}
\includegraphics[width=7cm,height=6cm]{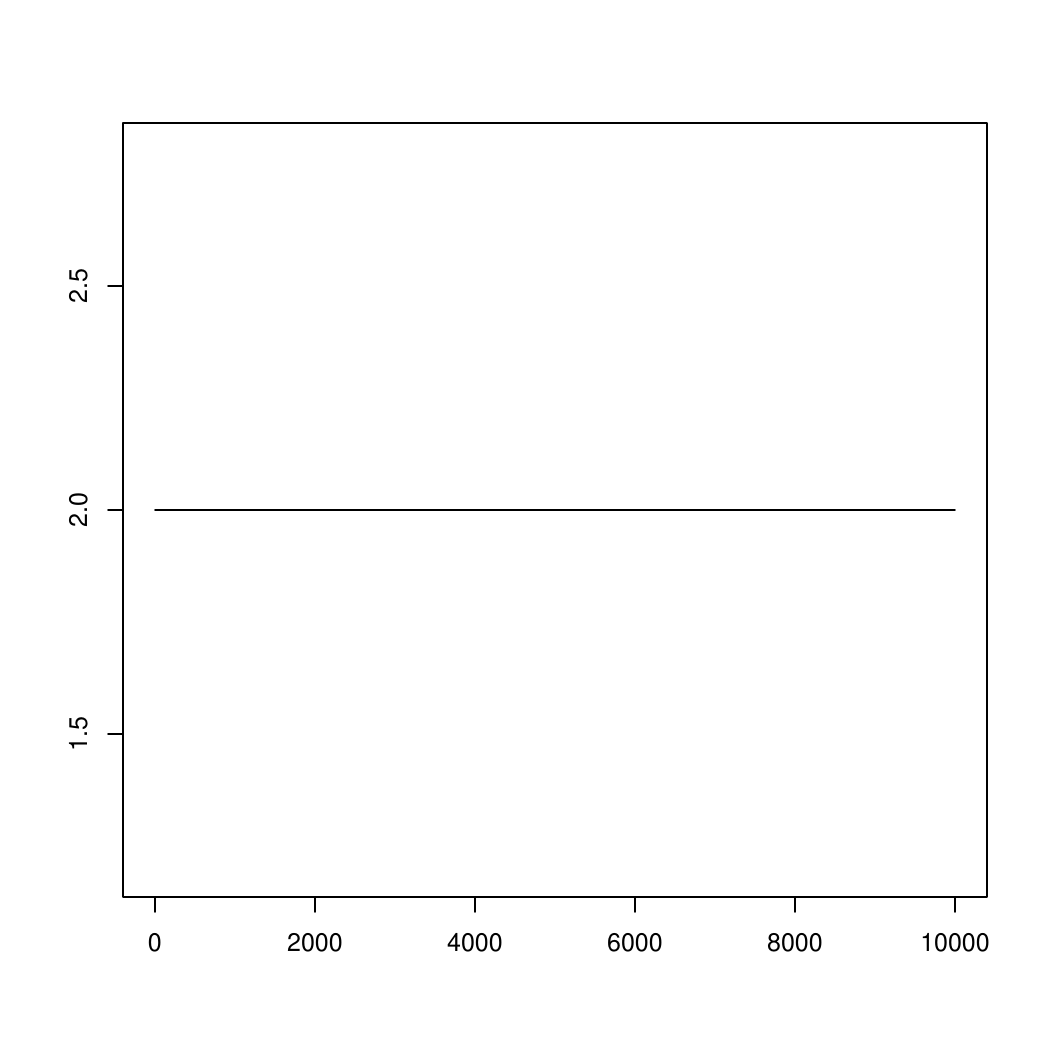}}
\hspace{2mm}
\subfigure[Trace plot of $\mu_{11}$.]{ \label{fig:20d_nu}
\includegraphics[width=7cm,height=6cm]{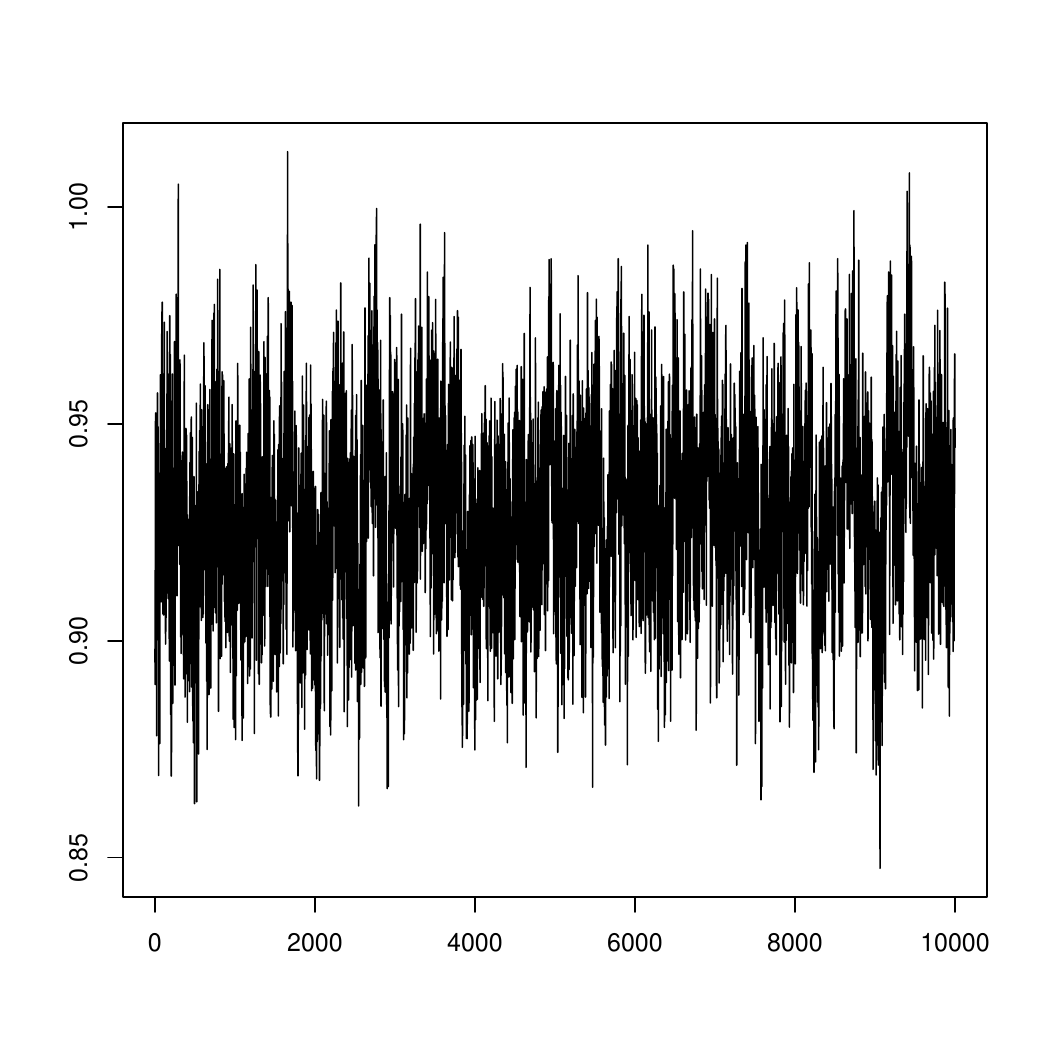}}\\
\vspace{2mm}
\subfigure[Trace plot of $L_{11}$.]{ \label{fig:20d_tau}
\includegraphics[width=7cm,height=6cm]{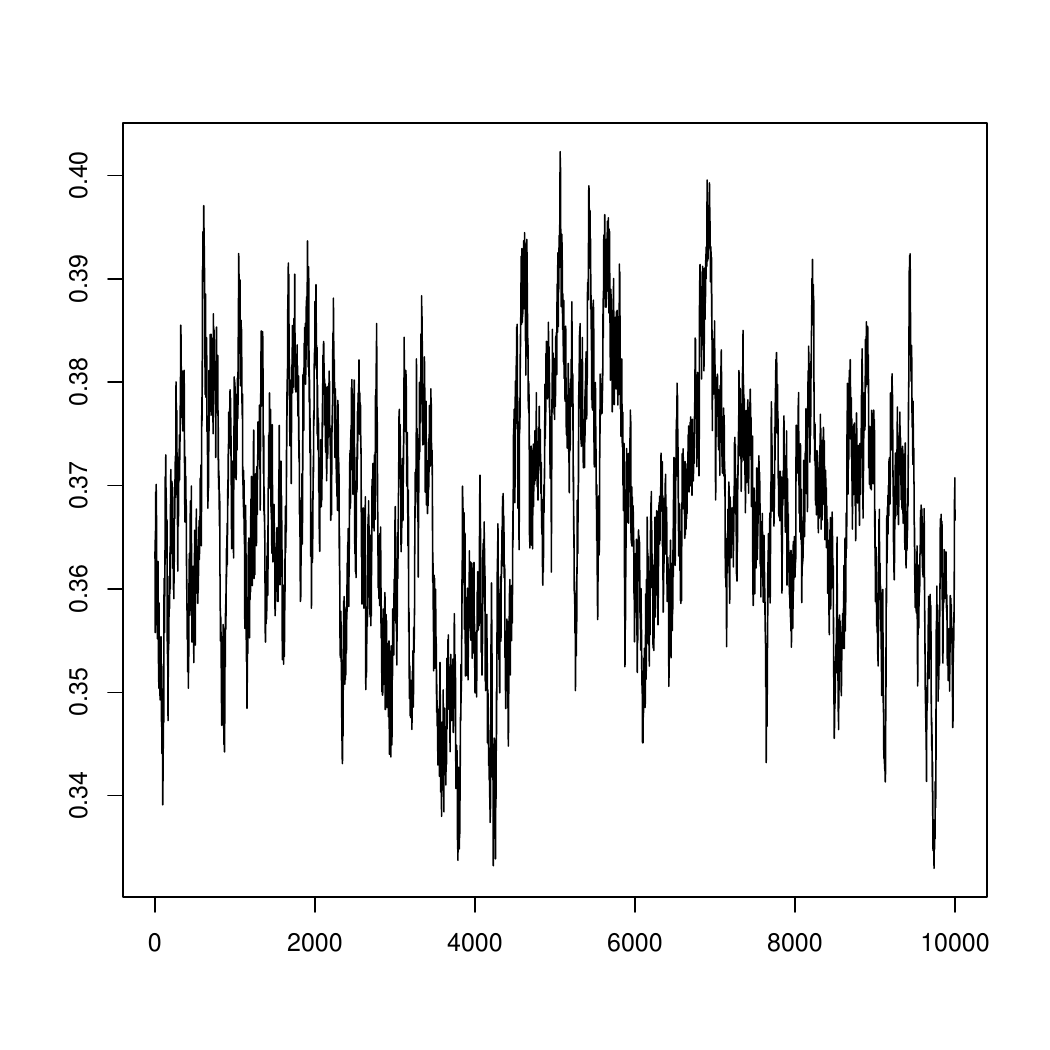}}
\hspace{2mm}
\subfigure[Trace plot of $\omega_1$.]{ \label{fig:20d_w}
\includegraphics[width=7cm,height=6cm]{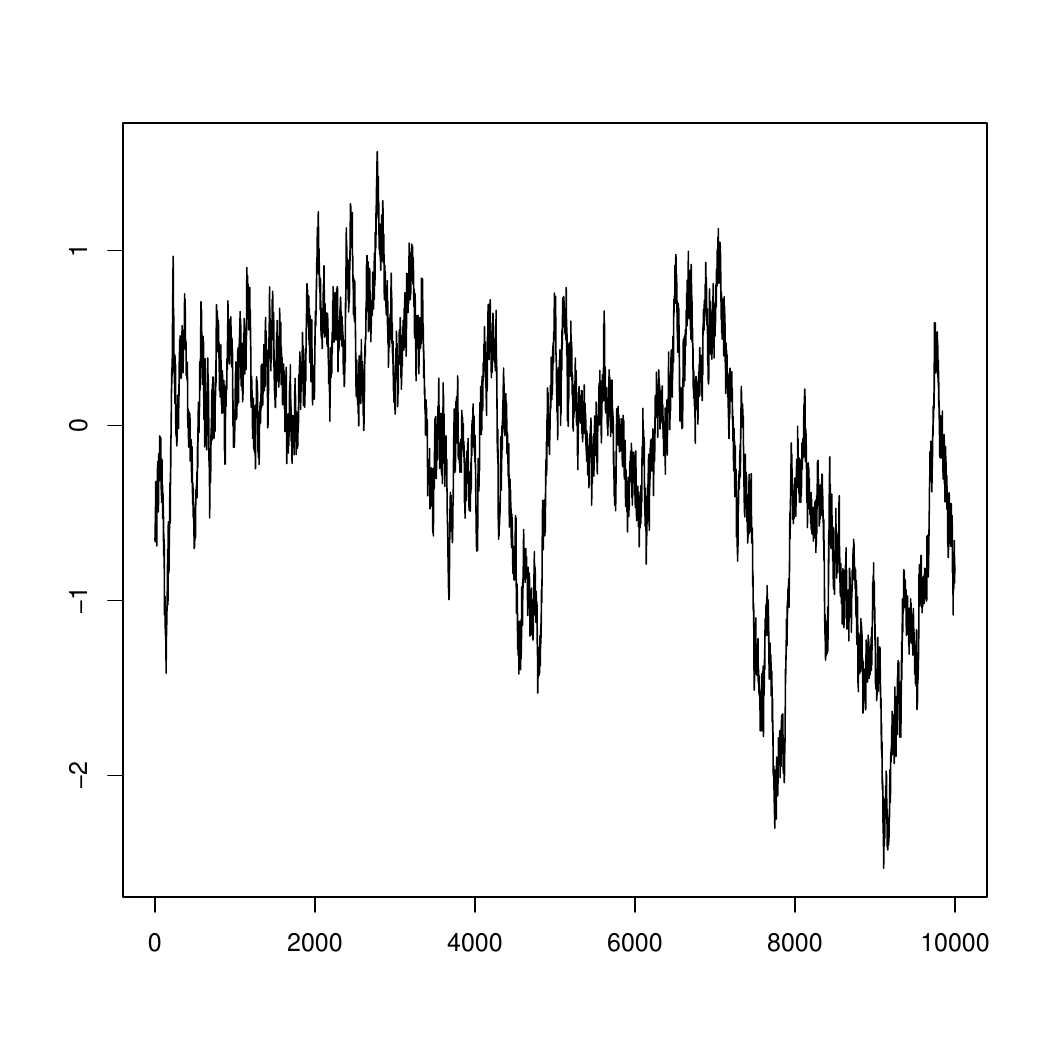}}
\caption{{\bf TTMCMC for 20-dimensional case:} Trace plots of $k$, $\mu_{11}$, $L_{11}$ and $\omega_1$. Adequate mixing
behavior of the TTMCMC chain is exhibited by the above panels.} 
\label{fig:20d_trace_plots}
\end{figure}

\begin{figure}
\includegraphics[width=7in,height=6.5in]{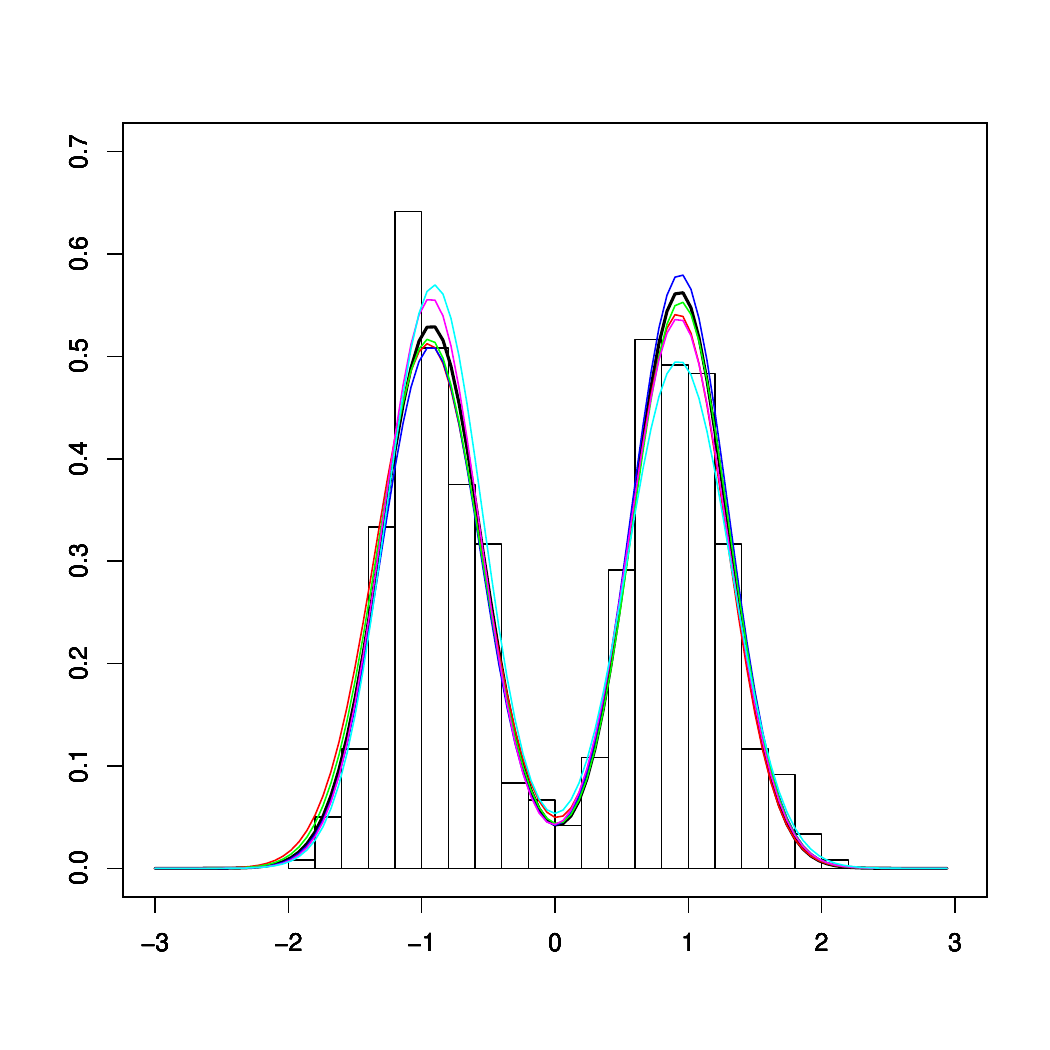}
\caption{{\bf TTMCMC for 20-dimensional case:} Goodness of fit of the posterior distribution of densities 
(coloured curves) to the histogram of the first component of the observed data $\left\{y_{i1};i=1,\ldots,600\right\}$. 
The thick black curve is the modal density and the other coloured curves are some densities contained in the 95\% HPD.}
\label{fig:20d_hpd}
\end{figure}

We conduct a further experiment, now with dimension $p=20$. Our data generation mechanism remains the same as
in Section \ref{subsubsec:simstudy_10d}, only the dimension is increased from $p=10$ to $p=20$. Our TTMCMC algorithm
also remains almost the same, with the same mixing improvement strategy. We again obtain $10,000$ samples by thinning
from a total of $15\times 10^7$ iterations. In this case, the overall acceptance rate, birth rate, death rate
and the no-change rate are $0.00741$, $0.00019$, $0.000421$ and $0.02163$, respectively. The time taken is 
$136$ hours and $44$ minutes. 
The trace plots and the goodness-of-fit diagram depicted in Figures \ref{fig:20d_trace_plots} and \ref{fig:20d_hpd}
once again speak in favour of our ideas, in particular, the great automation of our method, irrespective
of dimensions.

\section{{\bf Conclusion}}
\label{sec:conclusion}

The transformation based concepts of TMCMC in the fixed-dimensional set-up 
has led to the interesting variable-dimensional counterpart TTMCMC just as the 
traditional Metropolis-Hastings methodology has led to RJMCMC. 
Consequently, the advantages
of TMCMC over Metropolis-Hastings 
are expected to carry over to TTMCMC as compared to RJMCMC. Indeed, as we demonstrated in this paper, 
TTMCMC is simple to implement,
can update all the (variable number of) parameters in a single block while maintaining reasonable
acceptance rates thanks to drastic effective reduction of the dimensionality. 
In fact, TTMCMC effectively reduces the variable dimensional problem to a fixed dimensional problem
involving a single $\e$ or just a few, fixed number of $\e$'s, given any move type within
the birth, death or no-change moves.
The block updating strategy of TTMCMC using $\e$ or a few $\e$'s also ensures huge computational savings.
Furthermore, the mixture-type proposal distributions associated with TTMCMC ensures reasonable mixing properties.


There are three key features that manifested themselves in our comparative studies on TTMCMC and RJMCMC.
First, TTMCMC yields reasonable acceptance rates, which are larger than those of RJMCMC for the
same scales of the additive transformations. Importantly, in the gamma mixtures and the galaxy example, 
RJMCMC yields extremely poor acceptance rate, while that of TTMCMC is quite reasonable, for the same scales.

Second, ensuring reasonable mixing is a very challenging issue in variable dimensional problems. Here TTMCMC
outperforms RJMCMC very significantly in all the cases, as vindicated by the autocorrelation plots shown 
in Figure S-4 of the supplement. In other words, even in univariate situations, the random walk RJMCMC
completely fails to compete with TTMCMC. 

Third, it seems to be infeasible to devise appropriate RJMCMC move types in high-dimensional contexts. Indeed,
\ctn{Dell06} consider a maximum of only $5$-dimensional example for RJMCMC application. On the other hand, we have
demonstrated that our simple additive TTMCMC works even for dimensions as large as $20$. In this regard
it is useful to note that the split-merge proposals
of \ctn{Jain04} and \ctn{Jain07} are perhaps better candidates compared to those of \ctn{Richardson97} and \ctn{Dell06}
as they update all the allocation variables simultaneously, rather than Gibbs sampling. 
Since TTMCMC also generally updates all the variables in a single block, the general principles of their algorithm
and TTMCMC match. But a key difference is that we do not introduce allocation variables for mixture updation, and hence
have much less number of variables to update, which is expected to lead to better acceptance rate in our case.
It is also to be noted that the algorithms of \ctn{Jain04} and \ctn{Jain07} are devised for mixtures only, 
not for general variable-dimensional problems. In contrast, our default additive TTMCMC that we used for mixtures 
can be applied to all variable dimensional problems.  

A further issue with RJMCMC is that it tends to support more components than are expected.
The main issue responsible for this possible non-convergence 
is the requirement of dimension-matching for RJMCMC implementation.
This condition forces the acceptance ratio for the dimension-changing moves to depend upon the proposal density either via the
denominator (birth move) or through the numerator (death move). Thus, unlike fixed-dimensional Metropolis-Hastings, 
the acceptance ratio is not balanced by the presence of the proposal density in both numerator and denominator.
As already remarked in the discussion following Algorithm \ref{algo:ttmcmc}, 
this unbalanced nature of the RJMCMC acceptance ratio causes 
large number of birth moves if the proposal density is uniformly bounded by $1$, as in our examples. 
Since TTMCMC does not require dimension-matching, it has been possible to free the corresponding acceptance
ratio of the proposal density, which, in turn, completely solves the problem of bias towards large number of models
in finite number of iterations.

The wisdom that emerges from the investigations and the subsequent analyses 
is that even the simplest version of TTMCMC, namely, additive TTMCMC, is capable enough of exploring 
challenging variable-dimensional posteriors, providing ample support to our claim of automation
inherent within TTMCMC. On the other hand, as our implementations show, the corresponding random walk RJMCMC
do not measure up at all. In principle, there may exist 
RJMCMC algorithms which may perhaps perform reasonably in terms of convergence, but at the cost of being
problem-specific, complicated, hard-to-implement, and computationally burdensome. 

Also, very importantly, as we showed, our simple additive transformation exhibited very decent performance
even in dimension as large as $20$, thus providing a large boost to our claim of automation. 
To our knowledge, there exists no instance of RJMCMC that works in such high dimension. 

Thus, as per our experiments and knowledge, TTMCMC is remarkably close to automation, while 
automation for RJMCMC is nowhere in sight.

Apart from developing TTMCMC, we have also proposed, in a separate supplementary material, a general methodology
for summarizing the posterior distributions of densities. In particular, we have 
prescribed a procedure for obtaining the modes and desired HPD regions of the posterior distribution 
of density functions. Moreover, using these concepts as basis, we have proposed a convergence diagnostic criterion for the
underlying TTMCMC algorithm, which is again very generally applicable. 
The convergence diagnostic method seems to be particularly useful in variable-dimensional
contexts, where determining convergence is far more difficult than fixed-dimensional situations. Also, as 
we demonstrated with our applications, in the absence of optimal scaling theory in variable-dimensional situations, 
the criterion can provide guidance regarding choices of the scales of default additive TTMCMC.

Our results demonstrate that additive TTMCMC is promising enough to qualify as the default variable-dimensional
algorithm. This is also vindicated by the excellent performances of TTMCMC in challenging
spatio-temporal problems investigated by these authors and others. In this paper, we restricted ourselves
to mixture models because of their high standing in statistics and challenging nature of the associated
variable-dimensional problem. However, in a separate paper we shall present
detailed comparisons of TTMCMC and RJMCMC with respect to various other variable-dimensional problems. Our
investigations are on and we seek to establish TTMCMC as a far superior alternative compared to RJMCMC.

\section*{{\bf Acknowledgments}}
We are sincerely grateful to the two reviewers whose constructive comments 
have led to much improvement of this article.

\newpage

\renewcommand\thefigure{S-\arabic{figure}}
\renewcommand\thetable{S-\arabic{table}}
\renewcommand\thesection{S-\arabic{section}}

\setcounter{section}{0}
\setcounter{figure}{0}
\setcounter{table}{0}

\begin{center}
{\bf \Large Supplementary Material}
\end{center}

Throughout, we refer to our main paper \ctn{Das14c} as DB.

\section{{\bf Detailed balance for Algorithm 3.1 of DB}}

Before providing the proof of detailed balance in the general case, we first illustrate 
the proof with the example introduced in Section 3.1 of DB.

\subsection{{\bf Detailed balance for the simple example illustrated in Section 3.1 of DB}}
\label{subsec:detailed_balance1}

We assume the additive transformation and set $w_b=w_d=w_{nc}=\frac{1}{3}$. Also, we let $P(z_i=1)=P(z_i=-1)=p$
and the current state be $\bm x=(x_1,x_2) ~\in\mathbb R^{2}$.

\begin{figure}[H]

    \hspace{2mm}   

   \subfigure
   \centering
    \resizebox{\textwidth}{!}{
        \begin{tikzpicture}[font=\boldmath]

        \node[circle,fill=red!30,minimum size=5.5cm,font=\bfseries] (1) at (16,62) {\LARGE$x_1,x_2$};
        \node[circle,fill=gray!40,minimum size=5.5cm] (2) at (48,62) {\LARGE$x'_1,x'_2,x'_3$};
        
         \node[rectangle,fill=gray!20,minimum width=4cm,minimum height=3cm] (14) at (30,68) 
	 {\Large $x'_1=x_1+a_1\e , x'_2=x_1-a_1\e , x'_3= x_2+z_2a_2\e$};
         \node[rectangle,fill=gray!20,minimum width=4cm,minimum height=3cm] (14) at (30,55) 
	 {\Large $x_1=\frac{x'_1-a_1\e^{*}+ x'_2+a_1\e^{*}}{2},x_2=x'_3+a_1z^{c}_2\e, \e^{*}=\frac{x'_1-x'_2}{2}  $};

       \foreach \from/\to in {1/2}
        {\draw[->,ultra thick,  bend right= -22.5 fill=blue] (\from) to node[fill=white] {\large{$\epsilon$}} (\to);
            
        }
        
        \foreach \from/\to in {2/1}
        {\draw[->,ultra thick,  bend right= -22.5 fill=blue] (\from) to node[fill=white] {\large{$\epsilon$}} (\to);
        }


        \end{tikzpicture}
    }

  \caption{\bf Detailed balance condition.}
\label{fig:diagram_TTMCMC_supp}
\end{figure}
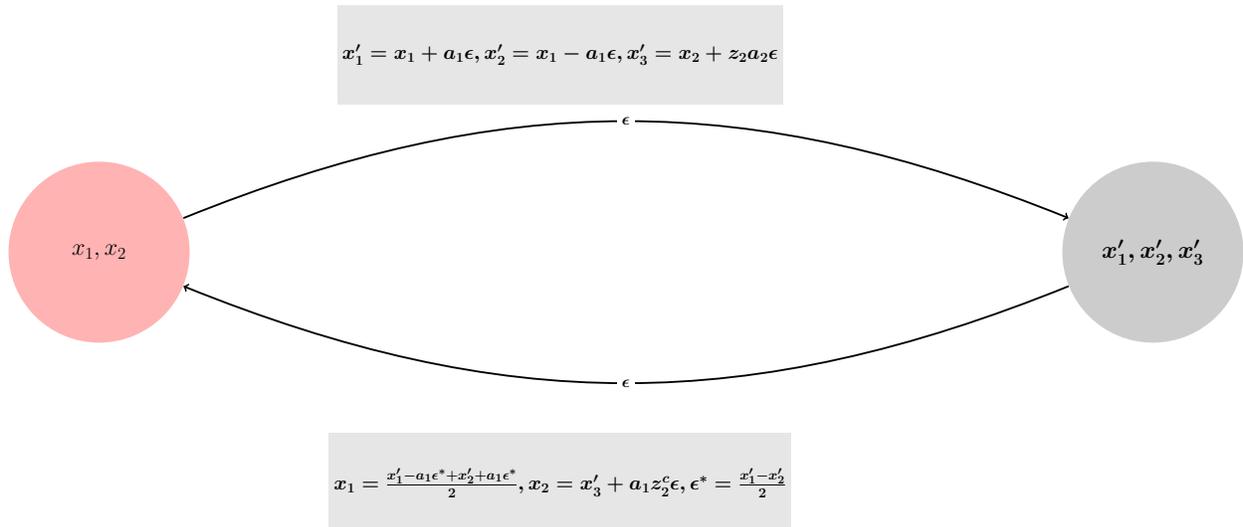

Figure \ref{fig:diagram_TTMCMC_supp} pictorially illustrates the detailed balance criterion.  
Specifically, according to our algorithm, the probability of transition $(x_1,x_2)\mapsto (x'_1,x'_2,x'_3)$ is given by:
\begin{align}
 &\pi(x_1,x_2)\times P(\text{birth move})\times P(\text{selecting one random coordinate from } x_1, x_2)\times \varrho(\e)\notag\\
 &\times P(z_2)\times a_b(\bm x,\e) \notag \\
 &=\pi(x_1,x_2)\times\frac{1}{3}\times \frac{1}{2} \times \varrho(\e)\times p\notag\\
 & \times\min\left\{1,\frac{1}{3}\times \frac{\pi(x'_1,x'_2,x'_3)}{\pi(x_1,x_2)}\times \left|\frac{\partial(x'_1,x'_2,x'_3)}{\partial(x_1,x_2)}\right|\right\}\notag\\
 &=\varrho(\e)\min\left\{\pi(x_1,x_2)\times p\times \frac{1}{6},\pi(x'_1,x'_2,x'_3)
 \times p \times \frac{1}{18}\times 2a_1\right\}.
 \label{eq:db_birth_simple}
 \end{align}
 
 For detailed balance to hold, we must be able to return from $(x'_1,x'_2,x'_3)$ to $(x_1,x_2)$. 
 The required transition, associated with the death move, has the following probability:
 \begin{align}
 &\pi(x'_1,x'_2,x'_3)\times P(\text{death move})\times P(\text{selecting } x'_1, x'_2)\times \varrho(\e)\notag\\
 &\times P(z^{c}_2)\times \left|\frac{\partial(x'_1,x'_2)}{\partial(x^{*}_1,\e^{*})}\right|\times a_d(\bm x',\e,\e^{*})\notag \\
 &=\pi(x'_1,x'_2,x'_3)\times\frac{1}{3}\times \frac{1}{3\times 2} \times \varrho(\e)\times p \times 2a_1\notag\\
 & \times\min\left\{1,3\times \frac{\pi(x_1,x_2)}{\pi(x'_1,x'_2,x'_3)}
 \times \left|\frac{\partial(x_1,x_2,\e^{*},\e)}{\partial(x'_1,x'_2,x'_3,\e)}\right|\right\}\notag\\
 &=\varrho(\e)\min\left\{\pi(x'_1,x'_2,x'_3)\times p 
 \times \frac{1}{18}\times 2a_1,\pi(x_1,x_2)\times p\times \frac{1}{6}\right\}.
\label{eq:db_death_simple}
\end{align}

So, (\ref{eq:db_birth_simple}) = (\ref{eq:db_death_simple}), implying that detailed balance holds for birth and death moves.
We now prove detailed balance for the general TTMCMC algorithm. 

\subsection{{\bf Proof of detailed balance for the general TTMCMC algorithm}}

To see that detailed balance is satisfied for the birth and death moves, note that associated with the birth move, 
the probability of transition $\bm x~(\in\mathbb R^k)\mapsto T_{b,\bz}(\bm x,\e)~(\in\mathbb R^{k+1})$ is given by:
\begin{align}
&\pi(\bm x)\times\frac{1}{k}\times w_{b,k}\times \varrho(\e)
\times\prod_{i\neq j=1}^kp^{I_{\{1\}}(z_i)}_iq^{I_{\{-1\}}(z_i)}_i\notag\\
&\times\min\left\{1, \frac{1}{k+1}\times\frac{w_{d,k+1}}{w_{b,k}}\times
\frac{\prod_{i\neq j=1}^k p^{I_{\{1\}}(z^c_i)}_iq^{I_{\{-1\}}(z^c_i)}_i}
{\prod_{i\neq j=1}^kp^{I_{\{1\}}(z_i)}_iq^{I_{\{-1\}}(z_i)}_i}
\times\frac{\pi(T_{b,\bz}(\bm x,\e))}{\pi(\bm x)}\times 
\left|\frac{\partial (T_{b,\bz}({\bm x}, \e))}{\partial({\bm x}, \e)}\right|\right\}\notag\\
&=\varrho(\e)\times\min\left\{\pi(\bm x)\times\frac{1}{k}\times w_{b,k}\times
\prod_{i\neq j=1}^kp^{I_{\{1\}}(z_i)}_iq^{I_{\{-1\}}(z_i)}_i,\right.\notag\\
&\quad\quad\quad\quad\left.\frac{1}{k(k+1)}\times w_{d,k+1}\times
\prod_{i\neq j=1}^k p^{I_{\{1\}}(z^c_i)}_iq^{I_{\{-1\}}(z^c_i)}_i
\pi(T_{b,\bz}(\bm x,\e))\times \left|\frac{\partial (T_{b,\bz}({\bm x}, \e))}{\partial({\bm x}, \e)}\right|
\right\}.
\label{eq:db_birth}
\end{align}
Here we assume that $x_j$ was selected, and was split into $g_{j,z_j=1}(x_j,\e)$ and $g_{j,z^c_j=-1}(x_j,\e)$. 
Hence, it is not necessary to simulate $z_j$.
For the remaining co-ordinates we need to simulate $z_i;~i\neq j=1,\ldots,k$.

At the reverse death move we must be able to return to $\bm x~(\in\mathbb R^k)$ from 
$T_{b,\bz}(\bm x,\e)~(\in\mathbb R^{k+1})$.
We select $g_{j,z_j=1}(x_j,\e)$ with probability $1/(k+1)$, then select $g_{j,z^c_j=-1}(x_j,\e)$ without
replacement with probability $1/k$, 
take their respective backward transformations after simulating $\epsilon\sim g$, and finally take the resultant average.
Thus, although it is not necessary to simulate $z_j$ here, we must simulate $z^c_i;~i\neq j=1,\ldots,k$
for the co-ordinates after re-labelling them appropriately to correspond to the remaining $(k+1)-2=k-1$ 
co-ordinates and $z_i;~i\neq j=1,\ldots,k$, the latter simulated in the balancing birth move.
The transition probability of the death move is hence given by:
\begin{align}
&\pi(T_{b,\bz}(\bm x,\e))\times w_{d,k+1}\times \varrho(\e)\times\prod_{i\neq j=1}^k p^{I_{\{1\}}(z^c_i)}_iq^{I_{\{-1\}}(z^c_i)}_i
\times\frac{1}{k+1}\times\frac{1}{k}
\times \left|\frac{\partial (T^{-1}_{d,\bz}({\bm x}, \e),\e^*)}{\partial({\bm x}, \e)}\right|
\notag\\
&\times\min\left\{1,(k+1)\times\frac{w_{b,k}}{w_{d,k+1}}\times\frac{\prod_{i\neq j=1}^kp^{I_{\{1\}}(z_i)}_iq^{I_{\{-1\}}(z_i)}_i}
{\prod_{i\neq j=1}^k p^{I_{\{1\}}(z^c_i)}_iq^{I_{\{-1\}}(z^c_i)}_i}
\times\frac{\pi(\bm x)}{\pi(T_{b,\bz}(\bm x,\e))}\times
\left|\frac{\partial (T_{d,\bz}({\bm x}, \e),\e^*)}{\partial({\bm x}, \e)}\right|\right\}
\notag\\
&=\varrho(\e)\times\min\left\{\pi(T_{b,\bz}(\bm x,\e))\times w_{d,k+1}\times
\prod_{i\neq j=1}^k p^{I_{\{1\}}(z^c_i)}_iq^{I_{\{-1\}}(z^c_i)}_i\times\frac{1}{k(k+1)}
\times \left|\frac{\partial (T^{-1}_{d,\bz}({\bm x}, \e),\e^*)}{\partial({\bm x}, \e)}\right|,\right.\notag\\
&\quad\quad\quad\quad \left. \frac{1}{k}\times w_{b,k}\times\prod_{i\neq j=1}^kp^{I_{\{1\}}(z_i)}_iq^{I_{\{-1\}}(z_i)}_i\times\pi(\bm x)
\right\}.
\label{eq:db_death}
\end{align}
Noting that $\left|\frac{\partial (T^{-1}_{d,\bz}({\bm x}, \e),\e^*)}{\partial({\bm x}, \e)}\right|
=\left|\frac{\partial (T_{b,\bz}({\bm x}, \e))}{\partial({\bm x}, \e)}\right|$, it follows that
(\ref{eq:db_birth}) = (\ref{eq:db_death}), showing that detailed balance holds for the birth and the death moves.
The proof of detailed balance for the no-change move type where the dimension remains unchanged is the same as that of
TMCMC, and has been been proved in the supplement of \ctn{Dutta14}.

\section{{\bf Irreducibility and aperiodicity of TTMCMC}}
\label{sec:ergodicity}
It is easy to see that our TTMCMC algorithm is irreducible and aperiodic. Assume that 
$\bx\in\mathbb R^k$, with $k\geq 1$. For $k'>0$ with $k'\neq k$, let $(k',A_{k'})$ have positive probability under
the target distribution, that is, $\pi(k',A_{k'})>0$; here $A_{k'}$ is a Borel set associated with $\mathbb R^{k'}$. 
Then $\mathbb R^{k'}$ can be reached from $\bx\in\mathbb R^k$ in a finite number of steps using the birth
and the death moves, accordingly as $k'>k$ or $k'<k$. 
Thus, if $k'>k$, $\mathbb R^{k'}$ can be reached in $(k'-k)$ steps by applying the birth move, and if $k'<k$,
then $\mathbb R^{k'}$ can be reached in $(k-k')$ steps using the death move.
Once $\mathbb R^{k'}$ is reached 
the no-change move-type and the transformations can be used to reach $A_{k'}$ in $k'$ steps. For the
proof of the latter see \ctn{Dutta14} and \ctn{Dey14}. 
Thus, $(k',A_{k'})$ can be reached from $\bx\in\mathbb R^k$ in $(|k'-k|+k')$ steps with positive probability.
Since the set $(k',A_{k'})$ is arbitrary, aperiodicity also follows.

\section{{\bf General TTMCMC algorithm for jumping $m$ dimensions}}
\label{sec:algo for jumping $m$ dimensions }
\begin{algo}\label{algo:ttmcmc2} \topline General TTMCMC algorithm for jumping $m$ dimensions.
\botline \normalfont \ttfamily
\begin{itemize}
 \item Let the initial value be $\supr{\bm x}{0}\in\mathbb R^k$, where $k\geq m$. 
 \item For $t=0,1,2,\ldots$
\begin{enumerate}
 \item Generate $u=(u_1,u_2,u_3)\sim Multinomial (1;w_{b,k},w_{d,k},w_{nc,k})$.
 \item If $u_1=1$ (increase dimension from $k$ to $k+m$), then
 \begin{enumerate}
 \item Randomly select $m$ co-ordinates from $\bm x^{(t)}=(x^{(t)}_1,\ldots,x^{(t)}_k)$ without replacement.
 Let $\bj_m=(j_1,\ldots,j_m)$ denote the chosen co-ordinates.
 \item Generate $\be_m=(\e_1,\ldots,\e_m)\stackrel{iid}{\sim} \varrho(\cdot)$ and for 
 $i=1,\ldots,k;~i\neq j_1,\ldots,j_m$, simulate  
 $z_i \sim Multinomial(1;p_i,q_i,1-p_i-q_i)$ independently. 
 \item Propose the birth move as follows: apply the transformation $x^{(t)}_i\rightarrow g_{i,z_i}(x^{(t)}_i,\e_1)$ 
 for $i\in\{1,\ldots,k\}\backslash\bj_m$ and, for each $\ell\in \bj_m$,
 split  $x^{(t)}_{\ell}$ into $g_{\ell,z_{\ell}=1}(x^{(t)}_{\ell},\e_{\ell})$ and 
 $g_{\ell,z^c_{\ell}=-1}(x^{(t)}_{\ell},\e_{\ell})$. In other words, the birth move is given by:
 \begin{align} 
 \bm x' &= T_{b,\bz}(\supr{\bm x}{t}, \be_m)=(g_{1,z_1}(x^{(t)}_1,\e_1),\ldots,
 g_{j_1-1,z_{j_1-1}}(x^{(t)}_{j_1-1},\e_1),\notag\\
& g_{j_1,{z_{j_1}=1}}(x^{(t)}_{j_1},\e_1),g_{j_1,{z^c_{j_1}=-1}}(x^{(t)}_{j_1},\e_1),
g_{j_1+1,z_{j_1+1}}(x^{(t)}_{j_1+1},\e_1),\ldots,
\notag\\
& g_{j_2-1,z_{j_2-1}}(x^{(t)}_{j_2-1},\e_1),
g_{j_2,{z_{j_2}=1}}(x^{(t)}_{j_2},\e_2),g_{j_2,{z^c_{j_2}=-1}}(x^{(t)}_{j_2},\e_2),
\notag\\
& g_{j_2+1,z_{j_2+1}}(x^{(t)}_{j_2+1},\e_1),\ldots,g_{j_m-1,z_{j_m-1}}(x^{(t)}_{j_m-1},\e_1),
g_{j_m,{z_{j_m}=1}}(x^{(t)}_{j_m},\e_m),
\notag\\
& g_{j_m,{z^c_{j_m}=-1}}(x^{(t)}_{j_m},\e_m),
g_{j_m+1,z_{j_m+1}}(x^{(t)}_{j_m+1},\e_1),\ldots,
g_{k,z_k}(x^{(t)}_k,\e_1)).\notag
\end{align}
Re-label the $k+m$ elements of $\bm x'$ as $(x'_1,x'_2,\ldots,x'_{k+m})$.
Notice that except for the co-ordinates $x^{(t)}_{j_1},x^{(t)}_{j_2},\ldots,x^{(t)}_{j_m}$, for which we use
$\epsilon_1,\epsilon_2,\ldots,\epsilon_m$ respectively for updating, for all the remaining co-ordinates 
we use only $\epsilon_1$.
\item Calculate the acceptance probability of the birth move $\bm x'$:
 \begin{align}
 a_b(\supr{\bm x}{t}, \be_m) &= 
 \min\left\{1, \frac{1}{(k+m)_m}\times\frac{w_{d,k+m}}{w_{b,k}}\times\dfrac{P_{(\bj_m)}(\bz^c)}{P_{(\bj_m)}(\bz)} 
 ~\dfrac{\pi(\bm x')}{\pi(\supr{\bm x}{t})} 
 ~\left|\frac{\partial (T_{b,\bz}(\supr{\bm x}{t}, \be_m))}{\partial(\supr{\bm x}{t}, \be_m)}\right| \right\},\notag
 \end{align}
where for integers $a>0$ and $r>0$ with $a>(r-1)$, we define $(a)_r = a\times (a-1)\times (a-r+1)$.
Also,
 \[ 
P_{(\bj_m)}(\bz)=\prod_{i\in\{1,\ldots,k\}\backslash\bj_m}p^{I_{\{1\}}(z_i)}_iq^{I_{\{-1\}}(z_i)}_i,
 \]
 and
 \[ 
P_{(\bj_m)}(\bz^c)=\prod_{i\in\{1,\ldots,k\}\backslash\bj_m}p^{I_{\{1\}(z^c_i)}}_iq^{I_{\{-1\}}(z^c_i)}_i.
 \]

\item Set \[ \supr{\bm x}{t+1}= \left\{\begin{array}{ccc}
 \bm x' & \mbox{ with probability } & a_b(\supr{\bm x}{t},\e) \\
 \supr{\bm x}{t}& \mbox{ with probability } & 1 - a_b(\supr{\bm x}{t},\e).
\end{array}\right.\]
\end{enumerate}
\item If $u_2=1$ (decrease dimension from $k$ to $k-m$, for $k\geq 2m$), then
 \begin{enumerate}
 \item Generate $\be_m=(\e_1,\ldots,\e_m)\stackrel{iid}{\sim} \varrho(\cdot)$. 
 \item Randomly, without replacement, select co-ordinates $\bj_m=(j_1,\ldots,j_m)$ and 
 $\bj'_m=(j'_1,\ldots,j'_m)$ from $\bx=(x_1,\ldots,x_k)$. For $\ell=1,\ldots,m$, 
 let $x^*_{j_{\ell}}=\left(g_{j_{\ell},z^c_{j_{\ell}}=-1}(x_{j_{\ell}},\e_{\ell})
 +g_{j'_{\ell},z_{j'_{\ell}}=1}(x_{j'_{\ell}},\e_{\ell})\right)/2$; 
 replace the co-ordinate $x_{j_{\ell}}$
 by the average $x^*_{j_{\ell}}$ and delete $x_{j'_{\ell}}$.
 \item Simulate $\bz$ by generating independently, for $i\in\{1,\ldots,k\}\backslash\bj_m$, 
 $z_i \sim Multinomial(1;p_i,q_i,1-p_i-q_i)$.
 \item For $i\in\{1,\ldots,k\}\backslash\bj_m$, apply the transformation $x'_i=g_{i,z_i}(x^{(t)}_i,\e_1)$.  
 \item Propose the following death move:
 \begin{align} 
 \bm x' &= T_{d,\bz}(\supr{\bm x}{t}, \be_m)\notag\\
 &=(g_{1,z_1}(x^{(t)}_1,\e_1),\ldots,
 g_{j_1-1,z_{j_1-1}}(x^{(t)}_{j_1-1},\e_1),x^*_{j_1},g_{j_1+1,z_{j_1+1}}(x^{(t)}_{j_1+1},\e_1),\notag\\
 &\ldots,g_{j_2-1,z_{j_2-1}}(x^{(t)}_{j_2-1},\e_1),x^*_{j_2},g_{j_2+1,z_{j_2+1}}(x^{(t)}_{j_2+1},\e_1),\notag\\
&\ldots,g_{j_m-1,z_{j_m-1}}(x^{(t)}_{j_m-1},\e_1), x^*_{j_m},
g_{j_m+1,z_{j_m+1}}(x^{(t)}_{j_m+1},\e_1),
\ldots,g_{k,z_k}(x^{(t)}_k,\e_1)).\notag
\end{align}
Re-label the elements of $\bm x'$ as $(x'_1,x'_2,\ldots,x'_{k-m})$.
 \item For $\ell=1,\ldots,m$, solve for $\e^*_{\ell}$ from the equations 
 $g_{\ell,z_{\ell}=1}(x^*_{\ell},\e^*_{\ell})=x_{j_{\ell}}$ and 
 $g_{\ell,z^c_{j_{\ell}}=-1}(x^*_{j_{\ell}},\e^*_{\ell})=x_{j'_{\ell}}$
and express $\e^*_{\ell}$ in terms of $x_{j_{\ell}}$ and $x_{j'_{\ell}}$.
Let $\be^*_m=(\e^*_1,\ldots,\e^*_m)$.
\item Calculate the acceptance probability of the death move:
$ a_d(\supr{\bm x}{t}, \be_m, \be^*_m)$ 
 \begin{align}
 &= 
 \min\left\{1,(k)_m\times\frac{w_{b,k-m}}{w_{d,k}}\times\dfrac{P_{(\bj_m,\bj'_m)}(\bz^c)}{P_{(\bj,\bj'_m)}(\bz)} 
 ~\dfrac{\pi(\bm x')}{\pi(\supr{\bm x}{t})} 
 ~\left|\frac{\partial (T_{d,\bz}(\supr{\bm x}{t}, \be_m),\be^*_m,\be_m)}
 {\partial(\supr{\bm x}{t},\be_m)}\right| \right\},\notag
 \end{align}
 where
 \[ 
P_{(\bj_m,\bj'_m)}(\bz)=\prod_{i\in\{1,\ldots,k\}\backslash\{\bj_m,\bj'_m\}}p^{I_{\{1\}}(z_i)}_iq^{I_{\{-1\}}(z_i)}_i,
 \]
 and
 \[ 
P_{(\bj_m,\bj'_m)}(\bz^c)=\prod_{i\in\{1,\ldots,k\}\backslash\{\bj_m,\bj'_m\}}p^{I_{\{1\}(z^c_i)}}_iq^{I_{\{-1\}}(z^c_i)}_i.
 \]
\item Set \[ \supr{\bm x}{t+1}= \left\{\begin{array}{ccc}
 \bm x' & \mbox{ with probability } & a_d(\supr{\bm x}{t},\be_m,\be^*_m) \\
 \supr{\bm x}{t}& \mbox{ with probability } & 1 - a_d(\supr{\bm x}{t},\be_m,\be^*_m).
\end{array}\right.\]
 \end{enumerate}
\item If $u_3=1$ (dimension remains unchanged), then implement steps (1), (2), (3) of Algorithm 3.1 
of \ctn{Dutta14}. 
\end{enumerate}
\item End for
\end{itemize}
\botline \rmfamily
\end{algo}

\section{{\bf Proof of detailed balance for General TTMCMC algorithm for jumping $m$ dimensions}}
\label{sec:detailed_balance}

To see that detailed balance is satisfied for the birth and death moves, note that associated with the birth move, 
the probability of transition $\bm x~(\in\mathbb R^k)\mapsto T_{b,\bz}(\bm x,\be_m)~(\in\mathbb R^{k+m})$, with
$k\geq m$, is given by:
\begin{align}
&\pi(\bm x)\times 
\frac{1}{(k)_m}
\times w_{b,k}\times \prod_{i=1}^m\varrho(\e_i)\times\prod_{i\in\{1,\ldots,k\}\backslash\bj_m}
p^{I_{\{1\}}(z_i)}_iq^{I_{\{-1\}}(z_i)}_i\notag\\
&\times\min\left\{1, \frac{1}{(k+m)_m}\times\frac{w_{d,k+m}}{w_{b,k}}\times
\frac{\prod_{i\in\{1,\ldots,k\}\backslash\bj_m} p^{I_{\{1\}}(z^c_i)}_iq^{I_{\{-1\}}(z^c_i)}_i}
{\prod_{i\in\{1,\ldots,k\}\backslash\bj_m}p^{I_{\{1\}}(z_i)}_iq^{I_{\{-1\}}(z_i)}_i}\right.\notag\\
&\hspace{6cm}\left.
\times\frac{\pi(T_{b,\bz}(\bm x,\be_m))}{\pi(\bm x)}\times 
\left|\frac{\partial (T_{b,\bz}(\supr{\bm x}{t}, \be_m))}{\partial(\supr{\bm x}{t}, \be_m)}\right|\right\}\notag\\
&=\prod_{i=1}^m\varrho(\e_i)\times\min\left\{\pi(\bm x)\times w_{b,k}\times
\frac{1}{(k)_m} 
\times\prod_{i\in\{1,\ldots,k\}\backslash\bj_m}p^{I_{\{1\}}(z_i)}_iq^{I_{\{-1\}}(z_i)}_i,
\frac{1}{(k)_m}\times\frac{1}{(k+m)_m} 
\right.\notag\\
&\hspace{1cm}\left.\times w_{d,k+m}\times\prod_{i\in\{1,\ldots,k\}\backslash\bj_m} 
p^{I_{\{1\}}(z^c_i)}_iq^{I_{\{-1\}}(z^c_i)}_i
\pi(T_{b,\bz}(\bm x,\be_m))\times \left|\frac{\partial (T_{b,\bz}(\supr{\bm x}{t}, \be_m))}{\partial(\supr{\bm x}{t}, \be_m)}\right|
\right\}.
\label{eq:db_birth2}
\end{align}
%


The transition probability of the reverse death move is given by:
\begin{align}
&\pi(\bm x)\times w_{d,k+m}\times \prod_{i=1}^m\varrho(\e_i)\times\prod_{i\in\{1,\ldots,k\}\backslash\bj_m} 
p^{I_{\{1\}}(z^c_i)}_iq^{I_{\{-1\}}(z^c_i)}_i\notag\\
&\hspace{4cm}\times\frac{1}{(k+m)_m}\times\frac{1}{(k)_m}
\times \left|\frac{\partial (T^{-1}_{d,\bz}(\supr{\bm x}{t}, \be_m),\be^*_m,\be_m)}{\partial(\supr{\bm x}{t}, \be_m)}\right|
\notag\\
&\times\min\left\{1,(k+m)_m\times\frac{w_{b,k}}{w_{d,k+m}}
\times\frac{\prod_{i\in\{1,\ldots,k\}\backslash\bj_m}p^{I_{\{1\}}(z_i)}_iq^{I_{\{-1\}}(z_i)}_i}
{\prod_{i\in\{1,\ldots,k\}\backslash\bj_m} p^{I_{\{1\}}(z^c_i)}_iq^{I_{\{-1\}}(z^c_i)}_i}\right.\notag\\
&\hspace{4cm}\left.
\times\frac{\pi(\bm x)}{\pi(T_{b,\bz}(\bm x,\be_m))}\times
\left|\frac{\partial (T_{d,\bz}(\supr{\bm x}{t}, \be_m),\be^*_m,\be_m)}{\partial(\supr{\bm x}{t}, \be_m)}\right|\right\}
\notag\\
&=\prod_{i=1}^m \varrho(\e_i)\times\min\left\{\pi(T_{b,\bz}(\bm x,\be_m))\times w_{d,k+m}\times
\prod_{i\in\{1,\ldots,k\}\backslash\bj_m} p^{I_{\{1\}}(z^c_i)}_iq^{I_{\{-1\}}(z^c_i)}_i\right.\notag\\
&\left. \hspace{4cm}\times\frac{1}{(k)_m}\times\frac{1}{(k+m)_m}
\times \left|\frac{\partial (T^{-1}_{d,\bz}(\supr{\bm x}{t}, \be_m),\be^*_m,\be_m)}
{\partial(\supr{\bm x}{t}, \be_m)}\right|,\right.\notag\\
&\hspace{4cm} \left. \frac{1}{(k)_m}\times w_{b,k}\times\prod_{i\in\{1,\ldots,k\}\backslash\bj_m}
p^{I_{\{1\}}(z_i)}_iq^{I_{\{-1\}}(z_i)}_i\times\pi(\bm x)
\right\}.
\label{eq:db_death2}
\end{align}
Noting that $\left|\frac{\partial (T^{-1}_{d,\bz}(\supr{\bm x}{t}, \be_m),\be^*_m,\be_m)}{\partial(\supr{\bm x}{t}, \be^*_m,\be_m)}\right|
=\left|\frac{\partial (T_{b,\bz}(\supr{\bm x}{t}, \be_m))}{\partial(\supr{\bm x}{t}, \be_m)}\right|$, it follows that
(\ref{eq:db_birth2}) = (\ref{eq:db_death2}), showing that detailed balance holds for the birth and the death moves.

\section{{\bf Jumping more than one dimensions at a time when there several sets of parameters are related}}
\label{sec:jump_more2}

It is often the case that changing dimension of one set of parameters forces changing dimension of the other
sets of parameters accordingly.
For instance, in a mixture problem with unknown
number of components, 
where the $i$-th component is characterized by the 
mean and standard deviation $(\mu_i,\sigma_i)$, when the dimension of the current $k$-dimensional
mean vector $(\mu_1,\ldots,\mu_k)$ is increased by one, then one must simultaneously increase the dimension
of the current $k$-dimensional vector of standard deviations $(\sigma_1,\ldots,\sigma_k)$ by one. 
In this section we extend TTMCMC to general situations of this kind. 

For an illustrative example, assume that the TTMCMC chain is currently at the state
$$\{(\mu_1,\log(\sigma_1)),(\mu_2,\log(\sigma_2))\}=
(\mu_1,\mu_2,\log(\sigma_1),\log(\sigma_2))\in\mathbb R^2\times\mathbb R^2.$$
Let $\bm x=(x_1,x_2,x_3,x_4)=(\mu_1,\mu_2,\log(\sigma_1),\log(\sigma_2))$.
Suppose that it is required to increase the dimension to $\mathbb R^3\times \mathbb R^3$ using
the additive transformation.

To achieve consistency with respect to dimensions such that the Jacobian is well-defined, 
we need to simulate two $\e$'s 
from $\varrho(\cdot)$: $\e_1$ for
splitting $x_1$ into $x_1+a_1\e_1$ and $x_1-a_1\e_1$, and $\e_2$ for splitting $x_3$ into $x_3+a_3\e_2$
and $x_3-a_3\e_2$. With the same $\e_1$ we can also update $x_2$ to $x_2+z_2a_2\e_1$, and 
$x_4$ to $x_4+z_4a_4\e_1$. 
Note that it is possible to use $\e_2$ to split $x_3$ into $x_3+a_3\e_2$
and $x_3-a_3\e_2$, and to update $x_4$ to $x_4+z_4a_4\e_2$, instead of using $\e_1$ to update $x_4$ to $x_4+z_4a_4\e_1$.
That is, we can use $\e_1$ and $\e_2$ for updating the sub-blocks $(\mu_1,\mu_2)$ and $(\log(\sigma_1),\log(\sigma_2))$,
respectively. However, using $\e_1$ for both the sub-blocks induces dependence between the updates through the common $\e_1$
and hence may be desirable since we are updating all the sub-blocks in a single block. Hence, in this article,
we confine ourselves to using a common $\e_1$ across the sub-blocks.

Hence, in this example, the birth move takes the form
$\bx'=T_{b,z_2,z_4}(\bx,\e_1,\e_2)=(x_1+a_1\e_1,x_1-a_1\e_1,x_2+z_2a_2\e_1,x_3+a_3\e_2,x_3-a_3\e_2,x_4+z_4a_4\e_1)
=(x'_1,x'_2,x'_3,x'_4,x'_5,x'_6)$.
Now the dimensions of both $\bx'=(x'_1,x'_2,x'_3,x'_4,x'_5,x'_6)$ and $(\bx,\e_1,\e_2)=(x_1,x_2,x_3,x_4,\e_1,\e_2)$
is 6, and so the Jacobian 
$$\left|\frac{\partial (T_{b,z_2,z_4}(\bx, \e_1,\e_2))}{\partial(\bx, \e_1,\e_2)}\right|
=\left|\frac{\partial (x_1+a_1\e_1,x_1-a_1\e_1,x_2+z_2a_2\e_1,x_3+a_3\e_2,x_3-a_3\e_2,x_4+z_4a_4\e_1)}
{\partial(x_1,x_2,x_3, x_4,\e_1,\e_2)}\right|=4a_1a_3,$$
is well-defined.
The acceptance probability of the birth move in this example is given by
\begin{align}
a_b(\bm x,\e_1,\e_2)
&=\min\left\{1, \frac{1}{3}\times\frac{w_{d,6}}{w_{b,4}}\times\prod_{i=2,4}\frac{p^{I_{\{1\}}(z^c_i)}_3q^{I_{\{-1\}}(z^c_i)}_i}
{p^{I_{\{1\}}(z_i)}_iq^{I_{\{-1\}}(z_i)}_i}
\times\frac{\pi(\bx')}
{\pi(\bx)}\times
\left|\frac{\partial (T_{b,\bz}(\bm x, \e_1,\e_2))}{\partial(\bm x, \e_1,\e_2)}\right|
\right\}\notag\\
&=\min\left\{1, \frac{1}{3}\times\frac{w_{d,6}}{w_{b,4}}\times\prod_{i=2,4}\frac{p^{I_{\{1\}}(z^c_i)}_iq^{I_{\{-1\}}(z^c_i)}_i}
{p^{I_{\{1\}}(z_i)}_iq^{I_{\{-1\}}(z_i)}_i}
\frac{\pi(\bx')}{\pi(\bx)}\times
4a_1a_3
\right\}.\notag\\
\label{eq:acc_birth_1}
\end{align}

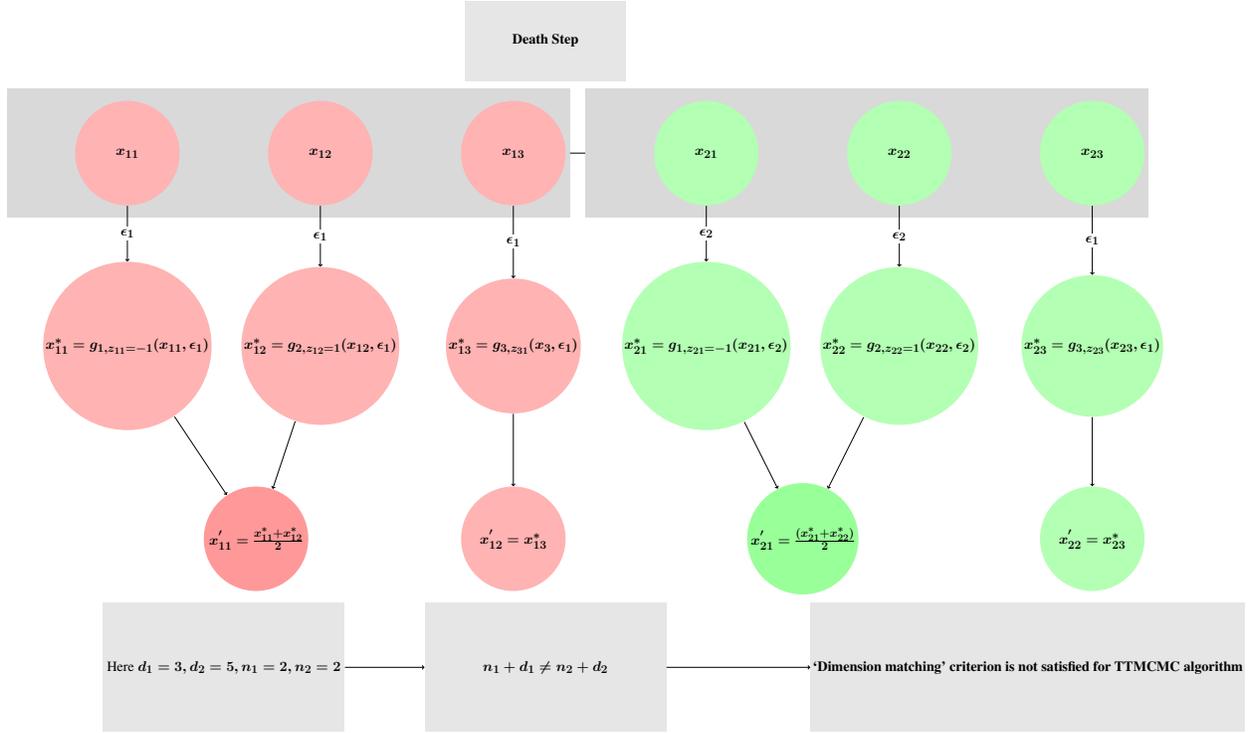
\begin{figure}[H]
    \centering
    \resizebox{\textwidth}{!}{
        \begin{tikzpicture}[ultra thick,font=\boldmath]

        \node[rectangle,fill=gray!20,minimum width =10cm,minimum height=5cm] (20) at (38,27) {\Huge \bf Death Step};
         \node[rectangle,fill=gray!30,minimum width =35cm,minimum height=8cm] (21) at (22,20) {\Huge $\pmb X_1$};
        \node[rectangle,fill=gray!30,minimum width =35cm,minimum height=8cm] (22) at (58,20) {\Huge \bf $X_2$};
        \node[circle,fill=red!30,minimum size=6.5cm] (1) at (12,20) {\Huge$ {x_{11}}$};
        \node[circle,fill=red!30,minimum size=6.5cm] (2) at (24,20) {\Huge$ {x_{12}}$};
        \node[circle,fill=red!30,minimum size=6.5cm] (3) at (36,20) {\Huge$ x_{13}$};
        \node[circle,fill=green!30,minimum size=6.5cm] (4) at (48,20) {\Huge$ x_{21}$};
        \node[circle,fill=green!30,minimum size=6.5cm] (5) at (60,20) {\Huge$ x_{22}$};
        \node[circle,fill=green!30,minimum size=6.5cm] (6) at (72,20) {\Huge$ x_{23}$};
        \node[circle,fill=red!30,minimum size=6.5cm] (7) at (12,8) {\Huge\Huge${x^{*}_{11}=g_{1,z_{11}=-1}(x_{11},\epsilon_1)}$};
        \node[circle,fill=red!30,minimum size=6.5cm] (8) at (24,8) {\Huge$ {x^{*}_{12}=g_{2,z_{12}=1}(x_{12},\epsilon_1)}$};
        \node[circle,fill=red!30,minimum size=6.5cm] (9) at (36,8) {\Huge$ {x^{*}_{13}=g_{3,z_{31}}(x_3,\epsilon_1)}$};
        \node[circle,fill=green!30,minimum size=6.5cm](10) at (48,8) {\Huge${x^{*}_{21}=g_{1,z_{21}=-1}(x_{21},\epsilon_2)}$};
        \node[circle,fill=green!30,minimum size=6.5cm] (11) at (60,8) {\Huge${x^{*}_{22}=g_{2,z_{22}=1}(x_{22},\epsilon_2)}$};
        \node[circle,fill=green!30,minimum size=8cm] (12) at (72,8) {\Huge$ {x^{*}_{23}=g_{3,z_{23}}(x_{23},\epsilon_1)}$};
        \node[circle,fill=red!40,minimum size=6.5cm] (13) at (20,-4) {\Huge${x^{'}_{11}=\frac{x^{*}_{11}+x^{*}_{12}}{2}}$};
         \node[circle,fill=red!30,minimum size=6.5cm] (14) at (36,-4) {\Huge $ {x^{'}_{12}=x^{*}_{13}}$};
        \node[circle,fill=green!40,minimum size=6.5cm] (15) at (54,-4) {\Huge$ {x^{'}_{21}=\frac{(x^{*}_{21}+x^{*}_{22})}{2}}$};
        \node[circle,fill=green!30,minimum size=6.5cm] (16) at (72,-4) {\Huge ${x^{'}_{22}=x^{*}_{23}}$};
        \node[rectangle,fill=gray!20,minimum width =15cm,minimum height=8cm] (17) at (18,-12) {\Huge Here $d_1=3, d_2=5,n_1=2,n_2=2$};
        \node[rectangle,fill=gray!20,minimum width =15cm,minimum height=8cm] (18) at (38,-12) { \Huge $n_1+d_1\neq n_2+d_2$};
        \node[rectangle,fill=gray!20,minimum width =15cm,minimum height=8cm] (19) at (68,-12) {\Huge \bf `Dimension matching' criterion is not satisfied for TTMCMC algorithm};
       
       \foreach \from/\to in {1/7,2/8,3/9,6/12}
        {\draw[->,ultra thick, fill=blue] (\from) to node[fill=white] {\Huge{$\epsilon_1$}} (\to);
        }
         \foreach \from/\to in {4/10,5/11}
        {\draw[->,ultra thick, fill=blue] (\from) to node[fill=white] {\Huge{$\epsilon_2$}} (\to);
        }
      
        {\draw[->,ultra thick, fill=blue] (7) -- (13);
          \draw[->,ultra thick, fill=blue] (8) -- (13);
          \draw[->,ultra thick, fill=blue] (9) -- (14);
           \draw[->,ultra thick, fill=blue] (10) -- (15);
           \draw[->,ultra thick, fill=blue] (11) -- (15);
           \draw[->,ultra thick, fill=blue] (12) -- (16);
           \draw[->,ultra thick, fill=blue] (17) -- (18);
           \draw[->,ultra thick,fill=blue] (18) -- (19);
           \draw[-,ultra thick,fill=blue] (21) -- (22);

        }
        
        \end{tikzpicture}
    }
     \caption{\bf Illustration of TTMCMC algorithm for jumping more than one dimension when 
     several sets of parameters are related.}

\label{fig:diagram_related}
\end{figure}

For the corresponding death move, that is, for moving from $\bx'=(x'_1,x'_2,x'_3,x'_4,x'_5,x'_6)$
to $\bx''=T_{d,\bz}(\bx',\e_1)=(\frac{x'_1+x'_2}{2},x'_3+z^c_2a_2\e_1,\frac{x'_4+x'_5}{2},x'_6+z^c_4a_4\e_1)
=(x''_1,x''_2,x''_3,x''_4)$, 
we must have, for the reverse of this death move,
$x''_1+a_1\e^*_1=x'_1$, $x''_1-a_1\e^*_1=x'_2$, $x''_3+a_3\e^*_2=x'_4$, 
$x''_3-a_3\e^*_2=x'_5$.
The first two equations yield $\e^*_1=\frac{x'_1-x'_2}{2a_1}$ and the last two equations yield 
$\e^*_2=\frac{x'_4-x'_5}{2a_3}$.
The Jacobian is given by
\begin{align}
&\left|\frac{\partial (T_{d,z_2,z_4}(\bx', \e_1);\e^*_1,\e^*_2,\e_1)}{\partial(\bx', \e_1)}\right|
=\left|\frac{\partial \left(\frac{x'_1+x'_2}{2},x'_3+z^c_2a_2\e_1,\frac{x'_4+x'_5}{2},x'_6+z^c_4a_4\e_1,\frac{x'_1-x'_2}{2a_1},
\frac{x'_4-x'_5}{2a_3},\e_1\right)}{\partial\left(x'_1,x'_2,x'_3,x'_4, x'_5,x'_6,\e_1\right)}\right|=\frac{1}{4a_1a_3}.
\end{align}
We accept this death move with probability
 \begin{align}
 a_d(\bm x'', \e_1, \e^*_1,\e^*_2) &= 
 \min\left\{1,3\times\frac{w_{b,4}}{w_{d,6}}\times\dfrac{P(\bz^c)}{P(\bz)} ~\dfrac{\pi(\bm x'')}{\pi(\bm x')} 
 ~\left|\frac{\partial (T_{d,\bz}(\bx', \e_1);\e^*_1,\e^*_2,\e_1)}{\partial(\bx',\e_1)}\right| \right\}\notag\\
 &= \min\left\{1,3\times\frac{w_{b,4}}{w_{d,6}}\times\prod_{i=2,4}\dfrac{p^{I_{\{1\}}(z_i)}_iq^{I_{\{-1\}}(z_i)}_i}
 {p^{I_{\{1\}}(z^c_i)}_iq^{I_{\{-1\}}(z^c_i)}_i} \times\dfrac{\pi(\bm x'')}{\pi(\bm x')} 
 \times\frac{1}{4a_1a_3}\right\}.
 \label{eq:acc_death_2}
 \end{align}
The key idea of the algorithm is described schematically in Figure \ref{fig:diagram_related}.

Note that for given $k$, in general mixture problems we would need to update 
$((\mu_1,\mu_2,\ldots,\mu_k)$, $(\log(\sigma_1),\log(\sigma_2),\ldots,\log(\sigma_k))$, $(\omega_1,\omega_2,\ldots,\omega_k))$,
where, for $j=1,\ldots,k$, $\omega_j$ correspond to the mixing proportion $\pi_j$, where $\sum_{j=1}^k\pi_j=1$, as
$\pi_j=\exp(\omega_j)/\sum_{\ell=1}^k\exp(\omega_j)$. If $(a_{\mu_1},\ldots,a_{\mu_k})$, 
$(a_{\sigma_1},\ldots,a_{\sigma_k})$, and $(a_{\omega_1},\ldots,a_{\omega_k})$ are the scales associated with
the three sub-blocks, then the Jacobian for the birth move, if the $j$-th component is selected, is given by
$8a_{\mu_j}a_{\sigma_j}a_{\omega_j}$, and that for the death move is $\left(8a_{\mu_j}a_{\sigma_j}a_{\omega_j}\right)^{-1}$.

\begin{figure}[H]
    \centering

    \resizebox{\textwidth}{!}
    {
        \begin{tikzpicture}[ultra thick,font=\boldmath]

        \node[rectangle,fill=gray!20,minimum size=5cm] (12) at (4,28) {\Huge \bf Birth Step};
        
        \node[rectangle,fill=gray!30,minimum width =35cm,minimum height=8cm] (13) at (-10,20) {\Huge \bf $X_1$};
        \node[rectangle,fill=gray!30,minimum width =35cm,minimum height=8cm] (13) at (30,20) {\Huge \bf $X_2$};
        \node[circle,fill=red!30,minimum size=6cm] (1) at (-20,20) {\bf \huge$x_{11}$};
        \node[circle,fill=red!30,minimum size=6cm] (2) at (0,20) {\huge$x_{12}$};
        \node[circle,fill=blue!30,minimum size=6cm] (3) at (20,20) {\huge$x_{21}$};
        \node[circle,fill=blue!30,minimum size=6cm] (4) at (40,20) {\huge$x_{22}$};
        
        \node[circle,fill=red!30,minimum size=2.5cm] (5) at (-28,4) {\huge$x^{'}_{11}=g_{1,z_{11}=1}(x_{11},\epsilon_1$)};
        \node[circle,fill=red!30,minimum size=2.5cm] (6) at (-12,4) {\huge$x^{'}_{12}=g_{1,z_{11}=-1}(x_{11},\epsilon_1  $)};
        \node[circle,fill=red!30,minimum size=4cm] (7) at (0,4) {\huge$x^{'}_{13}=g_{2,z_{12}}(x_{12},\epsilon_1)$};
        \node[circle,fill=blue!30,minimum size=4cm] (8) at (13,4) {\huge$x^{'}_{21}=g_{1,z_{21}=1}(x_{21},\epsilon_2)$};
        \node[circle,fill=blue!30,minimum size=2.5cm] (9) at (27,4) {\huge$x^{'}_{22}=g_{1,z_{21}=-1}(x_{21},\epsilon_2)$};
         \node[circle,fill=blue!30,minimum size=2.5cm] (10) at (40,4) {\huge$x^{'}_{23}=g_{2,z_{22}}(x_{22},\epsilon_1)$};


        
        \foreach \from/\to in {1/5,1/6,2/7,4/10}
        {\draw[->, ultra thick, fill=blue] (\from) to node[fill=white] {\Huge\Huge{$\epsilon_1$}} (\to);
        }
        \foreach \from/\to in {3/8,3/9}
        {\draw[->,ultra thick, fill=blue] (\from) to node[fill=white] {\Huge{$\epsilon_2$}} (\to);
 
        }

        \end{tikzpicture}

    }

\end{figure}

In general, $\bx\in\mathbb R^{mk}$ may be of the form $(\bx_1,\bx_2,\ldots,\bx_m)$, where 
$\bx_{\ell}=(x_{\ell,1},x_{\ell,2},\ldots,x_{\ell,k})$ for $\ell=1,2,\ldots,m$, where $m\geq 1$ is an integer. 
If the dimension of any one $\bx_{\ell}$ is changed, then the dimensions of all other 
$\bx_{\ell'};~\ell'\neq \ell$ must also change accordingly, as in the
above example. We provide the general TTMCMC algorithm as Algorithm \ref{algo:ttmcmc3} below.
It can be easily checked that detailed balance is satisfied for this algorithm.

\begin{algo}\label{algo:ttmcmc3} \topline General TTMCMC algorithm for jumping $m$ dimensions
with $m$ related sets of co-ordinates.
\botline \normalfont \ttfamily
\begin{itemize}
 \item Let the initial value be $\supr{\bm x}{0}\in\mathbb R^{mk}$, where $k\geq m$. 
 \item For $t=0,1,2,\ldots$
\begin{enumerate}
 \item Generate $u=(u_1,u_2,u_3)\sim Multinomial (1;w_{b,k},w_{d,k},w_{nc,k})$.
 \item If $u_1=1$ (increase dimension from $mk$ to $m(k+1)$), then
 \begin{enumerate}
 \item Randomly select one co-ordinate from $\bm x^{(t)}_1=(x^{(t)}_{11},\ldots,x^{(t)}_{1k})$ without replacement.
 Let $j$ denote the chosen co-ordinate.
 \item Generate $\be_m=(\e_1,\ldots,\e_m)\stackrel{iid}{\sim} \varrho(\cdot)$ and for $i\in\{1,\ldots,k\}\backslash\{j\}$ 
 simulate  
 $z_{\ell,i} \sim Multinomial(1;p_{\ell,i},q_{\ell,i},1-p_{\ell,i}-q_{\ell,i})$ independently, for every $\ell=1,\ldots,m$. 
 \item Propose the birth move as follows: for each $\ell=1,\ldots,m$, 
 apply the transformation $x^{(t)}_{\ell,i}\rightarrow g_{i,z_{\ell,i}}(x^{(t)}_{\ell,i},\e_1)$ 
 for $i\in\{1,\ldots,k\}\backslash\{j\}$ and, for each $\ell\in \{1,\ldots,m\}$,
 split  $x^{(t)}_{\ell,j}$ into $g_{\ell,z_{\ell,j}=1}(x^{(t)}_{\ell,j},\e_{\ell})$ and 
 $g_{\ell,z^c_{\ell,j}=-1}(x^{(t)}_{\ell,j},\e_{\ell})$. In other words, let $\bx'=
 T_{b,\bz}(\supr{\bm x}{t}, \be_m)=(\bx'_1,\ldots,\bx'_m)$
 denote the complete birth move, where, for $\ell=1,\ldots,m$, $\bx'_{\ell}$ is given by
 \begin{align} 
 \bm x'_{\ell} & 
 =(g_{\ell,z_{\ell,1}}(x^{(t)}_{\ell,1},\e_1),\ldots,
 g_{j-1,z_{\ell,j-1}}(x^{(t)}_{\ell,j-1},\e_1),\notag\\
& g_{j,{z_{\ell,j}=1}}(x^{(t)}_{\ell,j},\e_{\ell}),g_{j,{z^c_{\ell,j}=-1}}(x^{(t)}_{\ell,j},\e_{\ell}),
g_{j+1,z_{\ell,j+1}}(x^{(t)}_{\ell,j+1},\e_1),\ldots,
\notag\\
&\ldots,
g_{k,z_{\ell,k}}(x^{(t)}_{\ell,k},\e_1)).\notag
\end{align}
Re-label the $k+1$ elements of $\bm x'_{\ell}$ as $(x'_{\ell,1},x'_{\ell,2},\ldots,x'_{\ell,k+1})$.
Notice that, following the discussion presented in the illustrative example in the beginning of this section,
we use $\e_{\ell}$ only for splitting $x^{(t)}_{\ell,j}$ into 
$g_{j,{z_{\ell,j}=1}}(x^{(t)}_{\ell,j},\e_{\ell})$ and $g_{j,{z^c_{\ell,j}=-1}}(x^{(t)}_{\ell,j},\e_{\ell})$.
To update the remaining co-ordinates, we use $\e_1$ for all the blocks.

\item Calculate the acceptance probability of the birth move $\bm x'$:
 \begin{align}
 a_b(\supr{\bm x}{t}, \be_m) &= 
 \min\left\{1, \frac{1}{k+1}\times\frac{w_{d,k+1}}{w_{b,k}}\times\dfrac{P_{(j)}(\bz^c)}{P_{(j)}(\bz)} 
 ~\dfrac{\pi(\bm x')}{\pi(\supr{\bm x}{t})} 
 ~\left|\frac{\partial (T_{b,\bz}(\supr{\bm x}{t}, \be_m))}{\partial(\supr{\bm x}{t}, \be_m)}\right| \right\},\notag
 \end{align}
where 
 \[ 
P_{(j)}(\bz)=\prod_{\ell=1}^m\prod_{i\in\{1,\ldots,k\}\backslash\{j\}}
p^{I_{\{1\}}(z_{\ell,i})}_{\ell,i}q^{I_{\{-1\}}(z_{\ell,i})}_{\ell,i},
 \]
 and
 \[ 
P_{(j)}(\bz^c)=\prod_{\ell=1}^m\prod_{i\in\{1,\ldots,k\}\backslash\{j\}}
p^{I_{\{1\}(z^c_{\ell,i})}}_{\ell,i}q^{I_{\{-1\}}(z^c_{\ell,i})}_{\ell,i}.
 \]

\item Set \[ \supr{\bm x}{t+1}= \left\{\begin{array}{ccc}
 \bm x' & \mbox{ with probability } & a_b(\supr{\bm x}{t},\be_m) \\
 \supr{\bm x}{t}& \mbox{ with probability } & 1 - a_b(\supr{\bm x}{t},\be_m).
\end{array}\right.\]
\end{enumerate}
\item If $u_2=1$ (decrease dimension from $k$ to $k-m$, for $k\geq 2m$), then
 \begin{enumerate}
 \item Generate $\be_m=(\e_1,\ldots,\e_m)\stackrel{iid}{\sim} \varrho(\cdot)$. 
 \item Randomly, without replacement, select co-ordinates $j$ and 
 $j'$ from $\bx_1=(x_{1,1},\ldots,x_{1,k})$. For $\ell=1,\ldots,m$, 
 let $$x^*_{\ell,j}=\left(g_{j,z^c_{\ell,j}=-1}(x_{\ell,j},\e_{\ell})
 +g_{j',z_{\ell,j'}=1}(x_{\ell,j'},\e_{\ell})\right)/2;$$ 
 replace the co-ordinate $x_{\ell,j}$
 by the average $x^*_{\ell,j}$ and delete $x_{\ell,j'}$.
 \item Simulate $\bz$ by generating independently, for $\ell=1,\ldots,m$ and for $i\in\{1,\ldots,k\}\backslash\{j,j'\}$, 
 $z_{\ell,i} \sim Multinomial(1;p_{\ell,i},q_{\ell,i},1-p_{\ell,i}-q_{\ell,i})$.
 \item For $\ell=1,\ldots,m$ and for $i\in\{1,\ldots,k\}\backslash\{j,j'\}$, apply the transformation 
 $x'_{\ell,i}=g_{i,z_{\ell,i}}(x^{(t)}_{\ell,i},\e_1)$.  
 \item Propose the following death move $\bx'=T_{d,\bz}(\supr{\bm x}{t}, \be_m)=(\bx'_1,\ldots,\bx'_m)$ 
 where for $\ell=1,\ldots,m$, $\bx_{\ell}$ is given by
 \begin{align} 
 \bm x'_{\ell} 
 &=(g_{1,z_{\ell,1}}(x^{(t)}_{\ell,1},\e_1),\ldots,
 g_{j-1,z_{\ell,j-1}}(x^{(t)}_{\ell,j-1},\e_1),x^*_{\ell,j},g_{j+1,z_{\ell,j+1}}(x^{(t)}_{\ell,j+1},\e_1),\notag\\
& \ldots,g_{k,z_{\ell,k}}(x^{(t)}_{\ell,k},\e_1)).\notag
\end{align}
Re-label the elements of $\bm x'_{\ell}$ as $(x'_{\ell,1},x'_{\ell,2},\ldots,x'_{\ell,k-1})$.
 \item For $\ell=1,\ldots,m$, solve for $\e^*_{\ell}$ from the equations 
 $g_{\ell,z_{\ell,j}=1}(x^*_{\ell,j},\e^*_{\ell})=x_{\ell,j}$ and 
 $g_{\ell,z^c_{\ell,j}=-1}(x^*_{\ell,j},\e^*_{\ell})=x_{\ell,j'}$
and express $\e^*_{\ell}$ in terms of $x_{\ell,j}$ and $x_{\ell,j'}$.
Let $\be^*_m=(\e^*_1,\ldots,\e^*_m)$.
\item Calculate the acceptance probability of the death move:
 \begin{align}
 a_d(\supr{\bm x}{t}, \be_m, \be^*_m) &= 
 \min\left\{1,k\times\frac{w_{b,k-m}}{w_{d,k}}\times\dfrac{P_{(j,j')}(\bz^c)}{P_{(j,j')}(\bz)} 
 ~\dfrac{\pi(\bm x')}{\pi(\supr{\bm x}{t})} 
 ~\left|\frac{\partial (T_{d,\bz}(\supr{\bm x}{t}, \be_m),\be^*_m,\be_m)}
 {\partial(\supr{\bm x}{t},\be_m)}\right| \right\},\notag
 \end{align}
 where
 \[ 
P_{(j,j')}(\bz)=\prod_{\ell=1}^m\prod_{i\in\{1,\ldots,k\}\backslash\{j,j'\}}p^{I_{\{1\}}(z_{\ell,i})}_{\ell,i}
q^{I_{\{-1\}}(z_{\ell,i})}_{\ell,i},
 \]
 and
 \[ 
P_{(j,j')}(\bz^c)=\prod_{\ell=1}^m\prod_{i\in\{1,\ldots,k\}\backslash\{j,j'\}}p^{I_{\{1\}(z^c_{\ell,i})}}_{\ell,i}q^{I_{\{-1\}}(z^c_{\ell,i})}_{\ell,i}.
 \]
\item Set \[ \supr{\bm x}{t+1}= \left\{\begin{array}{ccc}
 \bm x' & \mbox{ with probability } & a_d(\supr{\bm x}{t},\be_m,\be^*_m) \\
 \supr{\bm x}{t}& \mbox{ with probability } & 1 - a_d(\supr{\bm x}{t},\be_m,\be^*_m).
\end{array}\right.\]
 \end{enumerate}
\item If $u_3=1$ (dimension remains unchanged), then implement steps (1), (2), (3) of Algorithm 3.1 
of \ctn{Dutta14}. 
\end{enumerate}
\item End for
\end{itemize}
\botline \rmfamily
\end{algo}

\section{{\bf Brief discussion on label switching}}
\label{sec:label_switching}
It is well-known that the mixture likelihood is invariant to permutations (labels) of the component 
parameters; hence, the mixture parameters are not identifiable. This problem is often
referred to as label-switching. So, if inference on the
parameters is of interest, then proper labeling of the components is necessary. \ctn{Richardson97}
considered ordering the mean parameters; see also \ctn{Stephens00} for other methods for tackling
label switching. 
However, \ctn{Lee09} argue and demonstrate that putting constraints on the prior parameter space
can have severe ill effects on both inference and computation.
Moreover, there seems to be a subtle question if identifiability is at all desirable
when inference regarding clustering of the data is of interest. To consider a simple example,
suppose that clustering the dataset $\{y_1,y_2,y_3,y_4\}$ using a two-component normal mixture model is of interest. 
Assume that $\{y_1,y_3\}$ are associated with $\nu_1$ and $\{y_2,y_4\}$ are associated with $\nu_2$, where
$\nu_1<\nu_2$. But because of this imposed constraint, the clusterings $\{\{y_1,y_3\},\{y_2,y_4\}\}$
and $\{\{y_2,y_4\},\{y_1,y_3\}\}$ can not be regarded as identical.

\section{{\bf Summarization of the posterior distribution of mixture densities}}
\label{sec:summary_density}

Note that the mixture setup induces a posterior distribution on mixture densities
of the form 
\begin{equation}
f(y_i\vert\bnu_k,\btau_k,\bpi_k,k)=\sum_{j=1}^k\pi_j\sqrt{\frac{\tau_j}{2\pi}}\exp\left\{-\frac{\tau_j}{2}(y_i-\nu_j)^2\right\}.
\label{eq:normix_supp}
\end{equation}
In other words, the set-up provides a way to 
make Bayesian inference regarding the unknown density of the observed data $y_1,\ldots,y_n$.
An obvious candidate of such density estimate is the unconditional posterior expectation
of the function
\begin{equation}
f(x\vert k,\bnu_k,\btau_k,\bpi_k)=\sum_{j=1}^k\pi_j\sqrt{\frac{\tau_j}{2\pi}}
\exp\left\{-\frac{\tau_j}{2}(x-\nu_j)^2\right\};\quad -\infty<x<\infty,
\label{eq:normix2}
\end{equation}
with respect to the posterior of 
$k,\bnu,\btau_k,\bpi_k$. For empirical purposes, one can just average 
$f(x\vert k,\bnu_k,\btau_k,\bpi_k)$ over TTMCMC samples of $k,\bnu,\btau_k,\bpi_k$.

Note, however, that the posterior expectation (or the corresponding empirical average)
fails to retain the finite mixture form of the resultant density estimate (see also \ctn{Richardson97}). 
More importantly, although this averaging yields a point density estimate, hitherto there
does not seem to be any attempt to quantify the uncertainty of the posterior distribution of the 
densities having the mixture form with unknown number of components.

Motivated by \ctn{Sabya11} who propose a methodology for obtaining the modes and any desired
highest posterior density credible regions associated with the posterior distribution of clusterings,
here we attempt the same for the posterior distribution of densities having form (\ref{eq:normix_supp}).
Following \ctn{Sabya11} here we propose a definition of ``central density":

\begin{definition}
\label{def:def1}
A density $f_0$ is ``central" which, for any $\epsilon>0$ satisfies the following equation:
\begin{align}
P\left(\left\{f:d(f_0,f)<\epsilon\right\}\right)=\sup_{g}P\left(\left\{f:d(g,f)<\epsilon\right\}\right),
\label{eq:central}
\end{align}
for some suitable metric $d$. 
\end{definition}
In this article, we consider the sup-norm metric between any two density functions $f$ and $h$, given by
$d(f,h)=\sup_{-\infty<x<\infty}|f(x)-h(x)|$. For empirical purpose we evaluate this metric at discrete
equidistant points $x_0,x_1,\ldots,x_m$ covering the effective support of the densities in question.

Observe that $f_0$ is the global mode of the posterior distribution of densities as $\epsilon\rightarrow 0$.
If the distribution of $f$ is unimodal, then the central density remains the same for all $\epsilon>0$.
However, for multimodal distributions, the central density varies with $\epsilon$, signifying 
existence of local modes, which we define as follows.

\begin{definition}
\label{def:def2}
We define $f_{\mbox{loc}}$ to be a local mode if
\begin{align}
\lim_{\epsilon\downarrow 0}\frac{\sup_{h\in\mathcal N(f_{\mbox{loc}},\eta)}
P\left(\left\{f\in\mathcal N(f_{\mbox{loc}},\eta):d(f,h)<\epsilon\right\}\right)}
{P\left(\left\{f\in\mathcal N(f_{\mbox{loc}},\eta):d(f,f_{\mbox{loc}})<\epsilon\right\}\right)}
&=1,
\label{eq:local_mode}
\end{align}
where $\mathcal N(f_{\mbox{loc}},\eta)=\left\{f:f(f_{\mbox{loc}},f)<\eta\right\}$ for some $\eta>0$.
\end{definition}

Note that unlike the distribution of clusterings considered by \ctn{Sabya11}, which is discrete,
the distribution of the mixture densities of the form (\ref{eq:normix2}) is continuous; this
is clear since although $k$, the number of mixture components is at most countable, the parameters
$\bnu$, $\btau$ and $\bpi$ are continuous. Hence, although obtaining the global mode in the case
of clusterings is an arduous task, here our problem is relatively easier.

It is nevertheless clear that without the aid of empirical methods the central density function defined
in (\ref{eq:central}) can not be obtained. Using available TTMCMC samples $\{f^{(j)};~j=1,\ldots,N\}$
of length $N$, the latter sufficiently large,
useful empirical methods can be devised, as we demonstrate in the next section.

\subsection{{\bf Empirical definition of central density function}}
\label{subsec:empirical_central_density}

We define that density $f^{(j)}$ as ``approximately central," which, for a given small $\epsilon>0$, satisfies 
the following equation:
\begin{equation}
f^{(j)}=\arg\max_{1\leq i\leq N}\frac{1}{N}\#\left\{f^{(\ell)};1\leq \ell\leq N:d(f^{(i)},f^{(\ell)})<\epsilon\right\}.
\label{eq:empirical_central_density}
\end{equation}
The central density $f^{(j)}$ is easily computable and the ergodic theorem ensures convergence
of $f^{(j)}$ almost surely to the true central density $f_0$.

\subsection{{\bf Construction of desired credible regions of densities}}
\label{subsec:credible}
Given a central density $f^{(j)}$, an approximate
95\% posterior density credible region is given by the set 
$\left\{f^{(\ell)};1\leq \ell\leq N:d(f^{(\ell)},f^{(j)})<\epsilon^*\right\}$, where $\epsilon^*$
is such that
\begin{equation}
\frac{1}{N}\#\left\{f^{(\ell)};1\leq \ell\leq N:d(f^{(\ell)},f^{(j)})<\epsilon^*\right\}\approx 0.95.
\label{eq:cred}
\end{equation}
In (\ref{eq:cred}) $\epsilon^*$ can be chosen adaptively by starting with $\epsilon^*=0$ and then slightly
increasing $\epsilon^*$ by a quantity $\zeta$ until (\ref{eq:cred}) is satisfied. 
In our applications, we chose $\zeta=10^{-5}$.
Approximate highest posterior density (HPD) regions
can be constructed by taking the union of the highest density regions. Following \ctn{Sabya11} 
we next discuss an adaptive methodology
for constructing HPD regions.

\subsection{{\bf Construction of desired HPD regions of densities}}
\label{subsec:hpd}
Assume that there are $\ell$ modes, $\{f^*_1,\ldots, f^*_{\ell}\}$, obtained by varying $\epsilon$ of 
the neighborhoods $\{f: d(f,f^{(i)})<\epsilon\};~i=1,\ldots,N$. 
Consider the regions $S_j=\{f:d(f^*_j,f)<\epsilon^*_j\};~ j=1,\ldots,\ell$. Set, initially, 
$\epsilon^*_1=\epsilon^*_2=\cdots =\epsilon^*_{\ell}=0$.
\begin{itemize}
\item[(i)] For $i=1,\ldots,N$, if the $i$-th TTMCMC realization $f^{(i)}$ does not fall in 
$S_j$ for some $j$, then increase $\epsilon^*_j$ by a small quantity, say, $\zeta$.
\item[(ii)] Calculate the probability of $\cup_{j=1}^{\ell}S_j$ as $P=\#\{\cup_{j=1}^{\ell}S_j\}/N$.
\item[(iii)] Repeat steps (i) and (ii) until $P\approx 0.95$ or any desired probability.
\end{itemize}
In step (i) we implicitly assume that, since $f^{(i)}\notin S_j$, $S_j$ must be a region with low probability, 
so its expansion is necessary to increase the probability. We achieve this expansion by increasing
$\epsilon^*_j$ by $\zeta$. Thus the sets $S_j$ are selected adaptively, by adaptively increasing $\epsilon^*_j$.
The desired approximate HPD region is then the final union of the $S_j$'s.

\section{{\bf TTMCMC convergence diagnostics for the mixture problem}}
\label{sec:conv_diag}

\subsection{{\bf Difficulties of convergence assessment in variable dimensional problems}}
\label{subsec:convergence_difficulties}
A particularly problematic area in variable dimensional problems is ascertaining 
whether or not the underlying MCMC algorithm has converged to the stationary distribution. 
The reason that the convergence assessment problem in transdimensional set-ups is more difficult in comparison with
the fixed-dimensional counterpart is that the dimensionality of the parameters, as well as their
interpretations, can vary with the iterations. 
The difficulty of the problem did motivate researchers to devise appropriate measures
of convergence diagnostics; however, to date, the developments are relatively few. \ctn{Sisson05}
provide a comprehensive review of such developments, along with their shortcomings. The shortcomings generally
pertain to marginal, rather than joint convergence assessment, computational burden that comes with 
implementing many independent runs of the sampler, and of course, various assumptions which may be difficult 
to validate in practice. 



\subsection{{\bf A new convergence diagnostic method for mixtures with known or unknown number of components}}
\label{subsec:new_conv_diag}

Armed with our metric-based methodology we now provide a convergence
diagnostic method for the challenging variable dimensional mixture problem.
Following the same principle as \ctn{Sabya11}, we divide our TTMCMC sample of size $N$ into 
$m$ equal parts, each part
having the same size $N/m$, assuming divisibility of $N$ by $m$. 
For each such subsample of size $N/m$, we compute a central density function 
and the corresponding approximate 95\% credible region.
If the $m$ credible regions thus obtained are close to each other, one can safely  infer that the 
$m$ subsamples arose from the same
stationary distribution. 

Analogous to the convergence diagnostic method \ctn{Sabya11}, our method can assess if two
credible regions corresponding to two separate subsamples are close to each other.
Let $(CR_{\epsilon_1},\epsilon_1)$ and $(CR_{\epsilon_2},\epsilon_2)$ denote the 95\% credible regions 
and the corresponding radii obtained from any two subsamples.
Suppose that $\eta_1>0$ is the {\it least positive value} such that $CR_{\epsilon_1+\eta_1}\supseteq CR_{\epsilon_2}$, 
and also suppose that $\eta_2>0$ is the {\it least positive value} such that $CR_{\epsilon_2+\eta_2}\supseteq CR_{\epsilon_1}$. 
Then, if both the increments $\eta_1,\eta_2$ are sufficiently small,
then the 95\% credible regions $CR_{\epsilon_1}$ and $CR_{\epsilon_2}$ can be said to be ``close".


Currently in this paper we restrict ourselves to mixture problems only. But from the construction 
it is clear that our proposed diagnostics is readily applicable to function estimation context. These developments, 
in our opinion, can play important roles in various
applications involving random basis function expansions, for instance, in nonparametric regression 
and functional data analysis. Since basis function expansions typically involve unknown number of summands,
TTMCMC based inference along with our procedure for summarizing posterior distribution of functions,
are expected to constitute a very interesting and important combination for such challenging
data analysis. Indeed, the functions may also be the modeled density (either discrete or continuous) associated
with the likelihood, indicating that our methods are very generally applicable. 
Since convergence in variable dimensional problems is particularly difficult to assess, our methodology,
which seems to provide a reliable convergence assessment criterion, perhaps provides a significant advance.

\section{{\bf Further simulation studies with the gamma mixtures with different data sizes}}
\label{sec:gamma_mixtures_varying_datasize}

In our simulation studies so far, we considered data sets of size $400$. We now experiment by
varying the data sizes for the four mixtures and note the changes for the TTMCMC based posteriors of $k$.

\subsection{{\bf 1-component mixture}}
\label{subsec:1comp_varying_datasize}
For a data set of size $60$, the posterior distribution of $k$ was concentrated on $k=1,2,3,4$
with probabilities $0.9308$, $0.0647$, $0.0041$ and $0.0004$, respectively, hence not differing too significantly
from our reported results when the data size was $400$. In this case, the overall acceptance rate turned out to be
$0.089759$, the birth rate was $0.016024$, the death rate was $0.357057$, and $0.151372$ was the no-change rate.
Further experiments with data sets larger than $60$ revealed that the posterior distribution of $k$ increasingly concentrated 
around $k=1$. For instance, with data size $1000$, the posterior of $k$ assigned probabilities
$0.9798$, $0.02$ and $0.0002$ to $k=1,2,3$, respectively. In this case, the overall acceptance rate was
$0.022835$, and the birth, death, no change rates were $0.00188$, $0.03401$ and $0.043119$, respectively. 
For data sizes smaller than 60, the information seemed to be insufficient to precisely capture $k=1$.

\subsection{{\bf 2-component mixture}}
\label{subsec:2comp_varying_datasize}
Since this is perhaps the most challenging example in that it is hard to distinguish two mixture components,
it is easy to anticipate that a somewhat large data set is necessary to capture the true information. As such,
we find that data sets of size $300$ or more produces good results. Indeed, for a dataset of size 300, we obtain
the posterior probabilities of $k=1,2,3$ to be $0.1917$, $0.8037$ and $0.0046$. The overall acceptance rate,
birth, death and the no-change rates are given by $0.064524$, $0.001132$, $0.002295$ and $0.158743$, respectively.
Thus, unlike the data of size $400$, we no longer obtain point posterior mass at $k=2$, although the truth
(namely, 2 components) has clearly been identified. 

\subsection{{\bf 3-component mixture}}
\label{subsec:3comp_varying_datasize}
In this example, we consider a consistency check by considering a dataset of size $1000$ and expecting 
our TTMCMC to give close to point posterior mass to $3$ components, given that it has given point 
posterior mass to $3$ components
for the data of size $400$. On implementation of TTMCMC, we find that consistency is indeed attained. 
The posterior probabilities for $k=3$ and $k=4$ are $0.9987$ and $0.0013$, respectively, while all other
values of $k$ received zero posterior mass.
The overall acceptance rate is $0.034308$, the overall birth and death rates are
$0.000028$ and $0.000035$, respectively, while the no-change rate is $0.102759$.

\subsection{{\bf 4-component mixture}}
\label{subsec:4comp_varying_datasize}
Since the 4-component mixture seems to be somewhat easy to identify, we investigate if TTMCMC can identify the
true number of components even for much smaller datasets. With our implementation
Indeed, for a dataset of size $170$, we 
find that $k=4,5,6,7$ receive posterior probabilities $0.7006$, $0.2932$, $0.0057$ and $0.0005$, respectively, while
the other values of $k$ receive zero posterior probability. In this case, the overall acceptance, 
birth, death and the no-change rates are $0.026976$, $0.000284$, $0.000289$ and $0.080268$, respectively.

\section{{\bf Comparison between additive TTMCMC and random walk RJMCMC in normal mixtures with respect
to the three real data sets}}
\label{sec:ttmcmc_vs_rjmcmc}
\subsection{{\bf Comparison in enzyme data}}
\label{subsec:enzyme_rjmcmc}
The implementation of random walk RJMCMC took 38 minutes and 29 seconds to yield $10,000$
realizations following a burn-in of 375,000 iterations, after storing one in 150 iterations out of
further $15,00,000$ iterations after the burn-in period. 

The RJMCMC algorithm yielded an overall acceptance rate
$0.05605344$, which is slightly larger than that of TTMCMC. The birth, death and no-change rates turned out to be
$0.007919$, $0.005228$ and $0.138663$, respectively. The birth and death rates are significantly larger than in TTMCMC,
while the no-change rate is smaller.

\begin{figure}
\centering
\subfigure[Trace plot of $k$.]{ \label{fig:rjmcmc_enzyme_k}
\includegraphics[width=7cm,height=6cm]{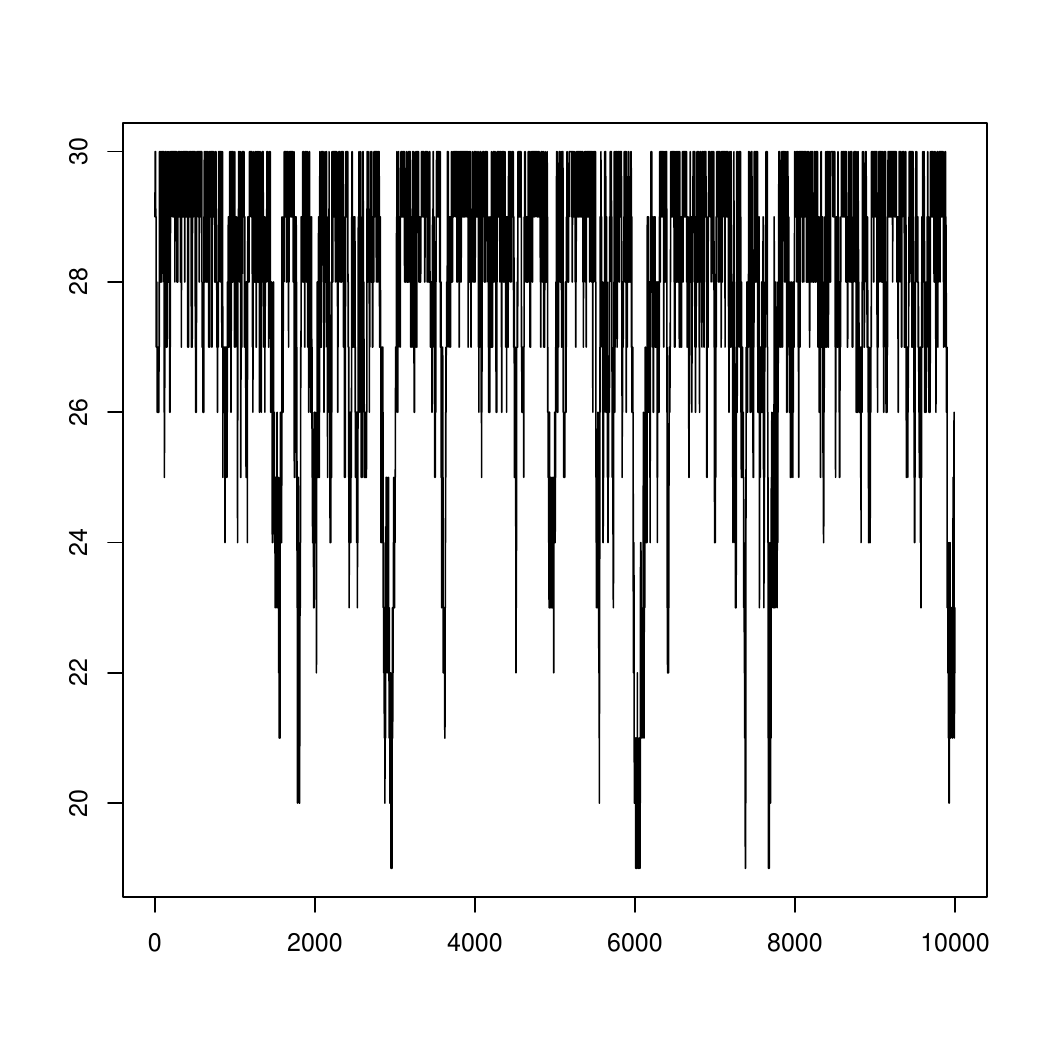}}
\hspace{2mm}
\subfigure[Trace plot of $\nu_1$.]{ \label{fig:rjmcmc_enzyme_nu}
\includegraphics[width=7cm,height=6cm]{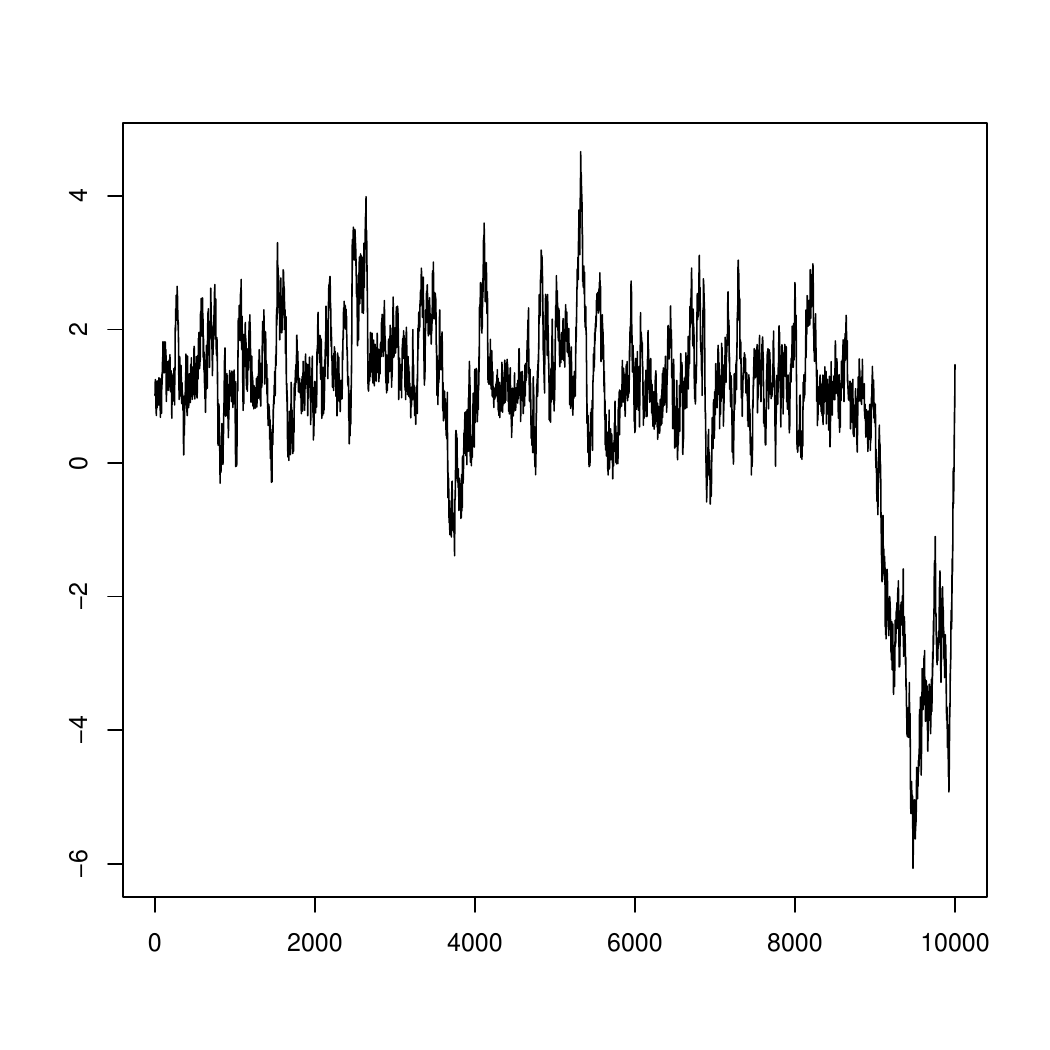}}\\
\vspace{2mm}
\subfigure[Trace plot of $\tau_1$.]{ \label{fig:rjmcmc_enzyme_tau}
\includegraphics[width=7cm,height=6cm]{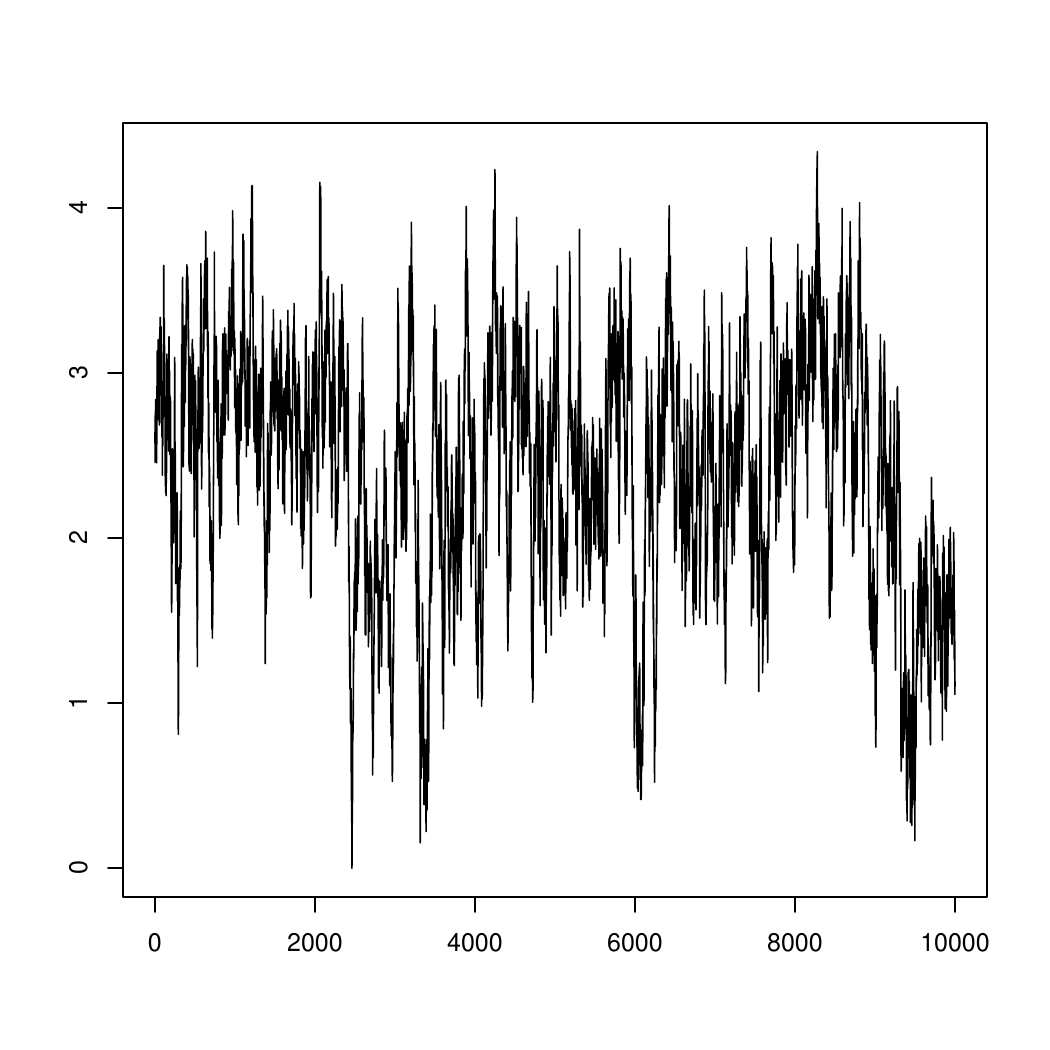}}
\hspace{2mm}
\subfigure[Trace plot of $\omega_1$.]{ \label{fig:rjmcmc_enzyme_w}
\includegraphics[width=7cm,height=6cm]{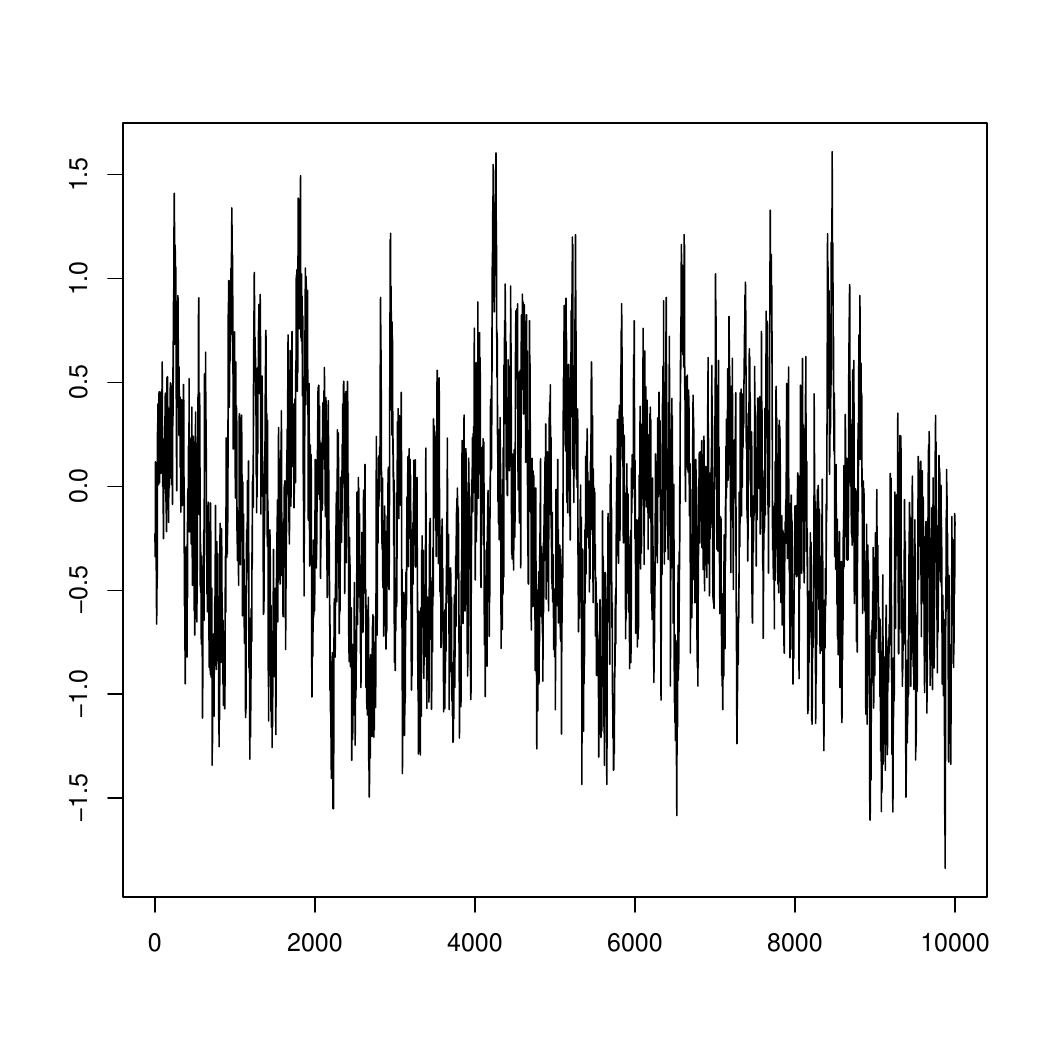}}
\caption{{\bf RJMCMC for the enzyme data:} Trace plots of $k$, $\nu^*_1$, $\tau^*_1$ and $\omega_1$.} 
\label{fig:rjmcmc_enzyme_trace_plots}
\end{figure}

However, we obtained $\eta_1=0.11463$ and $\eta_2=0.10610$, which are significantly larger than 
in TTMCMC, indicating better convergence of TTMCMC.
Moreover, the trace plots of $k$ displayed in Figure \ref{fig:rjmcmc_enzyme_trace_plots} show
that very large number of components are favored by RJMCMC, showing that the chain is far from convergence.
The main issue here seems to be the dependence of the acceptance rate on the proposal density $\prod_{i=1}^3\varrho(u_i)$. 
Since, $\prod_{i=1}^3\varrho(u_i)$, the product of left truncated standard normal densities, is less than one, it follows that
the acceptance probability of the birth move is higher than that of the death move. This explains
the large number of components favored by random walk RJMCMC, clearly impeding convergence. 

\subsection{{\bf Comparison in acidity data}}
\label{subsec:rjmcmc_acidity}

With RJMCMC based on random walk, the time for implementation is $11$ minutes and $25$ seconds,
much larger than that of TTMCMC. The overall acceptance rate turned out to be $0.196064$, 
smaller than that of TTMCMC. 
The birth, death and no-change rates are $0.012153$, $0.012177$ and $0.56334$, respectively,
that is, the birth and death rates are significantly larger than in TTMCMC while the no-change rate
is smaller.

As before, however, for this RJMCMC implementation, $\eta_1=0.03616$ and$\eta_2=0.03522$,
showing that the convergence is much inferior compared to TTMCMC.
Here, $k$ assigned positive posterior probabilities to large values and gave zero mass to $k=2$ and $k=3$, which
received full posterior mass from TTMCMC implementation, again
showing that in comparison with TMCMC, RJMCMC tends to assign larger posterior mass to larger number of components. 
\begin{figure}
\centering
\subfigure[Trace plot of $k$.]{ \label{fig:rjmcmc_acidity_k}
\includegraphics[width=7cm,height=6cm]{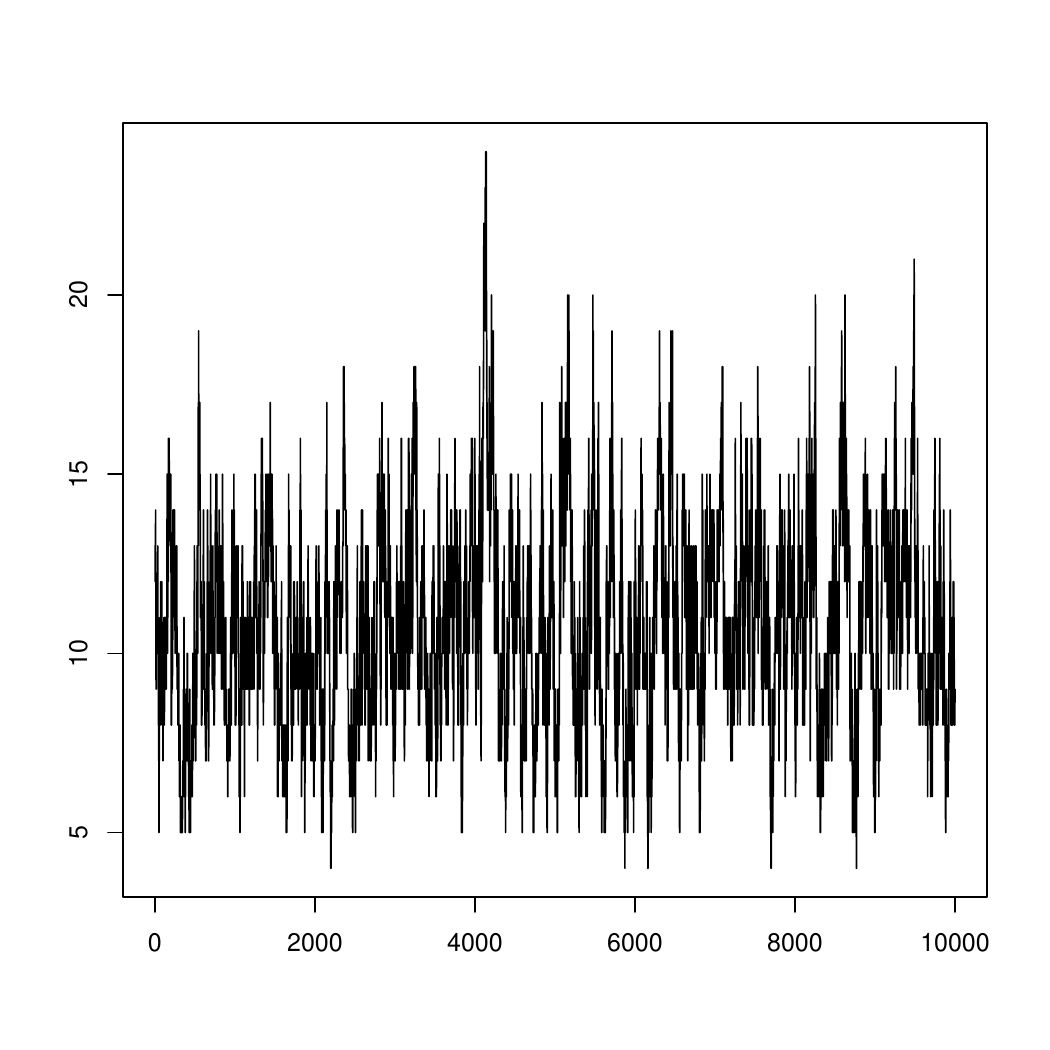}}
\hspace{2mm}
\subfigure[Trace plot of $\nu^*_1$.]{ \label{fig:rjmcmc_acidity_nu}
\includegraphics[width=7cm,height=6cm]{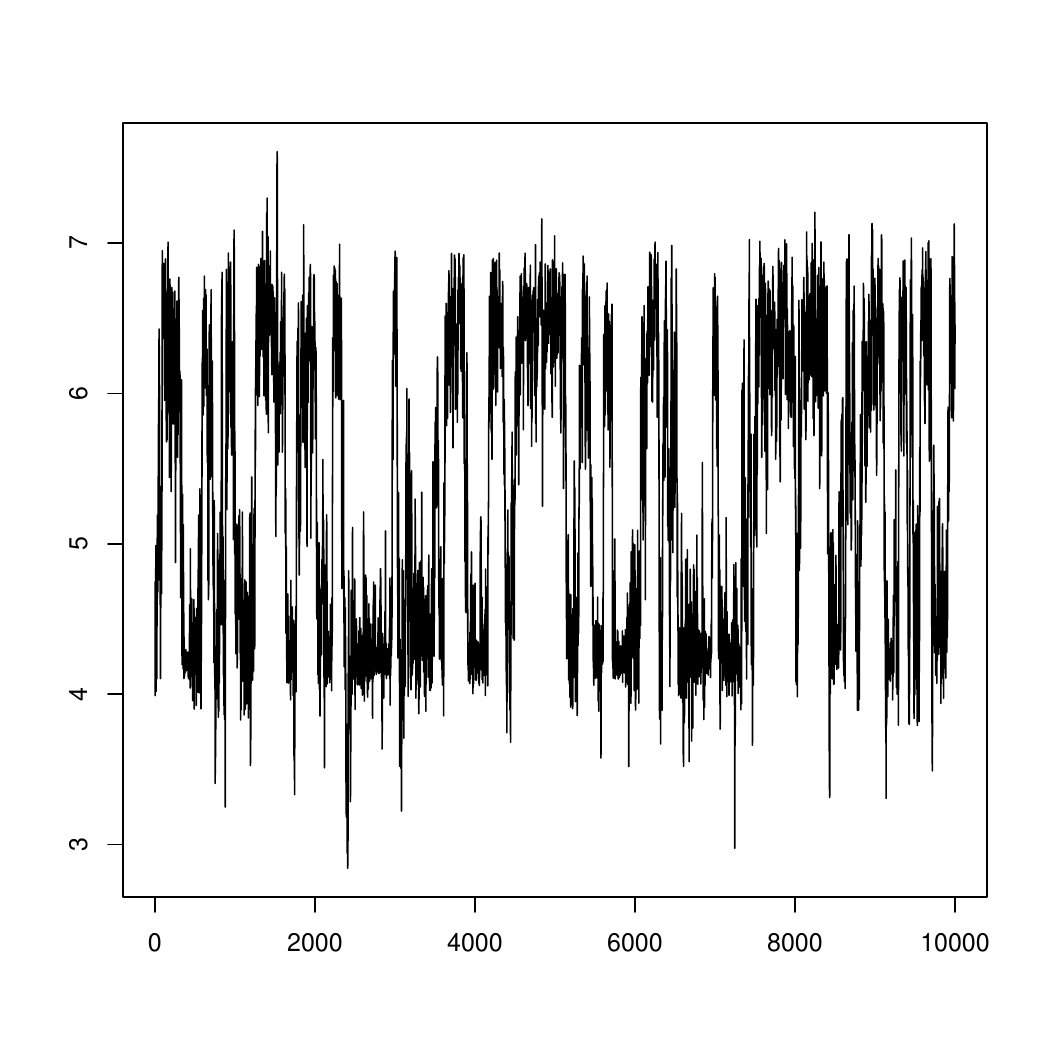}}\\
\vspace{2mm}
\subfigure[Trace plot of $\tau^*_1$.]{ \label{fig:rjmcmc_acidity_tau}
\includegraphics[width=7cm,height=6cm]{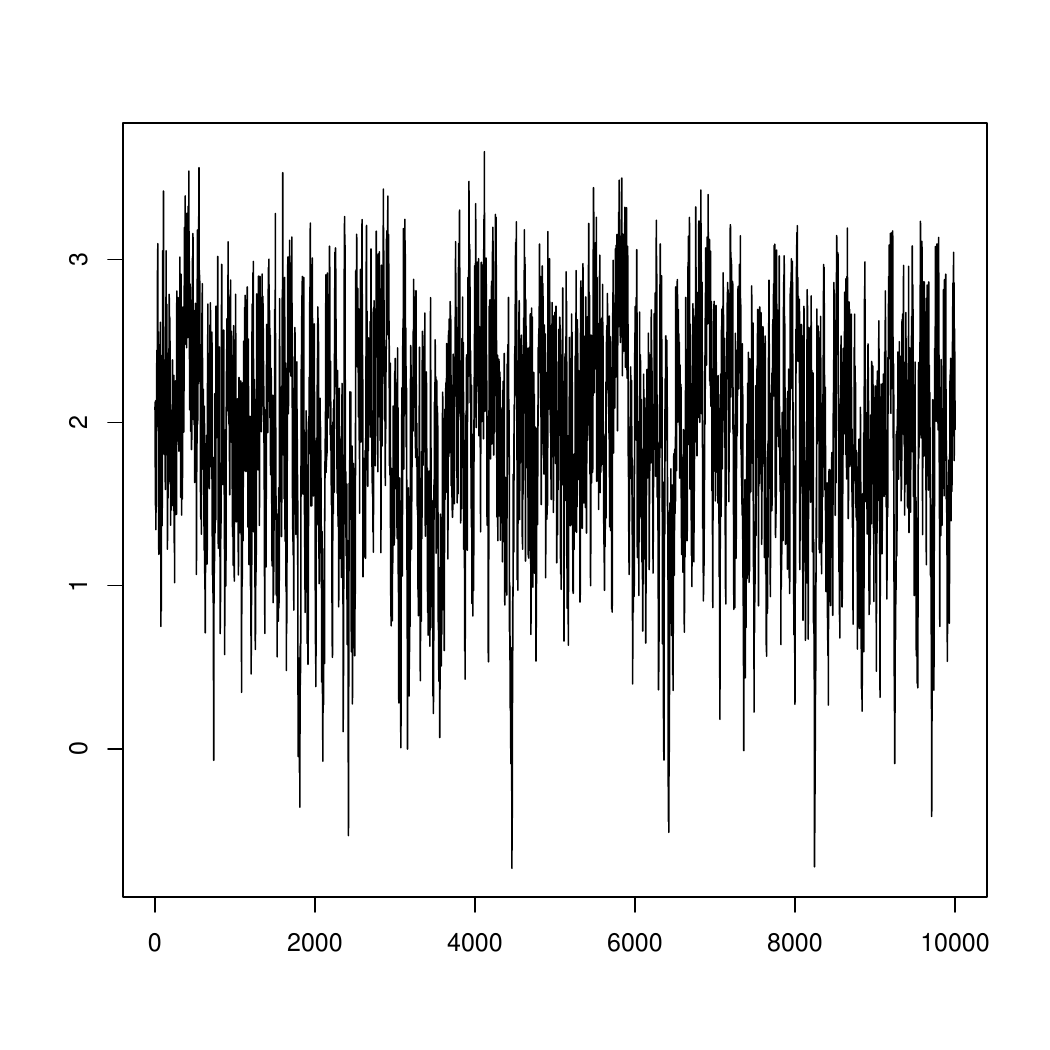}}
\hspace{2mm}
\subfigure[Trace plot of $\omega_1$.]{ \label{fig:rjmcmc_acidity_w}
\includegraphics[width=7cm,height=6cm]{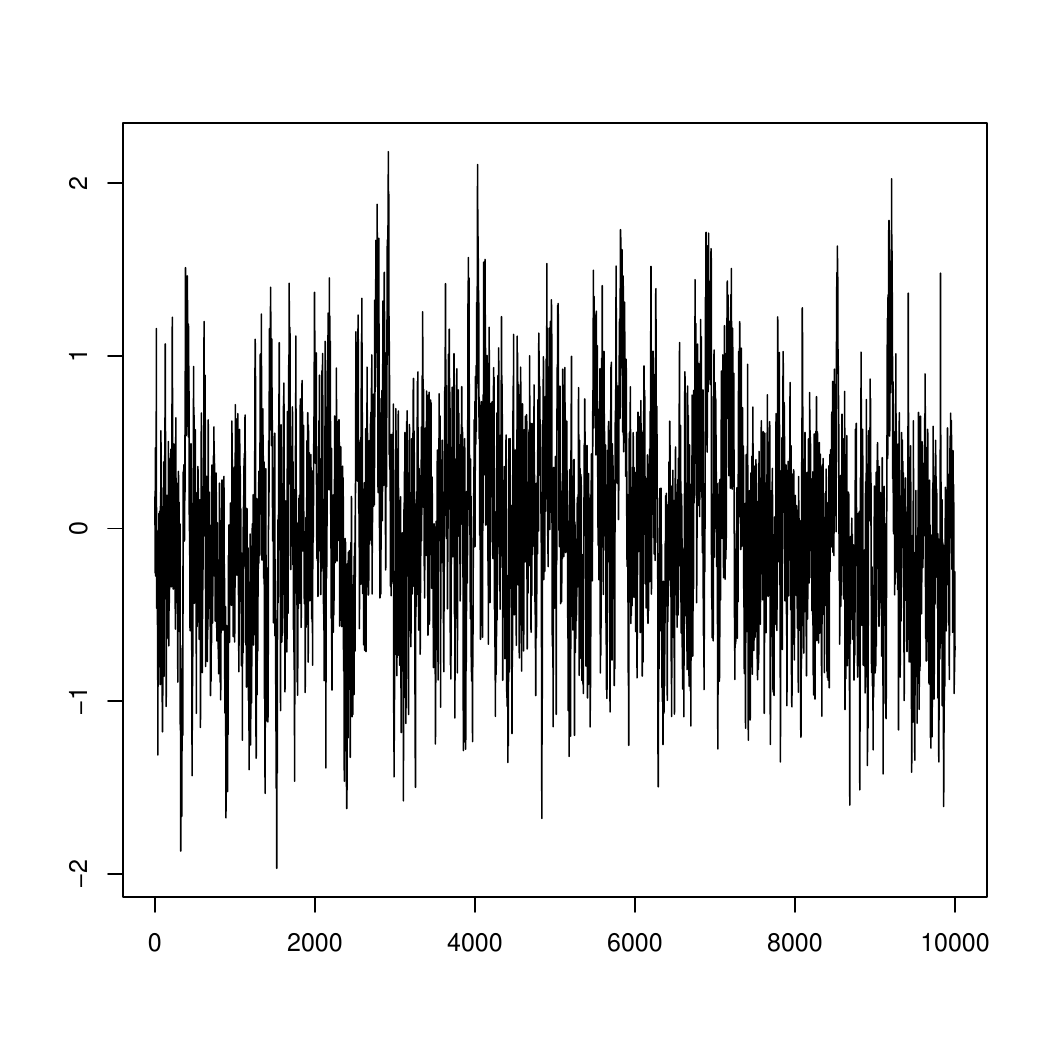}}
\caption{{\bf RJMCMC for the acidity data:} Trace plots of $k$, $\nu^*_1$, $\tau^*_1$ and $\omega_1$.} 
\label{fig:rjmcmc_acidity_trace_plots}
\end{figure}

\subsection{{\bf Comparison in galaxy data}}
\label{subsec:rjmcmc_galaxy}

\begin{figure}
\centering
\subfigure[Trace plot of $k$.]{ \label{fig:rjmcmc_galaxy_k}
\includegraphics[width=7cm,height=6cm]{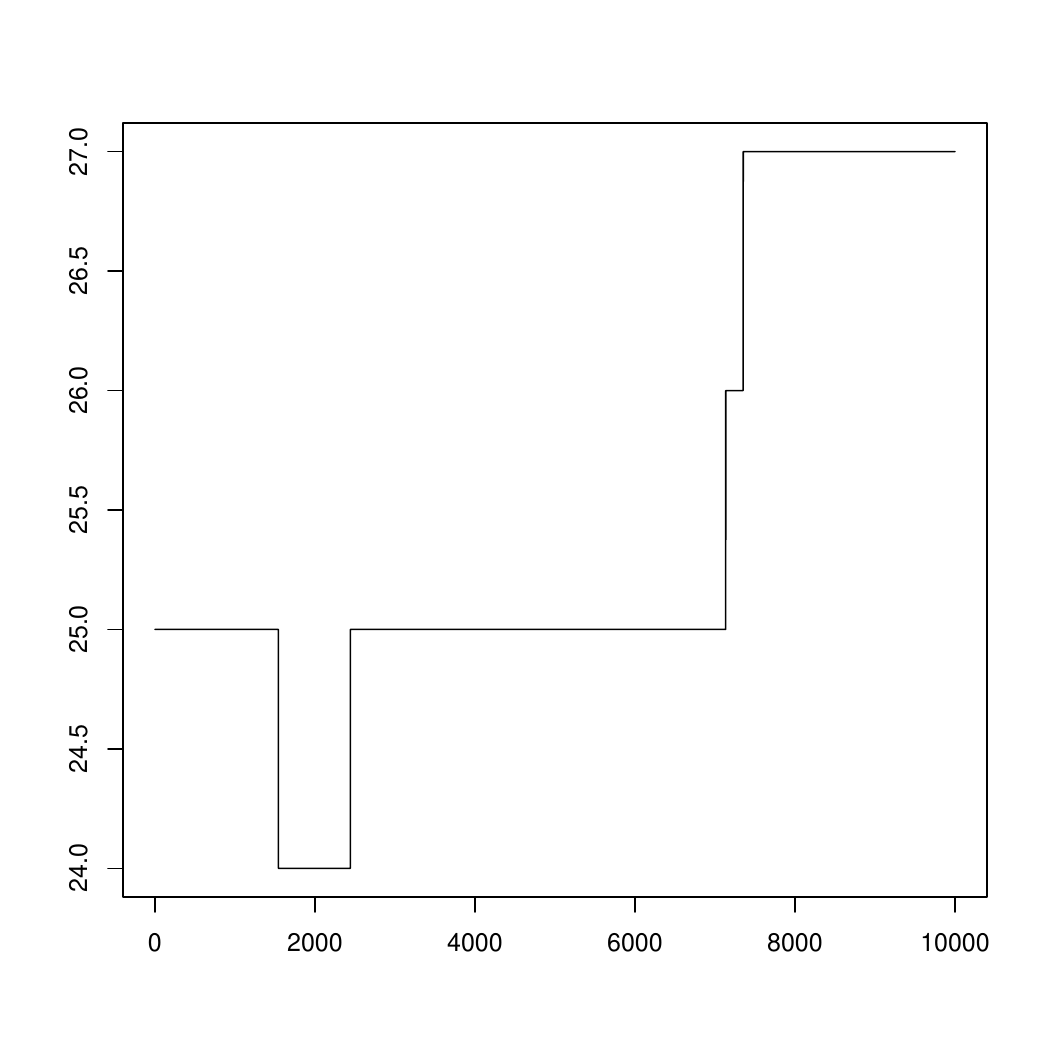}}
\hspace{2mm}
\subfigure[Trace plot of $\nu^*_1$.]{ \label{fig:rjmcmc_galaxy_nu}
\includegraphics[width=7cm,height=6cm]{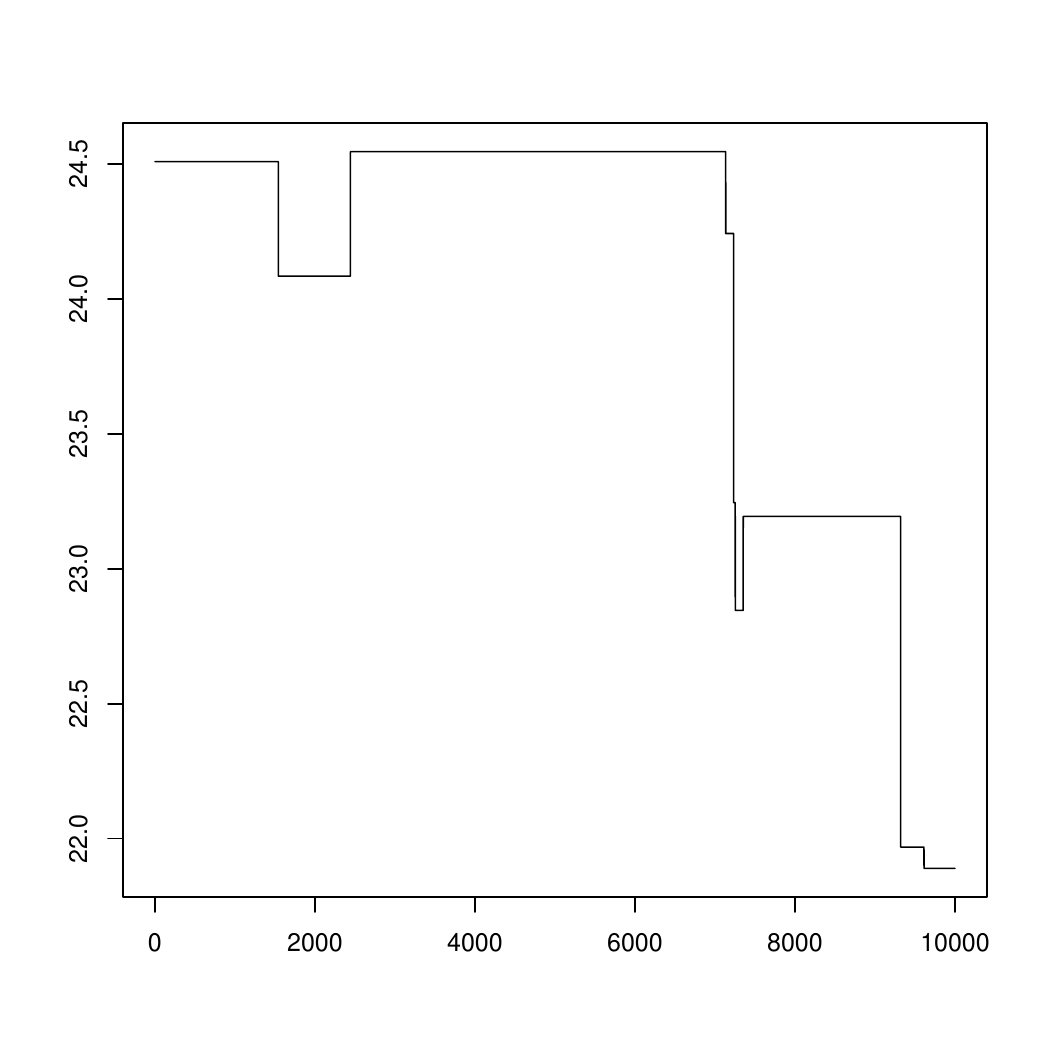}}\\
\vspace{2mm}
\subfigure[Trace plot of $\tau^*_1$.]{ \label{fig:rjmcmc_galaxy_tau}
\includegraphics[width=7cm,height=6cm]{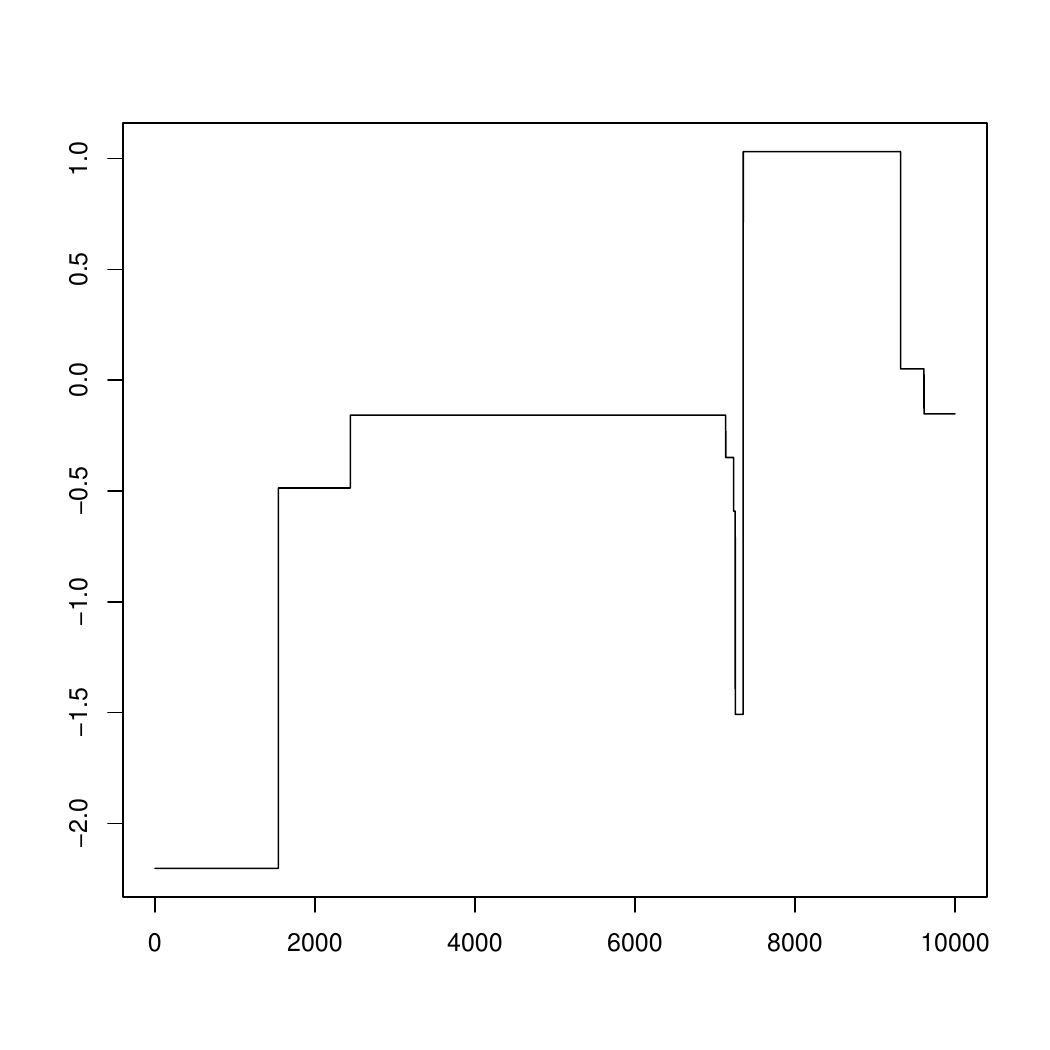}}
\hspace{2mm}
\subfigure[Trace plot of $\omega_1$.]{ \label{fig:rjmcmc_galaxy_w}
\includegraphics[width=7cm,height=6cm]{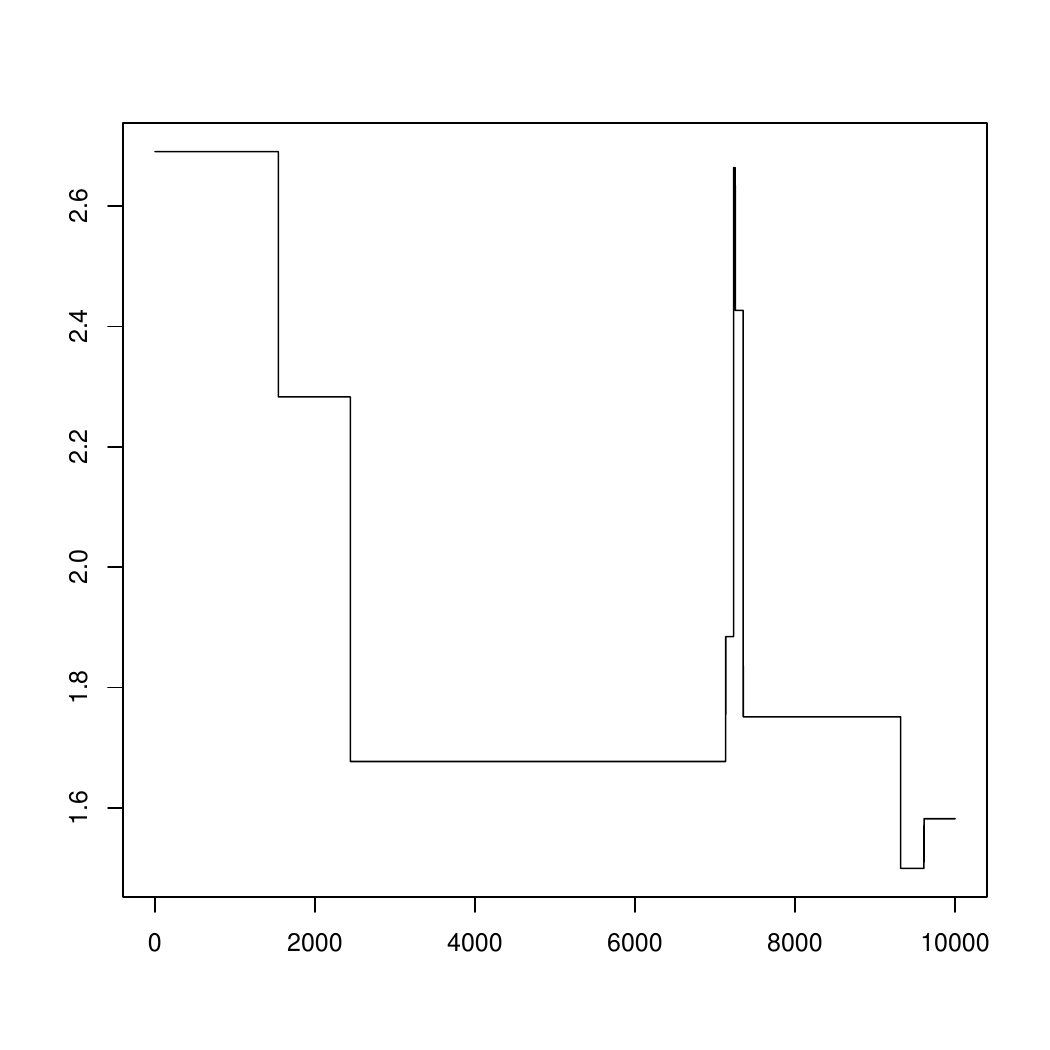}}
\caption{{\bf RJMCMC for the galaxy data:} Trace plots of $k$, $\nu^*_1$, $\tau^*_1$ and $\omega_1$. Poor performance
of the RJMCMC chain is exhibited by the above panels.} 
\label{fig:rjmcmc_galaxy_trace_plots}
\end{figure}

In this case, RJMCMC took 11 minutes and 3 seconds for implementation, which is significantly higher
than the computing time of TTMCMC. Here the acceptance rate of RJMCMC
turned out to be as low as $0.000008$, and the birth, death, no-change rates are $0.000007$, $0.000008$
and $0.00001$, respectively. 
Consequently, as the trace plots of Figure \ref{fig:rjmcmc_galaxy_trace_plots}
show, Bayesian inference based on RJMCMC would be absolutely hopeless! 
The reason for such miserable performance of RJMCMC particularly in this example is that here the local modes
are well-separated from one another and are concentrated on much smaller regions compared to the previous examples,
which, in accordance with high-dimensionality, render the jump size of the proposal too large for RJMCMC
for adequate performance. Low dimensionality on the other hand ensures excellent performance of TTMCMC.

\subsection{{\bf Comparison of the autocorrelations associated with additive TTMCMC and random walk RJMCMC
in the three real data examples}}
\begin{figure}
\centering
\subfigure[Enzyme.]{ \label{fig:rjmcmc_enzyme_acf_comp}
\includegraphics[width=12cm,height=7cm]{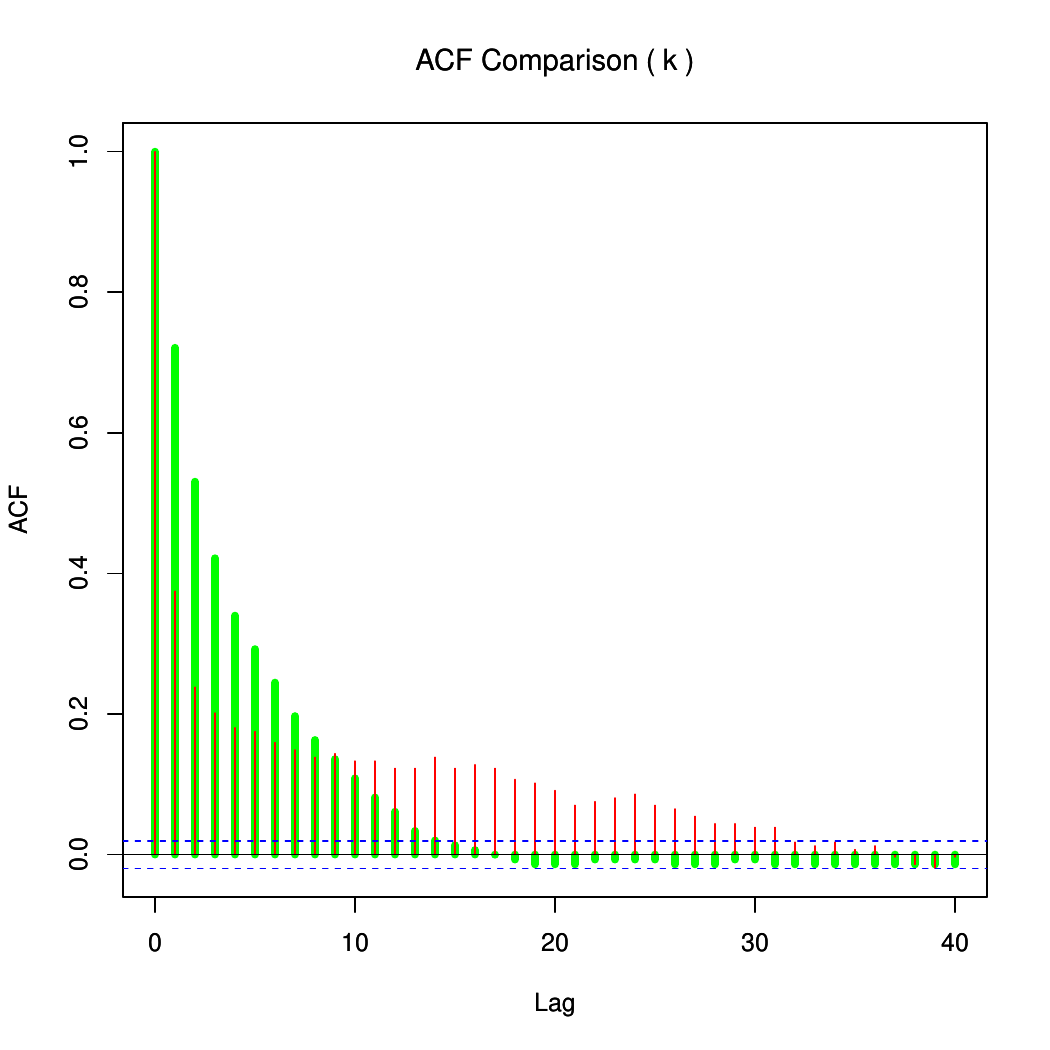}}\\
\subfigure[Acidity.]{ \label{fig:rjmcmc_acidity_acf_comp}
\includegraphics[width=12cm,height=7cm]{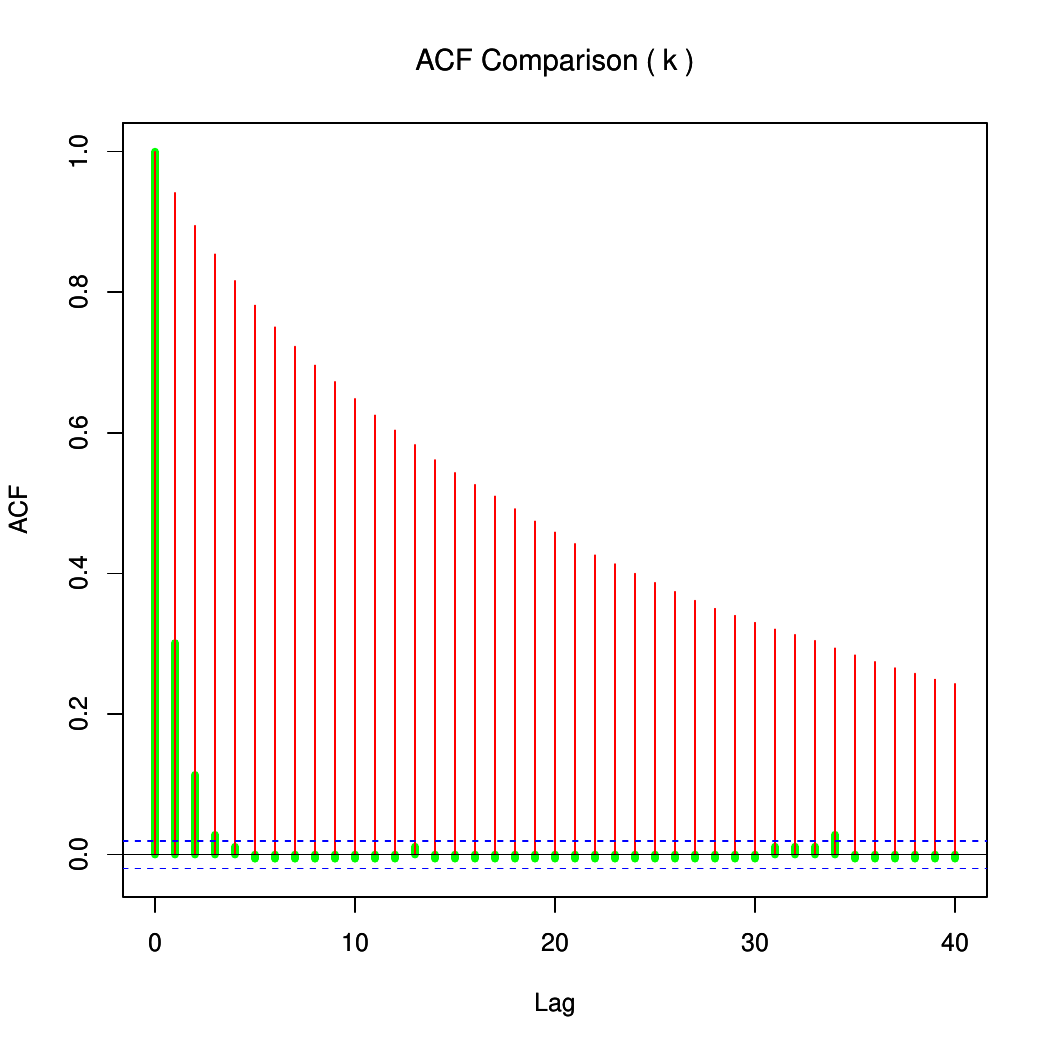}}\\
\subfigure[Galaxy.]{ \label{fig:rjmcmc_galaxy_acf_comp}
\includegraphics[width=12cm,height=7cm]{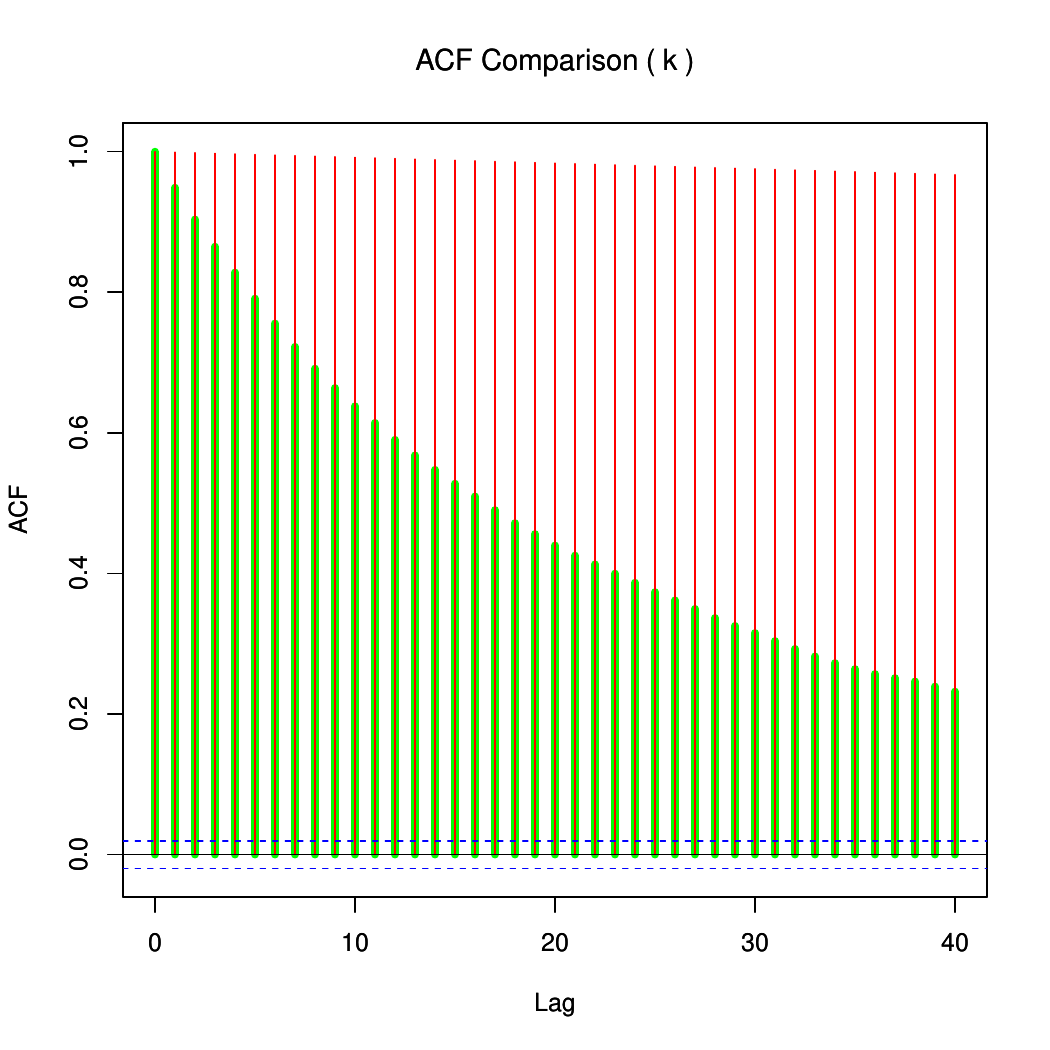}}
\caption{{\bf Autocorrelation comparisons between RJMCMC and TTMCMC for enzyme, acidity and galaxy data:} 
RJMCMC-based autocorrelations are depicted in red and TTMCMC-based autocorrelations are displayed in green.}
\label{fig:acf_plots}

\end{figure}

The comparisons of TTMCMC and RJMCMC with respect to the autocorrelations of $k$, associated with the three real data sets,
are provided in Figure \ref{fig:acf_plots}. TTMCMC outperforms RJMCMC very significantly.

\section{{\bf Comparisons between additive TTMCMC and RJMCMC with respect to the prior structure and 
the algorithm of Richardson and Green (1997) in the galaxy data context}}
\label{sec:further_comparisons}

In our main manuscript we have shown that for the galaxy data set, additive RJMCMC exhibits poor performance 
with respect to the prior we have chosen. We now consider the prior structure of \ctn{Richardson97} (henceforth, RG) 
and compare the results of our additive TTMCMC with the results reported in RG obtained by their RJMCMC algorithm.

Recall from Section 8.1 of DB that the data points $y_1,\ldots,y_n$ are assumed to be $iid$ 
as the normal mixture of the following form: for $i=1,\ldots,n$
\begin{equation*}
f(y_i\vert\bnu_k,\btau_k,\bpi_k,k)=\sum_{j=1}^k\pi_j\sqrt{\frac{\tau_j}{2\pi}}\exp\left\{-\frac{\tau_j}{2}(y_i-\nu_j)^2\right\},
\end{equation*}
where $\bnu_k=(\nu_1,\ldots,\nu_k)$, $\btau_k=(\tau_1,\ldots,\tau_k)$,and $\bpi_k=(\pi_1,\ldots,\pi_k)$.
Given $k>0$, for each $j$, $-\infty<\nu_j<\infty$, $\tau_j>0$, $0<\pi_j<1$ such that 
$\sum_{j=1}^k\pi_j=1$. 

\subsection{{\bf Prior structure}}
\label{subsec:normix_prior_supp}


Following RG, we assume that
\begin{align}
\beta &\sim \mathcal G\left(g,h\right);\label{eq:prior_beta}\\
[\tau_j] &\sim\mathcal G\left(\alpha,\beta\right);\label{eq:prior_tau_supp}\\
[\nu_j] &\sim N\left(\xi,\kappa^{-1}\right);\label{eq:prior_nu_supp}\\
[\pi_1,\ldots,\pi_k|k] &\sim \mathcal D\left(\delta,\ldots,\delta\right);\label{eq:prior_pi_supp}\\
[k]&\sim\mbox{Discrete Uniform}\left\{1,2,\ldots,k_{\max}\right\}.\label{eq:prior_k}
\end{align}
Furthermore, in order to somehow enforce identifiability, RG assume that $-\infty<\nu_1<\nu_2<\cdots<\nu_k<\infty$,
for all $k=1,\ldots,k_{\max}$.
RG consider $\alpha>1>g$ to express the belief that $\sigma^2_j=\tau^{-1}_j$ are similar, without being
informative about their absolute size. Specifically for the galaxy data, RG set $g=0.2$, $h=0.016$, $\alpha=2$,
$\kappa=0.0016$, $\xi=21.73$, $\delta=1$ and $k_{\max}=30$.

We consider two implementations of additive TTMCMC when the above prior structure is considered; in one
implementation we consider the above prior as it is and simulate $\beta$ in an additive TMCMC set-up
simultaneously with the joint additive TTMCMC step, and in the other case, we keep $\beta$ fixed as in \ctn{Cappe03}.
When $\beta$ is simulated, we reparameterize as $\exp\left(\beta^*\right)$, where 
$\beta^*\sim \log\left(\mathcal G\left(g,h\right)\right)$.

\subsection{Results of additive TTMCMC with RG's prior when $\beta$ is updated using additive TMCMC}
\label{subsec:case1}

\begin{figure}
\centering
\subfigure[Trace plot of $k$.]{ \label{fig:galaxy_k1}
\includegraphics[width=7cm,height=6cm]{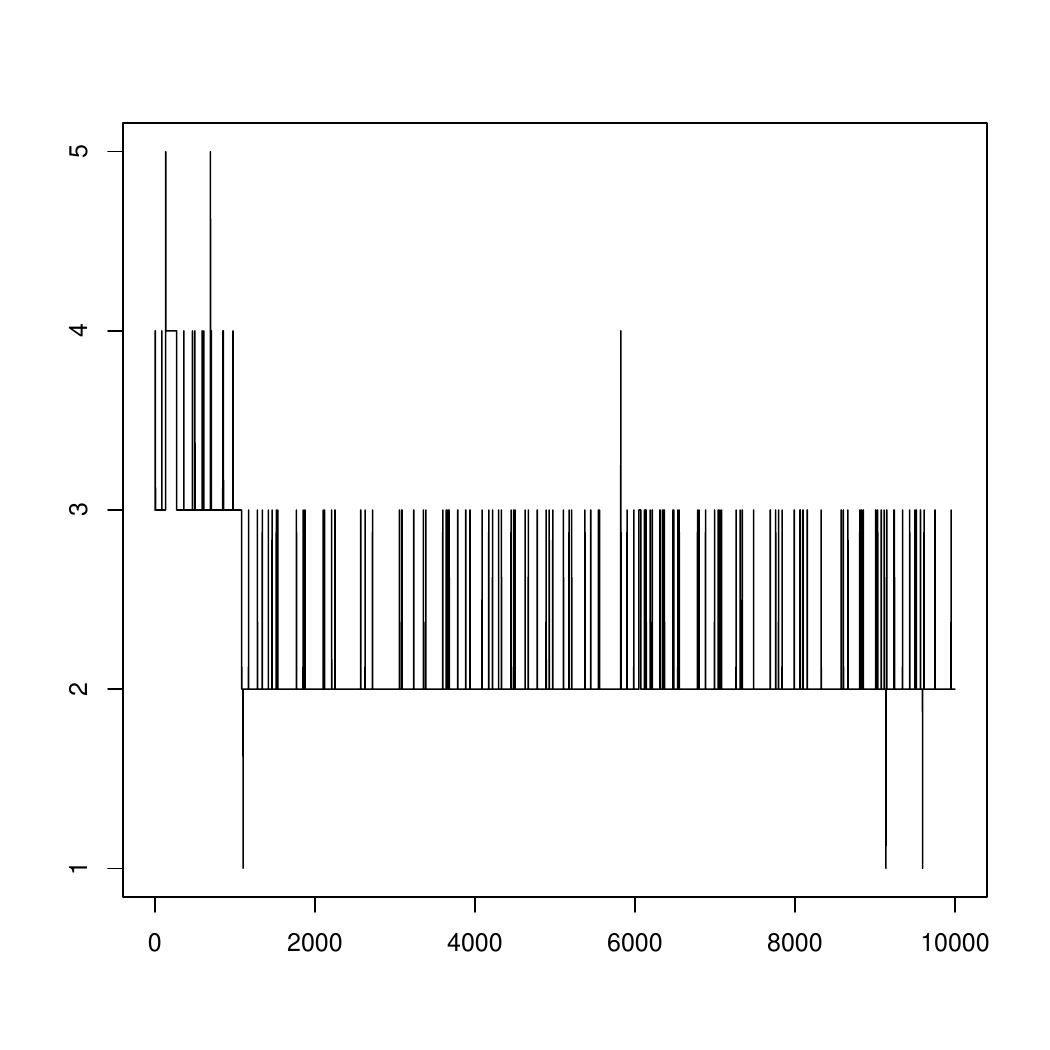}}
\hspace{2mm}
\subfigure[Trace plot of $\nu^*_1$.]{ \label{fig:galaxy_nu1}
\includegraphics[width=7cm,height=6cm]{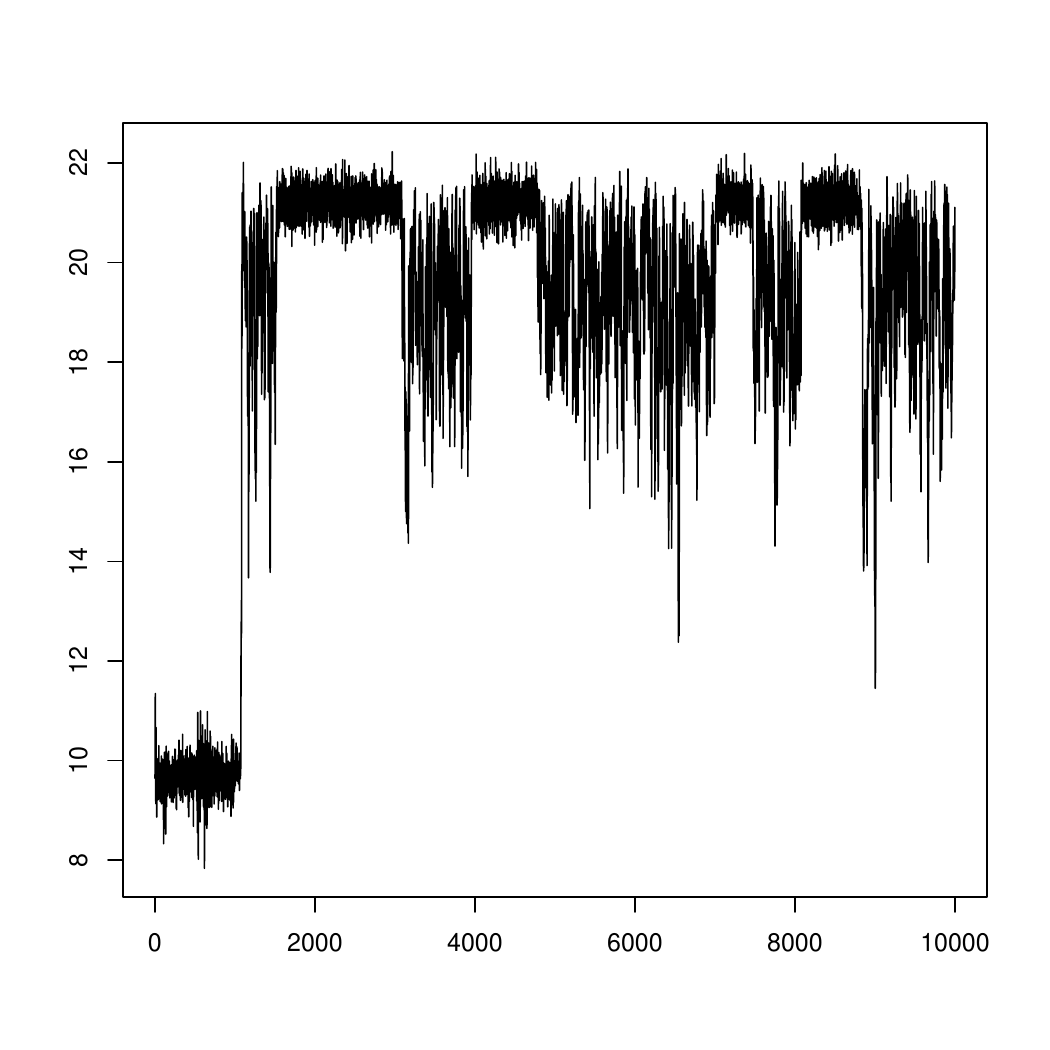}}\\
\vspace{2mm}
\subfigure[Trace plot of $\tau^*_1$.]{ \label{fig:galaxy_tau1}
\includegraphics[width=7cm,height=6cm]{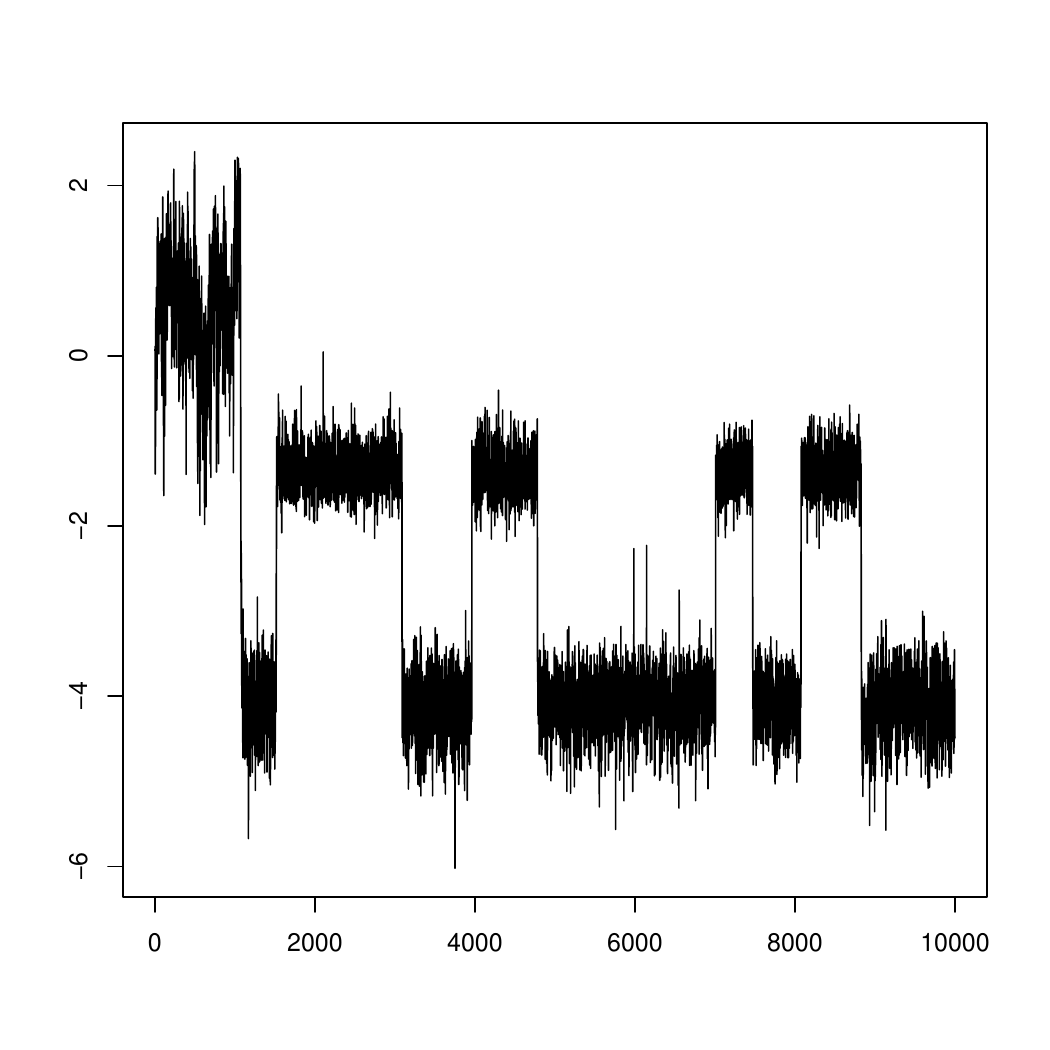}}
\hspace{2mm}
\subfigure[Trace plot of $\omega_1$.]{ \label{fig:galaxy_w1}
\includegraphics[width=7cm,height=6cm]{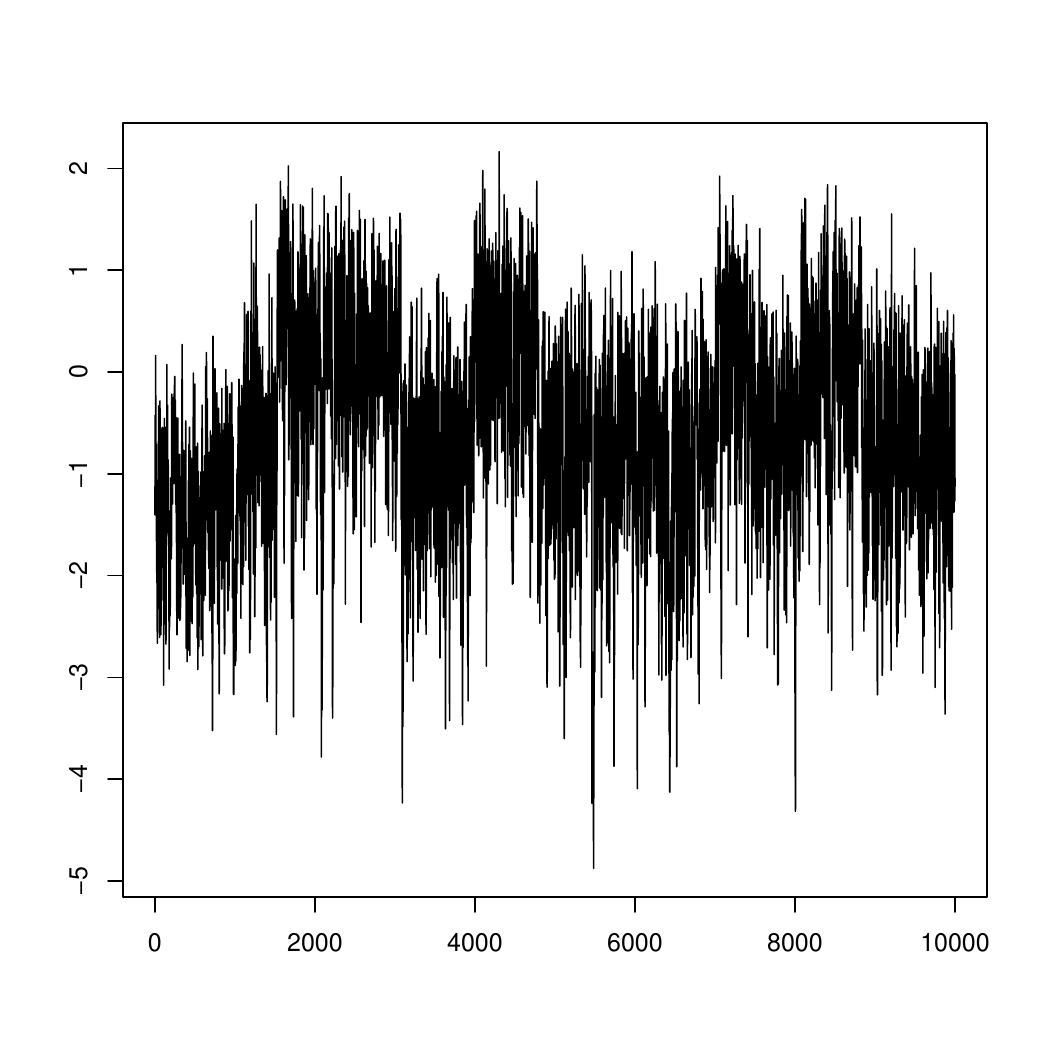}}\\
\vspace{2mm}
\subfigure[Trace plot of $\beta^*$.]{ \label{fig:galaxy_beta1}
\includegraphics[width=7cm,height=6cm]{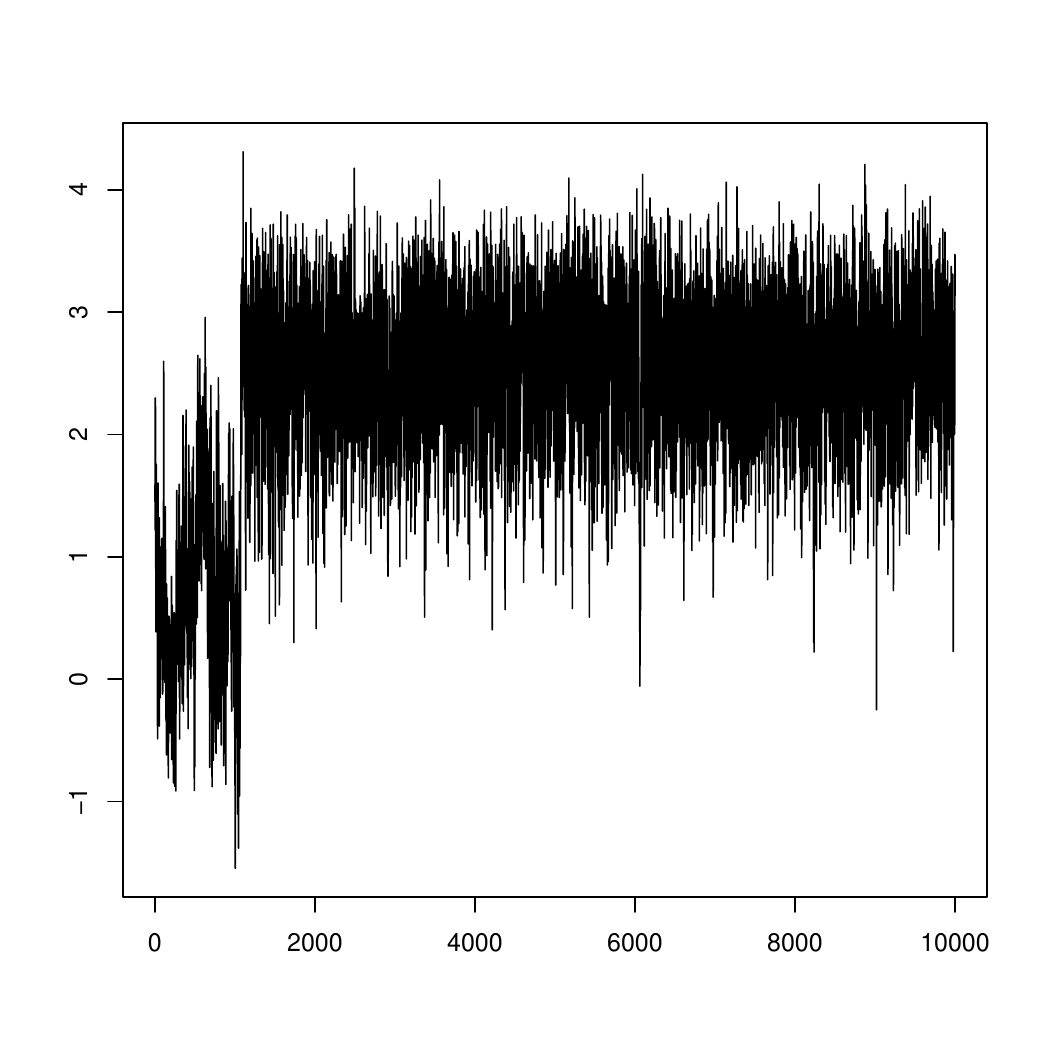}}
\caption{{\bf TTMCMC for the galaxy data with RG prior when $\beta$ is random and burn-in = 300,000:} 
Trace plots of $k$, $\nu^*_1$, $\tau^*_1$, $\omega_1$ and $\beta^*$.}
\label{fig:galaxy_trace_plots1}
\end{figure}

\begin{figure}
\includegraphics[width=7in,height=6.5in]{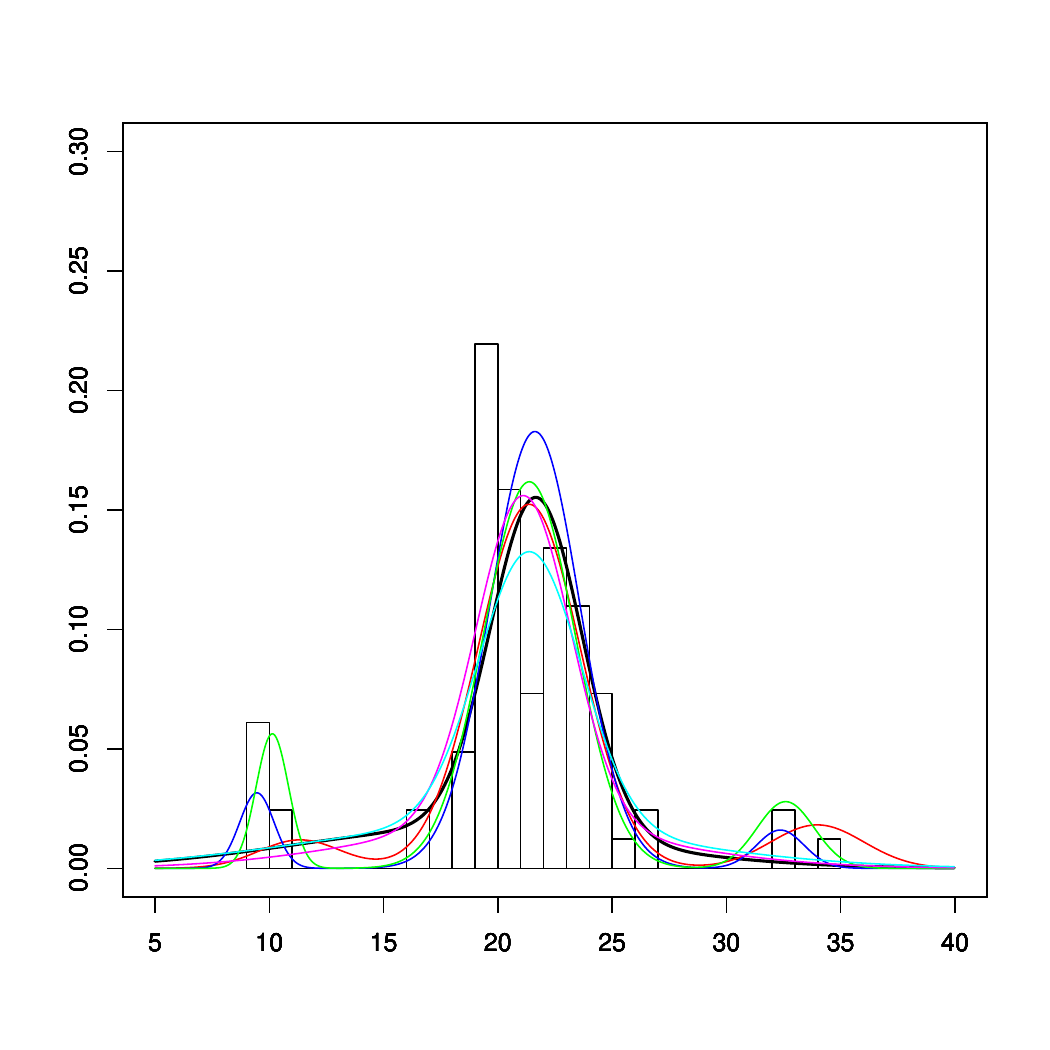}
\caption{{\bf TTMCMC for the galaxy data when $\beta$ is random and burn-in = 300,000:} 
Goodness of fit of the posterior distribution of densities
(colored curves) to the observed data (histogram). The thick black
curve is the modal density and the other colored curves are some densities contained in the 95\% HPD.}
\label{fig:galaxy_hpd1}
\end{figure}

For all the variables including $\beta$, we found the optimum scale for the additive transformation to be $0.5$.
As in our main manuscript, here also we assume a burn-in of 300,000 iterations, and a further 1,500,000 iterations, 
storing one in 150 iterations, thus obtaining a total of 10,000 realizations from the posterior distribution.
It took 1 minute and 20 seconds in our laptop and yielded an acceptance rate $0.089363$. The birth, death and no-change
rates are $0.002298$, $0.002336$ and $0.262948$, respectively.
The resulting trace plots
and the goodness-of-fit diagram are provided in Figures \ref{fig:galaxy_trace_plots1} and \ref{fig:galaxy_hpd1},
respectively. In this case, $k$ takes the values $1,2,3,4,5$ with probabilities $0.0003$, $0.8766$, $0.1079$,
$0.015$ and $0.0002$, respectively, which are quite different from the posterior distribution of $k$ obtained by RG.
Indeed, RG obtained much larger values of $k$, with significant posterior probabilities.
As we argued before, the inherent bias of RJMCMC methods for larger values of $k$ in finite samples 
seems to be responsible for this. The reason that we think that $k$ should not be large in this case is the following.
The prior on the $\tau$'s is set so that they are similar, and this does not seem to be a good strategy
for exploring relatively large number of modal regions with highly different local variabilities. Thus, 
the prior seems to be too smooth for the purpose, which is reflected in the results that we obtained.

\begin{figure}
\centering
\subfigure[Trace plot of $k$.]{ \label{fig:galaxy_k2}
\includegraphics[width=7cm,height=6cm]{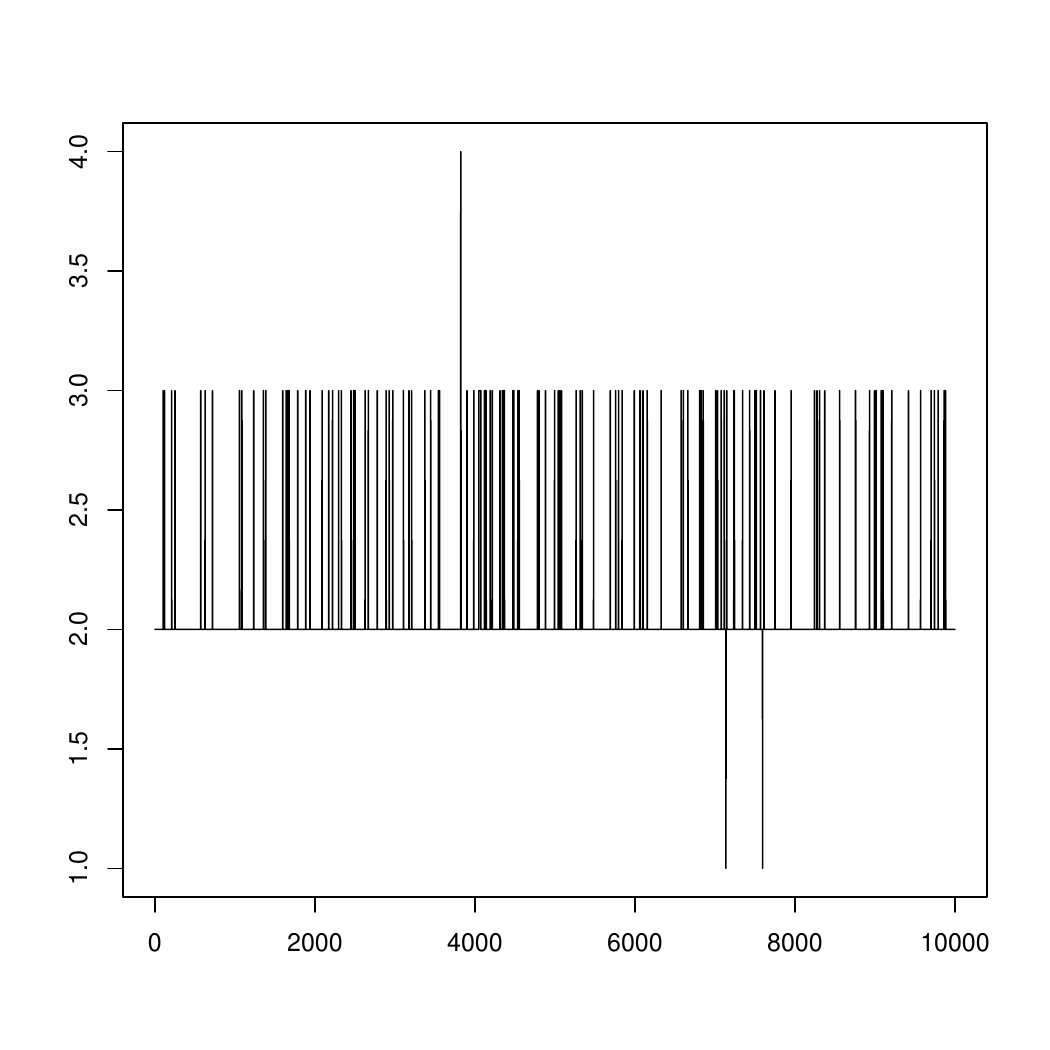}}
\hspace{2mm}
\subfigure[Trace plot of $\nu^*_1$.]{ \label{fig:galaxy_nu2}
\includegraphics[width=7cm,height=6cm]{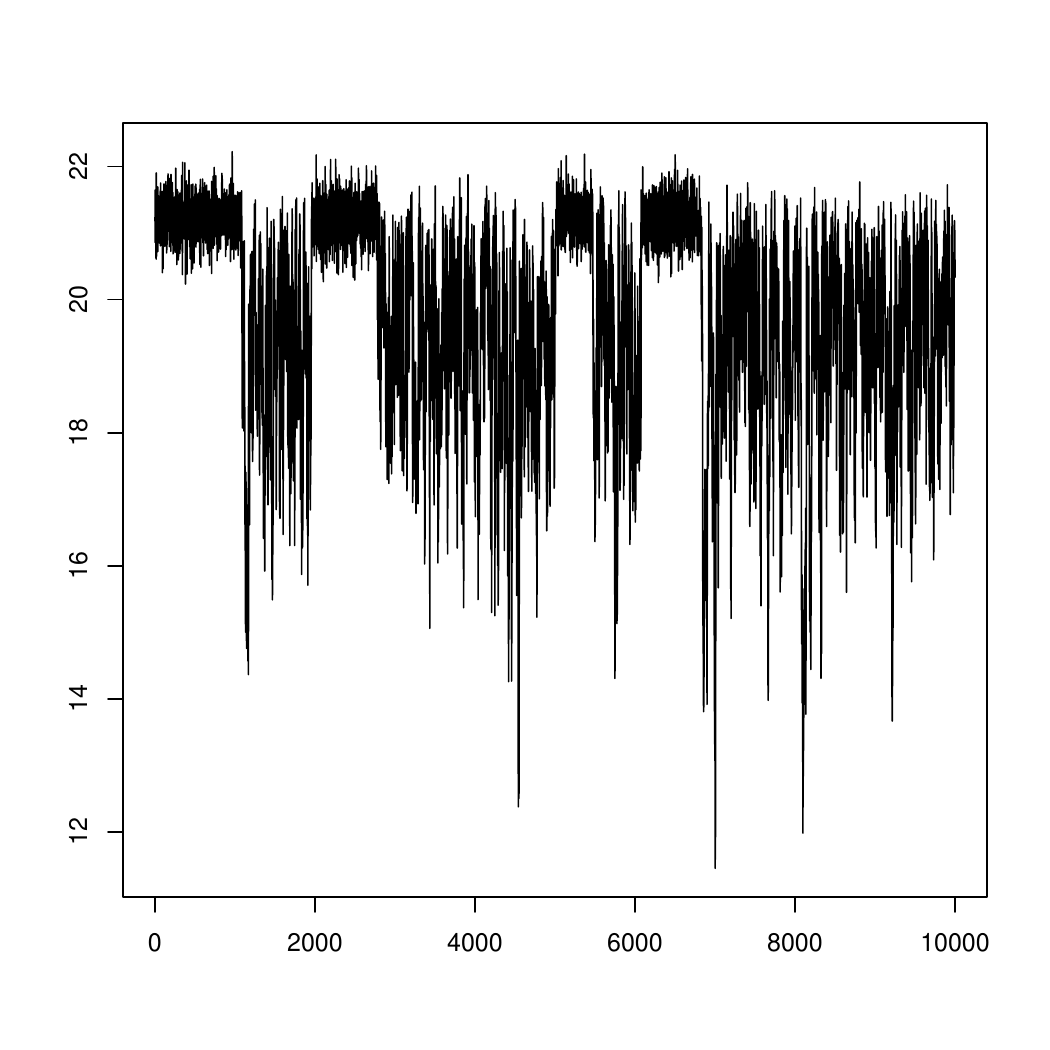}}\\
\vspace{2mm}
\subfigure[Trace plot of $\tau^*_1$.]{ \label{fig:galaxy_tau2}
\includegraphics[width=7cm,height=6cm]{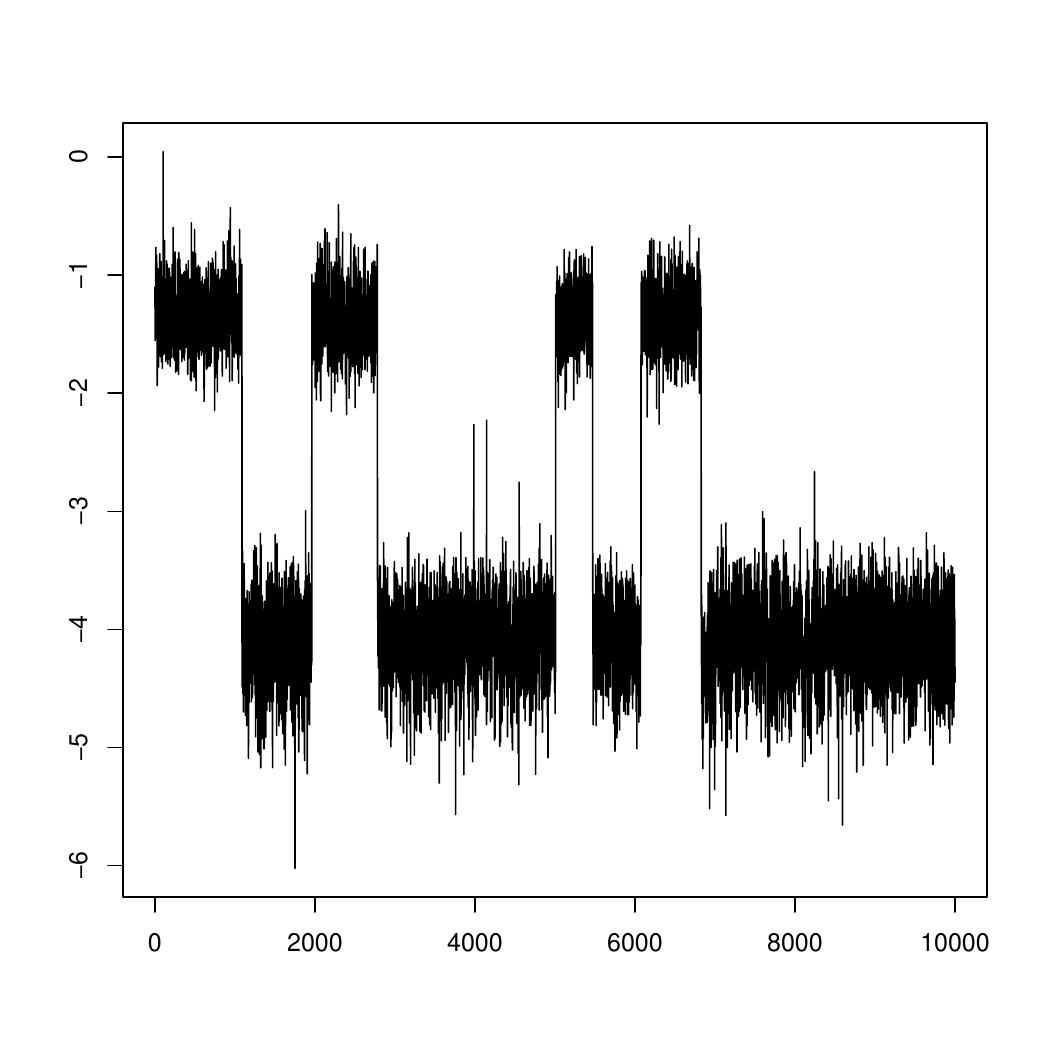}}
\hspace{2mm}
\subfigure[Trace plot of $\omega_1$.]{ \label{fig:galaxy_w2}
\includegraphics[width=7cm,height=6cm]{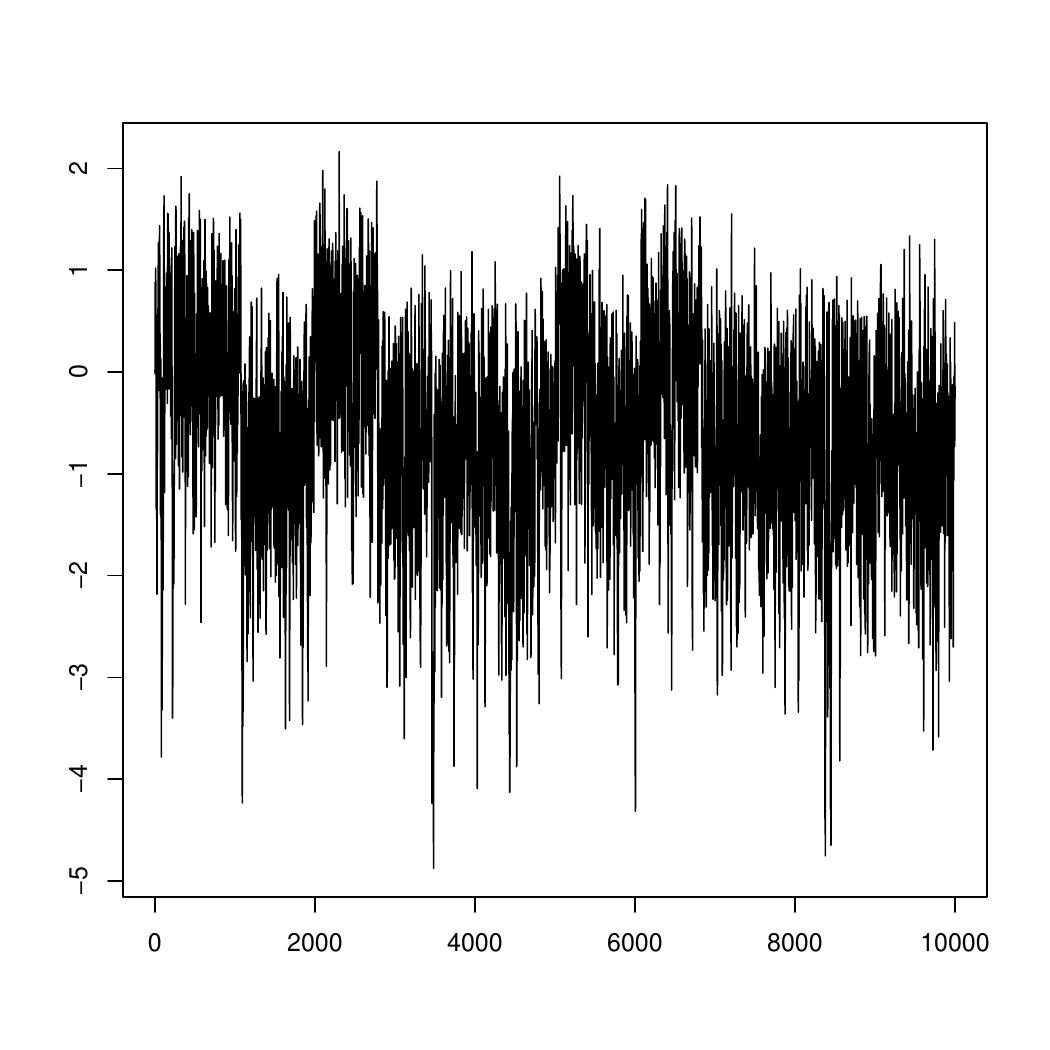}}\\
\vspace{2mm}
\subfigure[Trace plot of $\beta^*$.]{ \label{fig:galaxy_beta2}
\includegraphics[width=7cm,height=6cm]{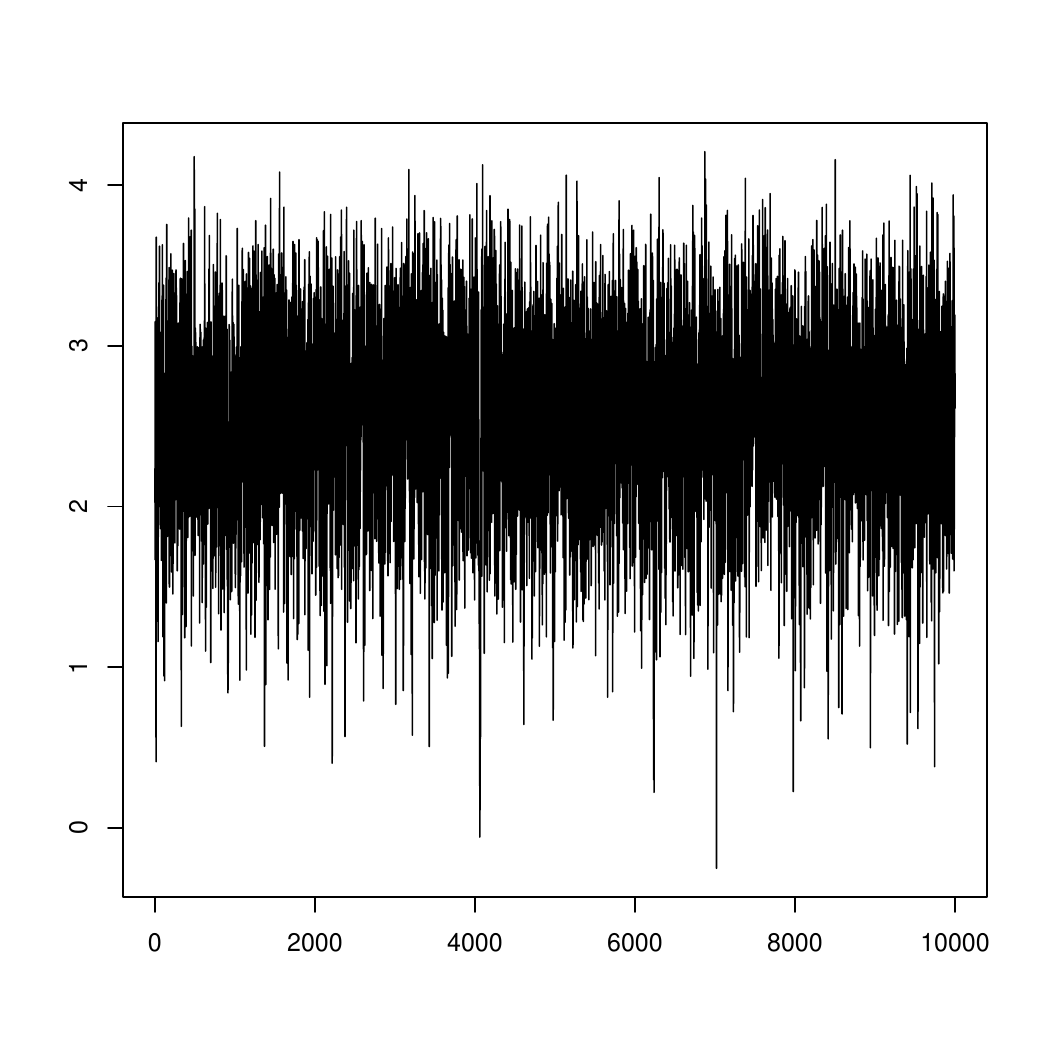}}
\caption{{\bf TTMCMC for the galaxy data with RG prior when $\beta^*$ is random and burn-in = 600,000:} 
Trace plots of $k$, $\nu^*_1$, $\tau^*_1$, $\omega_1$ and $\beta^*$.}
\label{fig:galaxy_trace_plots2}
\end{figure}

\begin{figure}
\includegraphics[width=7in,height=6.5in]{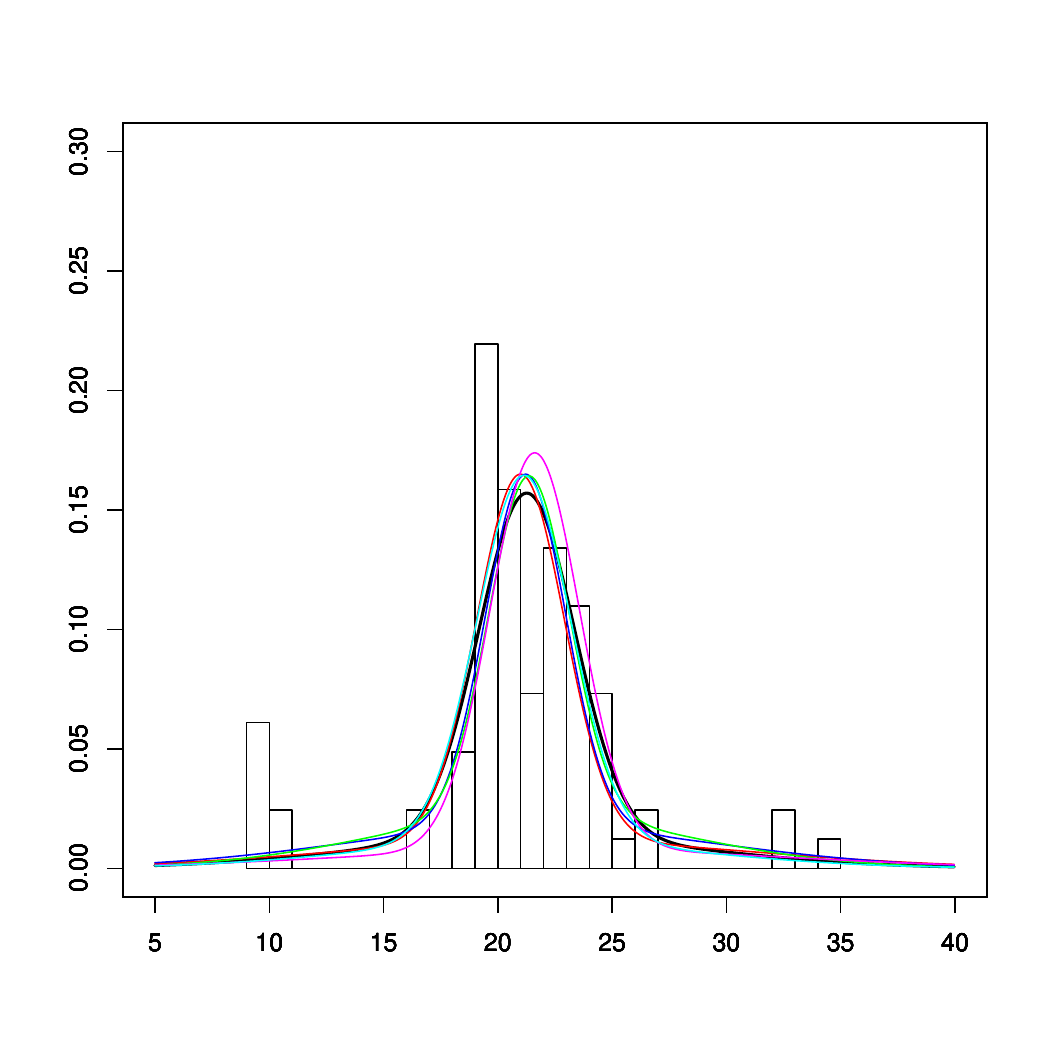}
\caption{{\bf TTMCMC for the galaxy data when $\beta$ is random and burn-in = 600,000:} Goodness of fit of the 
posterior distribution of densities (colored curves) to the observed data (histogram). The thick black
curve is the modal density and the other colored curves are some densities contained in the 95\% HPD.}
\label{fig:galaxy_hpd2}
\end{figure}

However, note that the trace plots visually indicate that the chain perhaps did not stabilize in the initial stages,
and so, to ensure proper convergence, we doubled the burn-in period from 300,000 to 600,000. 
In this case, the time taken is 1 minute 28 seconds and the overall acceptance rate turned out to be 0.09077,
while the birth, death and no-change rates are $0.002361$, $0.002394$ and $0.267125$, respectively.
The modified diagrams are provided in Figures \ref{fig:galaxy_trace_plots2} and \ref{fig:galaxy_hpd2}.
The current as well as the previous trace plots clearly indicate that the posteriors of $\nu$'s and $\tau$'s are
bi-modal; even the trace plots of the weights are suggestive of mild bi-modality. Importantly, the trace plots now 
indicate proper convergence and now $k$ takes the values $1,2,3,4$ with 
posterior probabilities $0.0002, 0.9842, 0.0155, 0.0001$ respectively. However, Figure \ref{fig:galaxy_hpd2} 
shows that the minor modes of the histogram are much ill-captured compared to that in Figure \ref{fig:galaxy_hpd1}.
In fact, the current posterior predictive densities are unimodal. Thus larger values of $k$ in 
Figure \ref{fig:galaxy_trace_plots1}
are not indicative of better exploration, but non-convergence of the chain, even after a large number of iterations.
Since larger number of mixture components can often {\it illusively} result in good fit of the minor modes of the
histogram, our exposition shows that one needs to exercise caution while analysing larger values of $k$.

The above exposition and arguments are applicable to RJMCMC as well. In fact,
RG use $\mathcal B(2,2)$, the Beta distribution with both parameters 2, as a proposal for their dimension changing move.
Since with probability approximately $0.21$ any realized value of $\mathcal B(2,2)$ has density less than one, 
it follows from the discussion in the third point following Algorithm 3.1 of our main manuscript that the RJMCMC algorithm of RG 
is influenced by its bias towards larger values of $k$ for finite samples, where the actual posterior
does not support more than 4 components. 

In other words, it seems that 
the algorithm of RG needed much longer run
to even attain convergence, and that the burn-in period of just 100,000 that RG considered (see page 742) 
seems to be too small given that even with 300,000 as burn-in, Figure \ref{fig:galaxy_trace_plots1} clearly showed 
lack of convergence of our TTMCMC algorithm. 


\subsection{Results of additive TTMCMC with RG's prior when $\beta$ is fixed}
\label{subsec:case3}

We now consider another experiment with $\beta$ fixed. This is motivated by \ctn{Cappe03} who set $\alpha=0.5$ and 
$\beta=0.001$. In this experiment we consider these values, keeping the remaining prior structure the same as RG 
for the galaxy data. With this prior and our
TTMCMC algorithm with all the scales of the additive transformations fixed at $0.5$ we consider a burn-in
of 15,00,000 iterations. Indeed, our chain did not converge even in 600,000 iterations, however,
the burn-in we chose turned out to be many more than sufficient for convergence. We thus implemented
our TTMCMC algorithm for 30,00,000 iterations, storing one in 150 iterations after the burn-in period.
We obtained an overall acceptance rate $0.054212$. The birth, death and the no-change rates turned out to be
$0.004234$, $0.004259$ and $0.154383$, respectively.
The time taken for the implementation is about 4 minutes.

\begin{figure}
\centering
\subfigure[Trace plot of $k$.]{ \label{fig:galaxy_k4}
\includegraphics[width=7cm,height=6cm]{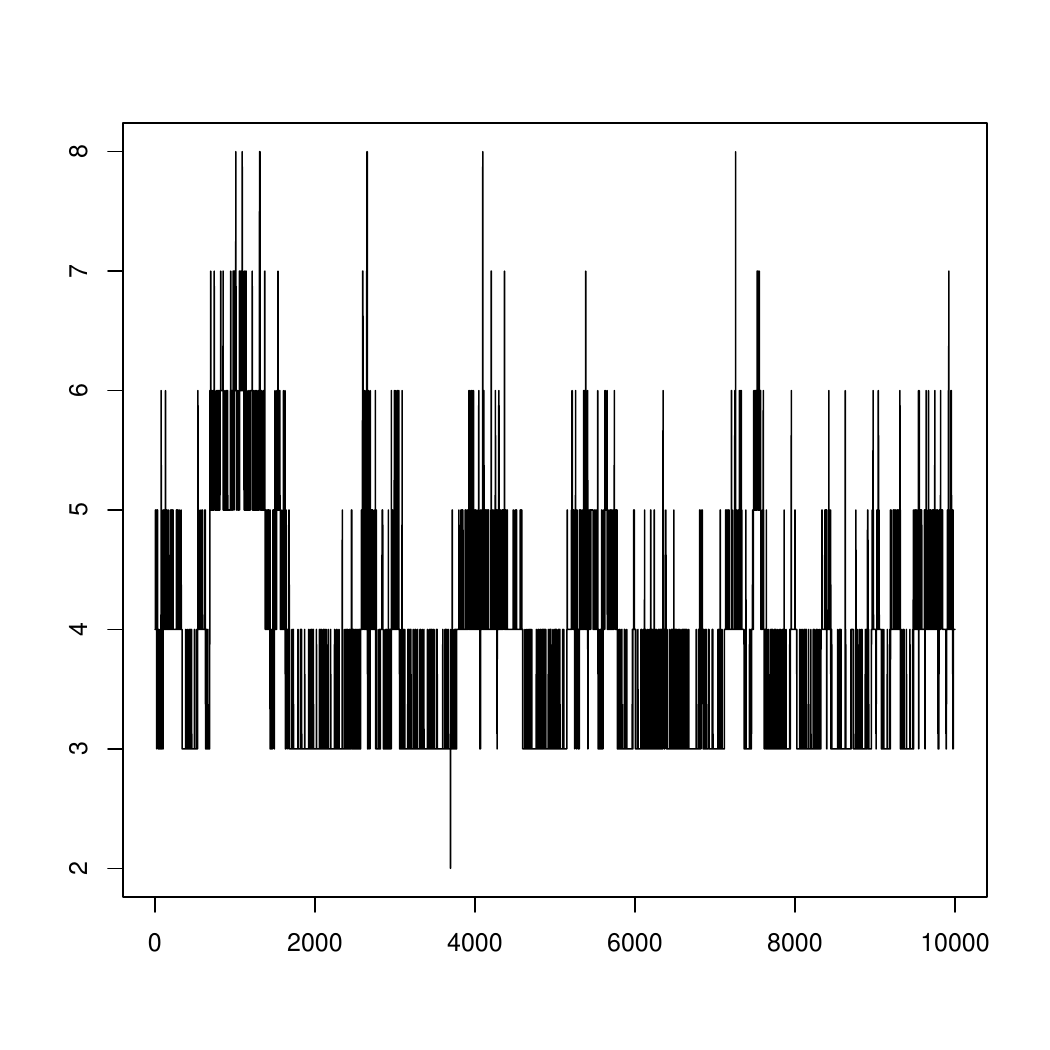}}
\hspace{2mm}
\subfigure[Trace plot of $\nu^*_1$.]{ \label{fig:galaxy_nu4}
\includegraphics[width=7cm,height=6cm]{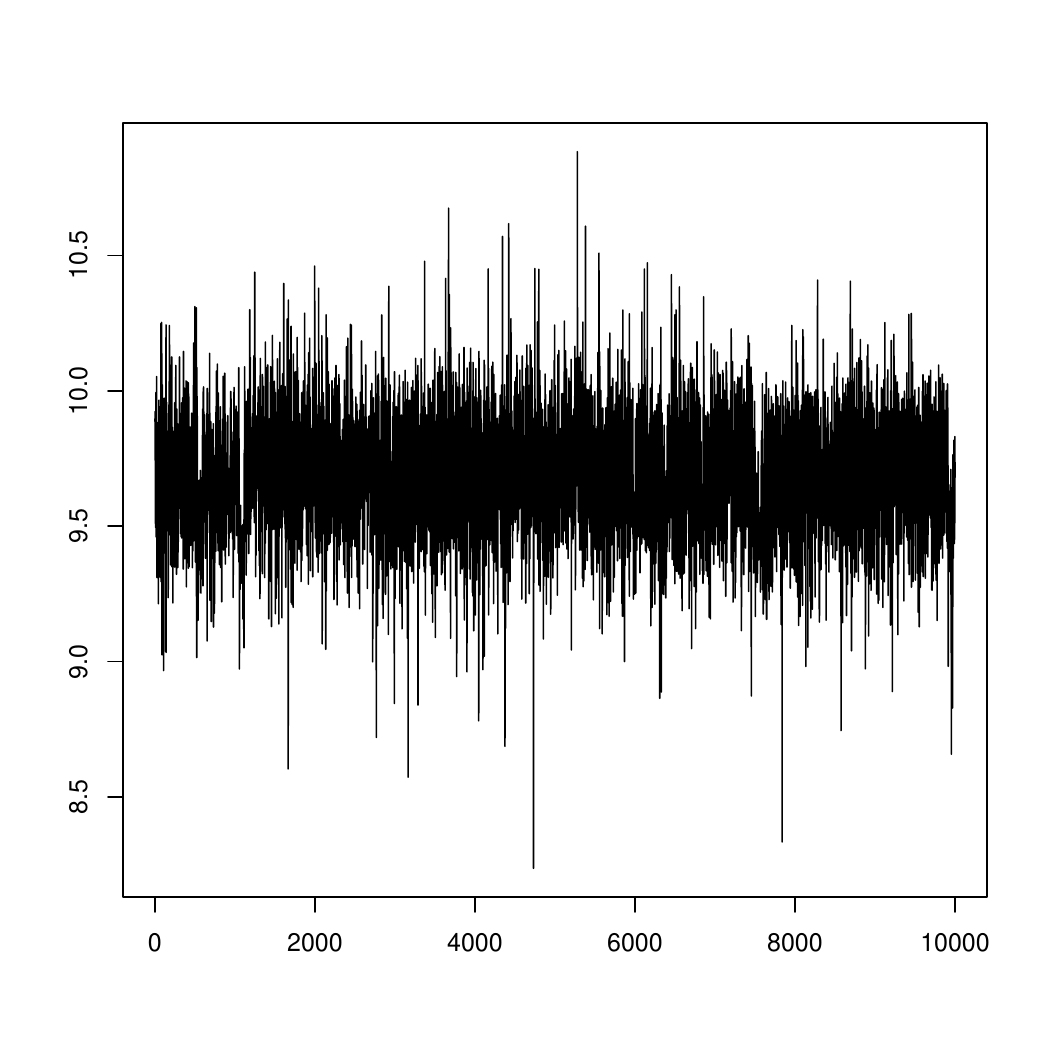}}\\
\vspace{2mm}
\subfigure[Trace plot of $\tau^*_1$.]{ \label{fig:galaxy_tau4}
\includegraphics[width=7cm,height=6cm]{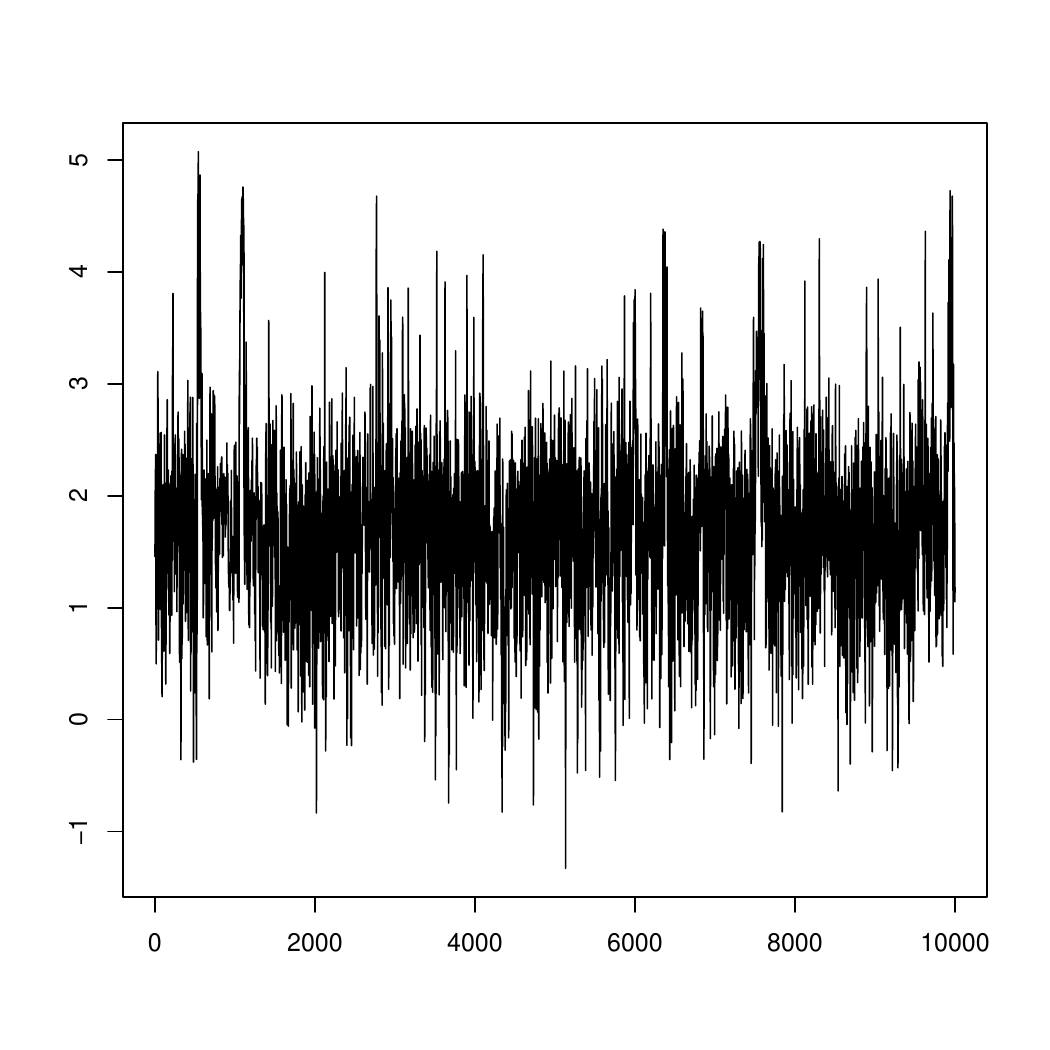}}
\hspace{2mm}
\subfigure[Trace plot of $\omega_1$.]{ \label{fig:galaxy_w4}
\includegraphics[width=7cm,height=6cm]{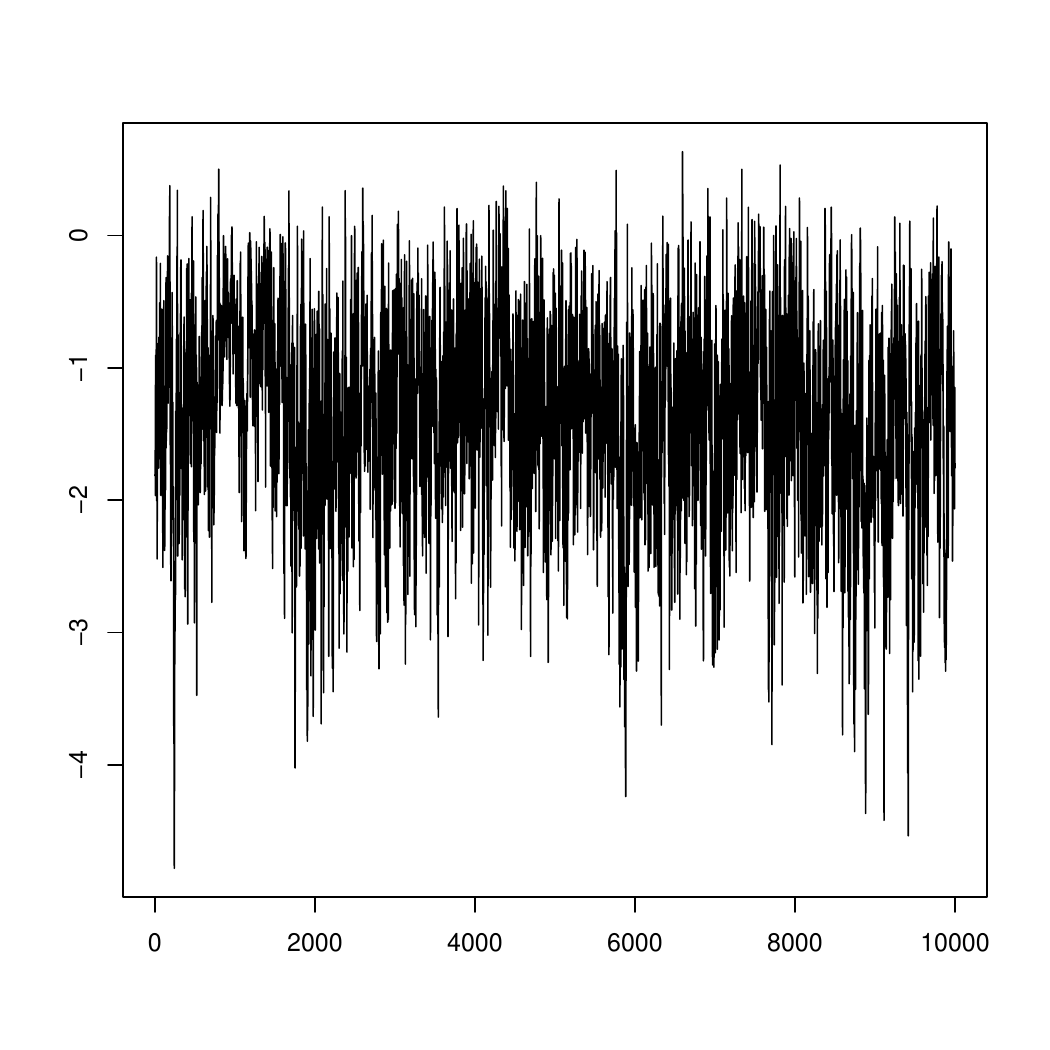}}
\caption{{\bf TTMCMC for the galaxy data with RG prior for fixed $\beta$:} Trace plots of $k$, $\nu^*_1$, $\tau^*_1$ and 
$\omega_1$ when $\beta$ is fixed.}
\label{fig:galaxy_trace_plots4}
\end{figure}

\begin{figure}
\includegraphics[width=7in,height=6.5in]{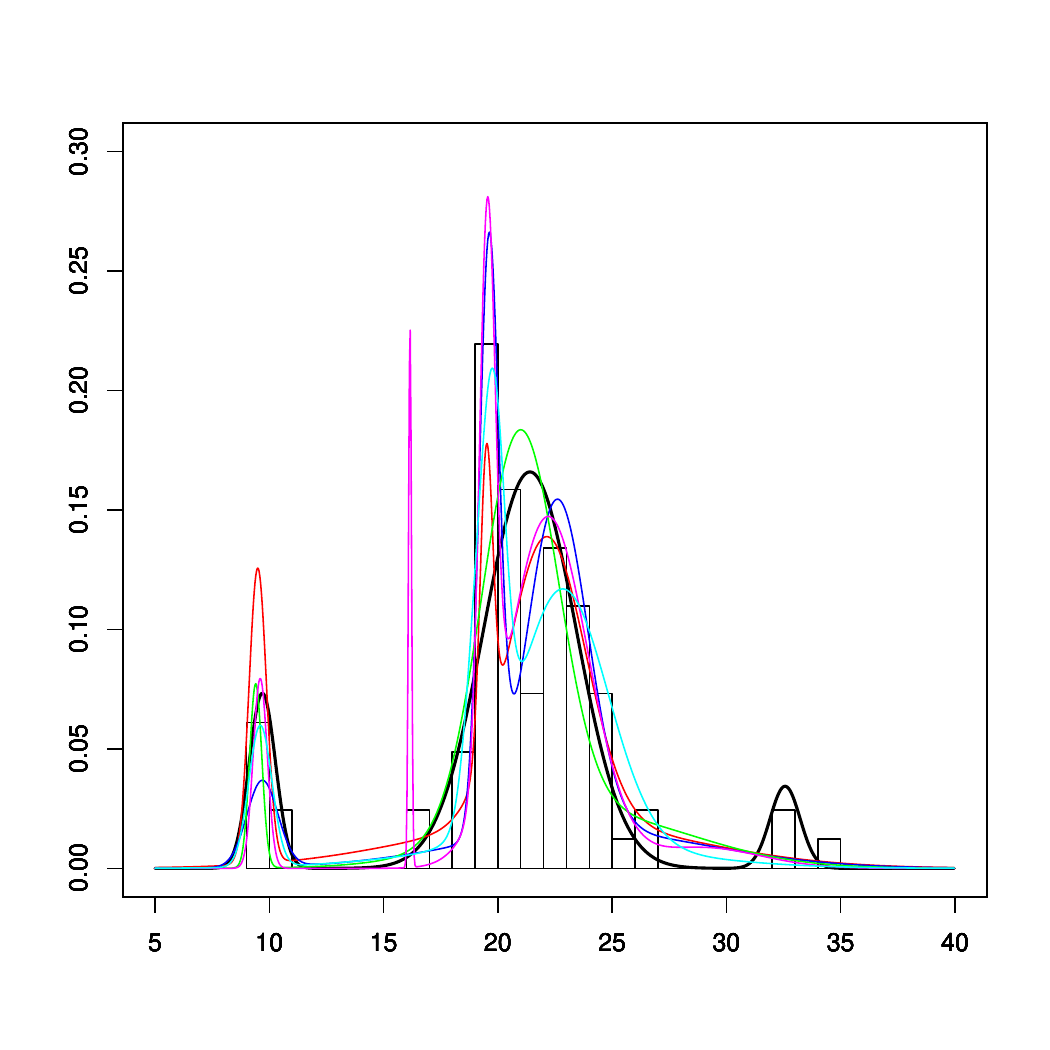}
\caption{{\bf TTMCMC for the galaxy data for fixed $\beta$:} Goodness of fit of the posterior distribution of densities
(colored curves) to the observed data (histogram). The thick black
curve is the modal density and the other colored curves are some densities contained in the 95\% HPD.}
\label{fig:galaxy_hpd4}
\end{figure}

The relevant plots are shown in Figures \ref{fig:galaxy_trace_plots4} and \ref{fig:galaxy_hpd4}.
Now, $k$ takes as large values as required, with significant posterior probabilities. Indeed, $k$ takes the values
2 to 8 with posterior probabilities $0.0002$, $0.4998$, $0.3249$, $0.1281$, $0.0385$, $0.0073$ and $0.0012$.
Expectedly, as shown in Figure \ref{fig:galaxy_hpd4}, the posterior predictive distribution provides reasonably good fit to the
histogram, capturing the minor modes much better than with the RG prior. The reason for much improved performance in this case with
fixed $\beta$ is that the $\tau$'s are now {\it a priori} independent and lets the data speak for itself, enabling
the posterior to adequately learn about the modal regions from the data.  

\newpage

\renewcommand\baselinestretch{1.3}
\normalsize
\bibliographystyle{ECA_jasa}
\bibliography{irmcmc}

\end{document}